\documentclass[12pt]{article}
\usepackage{jheppubmod}
\pdfoutput=1
\usepackage{hyperref}
\usepackage{psfrag}
\usepackage{array}
\usepackage{amssymb}
\usepackage{amsmath}
\usepackage{amsthm}
\usepackage{graphicx}
\usepackage{enumitem}
\usepackage{diagbox}
\usepackage[labelsep=quad]{subcaption}
\usepackage{cite}
\def\gnewt{G}
\def\qop{Q}

\def\fpast{f^{\text{past}}}

\def\fout{f^{\text{out}}}
\def\fin{f^{\text{in}}}
\def\fst{f^{\text{st}}}
\def\grey{\mathfrak{g}}
\def\white{\mathfrak{h}}
\def\hilbfull{{\cal H}}

\def\schwarzn{N}

\newcommand{\psine}{\Psi^{\text{ne}}} 

\def\psinesub[#1]{\psine_{#1}}

\newcommand{\tfd}{\Psi_{\text{tfd}}} 

\newcommand{\tfdtau}{\Psi_{\tau}} 

\newcommand{\cop}[1]{#1}
\newcommand{\al}{\cop{A}} 

\newcommand{\op}{{\cal O}} 

\def\tal{\widetilde{\al}} 

\def\Oright[#1]{\op_{#1}} 
\def\Oleft[#1]{\op_{L#1}} 
\def\anc{\cop{a}}

\def\an[#1]{\anc_{#1}} 
\def\aleft[#1]{\cop{a}_{L#1}} 
\def\arelr[#1]{\cop{a}^{\text{rel}}_{R#1}} 

\def\ta{\widetilde{\cop{a}}}

\def\alen[#1]{\al_{\text{L}, #1}} 

\def\enrange[#1,#2]{{\cal R}_{#1}}
\def\dimrange[#1,#2]{{\cal D}_{#1}}
\def\hilb[#1]{{\cal H}_{#1}}

\def\projrange[#1, #2]{\cop{P}_{#1}}
\def\projh[#1]{\cop{P}_{\hilb[#1]}}

\def\proj[#1]{\cop{P}_{#1}}

\def\tcaus{t_{\cal C}}

\def\cutoffT{\vartheta}

\def\tarelr[#1]{\widetilde{a}^{\text{rel}}_{R#1}}
\def\coeff[#1]{\alpha_{#1}}

\def\pb[#1,#2]{\{#1, #2\}}
\def\deb[#1,#2]{[#1,#2]_{\text{D.B.}}}

\def\tr{{\rm Tr}}

\def\Or[#1]{{\text{O}}\left({#1}\right)}
\def\dotl[#1,#2]{\left\langle #1,\, #2 \right\rangle}
\def\dotlb[#1,#2]{\left\langle #1,\, #2 \right\rangle}
\def\dotlm[#1,#2]{\left[ #1,\, #2 \right]}
\def\dotp[#1,#2]{(\vect{#1} \cdot\vect{#2})}
\def\aff[#1,#2]{\hat{#1}(#2)}

\def\n4sym{{\cal N}=4 SYM}
\def\>{\rangle}
\def\<{\langle}
\def\weight[#1,#2,#3]{\{(#1),#2,#3\}}
\def\ads[#1]{$\text{AdS}_{#1}$}

\def\rtor{{r_*}}

\def\rtors[#1]{r_{*#1}}

\newcommand{\be}{\begin{equation}}
\newcommand{\ee}{\end{equation}}
\newcommand{\ba}{\begin{align}}
\newcommand{\ea}{\end{align}}
\newcommand{\bs}{\begin{split}}

\def\sess\end{split}
\newcommand{\vect}[1]{{\vec{#1}}}

\def \anh {\mathfrak{a}}

\def \tildanh {\widetilde{\mathfrak{a}}}

\def\uent{\mathcal{U}}
\def\vent{\mathcal{V}}
\def\xent{\mathcal{\xi}}

\def\tune{{\cal T}}
\def\vol{ V_{\text{ol}}}

\def \sph{{\Omega}}
\def \scrip{{\cal I}^{+}}
\def \scrim{{\cal I}^{-}}
\def \scrippast{{\cal I}^{+}_{-}}
\def \scrimfuture{{\cal I}^{-}_{+}}

\def\alcut[#1]{{\cal A}_{#1, \epsilon}}
\def\alseg[#1,#2]{{\cal B}_{#1, #2}}

\def\projvac{{\cal P}_{\Omega}}
\def\supcharge[#1]{\{#1\}}
\def\suptrans{{\cal Q}}
\def\projsupeig[#1]{{\cal P}_{{\ell, m}}[{#1}]}
\def\transop[#1, #2]{T_{\{#1\}, \{#2\}}}
\def\supket[#1]{|\{#1\} \rangle}
\def\supbra[#1]{\langle \{#1\} | }

\def\projvac{P_0}

\def\maspect{m_{\text{B}}}
\newcommand{\ubulk}{U_{\text{bulk}}}

\newtheorem{lesson}{Conclusion}

\newtheorem*{wrongassumption}{Assumption}

\title{Lessons from the Information Paradox}
\author{Suvrat Raju}
\affiliation{International Centre for Theoretical Sciences, Tata Institute of Fundamental Research, Shivakote, Bengaluru 560089, India.}
\emailAdd{suvrat@icts.res.in}
\date{}
\abstract{
We review recent progress on the information paradox. We explain why exponentially small correlations in the  radiation emitted by a black hole  are sufficient to resolve the original paradox put forward by Hawking. We then describe a  refinement of the paradox that makes essential reference to the black-hole interior. This analysis leads to a broadly-applicable physical principle:  in a  theory of quantum gravity, a copy of all the information on a Cauchy slice is also available near the boundary of the slice. This principle can be made precise and established --- under weak assumptions, and using only low-energy  techniques  --- in asymptotically global AdS and in four dimensional asymptotically flat spacetime.  When applied to black holes, this principle tells us that the exterior of the black hole always retains a complete copy of the information in the interior. We show that  accounting for this redundancy provides a resolution of the information paradox for evaporating black holes and, conversely, that ignoring this redundancy leads to paradoxes even in the absence of black holes.  We relate this perspective to recent computations of the Page curve for holographic CFTs  coupled to nongravitational baths. But we argue that such models may provide an inaccurate picture of the rate at which information can be extracted from evaporating black holes in asymptotically flat space.  We discuss  large black holes dual to typical states in AdS/CFT and the new paradoxes that arise in this setting. These paradoxes also extend to the eternal black hole. They can be resolved by assuming that the map between the boundary CFT and the black-hole interior is state dependent. We discuss the consistency of state-dependent bulk reconstructions. We conclude by examining the viability of arguments for firewalls, fuzzballs and other kinds of structure at the horizon.
}

\begin{document}
\maketitle
\section{Introduction}
The objective of this article is to review some of the physical ideas that have emerged from discussions of the information paradox over the past few years. The emphasis of the article is not only on various aspects of the
paradox, but also on the broader lessons that the paradox holds for quantum gravity.

Although the information paradox is commonly traced back to Hawking's work \cite{Hawking:1976ra,Hawking:1974sw}  it has, in fact, long been clear that Hawking's calculation itself was not precise enough to lead to a paradox. We start by explaining this simple point, which follows from the fact that  exponentially small corrections can resolve the paradox when it is framed as Hawking did.

A significant advance was made by Mathur \cite{Mathur:2009hf} who pointed out that if one tries to maintain a description of a smooth interior as the black hole evaporates, then this leads to a seeming paradox with the monogamy of entanglement: degrees of freedom outside the horizon appear to be entangled with the interior and also with degrees of freedom far away from the black hole. This argument was, later, elaborated by AMPS  \cite{Almheiri:2012rt}.

We will describe how a consideration of this paradox leads to a robust physical lesson that we call the principle of ``holography of information'': in any theory of quantum gravity, a copy of all the information on a Cauchy slice also resides near the asymptotic boundary of the slice. This leads to a redundancy in description: one way to obtain information about a region is to make measurements in the region and another way to access the same information is to use gravitational effects and make measurements far away.

From the perspective of a local quantum field theory (LQFT) this is a very surprising statement since, in a LQFT, it is possible to change the state of one part of a Cauchy slice without affecting anything near the boundary. In gravity, however, the constraints of the theory force a modification of the metric outside a region if an excitation is introduced inside the region. The significance of these constraints is under-emphasized, perhaps because, classically, they do not carry much information. But in quantum mechanics, these constraints are much more significant.

We will show in some examples --- that when they are supplemented by a few additional weak assumptions --- these constraints directly lead to the principle of holography of information. The examples that we will consider are asymptotically anti-de Sitter (AdS) spacetimes  and four-dimensional asymptotically flat spacetimes with massless particles \cite{Laddha:2020kvp}. In both these cases, we will also demonstrate that the principle of holography of information is not an abstruse result about observables in quantum gravity.  Rather, it can be concretely verified in low-energy effective field theory, and is important even in the absence of black holes.

If one considers the life of a black hole in asymptotically flat space then as the black hole forms and then gradually evaporates, the principle of holography of information tells us that the region outside the black hole always retains a complete copy of what fell in. This renders the issue of ``information loss'' moot. Moreover, we show that ignoring these effects leads to a monogamy paradox of precisely the kind described earlier. We also describe why it is no harder --- and may, in fact, be easier --- to recover information by using these quantum-gravity effects than by ``collecting'' the Hawking radiation.

The physical effect above --- that degrees of freedom that appear to belong to one region may have a redundant description in terms of degrees of freedom in another region --- is similar to the ``island phenomenon'' that has received significant recent attention \cite{Penington:2019npb,Penington:2019kki,Almheiri:2019yqk,Almheiri:2019qdq}.  By studying quantum extremal surfaces, it is possible to derive the ``Page curve'' \cite{Page:1993df,lubkin1978entropy} for black holes in AdS  coupled to nongravitational baths. We briefly review this calculation. The Page curve suggests that information emerges gradually as the black hole evaporates but this happens because, in these systems, gravity switches off abruptly at some point. For more realistic black holes, such as black holes in asymptotically flat space, we argue that the correct picture is that the information is {\em always outside.}

In sections \ref{seclargeads} and \ref{secstatedep}, we turn to paradoxes involving large black holes in AdS. These black holes are unusual because they dominate the microcanonical ensemble and so do not evaporate. In this setting, the question is whether typical states have a smooth interior. We describe some paradoxes \cite{Almheiri:2013hfa,Marolf:2013dba} that were devised to argue  that typical black-hole microstates in AdS/CFT \cite{Maldacena:1997re,Gubser:1998bc,Witten:1998qj} could have firewalls lurking behind their horizon. We also describe a lesser-known version of the paradox that nevertheless crystallizes the essential issue and appears in the eternal black hole. 

To address these paradoxes, we review how the black-hole interior can be constructed in AdS/CFT. This construction, which was first described in \cite{Papadodimas:2013jku}, is dictated by effective field theory and so it must hold in any black-hole microstate that has a smooth interior. In particular, other prescriptions for understanding the black-hole interior must reduce to this construction if they are correct. This construction does not resolve paradoxes associated with large black holes by itself: it is logically possible that only a small fraction of CFT microstates correspond to black holes with a smooth interior for which the construction is correct, whereas other microstates correspond to firewalls for which the construction is inapplicable. This leads us to the phenomenon of state dependence. We show that if one allows the mapping between the black-hole interior and the boundary to be state dependent, then this neatly resolves several paradoxes about large black holes in AdS/CFT. But state dependence raises additional questions \cite{Marolf:2015dia,Raju:2016vsu}, and we describe
some interesting open questions about its consistency.

In section \ref{secisstructure}, we put together some perspectives on whether the region behind the horizon of a typical black hole is empty, as suggested by the conventional solution, or whether it corresponds to a fuzzball or a firewall. We point out that for black holes in asymptotically flat space and small black holes in AdS, there is no currently known good reason for us to give up the picture of an empty interior. We argue that the question of whether large black holes in AdS/CFT have empty interiors reduces to the previously-mentioned questions about state dependence.
The alternative to a state-dependent reconstruction of the interior is that typical microstates in AdS/CFT correspond to firewalls. We also explain, in this section, why the large and interesting family of classical solutions, known as ``microstate geometries'',  cannot be used to make any valid inferences about typical microstates of black holes.

This article covers a large number of topics. We will not attempt to review all the technical details of each topic.
Rather our idea is to to provide a bird's-eye perspective on several ideas and explain how they fit together in an accessible setting. At various points, we direct the reader to the appropriate sections in the original papers where details of the calculations can be found.  
 In keeping with an informal style, in some sections we have separated out some subtleties and addressed them in a ``question and answer'' format.

This article is divided into three main parts. In section \ref{secoldinfo}, we discuss Hawking's original formulation of the paradox, and how it can be easily resolved by the appropriate exponentially small corrections in Hawking radiation. In sections \ref{secintpar}, \ref{secholography}, and \ref{secresolveinfo}, we discuss paradoxes involving the monogamy of entanglement that appear for evaporating black holes, explain the principle of holography of information  and discuss how this principle  can resolve these paradoxes. In sections \ref{seclargeads} and \ref{secstatedep}, we discuss paradoxes involving large black holes in anti-de Sitter space. We explain how these paradoxes can be resolved by state-dependent mappings between the black-hole interior and the boundary but also describe some open questions regarding state dependence. Section \ref{secisstructure} discusses the question of whether horizon-structure is required, in light of the previous sections.

A short disclaimer is necessary. The literature on the information paradox is vast since the topic has attracted significant attention from the quantum-gravity community for decades. Therefore the preparation of this  article necessitated several choices about which topics to include, and which ones to drop and, moreover, these choices were guided, to some extent, by the author's own views and familiarity with these topics.  The author would like to apologize in advance for not devoting as much space to some ideas as they deserve. For some recent reviews that provide a complementary perspective, the reader is referred to \cite{Harlow:2014yka,Chakraborty:2017pmn,Perez:2017cmj}.

\section{Hawking's original paradox \label{secoldinfo}}
Popular descriptions of the information paradox often involve one of the following two statements.
\begin{enumerate}
\item[1.]
Hawking showed that a black hole emits radiation that depends {\em only} on its mass, charge and angular momentum.  This leads to a paradox with the unitarity of quantum mechanics since it suggests that the process of black hole formation and evaporation leads to loss of  information about the initial state, except for information about a few conserved charges.
\item[2.]
String theory shows that the evolution of the black hole must be unitary. But we do not know the precise ``mistake'' in Hawking's calculation.
\end{enumerate}
The purpose of this section is to explain why both the statements above have subtle errors. The correct statement, which will be the first significant conclusion that we will reach in this article, is as follows.
\begin{lesson}
\label{lessonone}
Hawking's original argument is not precise enough to lead to a paradox. This is because small corrections to Hawking's calculation, which are exponentially suppressed by the black hole entropy, are sufficient to ensure that information about the initial state is preserved.  \end{lesson}

To this end, in this section, we first review the derivation of Hawking radiation. This topic is, of course, reviewed in several textbooks \cite{birrell1984quantum,wald1994quantum}. Hawking's original derivation itself provides physical insight, and we review it in Appendix \ref{secraytracehawk}. In sections \ref{corrnull} and \ref{hawkingderiv},  we follow a different approach that  is more precise than the argument of Appendix \ref{secraytracehawk}.  We start by explaining a general result from QFT in curved spacetime about the universal form of correlations across any null surface. When applied to black holes in AdS and in flat space, this immediately yields the spectrum of Hawking radiation. This route to understanding Hawking radiation also allows us to introduce some techniques and notation that will be useful later in the article. 

We then review Hawking's original formulation of the information paradox. We follow this with another background subsection, where we explain a simple result from quantum statistical mechanics showing how most pure states yield physical observations that are exponentially close to mixed states. This will immediately lead us to the conclusion above.

We emphasize to the reader that, in this section, our objective is limited to demonstrating conclusion \ref{lessonone} and emphasizing the importance of exponentially small corrections.
There are more refined paradoxes about black hole evaporation that are addressed in subsequent sections. Moreover, one of the other objectives of this article is to provide evidence for a perspective on black hole information that suggests that the information is always available ``outside''.  This is discussed in sections \ref{secholography} and \ref{secresolveinfo},  and not in this section.

\subsection{Correlations across a null surface \label{corrnull}}

In this subsection, we establish a universal result about correlations between QFT degrees of freedom across any null surface. 
We will largely follow the treatment of \cite{Papadodimas:2019msp}.

Consider a point in a $(d+1)$-dimensional spacetime where the metric is smooth. Taking this point to be the origin, it is possible to locally introduce two null coordinates, $\uent$ and $\vent$, a set of $d-1$ transverse coordinates $\xent^{\alpha}$ and write the line element in the vicinity of this point as
\be
\label{nearpatchmetric}
ds^2 = -d \uent d \vent + \delta_{\alpha \beta} d \xent^{\alpha} d \xent^{\beta} + \ldots
\ee
The $\ldots$ indicate that away from the origin, the metric may deviate from this form as long as the deviations vanish smoothly as one approaches the origin. 

Consider a propagating scalar field, $\phi$,  in the vicinity of the origin. We are interested in theories with dynamical gravity. But we will assume  that gravity is weak, and focus on states $|\Psi \rangle$ to which we can associate a semiclassical metric. Then the short-distance structure of the two point function --- at length scales larger than the Planck scale but smaller than the characteristic curvature scale of the metric --- is expected to be universal. 
\be
\label{twoptfn}
\langle \Psi | \phi(x_1) \phi(x_2) | \Psi \rangle = {\Gamma(d-1) \over 2^d \pi^{d \over2} \Gamma({d \over 2})}  {1 \over s_{12}^{d-1 \over 2}}\left(1 + \Or[s_{12}]\right).
\ee
where $s_{12}$ is the square of the geodesic distance between the points $x_1$ and $x_2$. The idea is just that, at short distances, the space looks locally like Minkowski space, in which the two point function is as given above. This is discussed in more detail in \cite{Papadodimas:2019msp}. At smaller length scales, when the separation of $x_1$ and $x_2$ becomes comparable to an ultraviolet (UV) scale, we expect UV effects to be important. At long length scales, we expect that the effects of the curvature will be important. But for a large enough black hole, where the curvature length scales are large enough, there is an intermediate scale at which we expect \eqref{twoptfn} to be valid.

Now the idea is to integrate the field on both sides of the surface $\uent = 0$ so as to extract a set of ``local modes.'' 
To do this carefully, we introduce a function $\tune(\uent)$ that is supported for a very small interval of $\uent$ on the positive real axis near $\uent = 0$ and dies off smoothly at its endpoints. We normalize it so that $\int \tune(\uent)^2 {d \uent \over \uent} = 2 \pi$. We also introduce a small region, of volume $\vol$, in the transverse directions and then define the following operators integrated over the support of $\tune(\uent)$ and the small transverse region.
\be
\label{defnearhormodes}
\begin{split}
&\anh = \int   \partial_{\uent} \phi(\uent, \vent=0, \xent^a) \left(-\uent \right)^{-i \omega_0} \,  \tune(-\uent) d \uent {d^{d-1} \xent^a \over \sqrt{\pi \omega_0 \vol} } \\
&\tildanh =\int  \partial_\uent \phi(\uent,\vent=-\epsilon,\xent^a) \uent^{i \omega_0} \, \tune(\uent) d \uent {d^{d-1} \xent^a \over  \sqrt{\pi \omega_0 \vol}}.
\end{split}
\ee
Here $\epsilon$ is a small parameter that separates the support of the operators in $\vent$ coordinate. Its precise value is not important and it only ensures that the regions over which the operators $\anh$ and $\tildanh$ are defined are causally disconnected.   The operators are also implicitly dependent on the choice of the parameter  $\omega_0$.

\begin{figure}[!ht]
\begin{center}
\includegraphics[height=0.4\textheight]{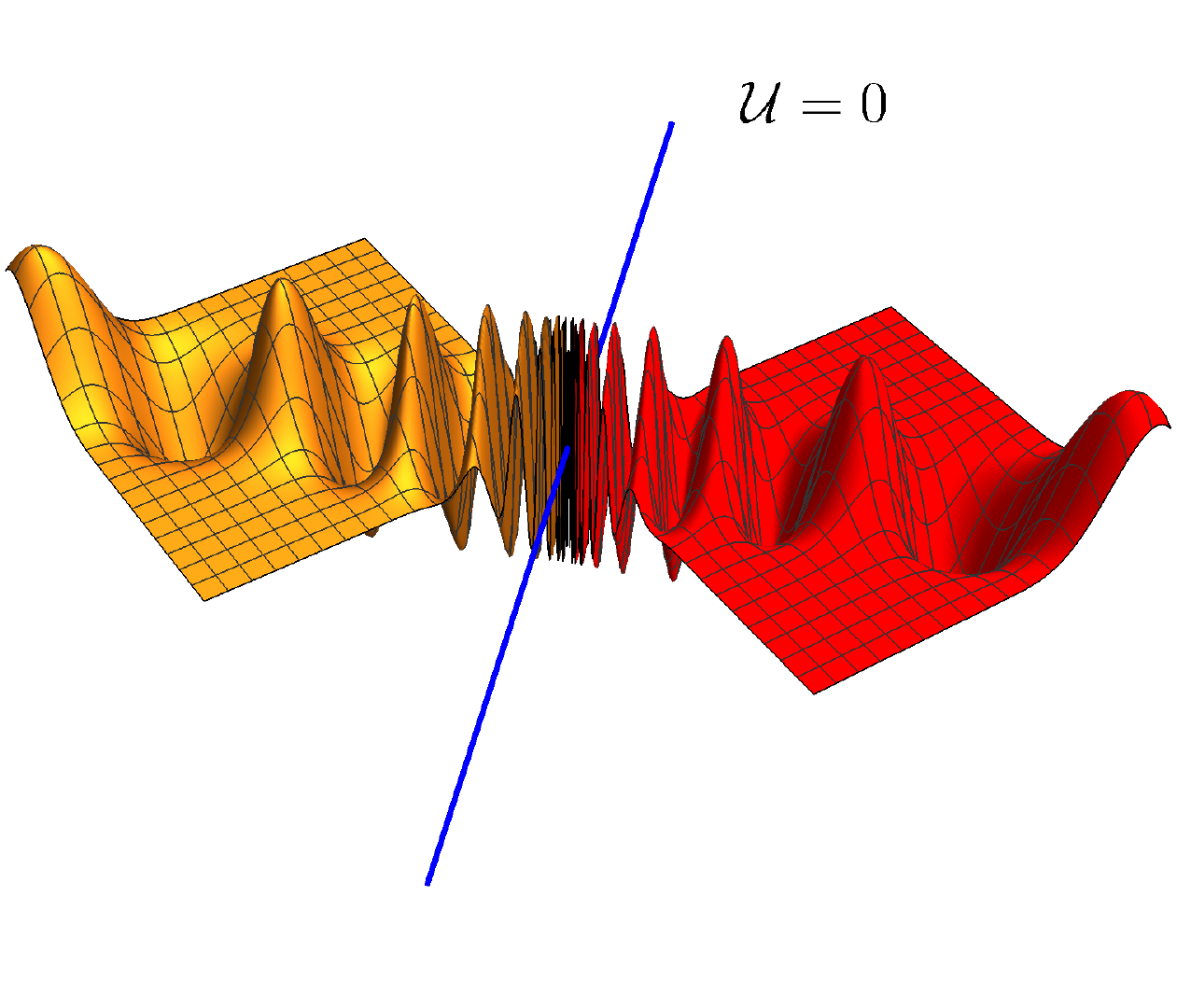}
\caption{\em The modes $\anh$ are extracted by integrating the field with the red smearing function to the right of $\uent = 0$. The modes $\tildanh$ are extracted by integrating the field with the yellow smearing function on the left. The oscillations of the smearing function tend to pile up near $\uent=0$, and these rapid oscillations provide the most significant contribution to the integral. The smearing functions are also separated by a small amount in the $\vent$ direction although this is not visible in the figure.\label{fignearhor}}
\end{center}
\end{figure}
The reader should not be intimidated by some of the notation that we have introduced above. What we are doing is very simple and is depicted in Figure \ref{fignearhor}. We have just taken a light ray that emanates from a point and integrated the field on both sides of the light ray with rapidly oscillating functions, where the frequency of oscillation is determined by $\omega_0$.  We are denoting these smeared field operators by $\anh$ and $\tildanh$. This process effectively extracts ``right moving modes'' (i.e. those that travel parallel to the surface $\uent = 0$) from the two sides of the surface at $\uent = 0$.

Since we know the short-distance behaviour of the two-point function of the field, and since the operators $\anh$ and $\tildanh$ are defined by smearing the field over a small region, this means we know everything about their two-point functions. A few short calculations --- for details we refer the reader to section 2 of \cite{Papadodimas:2019msp}--- then tell us that the two point functions of these modes must be
\be 
\label{samesidecorr}
\langle \Psi | \anh \anh^{\dagger} | \Psi \rangle = \langle \Psi | \tildanh \tildanh^{\dagger} | \Psi \rangle = {1 \over 1 - e^{-2 \pi \omega_0}}.
\ee
We can also work out the two-point correlators of the modes with each other, which turn out to be
\be
\label{oppsidecorr}
\langle \Psi | \anh \tildanh | \Psi \rangle = {e^{-\pi \omega_0}  \over 1 - e^{-2 \pi \omega_0}}.
\ee
Moreover, using the normalizations specified above and the canonical commutators of the field we find that
\be
\label{commutmodes}
[\anh, \anh^{\dagger}] = 1, \qquad
[\tildanh, \tildanh^{\dagger}] = 1.
\ee
And since the modes are defined on spacelike separated regions, as is clear from the definition above, we also have
\be
[\anh, \tildanh] =[\anh, \tildanh^{\dagger}] = 0.
\ee

The set of two point functions displayed in equations \eqref{samesidecorr} and \eqref{oppsidecorr} turns out to impose quite a strong requirement.  A little more analysis tells us that these equations can only hold if the action of $\anh$ on the state $|\Psi \rangle$ is proportional to the action of $\tildanh^{\dagger}$ and vice-versa. 
\be
\label{stateaction}
(\anh - e^{-\pi \omega_0} \tildanh^{\dagger})|\Psi \rangle = 0, \qquad
(\anh^{\dagger} - e^{{\pi \omega_0}} \tildanh)  |\Psi \rangle = 0. 
\ee

In almost all cases below we will be interested in spacetimes with spherical symmetry. The modes above are defined in the vicinity of a point but for spherically symmetric spacetimes they can also be defined in the vicinity of a sphere. The generalization is quite simple.
Say that we have a $(d-1)$-dimensional sphere of radius $r_0$ in the spacetime. We choose null coordinates, so that  $\uent=0, \vent = 0$ on the surface of the sphere.  Now say that in the vicinity of this sphere the metric is just
\be
\label{metricsphericalnull}
ds^2 = -d \uent d \vent + r_0^2 d \Omega_{d-1}^2 + \ldots
\ee
where the $\ldots$ again tell us that the metric may deviate smoothly from the form above as we move away from $\uent = 0, \vent = 0$. 

\begin{figure}[!ht]
\begin{center}
\includegraphics[height=0.4\textheight]{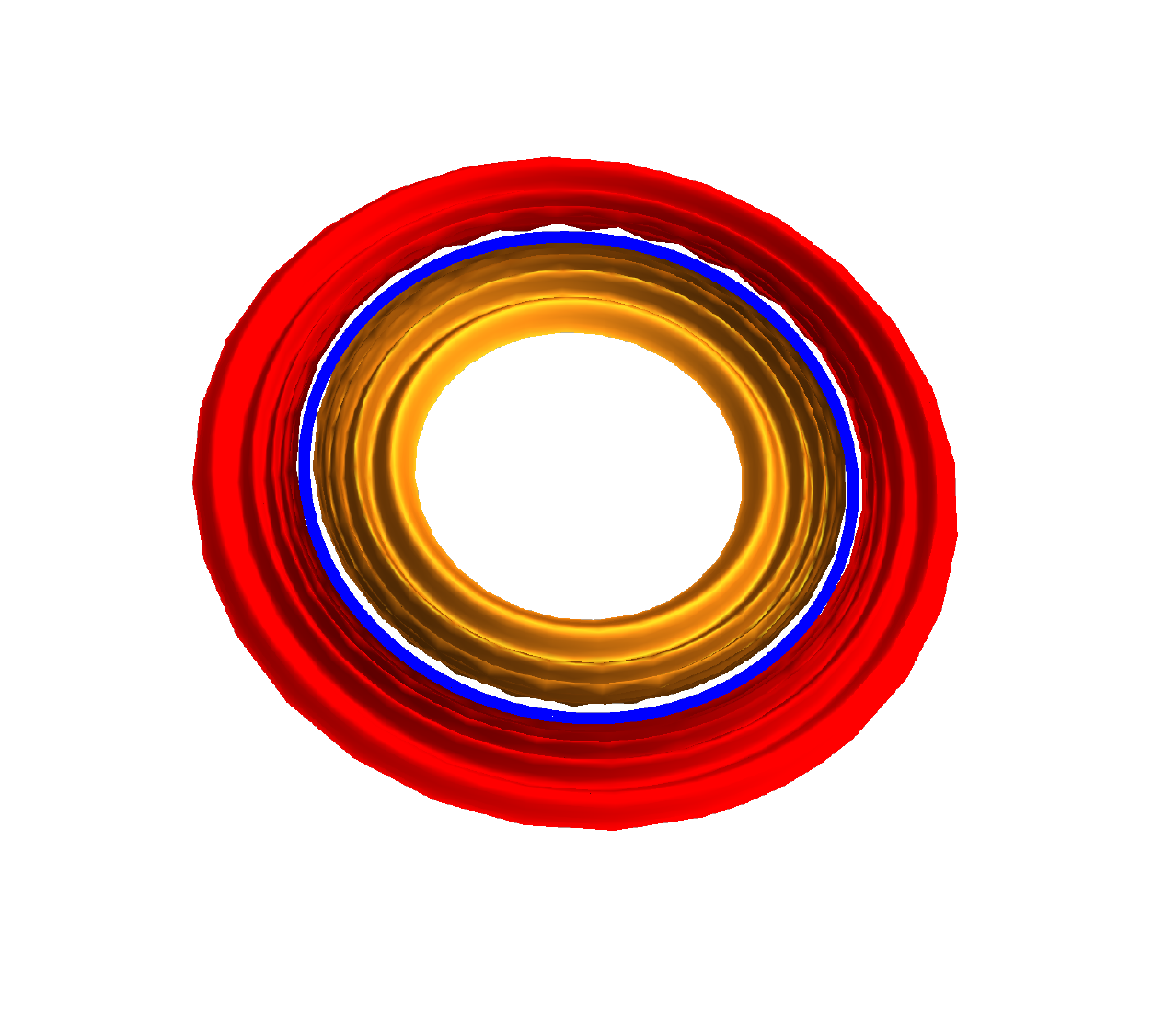}
\caption{\em If we have spherical symmetry, we can extract modes, $\anh$  by integrating the field along the trajectory of  an expanding light shell outside a sphere, and a contracting light shell inside the sphere. The blue circle in the middle is $\uent = 0$. The modes $\anh$ are obtained  by smearing the field with the red smearing function whereas $\tildanh$ are obtained by smearing the field with the yellow smearing function. The oscillations of both smearing functions pile up near $\uent = 0$. \label{figsphericalmode}}
\end{center}
\end{figure}
Then, just as above, we can define modes by integrating the field against a rapidly oscillating function in $\uent$ and against a spherical harmonic  on the sphere. We denote the set of quantum numbers that specify the spherical harmonic by $\ell$. The process is very similar to what we did earlier and is depicted in Figure \ref{figsphericalmode}.
\be
\label{sphericalmodedef}
\begin{split}
&\anh = {r_0^{{d-1 \over 2}}  \over  \sqrt{\pi \omega_0}  } \int\partial_{\uent} \phi(\uent, \vent=0, \Omega) \left(-\uent \right)^{-i \omega_0} \tune(-\uent) d \uent Y_{\ell}^*(\Omega) d^{d-1} \Omega, \\
&\tildanh = {r_0^{{d-1 \over 2}}  \over  \sqrt{\pi \omega_0}} \int \partial_\uent \phi(\uent,\vent=-\epsilon,\Omega) \uent^{i \omega_0} \tune(\uent) d \uent Y_{\ell}(\Omega) d^{d-1} \Omega,
\end{split}
\ee
The modes defined in this way now depend on a spherical harmonic rather than a point but some simple calculations show that they satisfy all of the properties displayed in equations \eqref{samesidecorr}, \eqref{oppsidecorr}, \eqref{commutmodes} and \eqref{stateaction}.

\subsection{A derivation of Hawking radiation \label{hawkingderiv}}
We now show how these results can be used to derive the form of Hawking radiation, both in AdS and in flat space. 

The main physical point is as follows.  For a black hole formed from collapse, at late times, the geometry develops an approximate time-translational symmetry. This symmetry can be used to relate the near-horizon modes defined above to modes that are defined near the asymptotic boundary.  Then the fact that the near-horizon modes are thermally occupied immediately tells us that the asymptotic modes are also thermally occupied to a good approximation.

\subsubsection{Flat space \label{secflathawkingrad}}
\begin{figure}[!ht]
\begin{center}
\includegraphics[height=0.4\textheight]{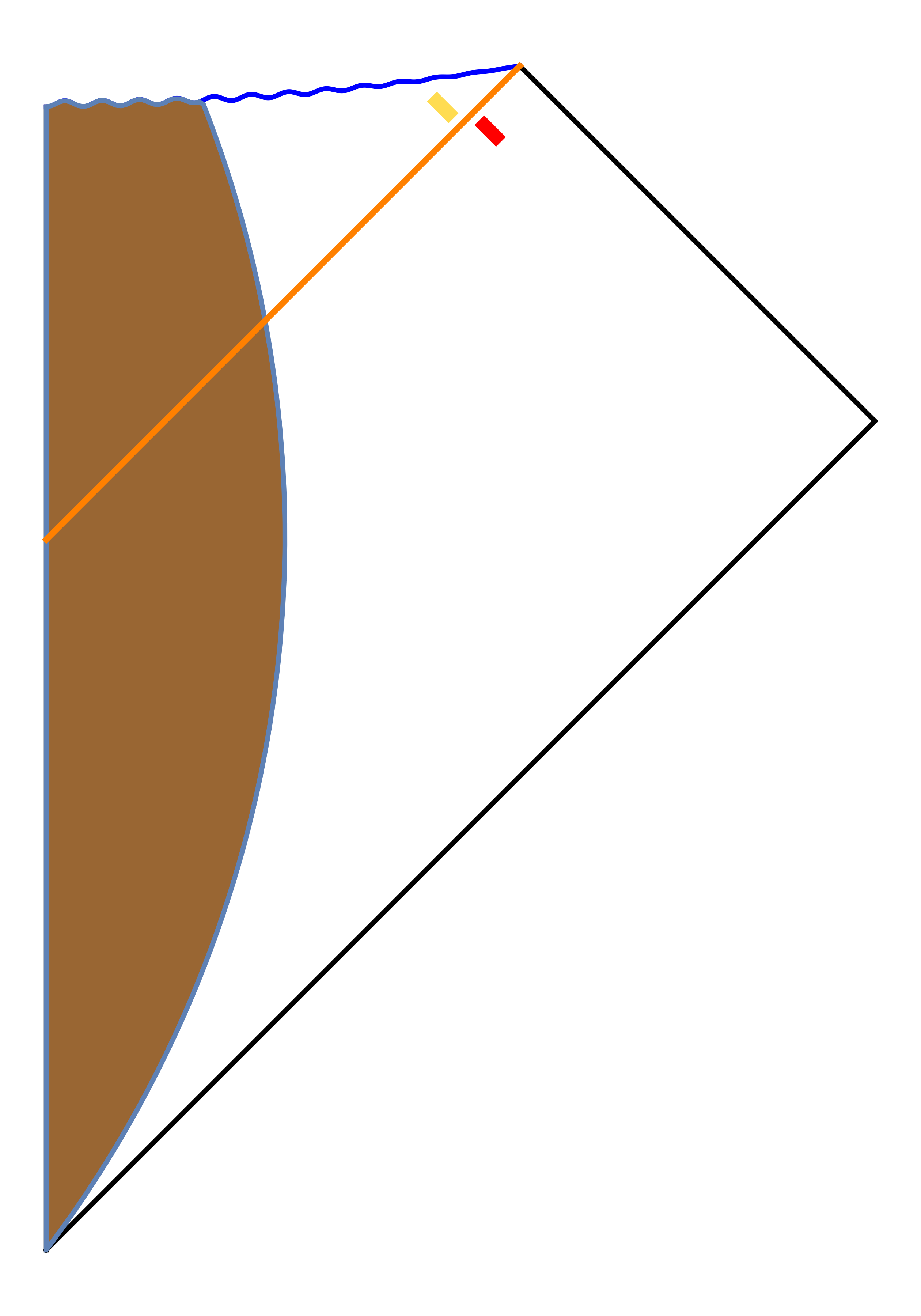}
\caption{\em A black hole formed by the collapse of infalling matter (brown). The horizon is a thick orange line. To derive the formula for Hawking radiation, we are interested in late times and in correlations of the field in the red and yellow regions marked near the horizon. \label{naivepenrose}}
\end{center}
\end{figure}

Consider a black hole formed from the collapse of some matter. The Penrose diagram, ignoring any backreaction of Hawking radiation, is displayed in Figure \ref{naivepenrose}. 

\paragraph{\bf The classical geometry \\}
Let us assume, for simplicity, that the black hole does not have charge or angular momentum. Then,  at {\em late times}, much after the infalling matter has fallen in,  the classical geometry tends to settle down to that of the Schwarzschild black hole. This follows because any perturbation near the horizon tends to die off exponentially. Therefore at late times we can write
\be
\label{latetimemetric}
ds^2 \underset{t \rightarrow \infty}{\longrightarrow} -f(r) dt^2 + dr^2 f^{-1}(r) + r^2 d \Omega_{d-1}^2.
\ee
where, for a black hole of mass $M$,
\be
\label{fexplicit}
f(r) = {1 - {\mu \over r^{d-2}}},
\ee
and the parameter, $\mu$ is related to the mass, $M$,  through $(d-1) \mu = 8 \pi^{{2-d \over 2}} \Gamma(d/2)  \gnewt M$ \cite{lrr-2008-6}. The black-hole horizon is at $r = r_h$ that satisfies $f(r_h) = 0$.
We emphasize that the limit displayed in equation \eqref{latetimemetric} is only an approximate statement about the line element for times that are large compared to the time scale of collapse but short compared to the evaporation time scale. For a black hole formed from collapse, the metric never exactly coincides with the right hand side of equation \eqref{latetimemetric}. Moreover since equation \eqref{latetimemetric} is a classical relation,  it does not tell us about the quantum state of the black hole. These caveats will be important below.

To analyze the propagation of fields in this geometry, it is often convenient to adopt tortoise coordinates
\be
d \rtor = {d r \over f(r)}.
\ee
Note that $\rtor \rightarrow -\infty$ near the horizon, both as we approach from outside ($r \rightarrow r_h^+$) and we approach it from inside ($r \rightarrow r_h^-$).

\paragraph{\bf The propagation of fields \\}
Now consider a minimally coupled scalar field, $\phi$,  in this geometry.  If we are considering physics over time scales where the back-reaction of evaporation does not cause appreciable changes in the geometry,  there is a convenient choice of solutions to the Klein-Gordon equation that can be used to expand the field. If we consider solutions with frequency, $\omega$, and angular quantum numbers denoted by $\ell$, then for the radial part of the solution, we choose one set of solutions, $\fin(\omega, \ell, \rtor)$, that are purely ``ingoing'' at the horizon and another set, $\fout(\omega, \ell, \rtor)$, that also has an outgoing component \cite{Candelas:1980zt,dewitt1975quantum}.
\be
\label{foutfinexpansion}
\begin{split}
&\fin(\omega, \ell, \rtor) \underset{r \rightarrow r_h^+}{\longrightarrow} \white_{\omega, \ell} e^{-i \omega \rtor};\\ 
&\fout(\omega, \ell, \rtor)  \underset{r \rightarrow r_h^+}{\longrightarrow} e^{i \omega \rtor} + \grey_{\omega, \ell} e^{-i \omega \rtor}.
\end{split}
\ee
Here $\grey_{\omega, \ell}$ and $\white_{\omega, \ell}$ are related to the so-called greybody factors of the black hole --- which can be computed by demanding that the solutions above be orthonormal in the Klein Gordon norm --- but will not be important for us.

At late times, the field can be expanded in the basis of solutions above  multiplied by operators $A_{\omega, \ell}$ and $B_{\omega, \ell}$ and their Hermitian conjugates.
\be
\label{fieldoutside}
\phi = \sum_{\ell} \int d \omega \left[A_{\omega, \ell} \fout(\omega, \ell, \rtor)  + B_{\omega, \ell} \fin(\omega, \ell, \rtor)\right] e^{-i \omega t} Y_{\ell}(\Omega) + \text{h.c.}
\ee

We can also quantize the field in the interior of the black hole. Here, the coordinate $\rtor$ plays the role of time, whereas the coordinate $t$ becomes spacelike. We do not understand what happens to the field near the singularity at $\rtor \rightarrow \infty$, and nor do we need to make any assumptions about what happens ``deep'' in the interior which is the region where $t \sim 0$.  We are interested only in the near-horizon behaviour, just behind the horizon, which still corresponds to the coordinate regime where $\rtor \rightarrow -\infty, t \rightarrow \infty$.  For a large black hole, the curvature continues to be small in this region in the  metric \eqref{latetimemetric}. Therefore we can 
expand the field in this region in terms of some new operators $\widetilde{A}_{\omega, \ell}$ and $C_{\omega, \ell}$ and a basis of functions, $\widetilde{\fout}$,
\be
\label{fieldinside}
\phi = \sum_{\ell} \int d \omega \left[\widetilde{A}_{\omega, \ell} e^{i \omega t} Y_{\ell}^*(\Omega) + C_{\omega, \ell} e^{-i \omega t} Y_{\ell}(\Omega)  \right]  \widetilde{\fout}(\omega, \ell,\rtor)   + \text{h.c},
\ee
where the new function is determined by its behaviour near the horizon,
\be
\label{ftexpansion}
\widetilde{\fout}(\omega, \ell, \rtor) \underset{r \rightarrow r_h^-}{\longrightarrow} e^{-i \omega \rtor},
\ee
and the fact that it solves the wave equation. The operators $C_{\omega, \ell}$ are related to the operators outside the horizon by the continuity of the field.\be
C_{\omega, \ell} = A_{\omega, \ell} \white_{\omega, \ell} + B_{\omega, \ell} \grey_{\omega, \ell}.
\ee
Note that continuity {\em cannot} be used to fix the operators $\widetilde{A}_{\omega, \ell}$ in terms of $A_{\omega, \ell}$ and $B_{\omega, \ell}$.

\paragraph{\bf Hawking radiation \\}
Now, we turn to the application of the result derived in section \ref{corrnull}. In the  so-called Kruskal coordinates, 
\be
\label{kruskaldef}
U = -{1 \over \kappa} e^{\kappa (\rtor - t)}, \quad \text{and} \quad V = {1 \over \kappa} e^{\kappa(\rtor + t)},
\ee
with $\kappa = f'(r_h)/2$ called the surface gravity, the metric 
automatically takes the form \eqref{metricsphericalnull} near the horizon, which is located at $U = 0$. Then we can define near-horizon modes in this geometry as described in section \ref{corrnull}.

The key point is to relate these near-horizon modes to the global modes described above. This is a matter of just taking the field expansion from equation \eqref{fieldoutside} and \eqref{fieldinside} and plugging it into the integrals shown in \eqref{sphericalmodedef}.  It turns out that this leads to a very simple relation between the global and near-horizon modes.
\be
\label{nearglobalrel}
\anh = a_{\omega_0 \kappa, \ell}; \qquad \tildanh = \widetilde{a}_{\omega_0 \kappa, \ell},
\ee
where $a_{\omega, \ell}$ and $\widetilde{a}_{\omega, \ell}$ are smeared versions of the operators $A_{\omega, \ell}$ and $\widetilde{A}_{\omega, \ell}$ respectively. 
\be
\label{smearing}
a_{\omega, \ell} = \int A_{\omega', \ell} q(\omega', \omega) d \omega'; \qquad \widetilde{a}_{\omega, \ell} = \int \widetilde{A}_{\omega', \ell} \widetilde{q}(\omega',\omega) d \omega'.
\ee
The precise smearing functions, $q$ and $\widetilde{q}$  can be read off using the results of section 3.3 of \cite{Papadodimas:2019msp}. We will not write them explicitly here but simply explain why they are sharply peaked about $\omega' = \omega$.  

From equations \eqref{foutfinexpansion} and \eqref{ftexpansion}, we see that the near-horizon expansion of the field is very simple when viewed in Kruskal coordinates. The oscillators $A_{\omega, \ell}$ and $\widetilde{A}_{\omega, \ell}$ multiply functions that take on the simple oscillatory form  $|U|^{i \omega \over \kappa}$.  The integral used to define the near-horizon modes in \eqref{sphericalmodedef} picks up precisely this oscillatory term and, moreover,  it is clear that the integral receives contributions from frequencies very close to $\kappa \omega_0$. The only detail about the smearing function that will be relevant is 
that the commutators of the modes above is unit normalized: 
\be
\label{normofcomm}
[a_{\omega, \ell}, a_{\omega, \ell}^{\dagger}] = 1; \qquad [\ta_{\omega, \ell}, \ta_{\omega, \ell}^{\dagger}] = 1.
\ee
Other than this, the  final result that the reader needs to keep in mind is just \eqref{nearglobalrel}, which is a simple and intuitive result.

Applying the result of equation \eqref{samesidecorr} from section \ref{corrnull}, we now immediately see that in a state, $|\Psi \rangle$, corresponding to a black hole, these modes are thermally occupied.
\be
\label{thermalexpectout}
\langle \Psi | a_{\omega, \ell} a_{\omega, \ell}^{\dagger} | \Psi  \rangle = {1 \over 1 - e^{-\beta \omega}},
\ee
where the inverse temperature is given by $\beta = {2 \pi \over \kappa}$.
We emphasize that the operators that appear in equation \eqref{thermalexpectout} are the ones that multiply the global mode functions. Therefore,  the expectation value above leads to a flux of particles at $\scrip$, which can be computed precisely using the asymptotic behaviour of $\fout$ as $\rtor \rightarrow \infty$.

We also emphasize that this derivation places no constraint on the ``ingoing'' modes, $B_{\omega, \ell}$.  Indeed, different choices can be made for the occupation of these operators. Two commonly used choices are the ``Unruh state'', where these modes are simply placed in their vacuum, and the ``Kruskal state'', where the ingoing modes are also thermally populated \cite{Candelas:1980zt}.

\subsubsection{AdS \label{adshawkingderiv}} 
We now describe how almost precisely the same procedure can be used to understand Hawking radiation in AdS. 

\paragraph{\bf Classical geometry \\}
We again assume that at late times, the metric tends to the form displayed in equation \eqref{latetimemetric}. With the AdS radius set to $1$, the only difference is that now the function $f(r)$ is given by
\be
\label{fadsexplicit}
f(r) = r^2 + 1 - {\mu \over r^{d-2}}.
\ee
The constant of proportionality between $\mu$ and the mass remains the same as in flat space. The black-hole horizon, $r_h$, is still located at the point where $f$ vanishes: $f(r_h) = 0$.  This means that, at late times, the metric tends to that of the global AdS Schwarzschild black hole.

We will again define a tortoise coordinate satisfying $d \rtor = {d r \over f(r)}$ which tends to $-\infty$ near the horizon and approaches a constant value near the asymptotic boundary.

\paragraph{\bf Propagation of fields \\}
In AdS, it is natural to impose normalizable boundary conditions on the field. Given a scalar field, $\phi$, of mass $m$, we demand that, as $r \rightarrow \infty$, the field dies off as $r^{-\Delta}$ where $2 \Delta = d + \sqrt{d^2 + 4 m^2}$. For this section, the relevance of this boundary condition is that it relates the left- and right-moving modes near the boundary.

Therefore, at late times, the field can be expanded as 
\be
\label{phiexpandads}
\phi = \sum_{\ell} \int d \omega A_{\omega, \ell} \fst(\omega, \ell, \rtor) e^{-i \omega t} Y_{\ell}(\Omega) + \text{h.c},
\ee
where, as opposed to flat space, there is now only one independent ``standing wave'' solution outside the horizon. Near the horizon, this function satisfies
\be
\fst(\omega, \ell, \rtor)  \underset{r \rightarrow r_h^+}{\longrightarrow}  \left( e^{i \omega \rtor} + e^{-i \delta_{\omega,\ell}} e^{-i \omega \rtor} \right). 
\ee
On the other hand, near the asymptotic boundary at $r \rightarrow \infty$, the function satisfies 
\be
\fst(\omega, \ell, \rtor) \underset{r \rightarrow \infty}{\longrightarrow} {\sqrt{G_{\omega, \ell}} \over r^{\Delta}}
\ee
In AdS/CFT this sets the relation between the modes $A_{\omega, \ell}$ and the modes of the generalized free field that is dual to this bulk scalar scalar field, $\op_{\omega, \ell}$, to be
\be
\label{bulkbdryrelation}
A_{\omega, \ell} \sqrt{G_{\omega, \ell}} = \op_{\omega, \ell}.
\ee

As in flat space when we cross the horizon, we can introduce some new modes, $\widetilde{A}_{\omega, \ell}$, and expand the field as
\be
\label{adsexpandbehindhor}
\phi = \sum_{\ell} \int d \omega \left[A_{\omega, \ell}  e^{-i \delta_{\omega, \ell}} e^{-i \omega t} Y_{\ell}(\Omega)  + \widetilde{A}_{\omega, \ell}  e^{i \omega t} Y_{\ell}^*(\Omega) \right] \widetilde{\fst}(\omega, \ell, \rtor)  + \text{h.c}
\ee
where $\widetilde{\fst}$ is specified by its behaviour as one approaches the horizon from inside
\be
\fst(\omega, \ell, \rtor) \underset{r \rightarrow r_h^-}{\longrightarrow} e^{-i \omega \rtor},
\ee
and the fact that it solves the wave equation.

We now again apply the result of section \ref{corrnull}. We again go to Kruskal coordinates defined precisely as in equation \eqref{kruskaldef}. The metric then again automatically takes the form shown in equation \ref{metricsphericalnull} near the horizon. Defining near horizon modes, as explained in section \ref{corrnull}, we find that they are related to the global modes through
\be
\anh = a_{\omega_0 \kappa, \ell}; \qquad \tildanh = \widetilde{a}_{\omega_0 \kappa, \ell}
\ee
where $a_{\omega, \ell}$ and $\widetilde{a}_{\omega, \ell}$ are, just as in equation \eqref{nearglobalrel},  smeared versions of the operators $A_{\omega, \ell}$ and $\widetilde{A}_{\omega, \ell}$. The smearing ensures that these operators have unit commutators, as displayed in \eqref{normofcomm}, rather than delta-function commutators.

So, once again, for a Schwarzschild black hole in AdS we find that the modes outside the horizon must be populated thermally exactly as displayed in equation \eqref{thermalexpectout}.
Note that this time there is no flux near the boundary since the global modes multiply standing waves.

\subsection{Hawking's argument for information loss \label{sechawkingparadox}}
We now turn to Hawking's argument that the formation and evaporation of black holes in asymptotically flat space would lead to information loss. This argument is laid out clearly in \cite{Hawking:1976ra}, and we will closely follow  sections 3 and 5 of that paper here.
The main point is as follows.  In section \ref{secflathawkingrad}, we considered the expectation value of the number operator, $\schwarzn_{\omega, \ell}$ for particles with a given frequency and angular momentum
\be
\label{schwarzndef}
\schwarzn_{\omega, \ell} = a_{\omega, \ell}^{\dagger} a_{\omega, \ell},
\ee
in a state corresponding to a black hole and found that it was the same as in a  thermal state.  But it is simple to see that this is also true of any power of $\schwarzn_{\omega, \ell}$, within the approximation that we have been using.
\be
\label{thermaldensity}
\langle \Psi | (\schwarzn_{\omega, \ell})^q|\Psi \rangle = \tr(\rho_{\omega, \ell} (\schwarzn_{\omega, \ell})^q) \quad \text{where} \quad \rho_{\omega, \ell} = {1 \over 1 - e^{-\beta \omega}}  e^{-\beta \omega \schwarzn_{\omega, \ell}}.
\ee

It is also possible to check that if one takes modes corresponding to different frequencies --- where in our treatment, ``different'' means larger than the width of the smearing function in equation \eqref{smearing} --- then there are no correlations at all.  For instance,
\be
\label{uncorrelated}
\langle \Psi | a_{\omega, \ell} a_{\omega', \ell}^{\dagger} |\Psi  \rangle = 0, \quad \text{for} \quad \omega \neq \omega'.
\ee

\begin{figure}[!ht]
\begin{center}
\includegraphics[height=0.4\textheight]{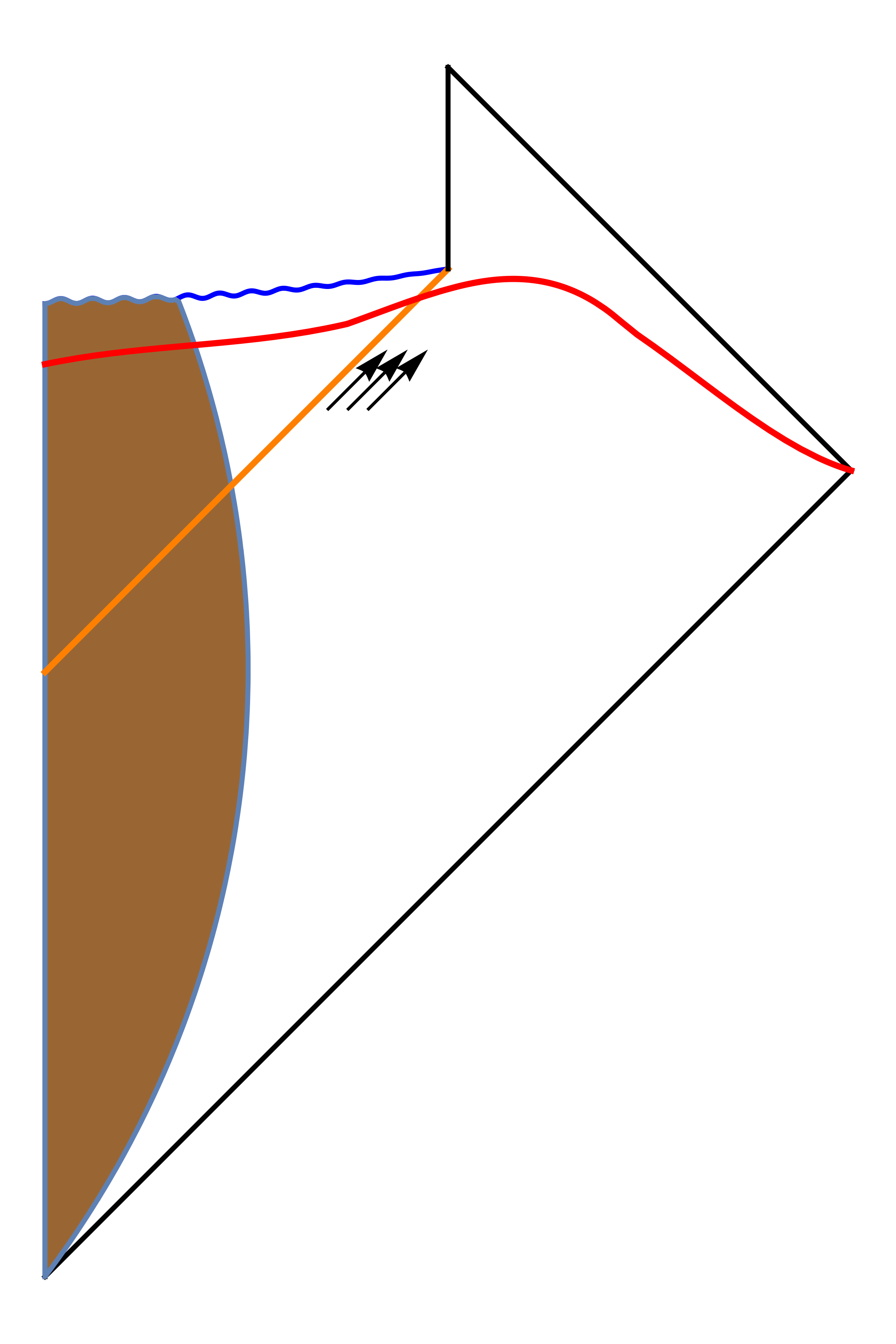}
\caption{\em A schematic Penrose diagram showing the formation of the black hole
and also its evaporation. Hawking radiation emerging from the horizon is marked with black arrows. The diagram also displays a ``nice slice'' that runs through the interior, stays away from the singularity, and also collects a significant fraction of the Hawking radiation.\label{backreactpenrose}}
\end{center}
\end{figure}

The precise spectrum of particles observed  on $\scrip$ also depends on greybody factors that are determined by the falloff of the mode $\fout$ as $r \rightarrow \infty$.  But these are functions of the frequency and angular momentum that depend on only the classical geometry at late times. In the example that we have been considering, the late-time geometry  depends on a single parameter, which we can take to be the mass or the temperature of the black hole. In 
the more general case, a black hole could have additional conserved charges such as angular momentum or electric charge. But, even then, it would appear that the probability to measure a certain number of particles, with a given frequency and angular momentum, at $\scrip$, depends only on a few numbers and not on the details of the collapse.  This is manifest in the calculation above:  if we place the 
ingoing $B_{\omega, \ell}$ modes in some different state, this does not affect the occupancy of the $A_{\omega,\ell}$ modes.

If we consider  a very large black hole, almost all of its evaporation is well captured by our calculation above since the curvature remains small at the horizon for most of its lifetime. This ceases to be true when the black hole
becomes very small but, Hawking argued that since ``information \ldots requires energy'', all the information about the initial state cannot emerge in the final stages of black hole evaporation.

It is possible to set up a similar paradox in AdS/CFT. Consider an arbitrary pure state that collapses to form a large Schwarzschild black hole.  Then the calculation of  section \ref{adshawkingderiv} suggests that irrespective of the details of the  initial state, the correlators of late-time generalized free fields are those of a thermal state. This may appear surprising. Where has the information about the initial state gone? 

Hawking suggested that this seeming lack of correlation between the initial state and the final state of the radiation would lead the observer outside
to adopt a ``principle of ignorance'' about the interior. Therefore the correct procedure was
for the observer outside to trace over all possibilities for the black-hole interior and thereby obtain a mixed state for the exterior.

If one considers the process  of gravitational collapse and black hole evaporation, then the central conclusion of \cite{Hawking:1976ra} was that  ``the final situation \ldots is described by a density matrix and not a pure state''.  This is what led Hawking to argue that the S-matrix in a theory of gravity cannot be unitary.  However, does this observation really constitute a paradox? 
In a large statistical system, how different are density matrices and pure states and is the calculation presented above accurate enough to distinguish between the two? In the next subsection we will take a small detour into quantum statistical mechanics to explore this issue before we return to this  paradox.

\subsection{How close are pure states to mixed states?\label{puremixed}}
In this subsection, we review some simple ideas about the physics of thermalization. The conclusions we will reach are elementary but these conclusions are, unfortunately,  not emphasized sufficiently in the high-energy or quantum-gravity
community.  We are interested in quantifying the differences between a mixed state, which is represented by a density matrix, and a pure state, when a system is probed with physical observables. To our knowledge the importance
of these results for black holes was first emphasized in  \cite{lloyd1988black}.

Consider the microcanonical ensemble in some system, which is given by the density
matrix, 
\be
\label{microcan}
\rho_{E} = {1 \over W} \sum_{E_0 - \Delta < E < E_0 + \Delta}  |E \rangle \langle E|,
\ee
where the sum is centered around some energy $E_0$ and has a width $2 \Delta$. All sums over E in this subsection will be over the same range, which we will henceforth omit. The normalization, $W$, just comes from the number of states in the band and is set by demanding that $\tr(\rho_{E}) = 1$.  We would like to understand how close this density matrix is, physically, to a pure state
\be
\label{puremicro}
|\Psi \rangle = \sum_{E}  a_{E} | E \rangle,
\ee
where we have a superposition of  states from the same energy band and the coefficients $a_{E}$ are picked randomly. 

We are looking for a physical notion of ``closeness''. In quantum mechanics, the physically meaningful observations have to do with the probabilities of various measurements. These are given by the expectation value of various projection operators. So we would like to understand how much the probability distribution for some measurement differs between the microcanonical density matrix and the pure state described above. This means that if $P$ is a projector we want to understand the typical size of 
\[
\delta = \left| \tr(\rho_{E} P) - \langle \Psi | P | \Psi \rangle \right|
\]

The natural probability distribution to place on the set of states is simply where all the complex coefficients $a_{E}$ are allowed to vary in an uncorrelated manner subject to the restriction that the state has unit norm.
\be
\label{haarmeasure}
d \mu_{\Psi} = {1 \over V} \delta(\sum_{E} |a_{E}|^2 - 1)\prod_{E}  d^2 a_{E}, 
\ee
and the normalization factor $V$ just ensures that $\int d \mu_{\Psi} = 1$. 

Then we can estimate the typical size of the deviation between a randomly picked pure state and the microcanonical mixed state using some simple integrals with this probability measure. This deviation is centered about $0$ since
\be
\begin{split}
\int d \mu_{\Psi} \langle \Psi | P | \Psi \rangle &= \int d^2 a_{E} \delta(\sum_{E} |a_{E}|^2 - 1) a_{E} a^*_{E'} \langle E' | P | E \rangle  \\
&= {1 \over W} \sum_{E} \langle E | P | E \rangle = \tr(\rho_{E} P)
\end{split}
\ee
But the average size of the deviation can also be computed through
\be
\label{projectordeviation}
\begin{split}
&\int d \mu_{\Psi} \big( \tr(\rho_{E} P) - \langle \Psi | P | \Psi \rangle \big)^2 = \int d \mu_{\Psi} \Big( \sum_{E, E'} (a_{E} a^*_{E'} - {1 \over W} \delta_{E, E'})  \langle E' | P | E \rangle \Big)^2 \\ &=  {1 \over (W + 1)} \sum_{E,E'} \Big( {1 \over W} \langle E | P | E' \rangle \langle E' | P | E \rangle - {1 \over W^2} \langle E | P | E \rangle \langle E' | P | E' \rangle \Big)   \leq {1 \over ( W + 1)}.
\end{split}
\ee
The steps above involve only some simple integrals and manipulations that can be found in section 2.1 of \cite{Raju:2018xue}.  In the last step above we have also used the fact that, as a projector, $P$ can have no eigenvalue larger than $1$ and also that $P^2=P$.

\begin{figure}[!ht]
\begin{center}
\includegraphics[height=0.5\textheight]{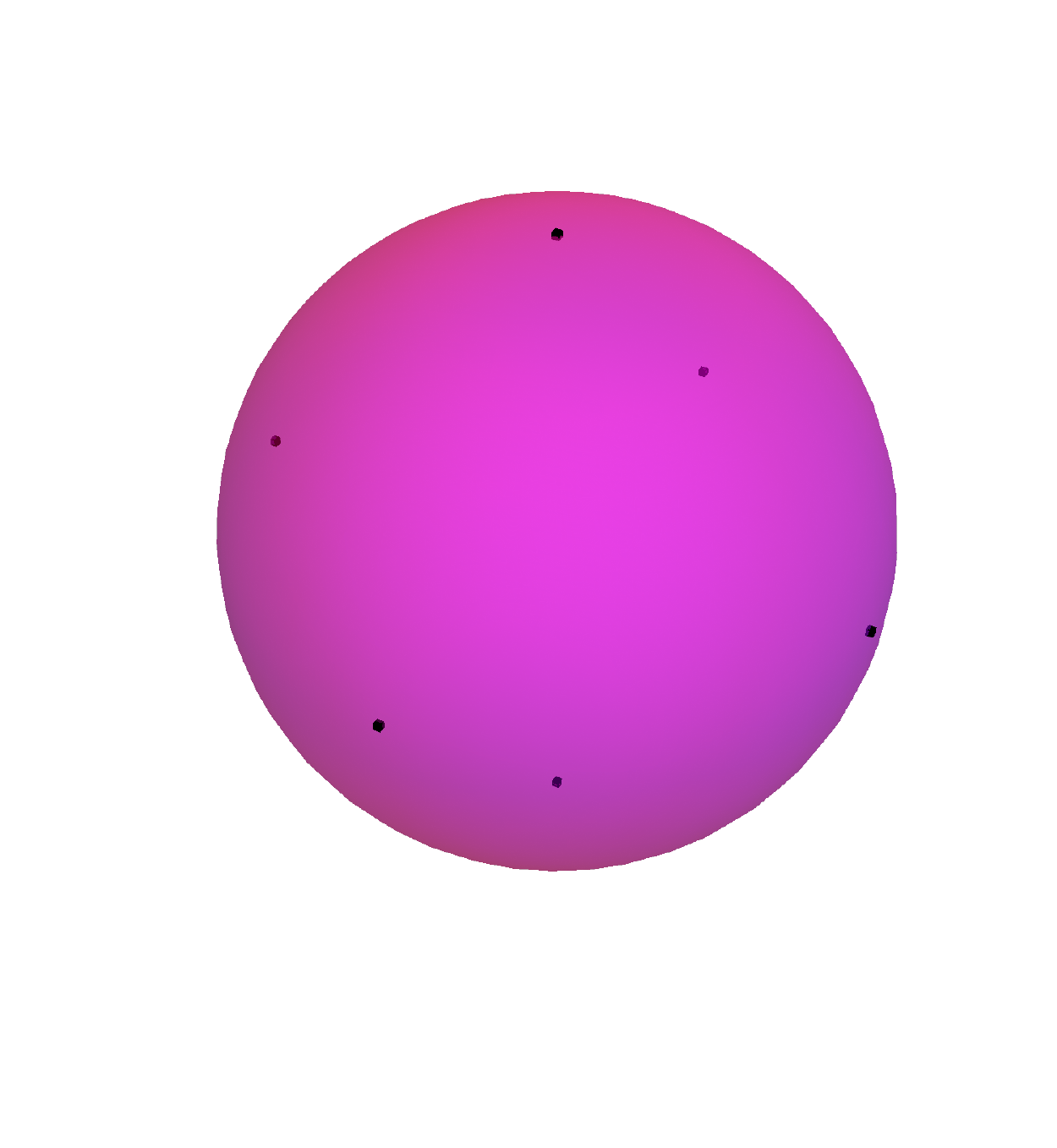}
\caption{\em In a high-dimensional Hilbert space, almost all pure states look exponentially close to the maximally mixed state. The volume occupied by atypical states (depicted by black dots) is exponentially small. \label{genericspace}}
\end{center}
\end{figure}

So far the results we have derived have been exact. However, if we now consider a system with a large number of degrees of freedom, then $W = e^{S}$, where $S$ is the entropy. So the result above tells us that random pure states are exponentially close to mixed states in a system with a large number of degrees of freedom. We emphasize that this result is so general that it does not require any dynamical assumptions, and follows purely from kinematic considerations. In particular, we emphasize that this result does {\em not} need the commonly discussed eigenstate thermalization hypothesis \cite{Deutsch,srednicki1994chaos,srednicki1999approach}. A visual cartoon of our result is shown in Figure \ref{genericspace}.

Note that these results do not apply to quantities like the von Neumann entropy. The von Neumann entropy has the property that it is zero for a pure state and non-zero for a mixed state. This is because it is not the expectation
value of any projector or a linear combination of such expectation values.\footnote{The argument is as follows. Say the von Neumann entropy had been the expectation value of some operator. Then since the von Neumann entropy is zero in all pure states, since any mixed density matrix can be written as a linear combination of pure density matrices,  and since expectation values are linear, the von Neumann entropy would be zero in all mixed states. This is absurd, and so this quantity cannot be the expectation value of any operator. \label{footvn}}  Consequently, it does not correspond to a direct physical observable. It can, of course, be reconstructed using a combination of many individual physical observations. But the results above tell us
that unless these individual observations are made to exponential precision,  the von Neumann entropy cannot be accurately reconstructed.

\subsection{Resolving Hawking's paradox \label{resolvehawking}}

The result above is directly relevant for Hawking's original paradox. We need to be careful in applying the result from section \ref{puremixed} to evaporating black holes since the statistical ensemble at a particular
energy is dominated by the Hawking radiation, which has more entropy than the original black hole. Moreover, in asymptotically flat space, the entropy is infinite due to a volume factor. However, we will assume that we can focus on some $e^{S}$ dimensional space of states, where $S$ is the Bekenstein-Hawking entropy, that can be used to describe the process of black-hole formation and evaporation.

Now,  we found above that the probability distribution for the number of particles emitted at a particular frequency and angular momentum coincided with a thermal distribution. Moreover, particles emitted at one frequency were uncorrelated with those emitted at another frequency. Now, if equations \eqref{thermalexpectout} and \eqref{uncorrelated} were to hold {\em exactly} then we would indeed be led to the conclusion that the final state was thermal. However, what the simple analysis of section \ref{puremixed} tells us is that if we just add $\Or[e^{-{S \over 2}}]$ corrections to the right hand side of equations \eqref{thermalexpectout} and \eqref{uncorrelated}, the results will be {\em perfectly consistent} with a pure state.

The derivation of Hawking radiation that we provided in section \ref{hawkingderiv} was more precise than Hawking's original derivation but it did not even keep track of leading  $\Or[{1 \over S}]$ effects, let alone those effects that are exponentially suppressed in $S$. Indeed, to keep track of power-law corrections is already challenging. For instance, imagine the following process. We form a black hole from collapse and before it starts evaporating, we perform a measurement of the energy of the black hole to within some accuracy $\Delta$. This partially collapses the wavefunction of the black hole, and forces it to have support on energies only within that range. Elementary considerations of statistical mechanics tell us that the black hole radiation should now be described by a state in the microcanonical ensemble and not the canonical ensemble. In the thermodynamic limit, these ensembles coincide, but for systems with a finite
entropy they differ at subleading order in $S$. In principle, by carefully keeping track of the fluctuations of the zero mode of the metric, which measures the energy, one should be able to correct \eqref{thermaldensity} and \eqref{uncorrelated} to derive the corresponding microcanonical results. But even this computation has never been performed. (See \cite{Brustein:2013ena,Brustein:2013qma,Saini:2015dea} for progress in this direction.)

We expect, from the results of section \ref{puremixed}, that the terms that are required to restore purity to Hawking radiation will be exponentially suppressed in $S$. In fact, it would contradiction the result of section \ref{puremixed} if final-state correlators were to differ from mixed-state correlators by a large amount, for a generic pure state.  
Since $S$ itself is inversely proportional to $\gnewt$, these terms are nonperturbative in the gravitational coupling. At this order, the notion of a metric may itself stop making sense. There is certainly no argument in the literature that bounds these terms.

Therefore we arrive at the following simple conclusion. Hawking was not justified in concluding from equations \eqref{thermaldensity} and \eqref{uncorrelated} that these results could only be consistent with a mixed state, since mixed and pure states are exponentially close, and since equations \eqref{thermaldensity} and \eqref{uncorrelated} are leading-order results derived in free-field theory. This leads us to conclusion \ref{lessonone}. 

\subsubsection{Some subtleties and possible questions \label{oldinfosubtle}}
We would like to address some possible subtleties by discussing the answers to commonly asked questions.
\begin{enumerate}[label={\bf Q\arabic*}, series=qseries]
\item
{\em Doesn't the no-hair theorem tell us that, even if Hawking's calculation is corrected at subleading orders, the outgoing radiation cannot carry any information except about a few conserved charges?}

The no-hair theorem  is a classical result,
 which applies to black hole spacetimes with a certain Killing symmetry. To set up the information paradox, we need to consider black holes formed by collapse. So to upgrade Hawking's argument to a true paradox, one would need some quantum analogue of the no hair theorem that tells us that, at late times, Hawking radiation is characterized by 
a few conserved charges up to exponential accuracy.
 
There are formidable difficulties in incorporating quantum effects into the no hair theorem. In fact, if one attempts to use the classical theorem, unchanged, in quantum mechanics, then the theorem is clearly wrong. Any black hole formed from collapse cannot be an energy eigenstate and must necessarily have some spread in its energy, $H$. (Recall that the physics of energy eigenstates is exactly time-translationally invariant in quantum mechanics.)  Therefore, the size of quantum fluctuations in the energy --- quantified by $\langle H^2 \rangle - \langle H \rangle^2$  --- contains independent information beyond the expectation value of the energy $\langle H \rangle$. We have already discussed this spread in section \ref{resolvehawking}.
So even this simple reasoning shows that specifying the state of a black hole  requires, at least,  information about the quantum mechanical probability distribution of the charges, and not just their expectation values.

But then we find that, in quantum mechanics, the observers outside also have information about correlations in the probability distributions of various charges. For instance, given an observable $O$ from the exterior of the black hole, there is  information in  the correlator  $\langle H O \rangle$ that goes beyond the information in the expectation values $\langle H \rangle$ and $\langle O \rangle$. Similarly, there are an infinite number of higher-point correlators one can write down outside, all of which give us independent pieces of information. There is no analogue of these correlators classically.\footnote{Here, we are comparing a classical pure state with a quantum mechanical pure state. Of course, even in classical mechanics one could consider an {\em ensemble} of black holes with probability distributions for these quantities, and compare that with a mixed state in quantum mechanics. Even then, quantum mechanical observables would have more information.} And so it is not true that the external observers have access to only a few numbers and ignorance about the details of the state in the interior. 

At best one might hope to prove that, in quantum mechanics, except for the moments of conserved charges, other correlators die off and approach some kind of universal value at late times. But to obtain a paradox one would need to prove that this exponential decay continues
till the deviation of the correlators from this universal value becomes smaller than $e^{-{S \over 2}}$ in size. But it is well known from the study of quantum statistical mechanics that such exponentially decaying signals retain a small tail at late times. This tail is precisely of the size $e^{-{S \over 2}}$ and therefore precisely of the right size to preserve information about infalling matter. 

To summarize briefly: in its current classical form, the no hair theorem cannot be used to frame a paradox. Not only has a quantum-mechanical generalization of this theorem not been proved, the arguments above imply that such a generalization does not exist.

\item
{\em Small corrections can resolve the paradox in principle. But can we show that such corrections do, indeed, retain information about the initial state?}

In this section, we have shown that Hawking's original argument is not precise enough to lead to a paradox. However, in sections \ref{secholography} and \ref{secresolveinfo}, we will do better and prove, under certain conditions, that correlators in the black hole exterior can uniquely identify the state of the interior.

 As we mentioned, Hawking suggested that the observer outside would have to adopt a a ``principle of ignorance'' about the black-hole interior. One of the main
takeaways of sections  \ref{secholography} and \ref{secresolveinfo} is that the correct situation in quantum gravity is exactly the opposite: the observer outside instead has complete information about the interior.

\item
{\em Have we found the mistake in Hawking's argument?}

The answer to this question depends on what one means by a ``mistake''. There is clearly no mistake in the computation of the leading-order correlators of operators on $\scrip$.

But there is a mistake in the claim that the leading order computation of these  correlators is  inconsistent with the final state being pure. As we have shown, in a generic pure state, we expect precisely the same leading-order result. There is also a mistake in the conclusion that just because these correlators are independent of the state of the infalling matter to leading order, this is enough to demonstrate a ``principle of ignorance.'' We will be able to formally prove, under weak assumptions, in section \ref{secholography} that this principle of ignorance is incorrect.

On the other hand, if the phrase  ``find a mistake''  is used to demand an explicit calculation of all the exponentially small corrections to Hawking radiation for a specified initial state then, indeed, such a calculation is not available.  But,  by this yardstick, there are no explicit checks of the unitarity of burning coal either. In fact, since black holes present a cleaner and more tractable system than coal we believe that, in the future, it will be easier to calculate these effects for evaporating black holes than for burning coal.

\end{enumerate}

\section{Paradoxes involving the interior \label{secintpar}}
In this section, we will explain how a more-formidable puzzle regarding black-hole evaporation can be derived if one also considers the interior of an evaporating black hole. This paradox uses a property of quantum entanglement, called ``monogamy''. This paradox was first outlined by Mathur \cite{Mathur:2009hf}, and then re-analyzed a few years later by AMPS \cite{Almheiri:2012rt}. 

The original paradox was framed using the strong subadditivity of the von Neumann entropy.  However, the von Neumann entropy is a subtle quantity in a quantum field theory, and even more so in quantum gravity.   The same physical paradox can be framed elegantly in terms of the monogamy of Bell correlators, which are completely well defined. This was done in  \cite{Raju:2018zpn} and we will follow this approach here.  

 The difference between the ``monogamy paradox''  of this section and the original information paradox is that, at first sight, it appears that the monogamy paradox cannot be resolved by small corrections. At the end of this section, we will explain that the monogamy paradox arises because of an incorrect physical assumption about how information is localized in quantum gravity. The point of using Bell correlators is only so that we can focus on this essential physics, and not worry about the technicalities of short-distance divergences in the von Neumann entropy.

In this section, we first explain some
of the necessary quantum-information background, and then return to the paradox. These quantum-information results will also be important for the discussion in later sections.

\subsection{Some results from quantum information}
\subsubsection{\bf Monogamy of entanglement \label{secmonogent}}
Two systems are said to be {\em entangled} if their observables have correlations that exceed possible classical correlations. Bell initially set out some inequalities \cite{bell1964einstein} that had to be satisfied by classical correlators but were not satisfied quantum mechanically. It is often convenient to think in terms of a slightly different observable called a ``CHSH'' operator. (The abbreviation refers to the initials of the authors of \cite{Clauser:1969ny}: Clauser, Horne, Shimony and Holt.)
 First we describe how this operator can serve as a diagnostic of entanglement, and then we describe how it can also be used to formulate the monogamy of entanglement.

\paragraph{\bf CHSH correlators and Tsirelson's bound \\}
 
Let $A_1, A_2$ be a pair of observables from one system  and  $B_1, B_2$ be a pair of observables from another system.    We assume that the eigenvalues of all the observables are in the range $[-1, 1]$. Both the systems can be part of a larger system and the precise mathematical assumption we require is just that, in this larger Hilbert space, $[A_i, B_j] = 0$. The CHSH operator, $C_{AB}$, is
defined as the joint observable
\be
\label{cabdefn}
C_{AB} =  A_1 (B_1 + B_2)  +  A_2 (B_1 - B_2).
\ee
If all the observables can be assigned values simultaneously --- which is the classical regime --- then it is easy to see that for any assignment of numerical values in the range $[-1,1]$ to $A_i$ and $B_j$, we will always have $|C_{AB}| \leq 2$.

On the other hand, quantum mechanically, this classical bound can be exceeded. Tsirelson \cite{cirel1980quantum} showed that, quantum mechanically, the expectation value of this observable can reach a larger value  in a state, $|\Psi \rangle$, of the combined system,
\be
|\langle \Psi |  C_{AB}  | \Psi \rangle| \leq 2 \sqrt{2}.
\ee
This is called Tsirelson's bound.

\begin{figure}[!ht]
\begin{center}
\includegraphics[height=0.3\textheight]{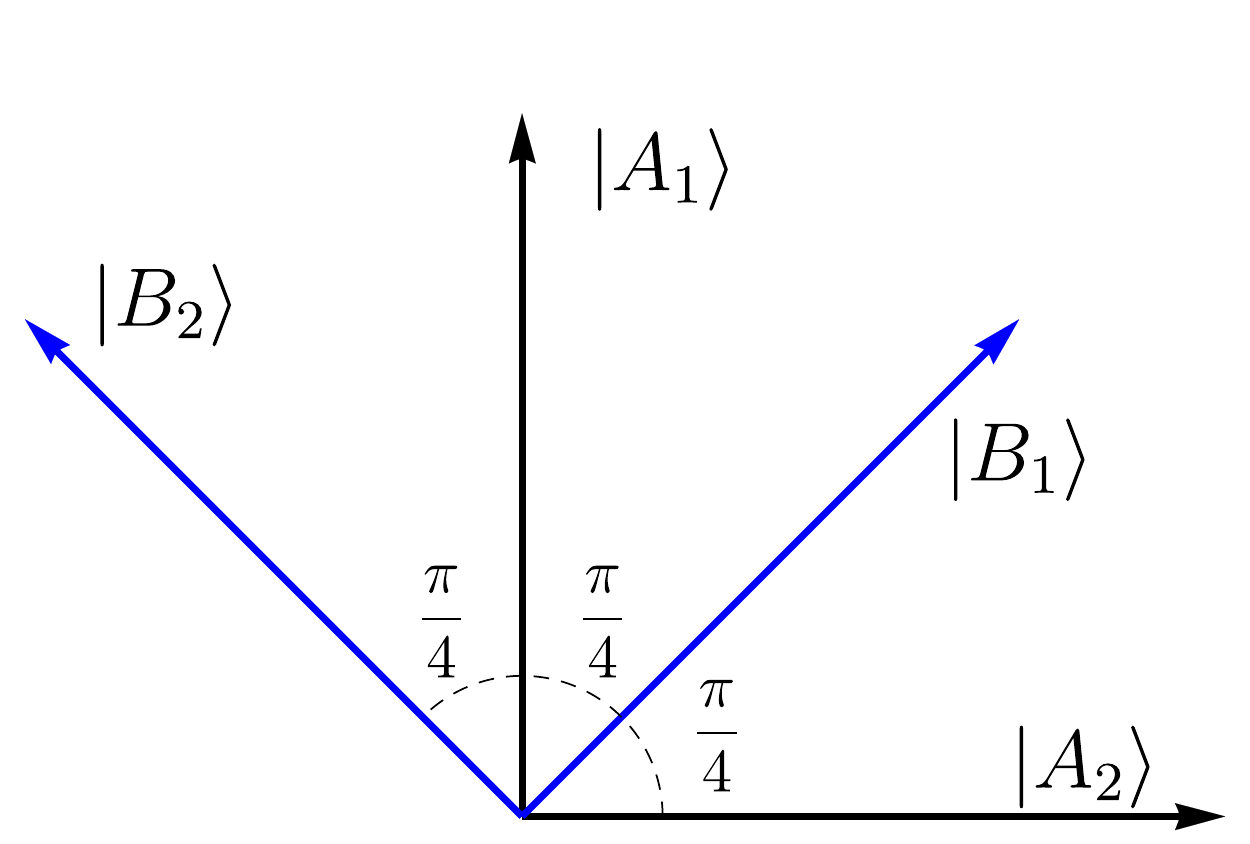}
\caption{\em If the vectors $|A_1 \rangle, |A_2 \rangle, |B_1 \rangle, |B_2 \rangle$ take the geometric configuration above, then Tsirelson's bound is saturated since $\langle A_1 | (|B_1 \rangle + |B_2 \rangle)  = \langle A_2| (|B_1 \rangle - |B_2 \rangle ) = \sqrt{2}$. \label{figcirelson}}
\end{center}
\end{figure}

It is easy to see how this bound can be saturated. Consider the four vectors 
\be
|A_1 \rangle = A_1 | \Psi \rangle; \quad  |A_2 \rangle = A_2 | \Psi \rangle; \quad  |B_1 \rangle = B_1 | \Psi \rangle; \quad  |B_2 \rangle = B_2 | \Psi \rangle.
\ee
Each of these vectors can, at most, have length $1$ since all the operators involved have no eigenvalue larger than $1$.   One configuration in which Tsirelson's bound is then saturated, is  when these four vectors all have length 1, all lie in a plane and take on the geometric configuration shown in Figure \ref{figcirelson}. In this configuration, $\langle B_1 | B_2 \rangle = \langle A_1 | A_2 \rangle = 0$ and  $|B_1 \rangle + |B_2 \rangle = \sqrt{2} |A_1 \rangle$ while $|B_1 \rangle - |B_2 \rangle = \sqrt{2} |A_2 \rangle$.

The systems probed by the $A$ and $B$ operators are said to be entangled if $2 \leq |\langle \Psi | C_{AB} |\Psi  \rangle| \leq 2 \sqrt{2}$, since this means that correlations between the two systems exceed the largest allowed classical value.

\paragraph{\bf Monogamy of entanglement \\}
If  two systems are entangled with each other, then ``monogamy'' means that neither of them can be entangled with a third system. There are various measures of entanglement, and many of them have an associated notion of monogamy. If we use CHSH
correlators, then the formulation of the monogamy of entanglement becomes particularly elegant. 

Let us take a third system probed by operators $C_1, C_2$, whose eigenvalues also lie in the range $[-1,1]$. Once again, the precise assumption is that  $[B_i, C_j] = [A_i, C_j] = 0$. Then one can define another CHSH operator, $C_{AC}$, just as in equation \eqref{cabdefn}. 
\be
\label{cacdefn}
C_{AC} =  A_1 (C_1 + C_2)  +  A_2 (C_1 - C_2).
\ee
Toner and Verstraete \cite{toner2006monogamy} (see also \cite{toner2009monogamy}) showed that in any state, $|\Psi \rangle$,
\be
\label{monogamyrel}
\langle \Psi| C_{A B} |\Psi \rangle^2 + \langle \Psi| C_{A C} |\Psi \rangle^2 \leq 8.
\ee

Equation \eqref{monogamyrel} is a rather remarkable inequality. Note if A and B are entangled, this means that $|\langle C_{AB} \rangle| > 2.$ But, if so then $\langle C_{AB} \rangle ^2 > 4$ and the inequality above tells us that $\langle C_{AC} \rangle^2 < 4$ or $|\langle C_{AC} \rangle| < 2$. Therefore if A and B are entangled, A and C cannot be entangled. So entanglement, as measured by CHSH correlators, is ``monogamous'': A can either be entangled with B or with C but not with both.

\subsubsection{Average entanglement between subsystems \label{secavent}}
Next, we turn to a slightly different problem: the average entanglement between two subsystems. We will present a simplified version of the results of \cite{Page:1993df,lubkin1978entropy}. Say that we have a large system that comprises two subsystems --- one of size $e^{S'}$ and the other of size $e^{S}$. We will work in the limit $e^{S'} \ll e^{S}$ or equivalently that $S - S' \gg 1$.  Say that
we put the combined system in a random state in its Hilbert space. (Here the word ``random'' is used in the sense of the probability distribution introduced in section \eqref{puremixed}.)  We would like to show that given any set
of operators in the smaller subsystem, satisfying some simple properties, it is always possible to find operators in the larger subsystem so that Tsirelson's bound is saturated.

First, lets start by exploring the density matrix of the smaller system. Let us consider a pure state in the larger system of the form
\be
|\Psi \rangle = \sum_{m=1}^{e^{S'}} \sum_{n=1}^{e^{S}} a_{m n} |m, n \rangle.
\ee
Here we have written this pure state as a linear combination of basis elements, $|m, n \rangle$ that are direct products of basis elements from the two systems. This leads to the different upper-bound in the sums over $m$ and $n$.

If the density matrix of the smaller system is denoted by $\rho$, its matrix elements are
\be
\rho_{m m'} = \sum_{n=1}^{e^{S}} a_{m n} a_{m',n}^*.
\ee
Consider the expectation value of the density matrix for a random state, using the measure introduced in section \ref{puremixed}
\be
d \mu_{\Psi} = {1 \over V} \delta(\sum_{m,n} |a_{m,n}|^2 - 1)\prod_{m,n}  d^2 a_{m,n}.
\ee
When the normalization factor is chosen to ensure  $\int d \mu_{\Psi} = 1$,  it is clear that
\be
\label{expectdensity}
\langle \rho_{m m'} \rangle = \int d \mu_{\Psi} \rho_{m m'} = {1 \over e^{S'}} \delta_{m m'}.
\ee
So the average density matrix in the smaller system is just proportional to the identity matrix. We can also compute the expected deviations of the density matrix from this average value by doing the integral
\be
\delta = {1 \over e^{S'}} \int d \mu_{\Psi} \sum_{m, m'} \left(\rho_{m m'} - {1 \over e^{S'}} \delta_{m m'} \right)^2 = {1 \over e^{S'}} \left[{e^{S} + e^{S'}  \over e^{S + S'} + 1} - {1 \over e^{S'}}\right].
\ee
We see that when $S,S' \gg 1$, we have  $\delta \approx {1 \over e^{S + S'}}$.
We can think of  $\sqrt{\delta}$ as  measuring the average deviation of each eigenvalue of the density matrix from $e^{-S'}$, and in the limit where $e^{S'} \ll e^{S}$  we see that this deviation is much smaller than the expected size of the eigenvalue itself. Therefore, for most of the volume of the Hilbert space, the density matrix of the smaller subsystem --- in the limit
where it is much smaller than its complement --- can reliably be taken to be the identity matrix.

The next result we show is that if the density matrix of a subsystem is proportional to the identity, it is very easy to saturate Tsirelson's bound.  Consider  two operators in the  subsystem that behave like ``pseudospin'' operators. This means that the operators $A_1, A_2$ share the following properties of the Pauli spin operators, $\sigma_z$ and $\sigma_x$: they satisfy  $A_1^2 = 1, A_2^2 = 1$ and $(A_1 + A_2)^2 = (A_1 - A_2)^2 = 2$.  It is always possible to find such operators by focusing on a two-dimensional subspace. Since the density matrix of the subsystem is proportional to the identity, there is  a choice of basis so that for every basis element of the subsystem, $|m \rangle$, we choose a basis element for the complementary subsystem, $|\widetilde{m} \rangle$, that allows us to write the state of the full system as
\be
|\Psi \rangle = {1 \over e^{S' \over 2}}\sum_{m=1}^{e^{S'}}|m , \widetilde{m} \rangle.
\ee
If the action of the two pseudospin operators on the first subsystem is specified by the matrix elements
\be
A_1 |m \rangle = \sum_q (A_1)_{m q} |q \rangle; \qquad A_2 |m \rangle = \sum_q (A_2)_{m q} |q \rangle,
\ee
then we consider those operators that act on the complementary subsystem as 
\be
\widetilde{A}_1 |\widetilde{m} \rangle = \sum_{q} (A_1)_{q m} |\widetilde{q} \rangle; \qquad \widetilde{A}_2 |\widetilde{m} \rangle = \sum_{q} (A_2)_{q m} |\widetilde{q} \rangle.
\ee
Since the matrix elements of these operators are just the transpose of the matrix elements of $A_1$ and $A_2$ it is clear that these operators also have operator-norm $1$. 
It is also clear that for the state above,
\be
A_1 | \Psi \rangle = \widetilde{A}_1 |\Psi \rangle; \qquad A_2 | \Psi \rangle = \widetilde{A}_2 |\Psi \rangle.
\ee
We now just set
\be
B_1 = {1 \over \sqrt{2}} (\widetilde{A}_1 + \widetilde{A}_2); \quad B_2 = {1 \over \sqrt{2}} (\widetilde{A}_1 - \widetilde{A}_2).
\ee
Therefore, using the property that the pseudospin operators square to the identity, we find
\be
\langle \Psi| A_1 (B_1 + B_2) |\Psi \rangle + \langle \Psi | A_2 (B_1 - B_2) |\Psi \rangle = 2 \sqrt{2}.
\ee
As promised, this means that if we can isolate any set of pseudospin operators acting on one subsystem --- and if the density matrix of this system is proportional to the identity --- then we can always find operators 
acting on the rest of the system so that Tsirelson's bound is saturated.

Finally, we mention that the statement about average entanglement can also be formulated in terms of the expected von Neumann entropy of one system. From the results above, we see that the expected value of the von Neumann entropy, when $S - S' \gg 1$, is just the logarithm of the dimension of the smaller subsystem.  Clearly an analogous result holds when $S' - S \gg 1$.  We can write this as
\be
-\langle \tr(\rho \log (\rho)) \rangle = \text{min}(S', S),
\ee
under the assumption that $|S' - S| \gg 1$. 

\begin{figure}[!ht]
\begin{center}
\includegraphics[width=\textwidth]{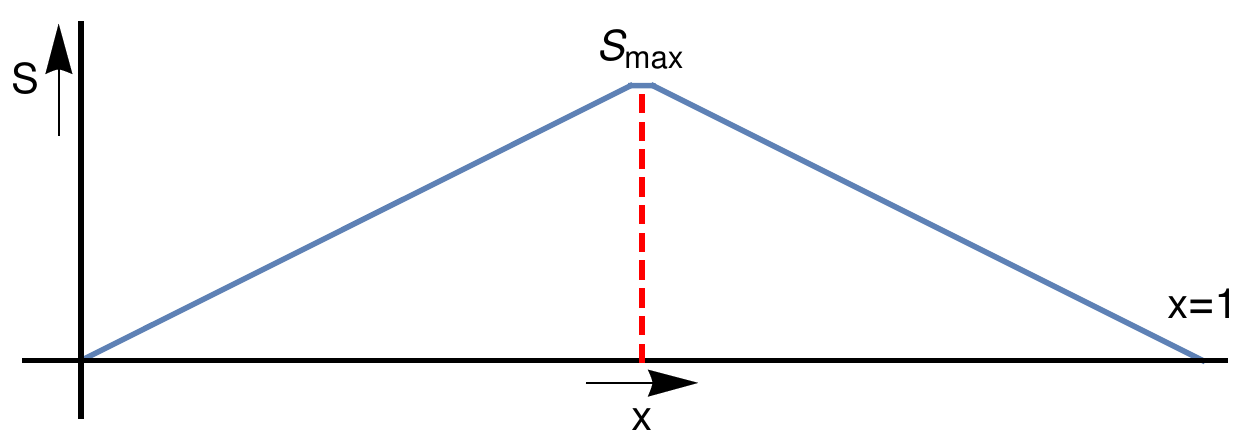}
\caption{\em If one divides an ordinary quantum mechanical system into two parts, where the ratio of the log of the dimension of one part to the log of the dimensions of the  entire system is $x$, then in a typical state, we expect that the entanglement entropy between the two parts will obey the Page curve shown above. \label{figpage}}
\end{center}
\end{figure}

If we plot this von Neumann entropy as a ratio of the size of one subsystem to the entire system we find that it obeys a characteristic curve, called a Page curve. This curve first rises, as the subsystem becomes larger as a fraction of the entire system. But when the size of the subsystem increases to more than half of the full system, then it is the complement of the subsystem that is smaller. And now the entropy is controlled by the size of the complement.  The Page curve is shown in Figure \ref{figpage}. Note that for a system in the thermodynamic limit, the condition $|S' - S| \gg 1$ is obeyed for almost all values of $S'$ and this leads to the sharp turnaround depicted in Figure \ref{figpage}.
\subsection{Entanglement across the horizon \label{entangacrosshor}}

The derivation of section \ref{hawkingderiv}, which led to the conclusion that a black hole emits particles
with a thermal spectrum, also tells us that the modes outside the horizon are entangled with ``mirror modes''
behind the horizon.  In this subsection, we will elucidate some of the properties of these modes behind the horizon, and use the results of the previous subsection to quantify the entanglement between degrees of freedom across the horizon.

Consider the slightly smeared modes, $a_{\omega, \ell}$ and $\ta_{\omega, \ell}$ that we discussed in section \ref{hawkingderiv}. From the analysis of section \ref{corrnull} we see that, on a black hole state $|\Psi \rangle$, these modes must satisfy
\be
\label{acrosshorcond}
\left(a_{\omega, \ell}  - e^{-{\beta \omega \over 2}} \ta_{\omega, \ell}^{\dagger}\right)|\Psi \rangle = 0, \qquad
\left( a_{\omega, \ell}^{\dagger} - e^{ {\beta \omega \over 2}} \ta_{\omega, \ell} \right) |\Psi \rangle = 0. 
\ee
Moreover, the modes on the two sides of the horizon must commute 
\be
[a_{\omega, \ell}, \ta_{\omega', \ell'}^{\dagger}] = [a_{\omega, \ell}, \ta_{\omega', \ell'}] = 0.
\ee
The correlators of the modes outside the horizon, when they are combined with \eqref{acrosshorcond} also imply 
correlators of modes across the horizon.
\be
\label{acrosshorcorr}
\langle \Psi | a_{\omega, \ell}  \ta_{\omega, \ell} |\Psi \rangle = {e^{-{\beta \omega \over 2}} \over 1 - e^{-\beta \omega}}.
\ee
These relations imply that the degrees of freedom probed by $a_{\omega, \ell}$ are {\em entangled} with the degrees of freedom probed by $\ta_{\omega, \ell}$.  To make this explicit, we now explain how one can define bounded composite operators that violate the Bell inequality.

First, consider the projector onto states that have a given value of the number operator, $\schwarzn_{\omega, \ell} = a_{\omega, \ell}^{\dagger} a_{\omega, \ell}$ that we already considered in equation \eqref{schwarzndef}. The projector onto eigenstates of this operator with eigenvalue $n$  be written down quite simply by exploiting the fact that the eigenvalues of this operator are integrally quantized.
\be
\label{projectdef}
|n \rangle \langle n | \equiv \int_0^{2 \pi} {d \theta \over 2 \pi} e^{i \theta (\schwarzn_{\omega, \ell} - n)}.
\ee
The bra-ket notation that we have chosen to denote the projector on the left hand side deliberately suggests that we are thinking of a small Hilbert space, which captures the simple harmonic degrees of freedom corresponding to this mode. But the more punctilious reader can simply consider the definition on the right hand side of \eqref{projectdef}, which is perfectly well defined in the full theory. Using this definition, we can also construct two additional operators,
\be
\label{raiselowerdef}
|n \rangle \langle n+1| \equiv  {1 \over \sqrt{n+1}} |n \rangle \langle n | a_{\omega, \ell}; \qquad 
|n + 1\rangle \langle n| \equiv  {1 \over \sqrt{n+1}} a_{\omega, \ell}^{\dagger} |n \rangle \langle n |.  
\ee

Similarly using the modes behind the horizon, and their number operator, $\widetilde{\schwarzn}_{\omega, \ell} = \ta_{\omega, \ell}^{\dagger} \ta_{\omega, \ell}$, one can construct the operators 
\[
|\widetilde{n} \rangle \langle \widetilde{n} |, \quad  |\widetilde{n} \rangle \langle \widetilde{n}+1|, \quad \text{and} \quad |\widetilde{n} + 1\rangle \langle \widetilde{n}|,
\]
which are defined in precise analogy to equations \eqref{projectdef} and \eqref{raiselowerdef}.

Now consider the following operators \cite{chen2002maximal}.
\be
\label{a1a2def}
\begin{split}
A_1 = \sum_{n=0}^{\infty} \left( |2 n + 1 \rangle \langle 2 n + 1| - |2 n \rangle \langle 2 n| \right), \\
A_2 = \sum_{n=0}^{\infty} \left( |2 n + 1 \rangle \langle 2 n| +  |2 n \rangle \langle 2 n + 1| \right),
\end{split}
\ee
which clearly satisfy $A_1^2 = A_2^2 = 1$.
There is an exactly analogous set of operators that one can construct using modes behind the horizon. 
\be
\begin{split}
\widetilde{A}_1 = \sum_{n=0}^{\infty} \left(|2 \widetilde{n} + 1 \rangle \langle 2 \widetilde{n} + 1| - |2 \widetilde{n} \rangle \langle 2 \widetilde{n}| \right), \\
\widetilde{A}_2 = \sum_{n=0}^{\infty} \left(|2 \widetilde{n} + 1 \rangle \langle 2 \widetilde{n}| +  |2 \widetilde{n} \rangle \langle 2 \widetilde{n} + 1| \right).
\end{split}
\ee

In a black hole state that satisfies the relations \eqref{acrosshorcond}, we can compute the correlators of these operators with some simple manipulations. We find that
\be
\langle \Psi | A_1 \widetilde{A}_1 | \Psi \rangle = 1; \qquad \langle \Psi | A_2 \widetilde{A}_2 | \Psi \rangle = {2 e^{-{\beta \omega \over 2}} \over 1 + e^{-\beta \omega}}.
\ee

We are almost done. We just need to recast the operators above in the CHSH form to see the nature of entanglement across the horizon. So we first define
\be
B_1 = \cos \theta \widetilde{A}_1 + \sin \theta \widetilde{A}_2  \quad \text{and}  \quad B_2 = \cos \theta \widetilde{A}_1 - \sin \theta \widetilde{A}_2,
\ee
and the CHSH operator $C_{AB}$ precisely as in \eqref{cabdefn}. Then using the correlators above, we find that
\be
\langle \Psi | C_{AB} |\Psi \rangle =  2 \cos \theta + {4 e^{-{\beta \omega \over 2}} \over 1 + e^{-{\beta \omega}}}\sin \theta .
\ee
This is maximized when $\tan \theta  = 2 {e^{-{\beta \omega \over 2}} \over 1 + e^{-{\beta \omega}}}$. The maximum value of the correlator, for this value of $\theta$,  is
\be
\label{maxentacross}
\langle \Psi | C_{AB} |\Psi  \rangle = {2 \over 1 + e^{-\beta \omega}} \left(1 + 6 e^{-{\beta \omega}} + e^{-2{\beta \omega}} \right)^{1 \over 2},
\ee
which is always larger than 2.

So we see that the modes on opposite sides of the horizon are entangled, in the sense that two point correlators measured across the horizon violate Bell inequalities. The degree of entanglement depends on the frequency, but the entanglement is present at all frequencies. For low frequencies, $\beta \omega \ll 1$, we see that Tsirelson's bound is almost saturated.

It is possible that the reader may have the following question.
\begin{enumerate}[qseries]
\item
{\em In the presence of interactions, the relations \eqref{acrosshorcond} may receive corrections. How does one know that the result for the CHSH correlator \eqref{maxentacross} is robust under such corrections?}\\
It was, in fact, to anticipate this issue that we constructed the projection operator \eqref{projectdef} and the transition operators \eqref{raiselowerdef} carefully within effective field theory. We did not simply assume that the larger Hilbert space contained a factor which supported the modes on the two sides of the horizon. As a result all the correlators that appear above can be written as correlators of the propagating fields on the black hole background. Moreover, these are all correlators of simple operators, which means that they can be well-approximated by correlators of an $\Or[1]$ number of field insertions.

In the presence of interactions, we expect that these correlators will all receive small corrections. This is true
even of the operators that appear in \eqref{a1a2def}. These appear to involve an infinite sum, but the higher order terms in this sum contribute a vanishingly small amount to the final expectation value of the CHSH operator. So this sum can be truncated to a finite order and computed reliably in perturbation theory.

Since the final violation of the Bell inequality that is displayed in \eqref{maxentacross} is an $\Or[1]$ violation, we expect that it will be robust, provided bulk interactions are weak. 
\end{enumerate}

\subsection{A monogamy paradox \label{secmonogamy}}

We now describe a paradox with the monogamy of entanglement that arises during black hole evaporation.

\begin{figure}[!ht]
\begin{center}
\includegraphics[height=0.4\textheight]{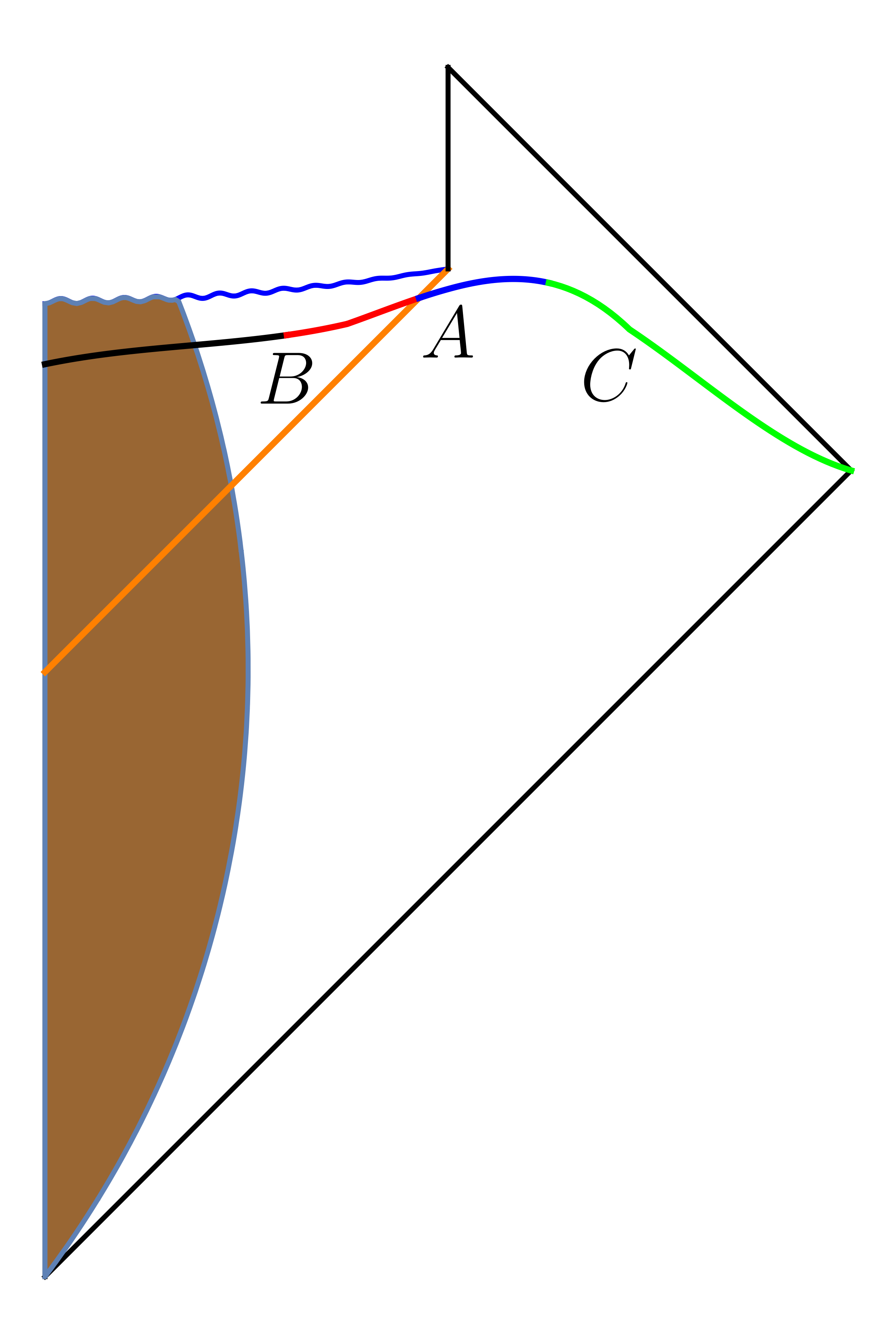}
\caption{\em A nice slice that runs through the geometry of an evaporating black hole, on which we have marked the regions A, B, C. \label{penroseabc}}
\end{center}
\end{figure}

Consider an evaporating black hole spacetime. As Figure \ref{penroseabc} shows, it is possible to draw a Cauchy slice
through the geometry that runs through the black-hole interior while the interior is still expected to be semiclassical, and {\em also} intersects a significant fraction of the Hawking radiation. On this slice, we can mark out three regions.
\begin{enumerate}
\item
Region A: A  region just outside the horizon of the black hole.
\item
Region B: A  region just inside the horizon of the black hole.
\item
Region C: A region near the boundary of the slice.
\end{enumerate}

To establish a paradox \cite{Mathur:2009hf} and \cite{Almheiri:2012rt} made the following assumption.
\begin{wrongassumption}{(Required for the monogamy paradox.)}
The Hilbert space of the theory
factorizes so that it is possible to associate distinct subfactors, $H_{A}, H_{B}, H_{C}$ with each of the three regions above. If $\hilbfull$ is the Hilbert space of the full theory then $H_{A} \otimes H_{B} \otimes H_{C} \subset \hilbfull$.
\end{wrongassumption}
In a LQFT, this assumption seems so obvious  that  \cite{Mathur:2009hf} and \cite{Almheiri:2012rt} did not even bother to make it explicit, although it was a tacit part of the argument. However, we will explain at the end of this section that, in a theory of gravity, this assumption fails.

We need some additional assumptions before we can set up a paradox using the results of section \ref{secavent}. Strictly speaking the results of that section are applicable to finite dimensional Hilbert spaces. Even in a LQFT, the Hilbert spaces, $H_{A}, H_{B}, H_{C}$ are infinite dimensional. However, the assumption is that, for all practical purposes, it should be possible to describe the evaporation of a black hole that starts with entropy $S$ in a Hilbert space whose dimension scales with $e^{S}$. Moreover, the assumption is that it should be possible to describe the physics in regions A and B using a subspace of this Hilbert space whose entropy is $\Or[1]$. For an old black hole, whose entropy has reduced from $S$ to $S'$, the assumption is that it should be possible to describe the physics in region C using a subspace whose dimension scales with $e^{S - S'}$. These assumptions arise
by thinking of black hole evaporation in terms of qubit-like models, where the evaporation process is modelled by
a qubit moving from the black hole region to the region C. These assumptions are somewhat plausible at first sight, and seem to be physically motivated. However, if the reader finds the reasoning somewhat sloppy this is because it is sloppy and has never been made more precise. 

In any case, proceeding with all these assumptions we now return to the operators that probe region A defined in \eqref{a1a2def}.  It is simple to see that these operators satisfy the criterion for ``pseudospin'' operators that we described in section \ref{secavent}. Namely, we have 
\be
A_1^2 = A_2^2 = {1 \over 2} (A_1 + A_2)^2 = {1 \over 2} (A_1 - A_2)^2 = 1.
\ee
Now these operators can be thought of as acting on the {\em complement} of $H_{C}$, which can denote by $\overline{H}_C$.  This complement includes $H_{A}, H_{B}$ and also the other degrees of freedom in the interior. Now consider an  old black hole, whose entropy has decreased to $S' < {S \over 2}$. Since black hole entropies are so large, we see that the moment the ratio ${S' \over S}$ drops even a little below ${1 \over 2}$,   the dimension of $H_{C}$ becomes much larger than the dimension of its complement. Therefore,
\be
\text{dim}(\overline{H}_C) \ll \text{dim}(H_C), \qquad \text{for~an~old~black~hole}.
\ee

We can now apply the results of section \ref{secavent}, which tell us that in a generic black hole state, $|\Psi \rangle$, we can find some operators $C_1, C_2$ that act entirely within $H_C$ 
so that if we consider the CHSH operator $C_{AC} = A_1 (C_1 + C_2) + A_2(C_1 - C_2)$, then 
\be
\langle \Psi | C_{AC} | \Psi \rangle = 2 \sqrt{2}.
\ee
Therefore, for an old black hole, degrees of freedom in region A are entangled with degrees of freedom in region C.

We showed in section \ref{entangacrosshor} that the same degrees of freedom in A were also entangled with partner modes in B. This is reflected by the result for $\langle \Psi | C_{AB} | \Psi \rangle$ in  \eqref{maxentacross}. But this leads to a contradiction with the result for the monogamy of entanglement since we have
\be
\langle \Psi | C_{AB} | \Psi \rangle^2 + \langle \Psi | C_{AC} | \Psi \rangle^2 = 8 + {4 \over (1 + e^{-\beta \omega})^2} \left(1 + 6 e^{-{\beta \omega}} + e^{-2{\beta \omega}} \right) > 8.
\ee
The surprising aspect of this contradiction, which was stressed in \cite{Mathur:2009hf}, is that it is an $\Or[1]$ contradiction and so it cannot be fixed by making small changes to the correlators that appear above.

\subsection{The firewall and fuzzball proposals}

The fuzzball resolution to the monogamy paradox can be understood if one asks a simple question: why does a similar paradox not appear if we consider a gas of black-body radiation that is slowly released from a box?  Even for this gas, we expect that the late radiation is entangled with the early radiation. But in this case, there are no ``mirror modes'' for the late radiation to be entangled with. So there is no question of a monogamy paradox for the gas. The original point of the  monogamy paradox, as it  was formulated in \cite{Mathur:2009hf}, was to argue that, in analogy with the gas, the black hole geometry also cannot support mirror modes behind the horizon that always remain correctly entangled with the
radiation emitted from the black hole.

In terms of the correlators above, the fuzzball proposal \cite{Skenderis:2008qn,Mathur:2005zp} suggests that we should drop the idea that $C_{AB} \geq 2$. But if the reader carefully traces our derivation above, this feature emerges simply from assuming
that the black hole geometry, at late times, becomes approximately time-translationally invariant and the horizon remains empty and {\em locally} resembles Minkowski space. Therefore, if one drops the idea that $C_{AB} \geq 2$,
one has to give up the idea that a black hole has an interior. The fuzzball proposal suggests that perhaps this is achieved because the spacetime ends  before the horizon is reached because of some topological effect,
such as an extra dimension that shrinks to zero size. The idea of ``fuzzball complementarity'' \cite{Mathur:2012jk,Avery:2012tf}  suggests that nevertheless, for an infalling observer, the geometry might look like the conventional black hole geometry.

The firewall proposal, on the other hand, was based on the idea that one does not expect any deviations from the conventional black hole geometry outside the horizon. If so, and if it is also the case that there is no entanglement between
the radiation outside the horizon and mirror modes behind the horizon, then an infalling observer will encounter a violent discontinuity at the horizon: ``fire'' rather than ``fuzz''. This is what led to the use of the term ``firewalls'' by AMPS \cite{Almheiri:2012rt}.

There are difficulties with both the firewall and fuzzball proposals. We discuss these at greater length in section \ref{secisstructure}. But, in a nutshell, a  firewall behind the horizon appears to violate effective field theory
and it is unclear what stabilizes it and prevents it from falling into the singularity.  The fuzzball proposal, on the other hand, which suggests that space should end before the horizon is reached runs into difficulties with simple but far-reaching results from statistical mechanics and Euclidean  gravity.

\subsection{A resolution to the paradox \label{resolvemonogamy}}

We will now point out a loophole in the logic that led to the monogamy paradox. After that, we will briefly outline a resolution that will be developed at length in the next two sections.

\paragraph{\bf Failure of factorization.}
We first  explain why the assumption made to set up the monogamy paradox --- namely that the  Hilbert space on a nice slice factorizes so that there are distinct factors $H_{A}, H_{B}, H_{C}$ --- is demonstrably wrong in quantum gravity.

The reason is that when gravity is treated quantum mechanically, it is impossible to write down an operator that modifies the state of the system  in the region A or the region B without modifying the state of region C. (Here we are referring to the regions as marked on Figure \ref{penroseabc}.)   This is a consequence of the fact that there are no local gauge-invariant operators in gravity \cite{dewitt1960quantization, kuchar1991problem, Giddings:2005id}. This may sound like an abstruse theorem, but it is simple to understand intuitively.  If one attempts to write down an operator that is strictly localized in region A or region B  then, since these regions have finite extent, the Heisenberg uncertainty principle tells us that this operator must be a superposition of an infinite number of operators of different energies. 

But such an operator cannot commute with the metric in region C, since this metric has information about the energy in the region that C encompasses, which includes A and B. For some simple operators, such as a smeared gauge-fixed quantum field operator, this commutator can be computed explicitly \cite{Donnelly:2018nbv}. But the argument for the non-zero commutator is quite general. 

One aspect of this argument that is not always appreciated is that there might be exactly degenerate energy eigenstates in the theory, or even states that are related by a global symmetry. For instance if one has two identical fields, $\phi_1$ and $\phi_2$, it is possible to consider two states such that one state has an excitation of $\phi_1$ in region B (in the interior) while another state has an excitation of $\phi_2$.  But this argument tells us that the operator that causes transitions between those eigenstates cannot be localized only in the black-hole interior, and so it must fail to commute with some other operator in the exterior. In this sense, gravity is very different from a gauge theory where an operator that induces a transition between two excitations of the same charge can be localized to a region.  The reader can easily verify this difference between gauge theories and gravity by attempting to write down the two states in effective field theory and also the operator that creates a transition between them.

There is another way to think about the same issue. Since the geometry itself fluctuates when gravity is quantized, we need some method of defining the location of quantum field operators. One common way of doing this is by means of a relational prescription where one specifies the location of an operator by means of geodesics that stretch to the asymptotic boundary. (See, for instance, section 3.1.1 \cite{Papadodimas:2015jra}.) These geodesics are the analogs of Wilson lines that are used to dress charged fields in gauge theories.  The operators obtained in this manner mimic local operators to a good approximation. So it makes sense to discuss their two-point correlators and the notion of entanglement across the horizon. But precisely because, in gravity, these operators must be dressed to infinity, they are not strictly local. This departure from exact locality is already visible within perturbation theory.

The monogamy paradox of \cite{Mathur:2009hf,Almheiri:2012rt} was predicated on the idea that one
should somehow be able to ignore these effects. Indeed, both in classical physics and in local quantum field
theory  --- where our intuitions are usually trained --- these effects do not appear. However, if one
starts asking fine-grained quantum-information questions about black-hole evaporation in a theory of gravity, the failure of factorization becomes important. This is therefore a loophole in the logic leading up the monogamy paradox. (Some recent discussions, from a differing perspective, can be found in \cite{Karlsson:2020uga,Karlsson:2019vlf}.)

\paragraph{Holography of Information.}
An examination of the argument above already informs us that the failure of the Hilbert space to factorize is 
 much more serious in gravity than it is in gauge theories. Even in a gauge theory, the algebra of operators
associated with a region has a nontrivial center \cite{Casini:2013rba,Casini:2014aia,Soni:2015yga,Ghosh:2015iwa}, and so there is some ambiguity in how to divide up the set of local operators between a region and its complement. But, in a gauge theory, there are some gauge invariant operators that are  unambiguously localized in a region like, for instance, a small Wilson loop operator that is completely confined
to one region. In gravity, on the other hand, if one considers a region that does not extend to infinity,
then the argument above suggests that it is impossible to write down even a {\em single} operator that is
supported on that region and commutes with all operators outside the region.

We will expand on this point in the next two sections. We will find that, in a theory of quantum gravity, the assumption of factorization fails as completely as possible: the degrees of freedom in the region C actually {\em contain} the degrees of freedom in both regions A and B. This is part of a more general characteristic of theories of quantum gravity that we will term the principle of holography of information. This immediately resolves the paradox above. The monogamy inequality \eqref{monogamyrel} holds only if the three pairs of operators from A,B,C commute with each other. However, the principle above tells us that the operators in C that are entangled with A do {\em not} commute with the operators in B and, in fact, operators in B can be equated with appropriate operators in C.  So it is not surprising that the monogamy inequality is violated.

We would like to emphasize a very important point. It is sometimes suggested in the literature \cite{Mathur:2017fnw} that the principle of holography of information is an assumption about ``new physics'', which is invented to save the conventional picture of the black hole. This conveys the impression that one has to make some sort of an aesthetic choice between giving up ``localized quantum information'' or giving up the conventional picture of the black-hole interior, and one can make either choice depending on one's predilections.  However, this is not correct.  As we will demonstrate below, the principle of holography of information is a general feature of theories of gravity that  follows from a careful analysis of low-energy physics. The only assumptions
that we need to make are that the full theory of quantum gravity shares some properties of the low-energy theory. This principle is also perfectly consistent with effective field theory, and moreover when we take $\gnewt \rightarrow 0$ we recover the usual picture of localized information that prevails in LQFTs.

As far as we know, this resolution to the monogamy paradox was first clearly stated in the literature in \cite{Papadodimas:2012aq}. (See page 43.) Similar, but somewhat distinct, ideas were considered earlier in \cite{Bousso:2012as} and \cite{Susskind:2012uw} but eventually rejected in favour of firewalls. Similar proposals were also explored in  \cite{Nomura:2013gna,Nomura:2012sw,Verlinde:2012cy}. Subsequent to these developments, these ideas were explored through the ER=EPR proposal and  \cite{Maldacena:2013xja} proposed a similar resolution of the monogamy paradox.  The idea that the black-hole interior can also be described by degrees of freedom far away in the exterior has since been verified in many different settings, including in recent work on quantum extremal islands as we will discuss in section \ref{secresolveinfo}.  These ideas are, of course, related to older ideas about black hole complementarity \cite{'tHooft:1984re,Susskind:1993mu,Susskind:1993if} and they were presented in \cite{Papadodimas:2012aq} in that context. In the next section we will present, what we feel, is the clearest and most direct argument for this effect that also shows how this effect is part of a more general feature of theories of quantum gravity and not just something that appears in the presence of black holes.

\section{Holography of information \label{secholography}}
In previous sections, we have alluded to the idea that operators far
away from the black hole could describe both the interior and also the near-horizon region. 
In this section,  we would like to take a step back and explain 
how this is part of a broader principle that should hold in any 
theory of quantum gravity. As we have  already mentioned, we call this the principle of {\em holography of information}, which can be articulated as follows.
\begin{lesson}
\label{lessonholography}
In a theory of quantum gravity, a copy of all the information available on a Cauchy slice is also available near the boundary of the Cauchy slice. This redundancy in description is already visible in the low-energy theory.
\end{lesson}

Before we embark on a technical analysis, we would like to present some perspective. From one point of view, the claim made in conclusion \ref{lessonholography} is  a surprising statement since it is in conflict with the intuition that one builds up by studying LQFTs or classical gravity. In a LQFT, apart from the Fock-space description of the Hilbert space, it is also possible to describe states precisely using the Schrodinger description, where the eigenkets of the field themselves are taken to be basis states. (See, for instance, chapter 10 of \cite{hatfield1991quantum}.) In this description it becomes clear that we can specify the state of different parts of a Cauchy slice entirely independently in a LQFT. This is sometimes called the ``split property'' \cite{Haag:1992hx} in the formal QFT literature. 

Even in classical gravity, it is clearly possible to specify the metric in one part of spacetime independently from the metric in another part of spacetime. The most familiar example of this comes from spherically symmetric configurations.  By Birkhoff's theorem, the metric outside a region is only sensitive to the {\em total mass} inside the region. The Corvino-Schoen theorem \cite{Corvino:2003sp} provides a generalization for the case without spherical symmetry. So the claim is that, in contrast to classical gravity and LQFTs, in quantum gravity, once one specifies the state near the boundary there is a unique way to ``fill it in.'' 

We sometimes find that, since our intuition about locality is commonly built on a study of  LQFTs and on classical gravity, the principle of holography of information elicits some discomfort at first sight.
But the good news is that the argument that follows is extremely simple, even from a technical perspective. So we hope that even the reader who is, at first, uncomfortable with conclusion \ref{lessonholography} will be able to check the correctness of this argument easily.

The principle of holography of information is related to previous formulations of the holographic principle \cite{Susskind:1994vu,Bousso:2002ju}. But what we would like to emphasize here is that it can be deduced, rather precisely, from within the low-energy theory of gravity itself.

We start by analyzing gravity in global AdS with normalizable boundary conditions.  Global AdS is qualitatively different from flat space only because it has a natural IR-cutoff, but this example already captures the essential physics. We will then show how, in this setting, the principle of the holography of information can be verified explicitly within low-energy effective field theory. We then explain how the result generalizes to asymptotically flat four-dimensional spacetimes. In this latter setting, we provide yet another procedure for
verifying this principle within low-energy perturbation theory.

\subsection{Holography of information in asymptotically anti-de Sitter spaces \label{holinfads}}
The argument in this subsection closely follows Appendix A of \cite{Laddha:2020kvp}. We will focus on the main physical ideas here and direct the reader to the original source for a more rigorous treatment.

\begin{figure}[!ht]
\begin{center}
\includegraphics[height=0.4\textheight]{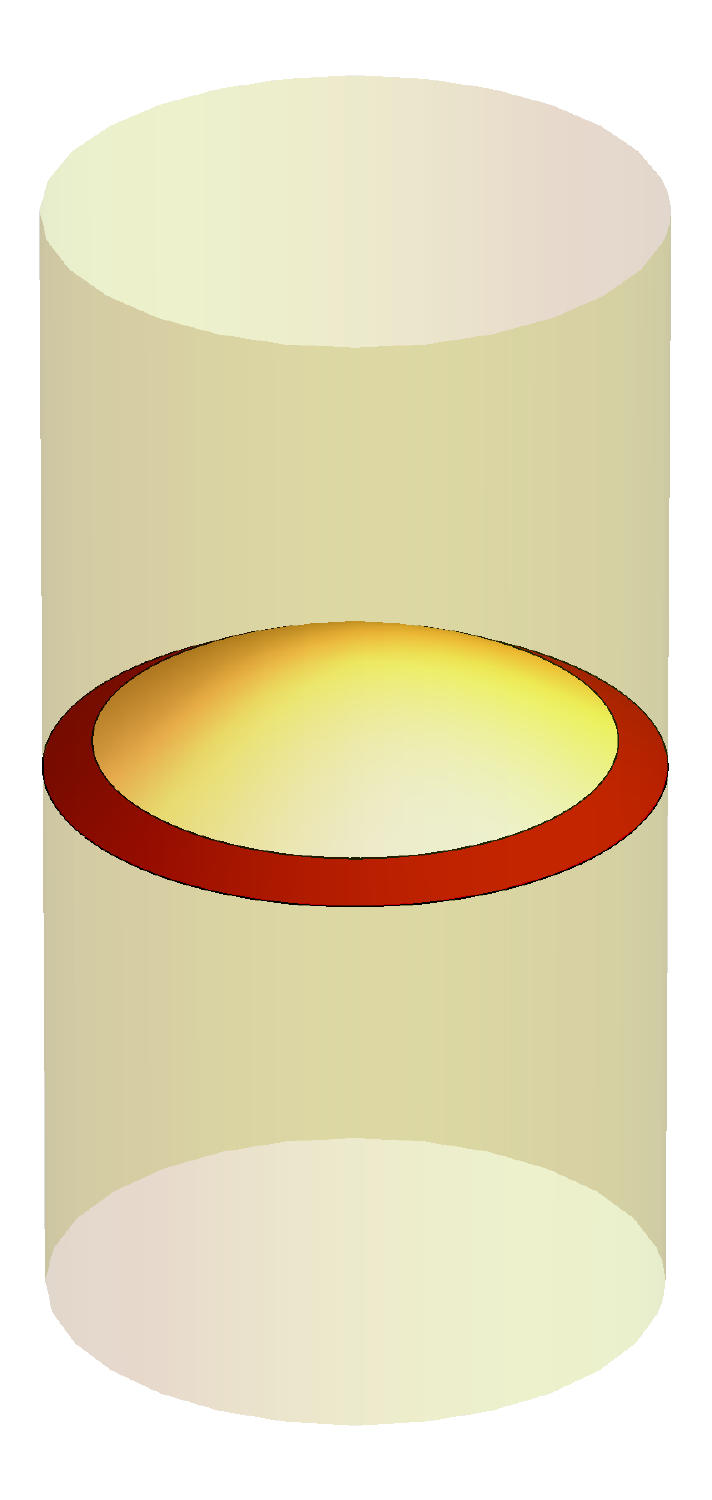}
\caption{\em  We would like to show that all information is available near the boundary (shaded) of a Cauchy slice. \label{figobservers}}
\end{center}
\end{figure}
Figure \ref{figobservers} shows a picture of global AdS with a Cauchy slice running through it. Let us choose coordinates so that asymptotic boundary is at $r \rightarrow \infty$, the asymptotic time is $t$ and the coordinates on the (d-1)-dimensional boundary sphere at a constant time are $\Omega$. This means that asymptotically the line element tends to
\be
ds^2 \underset{r \rightarrow \infty}{\longrightarrow} -(r^2 + 1) dt^2 + {d r^2 \over r^2 + 1} + r^2 d \Omega_{d-1}^2.
\ee

\begin{figure}[!ht]
\begin{center}
\includegraphics[height=0.4\textheight]{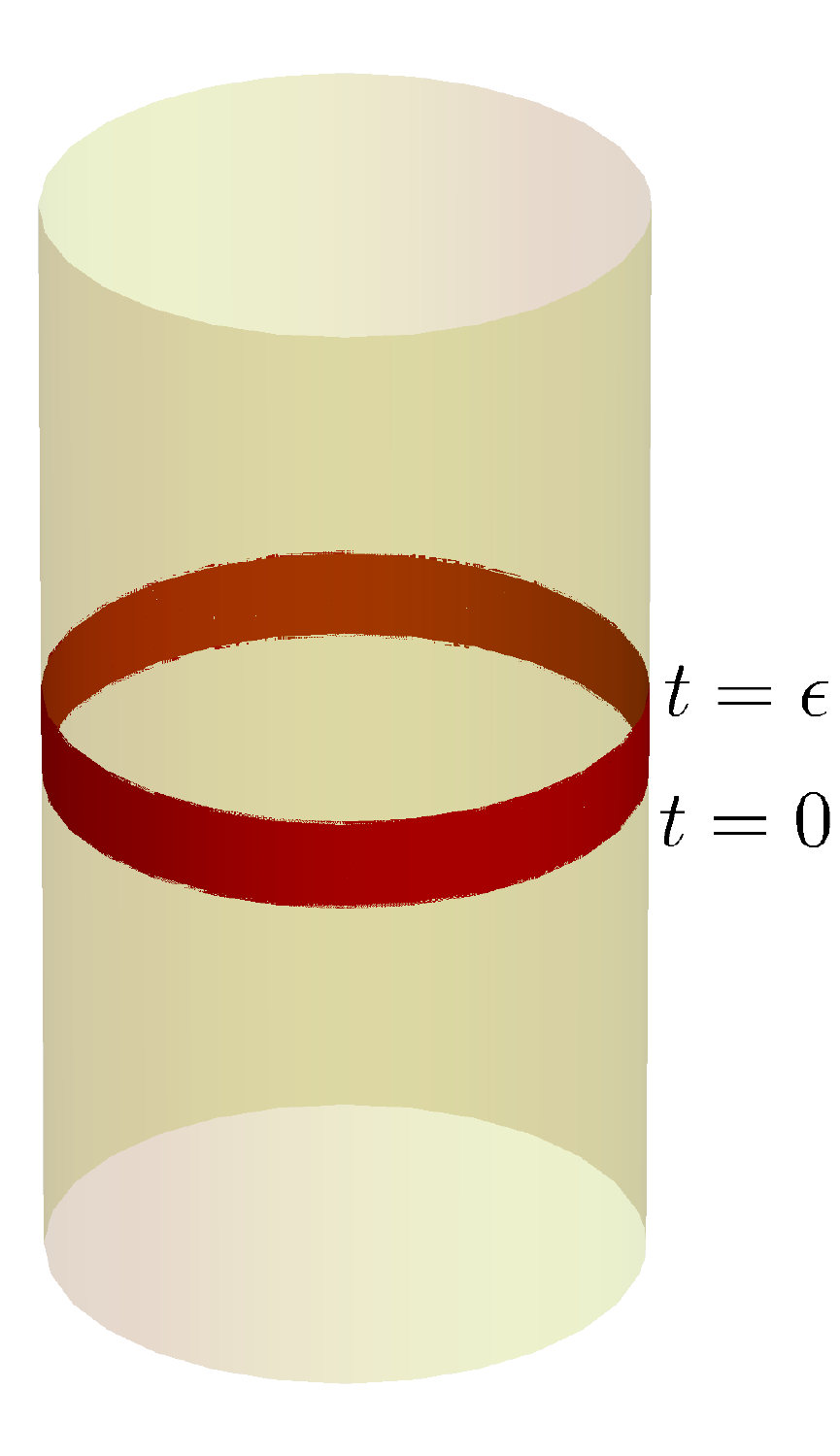}
\caption{\em  It is more precise to frame the principle of holography of information in terms of a time band on the boundary. \label{figband}}
\end{center}
\end{figure}
Physically, we would like to show that all the information on the Cauchy slice is available near its boundary, which is shaded in brown in Figure \ref{figobservers}. However, it is difficult to make this precise in a theory with a fluctuating metric since one has to worry about how to define this brown region in a gauge-invariant manner, and how its boundaries fluctuate in the presence of a large perturbation in the interior of the slice. So it is more convenient to think of operators that are pushed all the way to the asymptotic boundary, at 
$r \rightarrow \infty$. At the boundary, we take operators from the algebra of a small time band $[0, \epsilon]$ as shown in Figure \ref{figband}.   In a LQFT, the operators in this time band would correspond to bulk operators outside the ball with radius $\cot{{\epsilon \over 2}}$ at time  $t = {\epsilon \over 2}$. But, in gravity, we will show that observables from this algebra have information about the entire bulk.

We take the bulk theory to be a theory of gravity coupled to matter.  The matter sector could, for instance, correspond to a string theory. It is convenient to impose ``normalizable'' boundary conditions, which simply set all fluctuations to vanish near the asymptotic boundary. Then there is a well-defined procedure of taking the asymptotic limit of bulk operators, by taking the operator towards the boundary while simultaneously scaling it up with an appropriate power of $r$. 
For instance, say that we have a scalar field of mass $m$ described by the quantum field operator $\phi(r,t, \Omega)$. Then we obtain an asymptotic operator by taking the limit 
\be
O(t, \Omega) = \lim_{r \rightarrow \infty} r^{\Delta} \phi(r, t, \Omega)
\ee
where $\Delta = {d \over 2} + \left({d^2 \over 4} + m^2 \right)^{1 \over 2}$. This limit is well defined purely in the bulk theory.

\begin{figure}[!ht]
\begin{center}
\includegraphics[width=0.3\textwidth]{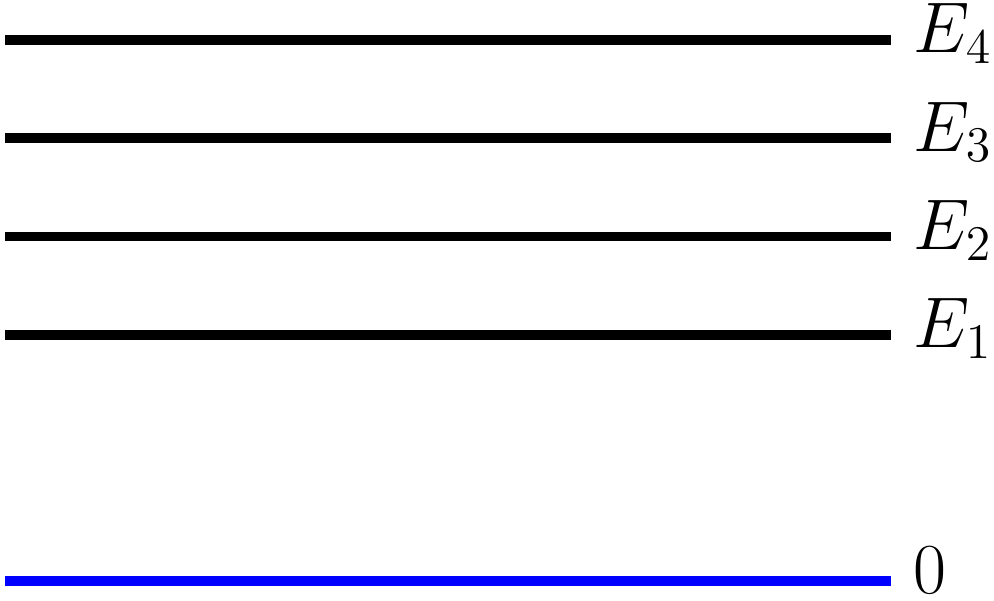}
\caption{\em A cartoon of the spectrum of states in global AdS. There is a unique vacuum, followed by a discrete spectrum.  This is just because the AdS scale provides an IR cutoff. The energy levels could be degenerate and there is, of course, an infinite tower of higher-energy states that we have not displayed.\label{figadsspect}} 
\end{center}
\end{figure}

The low-energy spectrum of a theory in AdS can be calculated reliably just by quantizing fields about global AdS. Since global AdS is like a box, the low-energy spectrum of the theory has a unique vacuum state, followed by gapped excitations. This is shown schematically in Figure \ref{figadsspect}. The reader who is unfamiliar with this fact should consult Appendix \ref{appadspert} for a primer on perturbation theory in global AdS.

Now, we turn to a physically crucial point. The energy of a state, in gravity, can be measured from near the boundary,  using the Gauss law. This is also true in the quantum theory \cite{Arnowitt:1962hi, Regge:1974zd} where the Gauss law is implemented by the Hamiltonian constraint \cite{DeWitt:1967yk}.  In AdS, just as in flat space,  we simply need to integrate a subleading term in the metric (in the appropriate gauge) to get the energy of the state. Let us call this integrated metric operator, $H$. (An explicit formula is given in Appendix \ref{appadspert}.)

Even if the UV-theory of quantum gravity in AdS is described in terms of some variables that look different from quantum fields, this operator, $H$ will continue to have a meaning provided that we can assign a meaning to the asymptotic metric operator. A key assumption in the argument is that when the low-energy theory is UV-completed, the spectrum of $H$  remains bounded below and moreover the the lowest eigenstate of this operator continues to correspond to the vacuum. 

 This latter assumption physically only states that one cannot ``completely hide'' energy in a theory of quantum gravity: if the metric near the AdS boundary is undisturbed then the assumption allows us to conclude, even in the full UV theory, that the global state is the vacuum. Moreover, we would like to emphasize that we are {\em not} making the assumption that $H$ is the Hamiltonian of the full theory.  We believe that this is a very reasonable assumption.

The operator $H$ is manifestly an observable in the boundary algebra and so it has a spectral decomposition $H = \sum E P_{E}$, where the projectors $P_{E}$ project onto the subspace of states where $H$ has eigenvalue $E$ and the sum runs over the spectrum of eigenvalues of this operator. By the standard rules of quantum mechanics, if an observer near the boundary measures $H$, the probability of obtaining $0$ in the state $|g \rangle$ is given by the expectation value $\langle g | \projvac | g \rangle$, where $\projvac$ is the projector on the vacuum above. Since this probability can clearly be determined by making observations of $H$ near the boundary, and calculating the relative frequency with which the observations yield $0$, we see that $\projvac$ is also an observable in the boundary algebra.\footnote{The projector on the vacuum was first discussed in the context of obtaining information from the time band in \cite{Banerjee:2016mhh}. However, the authors of \cite{Banerjee:2016mhh} wrote it in terms of a high-order polynomial of  the energy, making it appear to be a very complicated operator. The description of \cite{Banerjee:2016mhh} is not incorrect, but we see there is an alternative, simple and physical, interpretation of this operator.}

So far we have discussed the vacuum. Let us now consider the rest of the Hilbert space. The key point is that the Hilbert space of the theory can be taken to be the space of states obtained by exciting the vacuum with asymptotic operators
at arbitrary points of time. If $O(t)$ is such an asymptotic operator, then an example of such a state is 
\be
\label{examplestate}
O(t_1) \ldots O(t_n) | 0 \rangle.
\ee
We would like to consider the Hilbert space formed by taking the span of all states of the form \eqref{examplestate}, obtained by applying polynomials of $O$ or other asymptotic operators to the vacuum, with arbitrary choices of the time-coordinates $t_i$. The low-energy states shown in Figure \ref{figadsspect} can explicitly be written in this form. But our assumption that {\em all} states are of this form is a good assumption because the Hilbert space formed in this manner is manifestly closed under time evolution: time evolution by an amount $\tau$ just changes all the time coordinates above by $t_i \rightarrow t_i + \tau$ and so leads to another state formed by the action of asymptotic operators on the vacuum. So it is clear that just the requirement of unitarity cannot forced us to include more states in the Hilbert space.

A simple result \cite{Banerjee:2016mhh} then tells us that any state in this Hilbert space can be generated by applying only operators from the time band $[0, \epsilon]$ to the vacuum.  This means that any state of the form \eqref{examplestate}, where the $t_i$ are arbitrary, can be rewritten as a linear combination of states of the form  $O(t_1) \ldots O(t_n) | 0 \rangle$, where $t_i$ are {\em restricted} to $[0, \epsilon]$. 
To prove this, let us start by considering those states obtained by the application of a single operator to the vacuum. If a state in this space is orthogonal to all states that can be produced by applying operators from the band $[0,\epsilon]$ on the vacuum, then there is a vector of finite norm, $|\Psi \rangle$ so that $\langle \Psi | O(t) | 0 \rangle = 0$ for all $t \in [0, \epsilon]$. But this matrix element is analytic when $t$ is continued in the upper-half plane since it can be evaluated by inserting a complete set of intermediate energy eigenstates, all of which have positive energy.
\be
\langle \Psi | O(t) | 0 \rangle = \sum_{E} \langle \Psi | E \rangle \langle E| O(0) | 0 \rangle e^{i E t}.
\ee
But then a standard result about analytic continuation tells us that if such a function vanishes for $t \in [0, \epsilon]$ then it must vanish on all the real line. But this is impossible since $|\Psi \rangle$ was itself produced by applying operators from somewhere on the real line to the vacuum. This argument can be extended to states produced by applying higher-order polynomials on the vacuum and using the edge-of-the-wedge theorem \cite{streater2016pct}.

What we have shown is that for every state $|n \rangle$ in the Hilbert space, there is an operator in the algebra of the time band, $X_n$ so that
\be
\label{rs}
|n \rangle = X_n |0 \rangle
\ee

The result above, in the form that we have stated may seem unfamiliar. So we would like to demystify the result by focusing on a simple qubit example, which explains what is going on.
Consider the EPR state, 
\be
|\text{EPR} \rangle = a |0 0 \rangle + b |1 1 \rangle
\ee
The full Hilbert space is 4-dimensional. This is the same as the dimension of the vector space of {\em operators} acting on the first qubit. In fact, by acting with all operators on the first qubit, starting with the state $|\text{EPR} \rangle$, one can reach any state in the Hilbert space. For instance, 
\be
\label{eprbasis}
\begin{split}
&(1 + \sigma^{z}) |\text{EPR} \rangle = 2 a |0 0 \rangle; \quad (1 - \sigma^{z}) |\text{EPR} \rangle = 2 b |1 1 \rangle; \\
&(\sigma^x + i \sigma^y) |\text{EPR} \rangle = 2 a |1 0 \rangle; \quad (\sigma^x - i \sigma^y)  |\text{EPR} \rangle = 2 b |0 0 \rangle; 
\end{split}
\ee
where the $\sigma$-matrices act {\em only} on the first qubit. The relations above show how can obtain a basis for the Hilbert space, but by taking other linear combinations of the $\sigma$-matrices, clearly one can obtain any vector in the Hilbert space. One can of course generate the same space by acting with operators from both qubits, but the operators on the second qubit are redundant.

There are two important physical points that are already evident above. 
\begin{enumerate}
\item
This construction, which just uses straightforward quantum mechanics, obviously does {\em not} imply any loss of locality. This is because, physically, an observer who has access to only the first qubit can only act with {\em unitary} excitations on the first qubit and cannot ``act'' with arbitrary operators. There is an unfortunate difficulty with language here. In equation \eqref{eprbasis}, one uses the verb ``act'' to represent the mathematical application of the $\sigma$-matrices to the EPR state. But this does not correspond to a physically allowed ``action.'' Of course, unitary operators 
on the first qubit do not change the density matrix of the second qubit at all. 
\item
In spite of Equation \eqref{eprbasis}, an observer with access to only the first qubit {\em cannot} determine the full state of the system. In particular, any measurement on the first qubit --- including expectation values of the operators shown in Equation \eqref{eprbasis} --- is insensitive to the action of a unitary operator on the {\em second} qubit. 
\end{enumerate}

The result \eqref{rs} is just the infinite dimensional analogue of the simple qubit example. In particular, both the physical caveats that we have mentioned apply to this result. First, this result would hold even in a LQFT and so does not imply any violation of locality. Second, an observer near the boundary {\em cannot} use the result \eqref{rs} by itself to obtain information about the bulk. In a LQFT, for instance, there would be exactly local operators that would commute with all observations near the boundary time band, in spite of \eqref{rs}.  These are analogous to unitaries acting on the second qubit in our qubit-example.
 
We have proved equation \eqref{rs} independently, but it can also be understood as a manifestation of the so-called ``Reeh-Schlieder'' theorem \cite{Haag:1992hx}. We try to eschew this terminology because it might appear intimidating and  because the context in which we are using the result, where we are allowed to use operators smeared all over the boundary, is rather trivial.

But now notice what happens when the equation \eqref{rs} is combined with the fact that the energy can be measured from the boundary. If we consider the algebra of operators near the boundary from the time band $[0, \epsilon]$, then since the operators that enter equation \eqref{rs} and the projector on the vacuum are both elements of this algebra, so is the operator
\be
\label{transnm}
|n \rangle \langle m| = X_n \projvac X_m^{\dagger}.
\ee
But since all operators in the Hilbert space can be written as linear combinations of the operators that appear in equation \eqref{transnm}, this means that {\em all} operators in the theory can be found in the algebra of operators from the boundary time band.

This is a remarkable and simple result 
that  directly implies that all ``information'' about the bulk is available in this algebra. By ``information'' about a quantum state, we mean the expectation value of all possible observables in the theory. Since we have shown that  any observable can be written as a linear combination of elements of this algebra, this means that all physical expectation values can be 
obtained through measurements of observables in this algebra. Another way to state this result is that ``any two distinct states in the theory can be distinguished by observations from the algebra of boundary operators in the time band $[0, \epsilon]$.''

Marolf earlier also argued \cite{Marolf:2008mf,Marolf:2006bk,Marolf:2013iba} that gravitational theories are holographic since the Hamiltonian is a boundary term. The  argument is that, as a consequence, the boundary algebra is closed under time evolution.  Therefore, information on the boundary is conserved. If one accepts that information in the bulk must always reach the boundary at some point of time --- whether in the future or the past --- then this argument implies that information about the bulk must be available near the boundary at all times. The argument that we have presented above is somewhat different since we only assumed that the boundary metric correctly identifies the vacuum, and combined that with a precise construction of the Hilbert space. 

\subsection{Low-energy physical tests of the holography of information \label{lowenergy}}
We now want to explain that the holography of information is not some abstract result about operator algebras. Rather, it is a very physical phenomenon that can be tested within perturbation theory using low-energy thought experiments. 

\begin{figure}[!ht]
\begin{center}
\includegraphics[height=0.5\textheight]{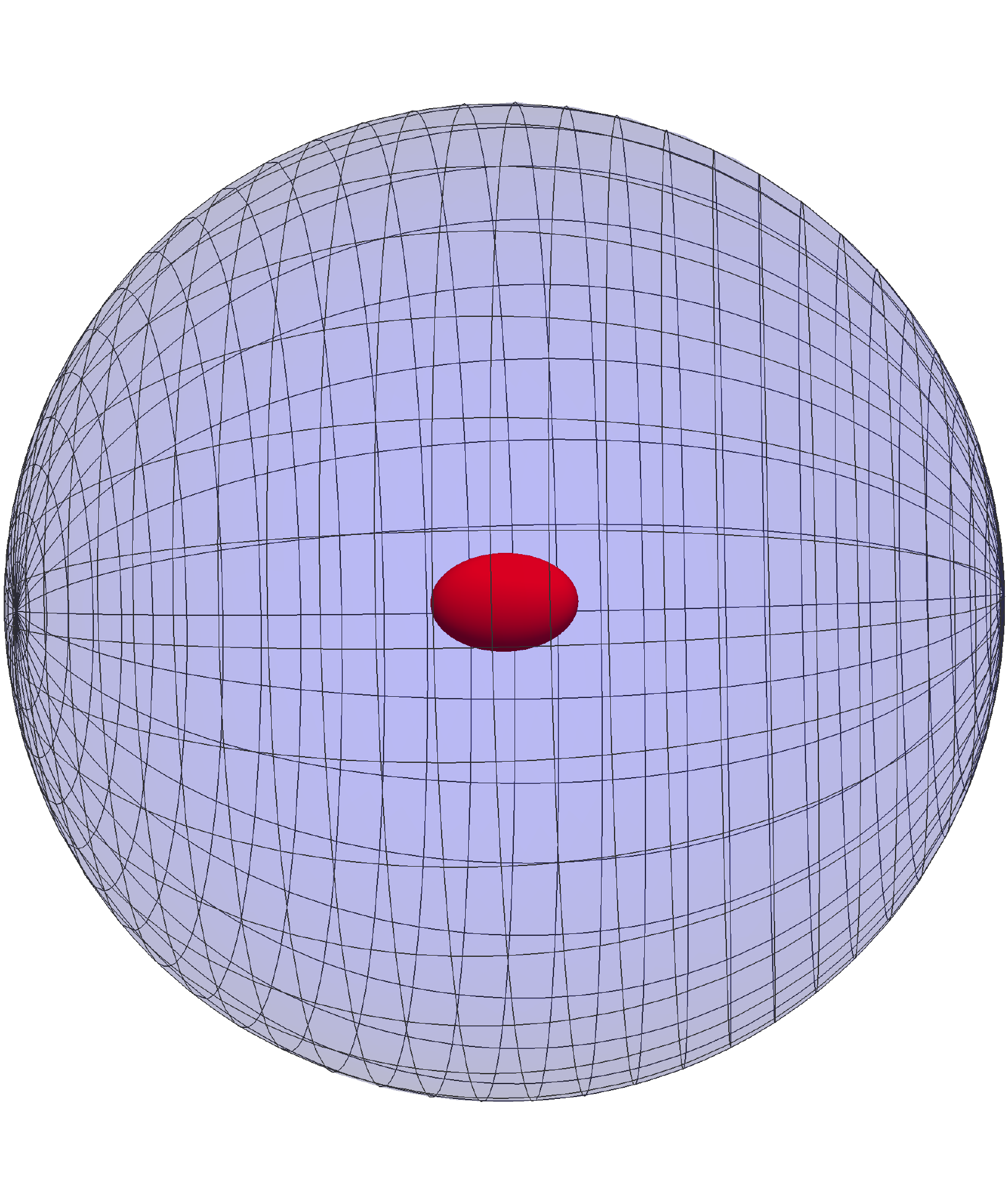}
\caption{\em The configuration that is relevant for the test of the holography of information. The task is to identify the excitation (red ellipsoid) in the middle of AdS by making observations on a sphere near infinity for a small time interval. \label{figdetectors}}
\end{center}
\end{figure}
Consider some observers who live near the boundary of global AdS. This setup is shown in Figure \ref{figdetectors}. The observers are put in a {\em low} energy state of the theory that we denote by $| g \rangle$  and they are  asked to determine what the state is. We allow the observers access to a number of identically prepared systems, and provide them with the ability to act with low-energy unitaries near the boundary, and make projective measurements of the energy near the boundary. In summary, this is the {\em textbook} framework for understanding quantum information questions.

\begin{figure}[!ht]
\centering
\begin{subfigure}{0.4\textwidth}
\centering
\includegraphics[height=0.3\textheight]{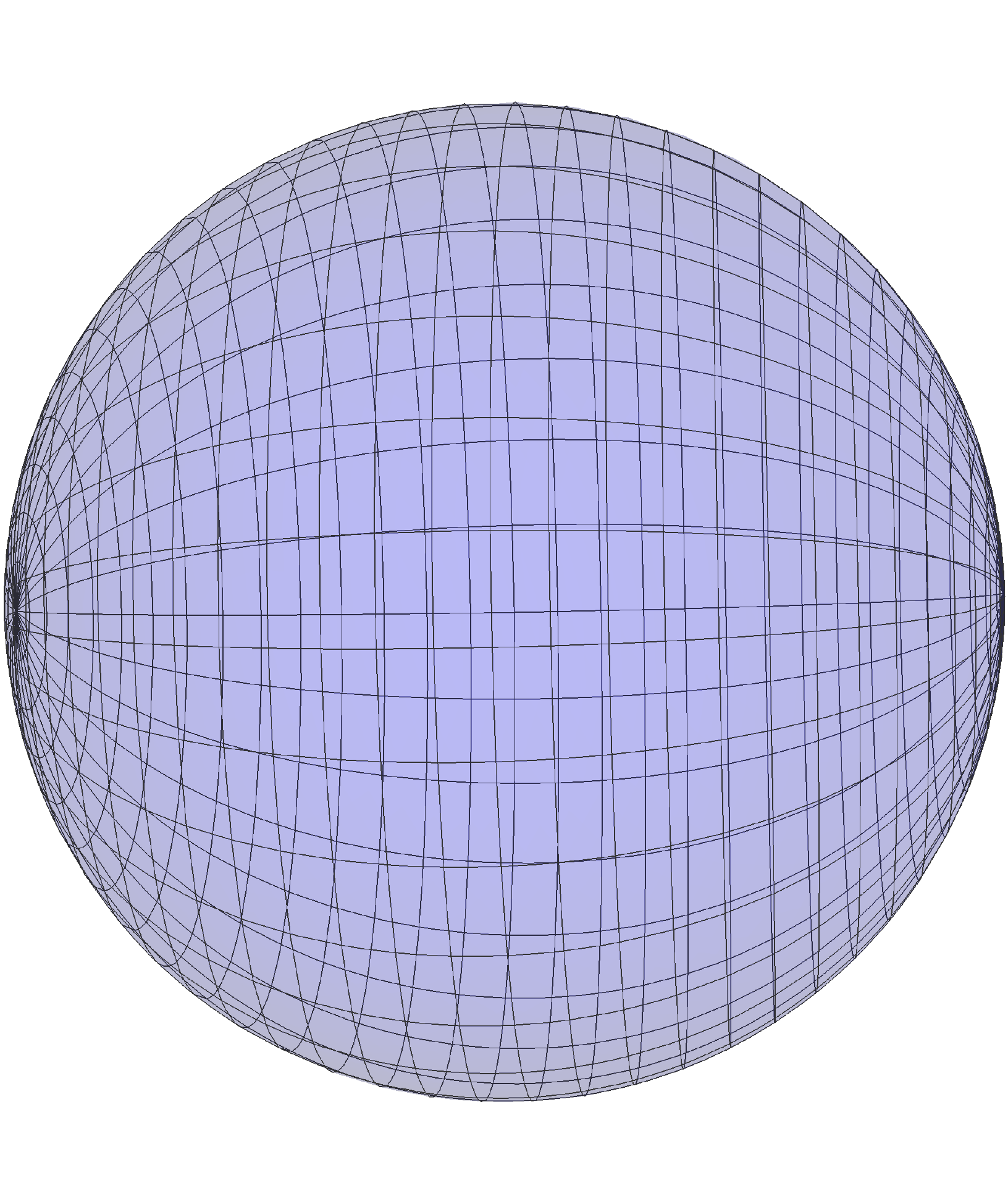}
\caption{\label{leftfig}}
\end{subfigure}
\hspace{0.15\textwidth}
\begin{subfigure}{0.4\textwidth}
\centering
\includegraphics[height=0.3\textheight]{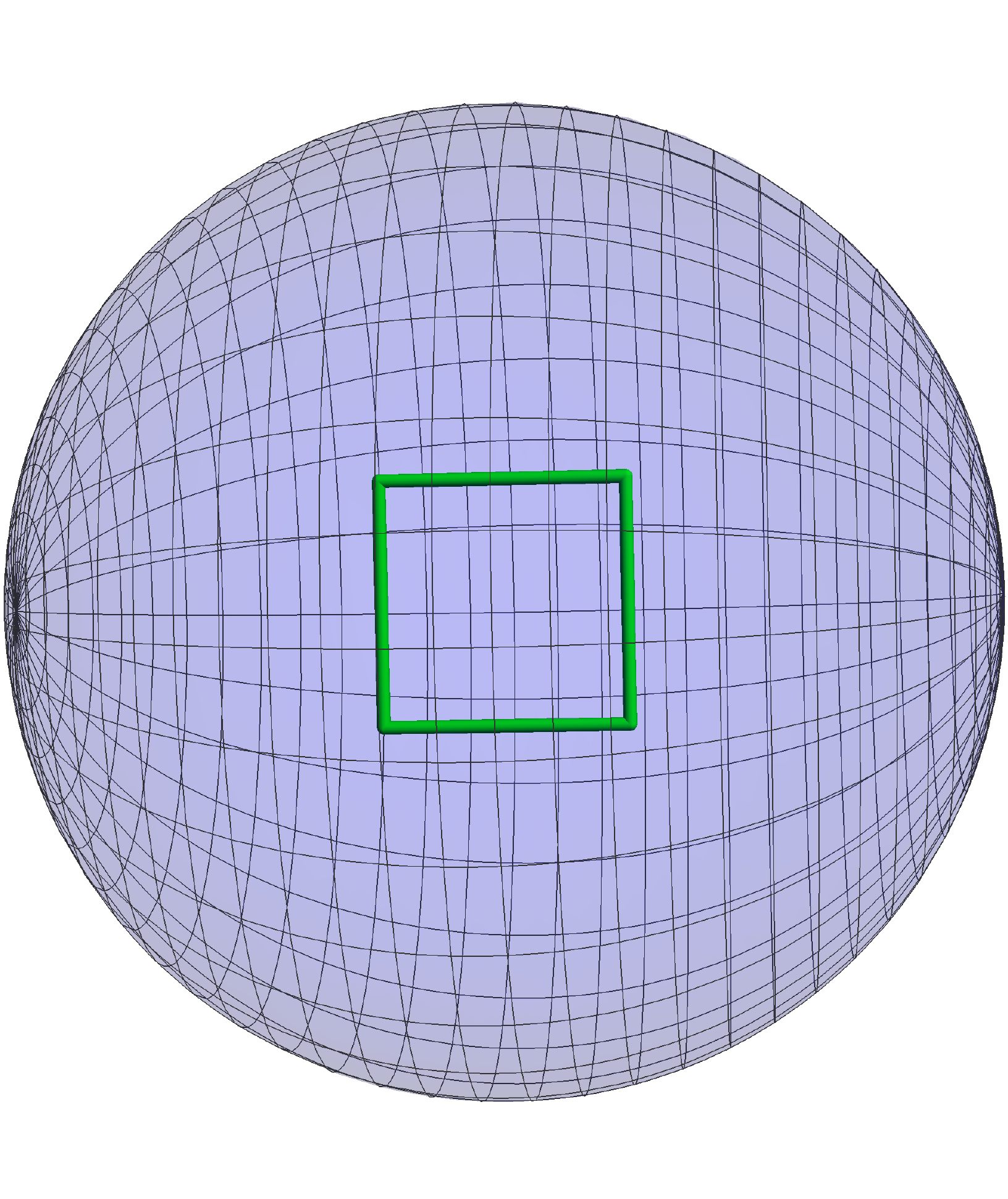}
\caption{\label{rightfig}}
\end{subfigure}
\caption{\em In a nongravitational gauge theory, observations made on a sphere far away  cannot distinguish the vacuum, shown on the left, from the vacuum after a Wilson loop operator has acted on it (shown on the right).  \label{figloop}}
\end{figure}

The simplest example that shows how gravity localizes information differently from an LQFT is to give the observers the task of determining whether the state, $|g \rangle$ coincides with the vacuum, $|0 \rangle$, or not. In a LQFT, it is always possible to find unitary operators, $\ubulk$, that  commute with all observables outside a ball on a Cauchy slice.  For instance, in a gauge theory we could consider a Wilson loop operator near the center of AdS as shown in Figure \ref{figloop}.  Therefore, in a LQFT the observers cannot distinguish between $| 0 \rangle$ and $\ubulk | 0 \rangle$. 

But in a theory of gravity, the observers can perform this task just by measuring the energy repeatedly on identically prepared systems. The probability that this measurement yields $0$ is just given by  $|\langle g| 0 \rangle|^2$. And this quantity immediately distinguishes between the case $|g \rangle = |0 \rangle$ and any other case.

For more refined questions, the observers need to measure correlators of the energy and other observables.  Physically this can be done by first manipulating the system with a unitary manipulation and subsequently measuring the energy. A prototypical task is as follows. Let $X$ be a low-energy operator near the boundary and we denote $|X \rangle \equiv X | 0 \rangle$. The observers are now placed in a state $|g \rangle$. Say that the observers, using the method above, already found out that $\langle 0 | g \rangle = 0$ and now want to determine if $|g \rangle = | X \rangle$ or not. For most states, measurements of the energy alone are insufficient to answer this question.

However, recall that the observers are allowed to act with the unitary operator $e^{i J X}$. If they first do this and  {\em subsequently} measure the energy then the probability that they obtain $0$ is just given by 
\be
\langle g | e^{-i J X} \projvac e^{i J X} | g \rangle =  J^2 |\langle g | X \rangle|^2 + \Or[J^3]
\ee
which can be checked just by expanding the exponentials out to second order, and using $\langle 0 | g \rangle = 0$ and $X|0 \rangle = |X \rangle$. By acting with this unitary operator repeatedly, with different small values of $J$ and then measuring the energy, the observers can determine the coefficient of the $\Or[J^2]$ term above. This coefficient directly gives them the answer that they seek!

As we explained above, states of the form $| X \rangle$ already provide a basis for the Hilbert space. At low energies, this does not even require the formal proof above but can be checked {\em numerically}. This numerical check is performed in Appendix A of \cite{Chowdhury:2020hse}. Therefore just the simple procedure above allows the observers to measure the {\em magnitude} of the overlap of $|g \rangle$ with any basis element. 

The phase can be obtained as follows.  Since the overall phase of the state is arbitrary, the observers can just pick a Hermitian operator $X_r$ near the boundary.  With $|X_r \rangle \equiv X_r | 0 \rangle$, they can determine $|\langle g| X_r \rangle|^2$ using the procedure above and then declare that, for this particular state,  $\langle g | X_r \rangle = |\langle g | X_r \rangle|$. Now by acting with unitaries of the form $e^{i J(X + X_r)}$, the observers can also determine $|\langle g| X \rangle + \langle g | X_r \rangle|^2$. But this quantity, when combined with the knowledge of $|\langle g| X \rangle|^2$, immediately gives them  $\text{Re} \left(\langle g | X \rangle\right)$. This leaves the observers with only a {\em sign} ambiguity in $\text{Im}(\langle g | X \rangle)$. This sign ambiguity can also be removed and this is  explained in \cite{Chowdhury:2020hse}.

\subsubsection{Subtleties}
We now discuss some subtleties that may not have been apparent in our discussion. We have framed this subsection, somewhat informally,  in terms of a set of questions that readers may have.
\begin{enumerate}[qseries]
\item
{\em The fact that the energy can be measured at infinity is a property of canonical quantum gravity. But how do we know that it continues to hold in the full theory of quantum gravity, which may have a description in terms of some ``other'' degrees of freedom? } 

The precise property that was used above was that the  integrated boundary  metric could be used to accurately identify the vacuum. If one considers high-energy states, such as black-hole microstates, then nearby states are separated by 
energy differences that are exponentially small in the black hole entropy.  But we are {\em not} interested in using the boundary metric to distinguish between such states. So the precise assumption is that the  only state that is annihilated by the integral of the subleading component of the metric is the vacuum. In more physical language, the assumption is that one does not have a state of nonzero energy, in quantum gravity, where the energy is entirely shielded from detection by the boundary metric.
\item
{\em  How does one know that ``non-perturbative'' states, such as those involving excitations of D-branes, are described by states produced by applying local operators near the boundary to the vacuum?} 

These states will be included in the Hilbert space provided that they can be produced by applying polynomials of asymptotic quantum-field operators on the boundary followed by time evolution.  Note that we don't need some kind of ``D-brane field operator''  near the boundary to produce such states. For instance,  black holes that can be formed from the collapse of ordinary matter are included in this Hilbert space even though there is no ``black hole field''.  This is because one can start with matter that is well described by an excitation of quantum fields and allow it to collapse by the dynamics. The same process will also create other nonperturbative states. The second important point is that, by construction, Hamiltonian evolution keeps us within the space of states formed by acting with asymptotic operators as explained in the main text. So this space forms a superselection sector and one can safely formulate the theory in such a sector.

\item
{\em The Reeh-Schlieder theorem holds even in ordinary LQFTs. Why doesn't one see this redundancy of information there?} 

It is important to distinguish between two statements: (a) it is possible to produce arbitrary states by applying near-boundary operators to the vacuum and (b) it is possible to unambiguously identify arbitrary states
using operators near the boundary. The Reeh-Schlieder theorem only tells us that statement (a) is true and this statement is true even in a LQFT without gravity. But statement (b) is very different from statement (a) and is not true in a LQFT. The important gravitational input in our argument was that the vacuum could be identified from infinity. This is what makes statement (b) true as well. Consider the construction of the transition operator displayed in equation \eqref{transnm}. Without the insertion of $\projvac$, which is a boundary operator only in a theory of gravity, the operators $X_n$ and $X_m$ are of absolutely no use in constructing the left hand side.
\item
{\em Gauge theories also have a Gauss law. Why doesn't one see this redundancy of information there?} 

The key difference between gauge theories and gravity is that gauge theories have both positive and negative charges. Consequently, it is possible to construct {\em exactly local gauge invariant operators} in a nongravitational 
gauge theory. For instance consider a Wilson-loop operator that is localized near the center of AdS. This operator commutes with all operators near the boundary, and so an observer near the boundary cannot differentiate one state
from another state where this Wilson loop operator has acted as shown in Figure \ref{figloop}.

A more formal statement is that gauge theories obey the so-called split property. It is possible to prepare the state of the theory so that it looks like any desired state, say state 1, for all observations made near the boundary
and any other desired state, say state 2, for all observations made near the inter of the space. In fact, if one separates the interior of the space from the near-boundary region by a small ``collar'' then the near-boundary observations cannot even be used to tell the amount of charge in the interior, since the charge can be hidden from the boundary using compensating charges in the collar.
\item
{\em Is this phenomenon consistent with effective field theory?} 

In deriving the effect above, we used {\em only} effective field theory. We then  assumed that the full theory of quantum gravity shared some of the robust properties of effective field theory at low energies.  The low energy tests outlined in section \ref{lowenergy} are also carried out within the realm of effective field theory.  Therefore this phenomenon---of a redundancy in description --- is not only consistent with effective field theory, it is a consequence of effective field theory!

\end{enumerate}

\subsection{Holography of information in asymptotically flat spacetimes \label{holinfoflat}}
We now explain how this result can be extended to asymptotically flat spacetimes. So far, this extension is well understood for four dimensional spacetimes although we will make some comments about other dimensions below. Our treatment in this section will closely follow \cite{Laddha:2020kvp}. 

It is convenient to work in retarded Bondi coordinates. In these coordinates, we demand that, asymptotically the metric coincide with Minkowski space.
\be
ds^2 \underset{r \rightarrow \infty}{\longrightarrow}  - du^2 - 2 du dr + r^2 \gamma_{AB}d \sph^{A} d \sph^{B}  + h_{\mu \nu} d x^{\mu} d x^{\nu},
\ee
where $h_{\mu \nu}$ indicates the subleading fluctuations of the metric. Here, $A,B$ run over the two spherical coordinates $\theta, \phi$ and $\gamma_{AB}$ is the usual round metric on the 2-sphere. The limit above, where one keeps $u$ fixed and takes $r \rightarrow \infty$ takes us to future null infinity, denoted by $\scrip$.  The past boundary of null infinity, $\scrippast$, is reached by taking $u \rightarrow -\infty$ on $\scrip$.

In the so-called ``Bondi gauge'' (see section 3.1.2 of \cite{Compere:2018aar} for a nice review), the interesting subleading components of the metric are
\be
C_{AB}(u, \Omega) = \lim_{r \rightarrow \infty} {1 \over r} h_{AB}(r, u, \Omega),
\ee
which is called the ``shear'' and
\be
\maspect(u, \Omega) = {1 \over 2} \lim_{r \rightarrow \infty} r h_{u u}(r, u, \Omega),
\ee
which is called the Bondi mass aspect. If we have other massless fields, then there is a similar well-defined limit to extract their boundary values at null infinity.\footnote{This limit must be taken with some care. See the discussion in \cite{Strominger:2017zoo} or Appendix B of \cite{Ghosh:2017pel}.} For a massless scalar, $\phi$, we
will denote the boundary value by $O$ where
\be
O(u, \Omega) = \lim_{r \rightarrow \infty} r \phi(r, u, \Omega).
\ee

\begin{figure}[!ht]
\begin{center}
\includegraphics[height=0.5\textheight]{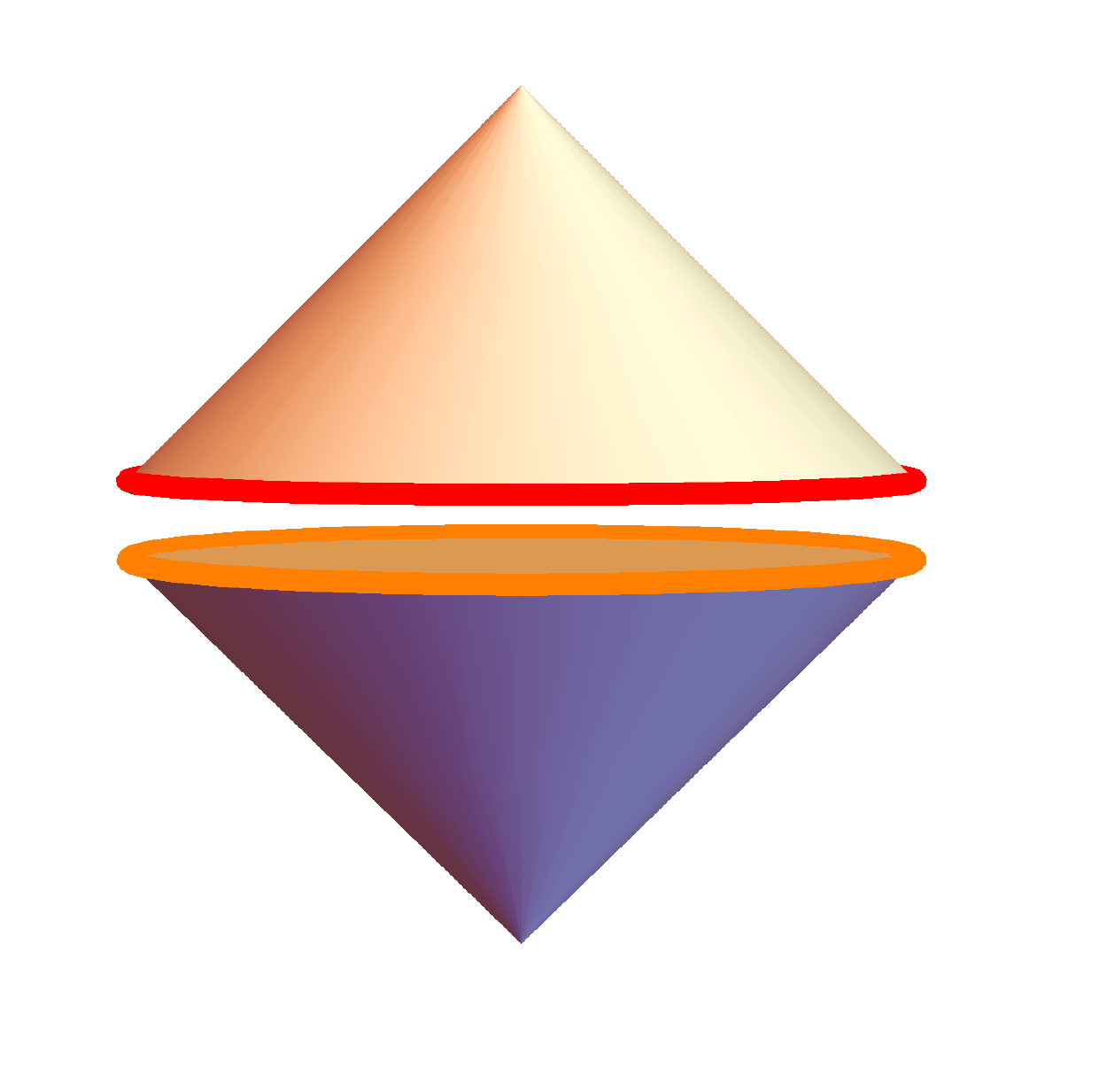}
\caption{\em In asymptotically flat four dimensional spacetimes, all information about massless particles is available near the past boundary of future null infinity (showed as a red band) or the future boundary of past null infinity (orange). \label{allinfo}}
\end{center}
\end{figure}

Due to infrared effects, it turns out that the vacuum is infinitely degenerate in four dimensional asymptotically flat space \cite{Strominger:2013jfa}.  The various vacua are labelled by the eigenvalues of conserved charges from the asymptotic symmetry group \cite{Bondi:1962px} that are called ``supertranslation charges''. These supertranslation charges can be labeled by the same indices as a spherical harmonic. We denote them by $\suptrans_{\ell, m}$ and a vacuum, $|\supcharge[s] \rangle$  is specified by giving an infinite number of real numbers, $s_{\ell, m}$
\be
\suptrans_{\ell, m} |\supcharge[s] \rangle = s_{\ell, m} |\supcharge[s] \rangle.
\ee

While the vacuum structure in 4d flat space is very intricate, the description of excited states on $\scrip$ is very simple, when compared to AdS. This was first shown by Ashtekar, as part of the ``asymptotic quantization'' program  \cite{Ashtekar:1981sf,Ashtekar:1981hw,Ashtekar:1987tt,Ashtekar:2018lor}. A remarkable outcome of this program, which is perhaps somewhat underappreciated, is that even in the full nonlinear theory of general relativity, the structure of excitations is simply that of a Fock space. This is also true of other dynamical fields; whatever their interactions in the bulk, they become free near the boundary. Therefore,
the set of all massless excitations can be described in the following Hilbert space.  Let $O(h)$ denote a massless quantum field operator smeared with a smooth function $h$ in $u$ and on the sphere. We can then generate a Fock space by repeatedly acting with such operators.
\be
\label{hilbsexc}
\hilb[s] = \text{span~of~}\big\{O(h_1) ... O(h_n) | \{s \} \rangle \big\}.
\ee
Our notation is the same as equation 2.16 of \cite{Laddha:2020kvp} except that we have not displayed the radiative modes of the graviton and included them implicitly above. The full Hilbert space of massless particles is obtained by taking the direct sum of all of these individual Fock spaces
\be
\label{hfulldef}
{\cal H} = \bigoplus_{\supcharge[s]} {\cal H}_{\supcharge[s]},
\ee

The Hilbert space above does not include massive particles because the operators that describe massive particles have a good limit at future and past infinity and not at null infinity. Physically, the reason for this is that, at a distance $r$ the force mediated by a massive particle of mass $m$ dies off as $e^{-m r}$ and so it dies off faster than any power of $r$ as one approaches null infinity \cite{winicour1988massive}. We believe it should be possible to include these particles in the ambit of the principle of holography of information but, for now, we will restrict ourselves to excitations of massless particles. Note that even if a theory has massive particles, the Hilbert space can be written as a factorized space of excitations of the massive and massless particles \cite{Campiglia:2015kxa} and our statements below then apply after massive particles have been traced out to yield a possibly mixed state on the massless Hilbert space, ${\cal H}$.

So far we have reviewed standard material but we now proceed to our argument. We will now show that all operators ${\cal H} \rightarrow {\cal H}$ can be represented as operators supported in the region $u \in (-\infty, -{1 \over \epsilon})$. This region is ``near'' the past boundary of future null infinity,  $\scrippast$, and is marked on  Figure \ref{allinfo}. This is the precise sense in which the principle of holography of information holds in this setting. There are two steps in the argument, just as in the case of AdS.

\paragraph{\bf Step 1: Vacuum operators are located near $\scrippast$.}
First we show that the projector onto the manifold of vacua and transition operators between different vacua are operators that are accessible near $\scrippast$.  This is the physically important step, and the one that is special to gravity.

This follows because both the Hamiltonian and the supertranslation charges are elements of the algebra in $u \in (-\infty, -{1 \over \epsilon})$.  Physically, if an observer were to attempt to measure these charges near $\scrippast$, the probability, in quantum mechanics, for the observer to obtain a given result is given by the expectation value of the projector onto eigenspaces of these operators. Therefore, these projectors are themselves elements of the algebra near $\scrippast$. By taking a sharp limit of these projectors we conclude that the operator $|\supcharge[s] \rangle \langle \supcharge[s] |$ is an element of the algebra near $u \rightarrow -\infty$.\footnote{We are being a little sloppy here, since the vacua are normalized as $\langle \supcharge[s] | \supcharge[s'] \rangle = \delta(\supcharge[s] - \supcharge[s'])$ and so to obtain a projector we need to smear this operator over some finite range of charges.}

Second, within the low-energy theory, one can find operators that change the value of the supertranslation charge. By starting with a vacuum with a given supertranslation charge, $|\supcharge[s] \rangle$, acting with such an operator, and projecting back to the manifold of zero-energy states we can also construct the operators, 
\be
\transop[s, s'] = |\supcharge[s] \rangle \langle \supcharge[s'] |,
\ee
for an arbitrary pair of vacua.
We conclude that these operators are also physically accessible near $\scrippast$. 

\paragraph{\bf Step 2: Arbitrary operators, ${\cal H} \rightarrow {\cal H}$ are located near $\scrip$.}
Armed with the result above, we now note that if $|n_{\supcharge[s]} \rangle$ is a state in the Fock space $\hilb[s]$ then it can be approximated 
arbitrary well by a state from the space
\be
\text{span~of~}\big\{O(\widetilde{h}_1) ... O(\widetilde{h}_n)\big\} | \supcharge[s] \rangle,
\ee
where the smearing functions $\widetilde{h}_i$ now have support only for $u \in (-\infty, -{1 \over \epsilon})$. The proof of this statement is exactly the same as in AdS, and so we will not repeat it here. As we emphasized repeatedly, in the discussion above, this result holds even without gravity and by itself does not allow an observer near $\scrippast$ to obtain any information about the rest of $\scrip$.

However, when this result is combined with the result from Step 1 above, we find something remarkable. Let $|n_{\supcharge[s]} \rangle \langle m_{\supcharge[s']} |$ be the operator that maps an arbitrary state which lives in $\hilb[s]$ to another arbitrary state that lives in $\hilb[s']$.  Since for both states we have 
\be
|n_{\supcharge[s]} \rangle = X_{n} |\supcharge[s] \rangle;\qquad |m_{\supcharge[s']} \rangle = X_{m} |\supcharge[s'] \rangle,
\ee
where $X_n$ and $X_m$ are operators accessible near $\scrip$ we find that 
\be
|n_{\supcharge[s]} \rangle \langle m_{\supcharge[s']} |= X_n \transop[s, s'] X_m.
\ee
The operator on the right is just a product of three operators, all of which are accessible near $\scrip$, and therefore it is also accessible near $\scrip$. Since {\em every operator} in ${\cal H}$ can be written as a linear combination of the operators that appear above, we conclude that every operator that acts on this Hilbert space lives in the algebra near $\scrip$.

The result above tells us that all information about massless particles  can be obtained from near $\scrip$. Another way to state the result that we have found is that any two distinct states --- pure or mixed --- on ${\cal H}$ can be distinguished just by observations near $\scrippast$. This is surprising, when viewed from the perspective of LQFT, since one usually imagines that it would require observations on all of $\scrip$ to uniquely identify a state. But the remarkable feature of the holography of information is that the same information is also available just near $\scrippast$.

\paragraph{\bf A possible refinement. \\}
\begin{figure}[!ht]
\begin{center}
\includegraphics[height=0.5\textheight]{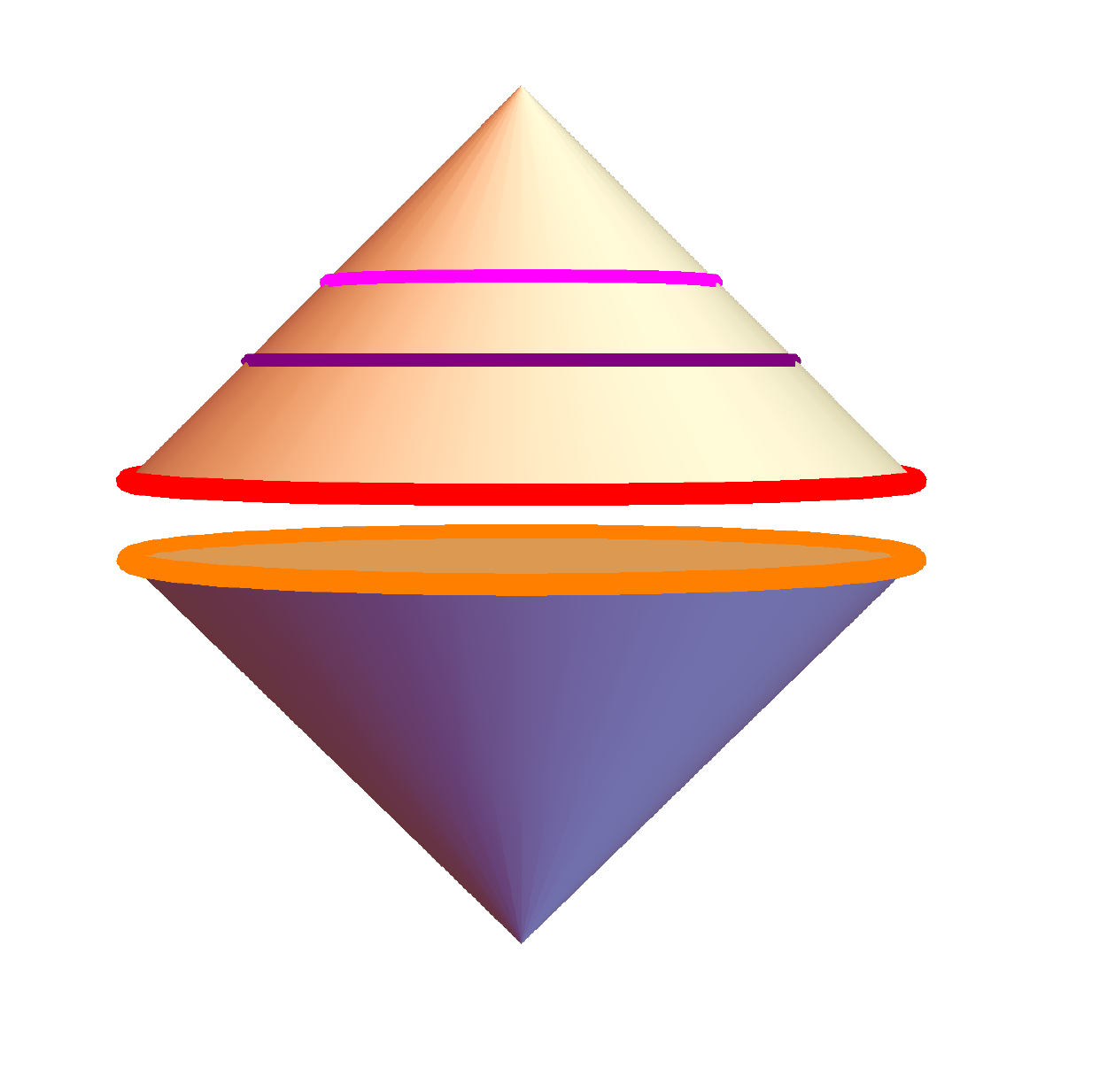}
\caption{\em With stronger assumptions we find that, on future null infinity, information available on any cut is available on any cut to its past. But the converse is not true. \label{cutinfo}}
\end{center}
\end{figure}
It is possible to refine this result a little to show that all the information to the future of a cut on $\scrip$ is available on the cut itself. This is shown in Figure \ref{cutinfo}. 

 When the Bondi mass aspect is integrated over the celestial sphere, it yields the ``Bondi mass'' $M(u)$ at a cut at $u$. Now, in the full nonlinear theory of general relativity, it is possible to derive the Dirac brackets between the Bondi mass and any other massless field on null infinity. These commutators are given by
\be
\label{commgr}
[M(u), O(u', \sph)] = -4 \pi \gnewt i  \partial_{u'} O(u', \sph) \theta(u' - u).
\ee
One could assume that these commutators continue to hold in the full UV theory of quantum gravity. This is not as unreasonable an assumption as it might first appear because these commutators depend only on the very weak-field limit of the action. So they are unchanged if one adds an arbitrary number of higher-derivative terms to the Einstein-Hilbert action.

In any case, assuming that this relation continues to hold,  we obtain the remarkable result. 
\be
\label{restwo}
O(u_0 + U, \sph) = e^{{i M(u_0 - \epsilon) \over 4 \pi \gnewt}  U} O(u_0, \sph) e^{{-i M (u_0 - \epsilon) \over 4 \pi \gnewt} U}, ~~~\text{for}~~~ U> 0.
\ee
This means that the Bondi mass at any cut at $u_0 - \epsilon$ can be used to evolve observables from $u_0$ to any value in the future.\footnote{We note there are some subtleties in defining the Bondi mass at a finite cut as an operator in the quantum theory \cite{Bousso:2017xyo}, which do not appear if we only consider the conserved charges that are defined in the limit as $u \rightarrow -\infty$. See Appendix B of \cite{Laddha:2020kvp}.  In showing that all information is available near $\scrippast$ we only used these conserved charges.  But equation \eqref{commgr} assumes that, at least {\em conjugation} by the bounded operator $e^{i M(u)}$  is well defined at finite $u$.}

This result can be interpreted as a possible refinement of the principle of the holography of information.  Think of $\scrip$ as the asymptotic limit of a sequence of Cauchy slices that are pushed forward in time.  Then, referring to Figure \ref{cutinfo}, there is an appealing physical interpretation of result \ref{restwo}: the information in a region (the future of a cut) is available at its ``boundary'' (the cut) without having to go all the way to the boundary of the full slice, which is at $u \rightarrow -\infty$.  We emphasize that even if the stronger assumptions made for this result fail to hold, this does not affect the validity of the principle of holography of information proved previously.

The focus in this section has been on the question of where information is available. This is just part of a broader program of understanding how holography works in flat space, including the question of how asymptotic symmetries constrain asymptotic correlators and other physical quantities \cite{Bagchi:2012cy,Bagchi:2010eg,Bagchi:2014iea,Banerjee:2019prz,Pasterski:2016qvg,He:2017fsb,Mishra:2017zan} on which there is significant literature starting with the work of de Boer and Solodukhin \cite{deBoer:2003vf}.

\subsubsection{Subtleties}
We again turn to some subtleties that are implicit in the argument above.
\begin{enumerate}[qseries]
\item {\em Does the fact that supertranslations seem to play a key role somehow suggest that four dimensions are special?}

In fact supertranslations are a {\em complication} for our argument. As the reader can check by comparing the argument above with the argument in AdS, the argument made in the presence of the IR cutoff in AdS is simpler. In this sense, our argument is quite different from \cite{Hawking:2016sgy,Hawking:2016msc}, where the soft charges played a key role.

The reason for restricting to four dimensions is that  the vacuum structure of the theory is not well understood in higher dimensions. We direct the reader to \cite{Kapec:2015vwa,Hollands:2016oma,Aggarwal:2018ilg,Campiglia:2017xkp,He:2019pll} for some recent discussion and differing perspectives. However, whether or not one needs to include an infinite set of vacua, distinguished by
the supertranslation charges, it seems likely that the argument above will generalize to higher dimensions. For instance, if the asymptotic symmetry group for higher d reduces to the Poincare group, then we expect to have a unique vacuum that is specified by the Hamiltonian. 
Since the Hamiltonian is defined at $\scrippast$ even in higher dimensions, the argument in higher dimensions will then resemble the argument in AdS. On the other hand, if one does need to include a family of vacua then, provided these vacua are distinguished by charges at $\scrippast$, the argument that we have given in four dimensions  above will go through.

\item{\em Isn't the Penrose diagram different from Figure \ref{allinfo}  in the presence of black holes?}

  Our assumption is that in the quantum theory, black holes always evaporate and so the Penrose diagram is always trivial. Therefore the total Hamiltonian, and the supertranslation charges are always defined in $\scrippast$ and do not have a separate contribution that comes from the horizon. 

\item {\em What if the vacuum structure is more complicated, and additional charges beyond  supertranslation charges are required to specify a unique vacuum?}

There has been some discussion in the literature \cite{Campiglia:2016jdj,Sahoo:2018lxl} about the possibility of requiring 
additional charges. But provided these additional charges are also defined at $\scrippast$, our argument will generalize trivially. The requirement that the charges should be defined at $\scrippast$ is very mild since if the charges 
can be defined on a large Gaussian surface at large-$r$, then it is natural that the same charges will also carry over to $\scrippast$. 
\end{enumerate}

\subsection{More low-energy tests of the holography of information \label{morelowenergy}}
We conclude this description by providing yet another procedure to check the principle of holography of information within perturbation theory. This time, we will present the check in flat space rather than in AdS.  We believe that it is possible to generalize the protocol of section \ref{lowenergy} but in this subsection, we will adopt a slightly different route in order to give the reader a different view on these results.

\begin{figure}[!ht]
\begin{center}
\includegraphics[width=0.7\textwidth]{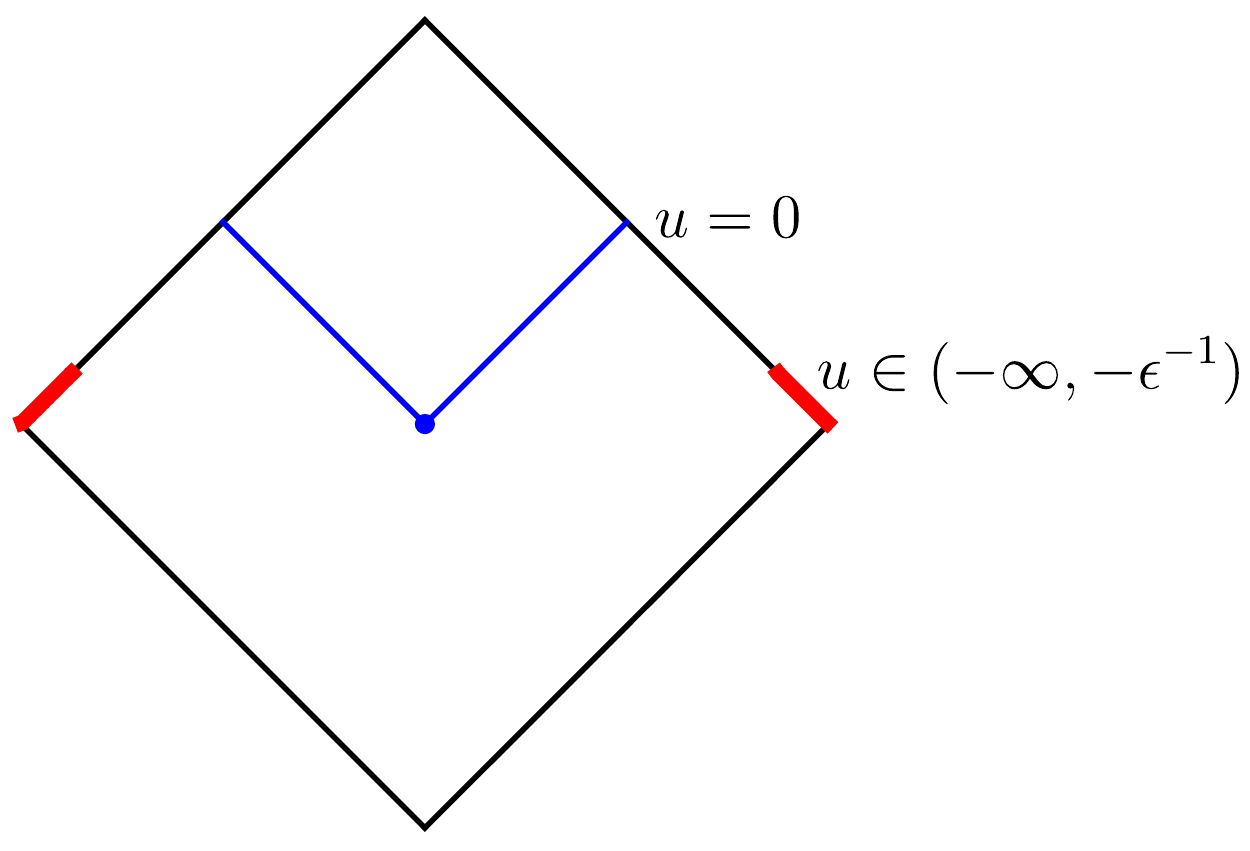}
\caption{\em The state $|f \rangle$ contains a pulse that hits $\scrip$ near $u = 0$. Nevertheless, in a theory of gravity by making observations in $u \in (-\infty, -{1 \over \epsilon})$ (thick red segment), it is possible
to determine the precise profile of the pulse. It is also possible to differentiate a pulse of one field from a pulse of another field that is related to the first one by a global symmetry. \label{figpulse}}
\end{center}
\end{figure}

Consider the following simple state, which comprises a simple excitation of a dynamical scalar field on top of some vacuum.
\be
\label{fdef}
|f \rangle = e^{-i \lambda \int f(u, \Omega) O(u, \Omega)} |0 \rangle.
\ee
Here the state is specified by a single ``profile function'' $f(u, \Omega)$. Let us take $f(u, \Omega)$ to have compact support near $u = 0$. Then, physically, this state corresponds to a pulse of the scalar field with some thickness specified by $f$ that comes and hits null infinity near $u = 0$. (See Figure \ref{figpulse}.) In the state above, $\lambda$ is a perturbative parameter and $|0 \rangle$ is some normalizable vacuum state, comprised of the different vacua above. 

Now, consider a set of observers, who are constrained to make observations on $\scrip$ only for retarded times in the range $u \in (-\infty, -{1 \over \epsilon})$. These observers are allowed to probe the state in \eqref{fdef} for different values of $\lambda$, so that they can work in perturbation theory in $\lambda$. Can these observers determine the complete form of the profile function $f(u)$?  

As Figure \ref{figpulse} makes clear, this task requires the observers to extract information about the pulse {\em before} the pulse actually reaches $\scrip$. We have taken the function $f(u, \Omega)$ to have compact support so there are {\em no tails} of the function that reach the region $u \in (-\infty, -{1 \over \epsilon})$. Therefore, this task is clearly impossible in a LQFT. 

Nevertheless, in a theory of gravity, the observers can complete this task as follows. The observers measure a  two-point correlation function of the energy and the fluctuations of the field itself for $u \in (-\infty, -{1 \over \epsilon})$. The entire correlation function can be determined using observations in this range of $u$, since the energy itself is the limit of the Bondi mass as $u \rightarrow -\infty$;  the fact that it can be measured on the past boundary of $\scrip$ is just a simple consequence of the Gauss law. A simple calculation in effective field theory then tells us that
\be
\label{fresult}
\langle f | M(-\infty) O(u, \sph') | f \rangle =  \gnewt \lambda \int {f(x, \sph') \over (x - u - i \epsilon^{+})} d x + \Or[\lambda^2].
\ee
Since the observers are allowed to probe the state for different values of $\lambda$ for small $\lambda$ they can unambiguously extract the first order term above.

The important aspect of \eqref{fresult} is that even though the function $f$ itself does not have any support for large negative $u$, the correlator is nonzero in this interval.  Moreover, the right hand side of \eqref{fresult} allows us to unambiguously extract the function $f$. A formal proof is as follows: the right hand side of \eqref{fresult} is the analytic when $u$ is extended in the upper half plane. Therefore if two functions $f_1$ and $f_2$ yield the same right hand side for  $u \in (-\infty, -{1 \over \epsilon})$, these two functions must coincide for all $u$.

But this can also be seen from a more hands-on perspective. Since $f(x)$ has compact support we have
\be
\label{fseries}
\int {f(x, \sph') \over (x - u - i \epsilon^{+})^3} d x = -\sum_{n=0}^{\infty} {1 \over  u^{n+1}} \int x^n f(x, \sph') d x.
\ee
So by determining the series expansion of the correlation function for different values of $u$ --- which the observers can do since they have access to $u$ in a range  --- the observers can extract all the moments of the function $f$.

This perspective, of thinking about information accessible near $\scrippast$ in terms of correlators of fluctuations
of metric with other dynamical fields, also provides insight into other subtleties that we now describe briefly.
\begin{enumerate}
\item
{\bf The price of being far away.} 
The expression \eqref{fseries} also gives us an estimate of the ``price'' that the observers must pay to extract information about the pulse before the pulse actually reaches $\scrip$. If $f$ has compact support, then  higher-order terms in the power series are suppressed since the ratio ${x \over u}$ is small. Therefore, it is hard for the observers to extract the higher moments of $f$, which multiply higher terms in this decaying power series. 
\item
{\bf Classical and nongravitational limit.} The correlator above is a quantum-gravity effect. This can be seen by restoring the fundamental constants in equation \eqref{fresult}. We find, through dimensional analysis that $\gnewt$ in equation \eqref{fresult} must be replaced with
\be
\gnewt \rightarrow {\hbar \gnewt \over c^3}
\ee
We now see that in the limit $\hbar \rightarrow 0$, the right hand side of \eqref{fresult} vanishes. This also happens in the limit $\gnewt \rightarrow 0$. This is consistent with the expectation that both in classical gravity, and in LQFT,  the observers cannot determine the form of the pulse before it reaches $\scrip$. The difference between the classical and quantum case is also discussed in \cite{Jacobson:2019gnm}.
\item
{\bf Global symmetries.} It may be that, for other reasons, global symmetries do not exist in a theory of quantum gravity. (See \cite{Harlow:2018tng} and references there for a recent discussion.) But, at the level of our perturbative protocol, the existence of global symmetries does not pose any obstruction, and does not prevent the observers near $\scrippast$ from identifying the state. In particular, say that there is another scalar field $\widetilde{O}$, which is related to $O$ by a global symmetry. And say that the observers are tasked with distinguishing the state $|f \rangle$ from the state $|\widetilde{f} \rangle$ where
\be
\label{ftildedef}
|\widetilde{f} \rangle = e^{i \lambda \int f(u, \Omega) \widetilde{O}(u, \Omega)} |0 \rangle.
\ee
These two states can be distinguished by comparing correlators of $M(-\infty) O(u)$ with correlators of $M(-\infty) \widetilde{O}(u)$. The table below gives the values of the two possible correlators in the two possible states. The reason that the observers can distinguish these two states is because they
are not just measuring the energy far away (which is the same for both states) but measuring how fluctuations in the energy are correlated with fluctuations in other dynamical fields. 
\begin{center}
\begin{tabular}{|c|c|c|}
\hline
\diagbox[width=10em]{State}{Correlator} & $\langle M(-\infty) O(u) \rangle$ & $\langle M(-\infty) \widetilde{O}(u)$ \\ \hline
$|f \rangle$ & Same as equation \eqref{fresult} & 0 \\ \hline 
\parbox[c]{1.6em}{\vspace{0.02in}$|\widetilde{f} \rangle$} & 0 & Same as equation \eqref{fresult} \\ \hline
\end{tabular}
\end{center}
\end{enumerate}
In this subsection, we considered the specific family of states displayed in equation \eqref{fdef}. But when viewed from the perspective of null infinity,  the structure of massless excitations in flat space is very simple and the Hilbert space is just a Fock space. So even this simple example already strongly suggests that if one considers more complicated states then, while two point correlators may be insufficient, higher point correlators near $u \rightarrow -\infty$ can be used to uniquely identify the state. The argument in section \ref{holinfoflat} provides a proof of this expectation.

\section{A new perspective on black hole information \label{secresolveinfo}}
The principle of holography of information provides a fresh perspective of black hole information. Recall that the discussion above, both in the case of AdS and in the case of flat space, was
quite general and included black hole states. Usually, the question that is commonly asked about black holes is  ``how does the information comes out of the horizon?''. But if the principle of holography of information is correct --- and we have provided a proof of this principle in some cases in section \ref{secholography} --- it immediately leads to the following conclusion.
\begin{lesson}
\label{lessonalwaysoutside}
The principle of holography of information  implies that the exterior of a black hole always contains a complete copy of the information in the interior. 
\end{lesson}
This conclusion  immediately renders the issue of information loss moot. Even as the black-hole interior evaporates away, the copy of the information outside remains.  Second, as we discussed briefly in section \ref{oldinfosubtle}, the conclusion above also reveals a clear error in Hawking's argument.
\begin{lesson}
\label{lessonhawkerror}
An error in Hawking's argument is that it assumes that quantum gravity localizes information like local quantum field theories.
\end{lesson}
This is an error because in a LQFT, one can specify the state of the black-hole interior independently of the exterior, which leads to Hawking's ``principle of ignorance''. But in a theory of quantum gravity, the principle of holography of information tells us that the situation is reversed: the state of the exterior completely specifies the state of the interior. 

The conclusions above can be viewed as resolutions of the information paradox for evaporating black holes. 
In this section, we discuss some further implications of this resolution, first for flat-space black holes and then for small black holes in AdS. For this latter case, we also briefly discuss how the difficulty of extracting information from the black hole changes with time from when the black hole forms to when it completely evaporates. We also revisit the monogamy paradox described in section \ref{secmonogamy}. 
Here, we show how neglecting the principle of holography of information leads to a monogamy paradox even in empty space, in the absence of black holes.  We also explain why the operators in the interior of a black hole are precisely of the kind that appear in the empty-space paradox.  This leads to the conclusion that the seeming violation of the monogamy of entanglement that appears during black hole evaporation is not a paradox and is, in fact, expected in a theory of quantum gravity. In the last part of this section, subsection \ref{secpage}, we discuss the recent computations of the Page curve for AdS black holes coupled to nongravitational baths and contrast the results obtained in these papers with the results obtained above.

\subsection{Black holes in asymptotically flat space}
We remind the reader of the caveat that our result for the holography of information in flat space applies, strictly speaking, for theories with massless particles. We believe that, qualitatively, these results can be generalized to incorporate massive particles but this has not been done so far.

\begin{figure}[!ht]
\begin{center}
\includegraphics[height=0.5\textheight]{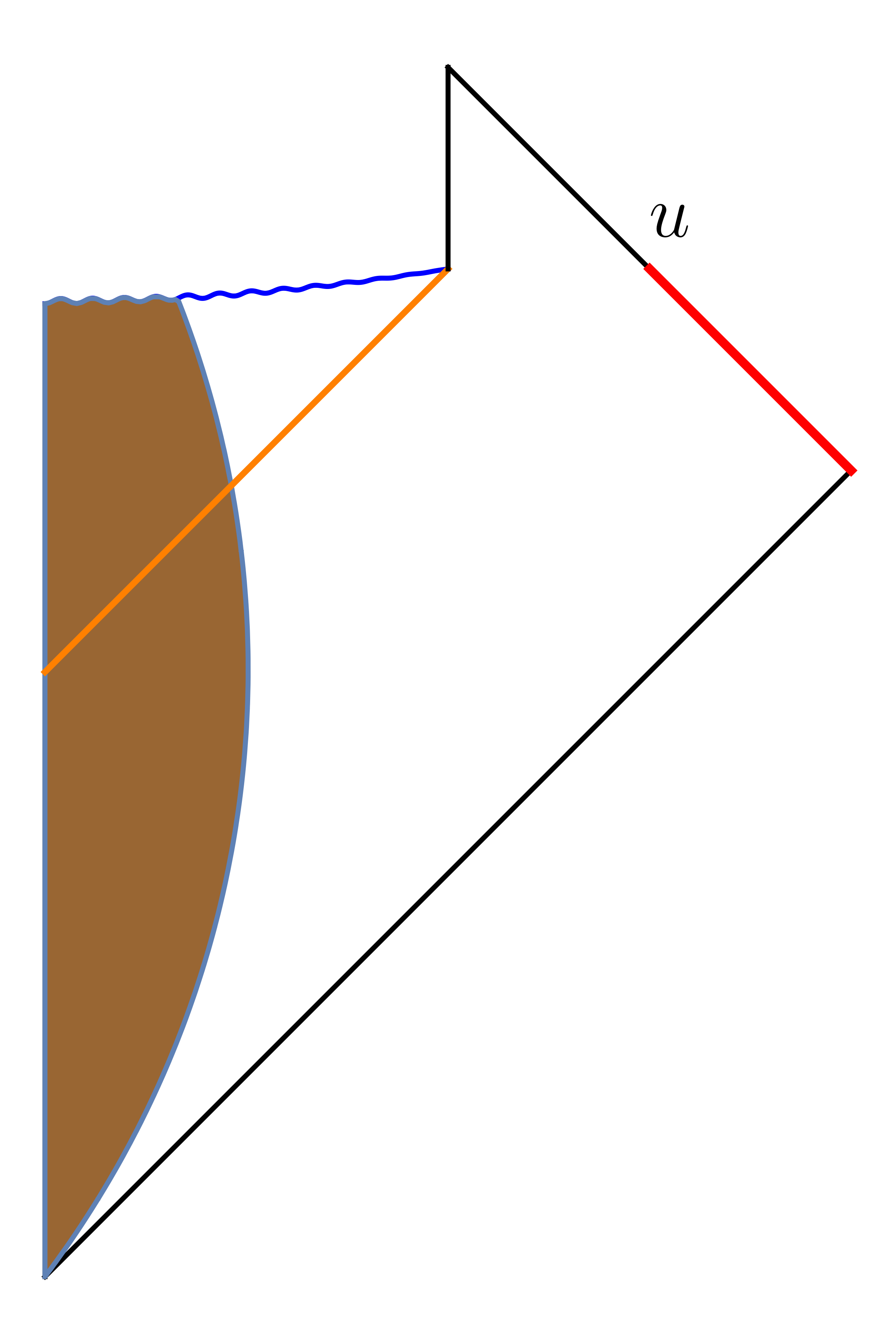}
\caption{\em One of our results is that the fine-grained entropy of the segment $(-\infty, u)$ of null infinity (marked in red) is independent of $u$.}
\end{center}
\end{figure}
\begin{figure}[!ht]
\begin{center}
\includegraphics[width=0.7\textwidth]{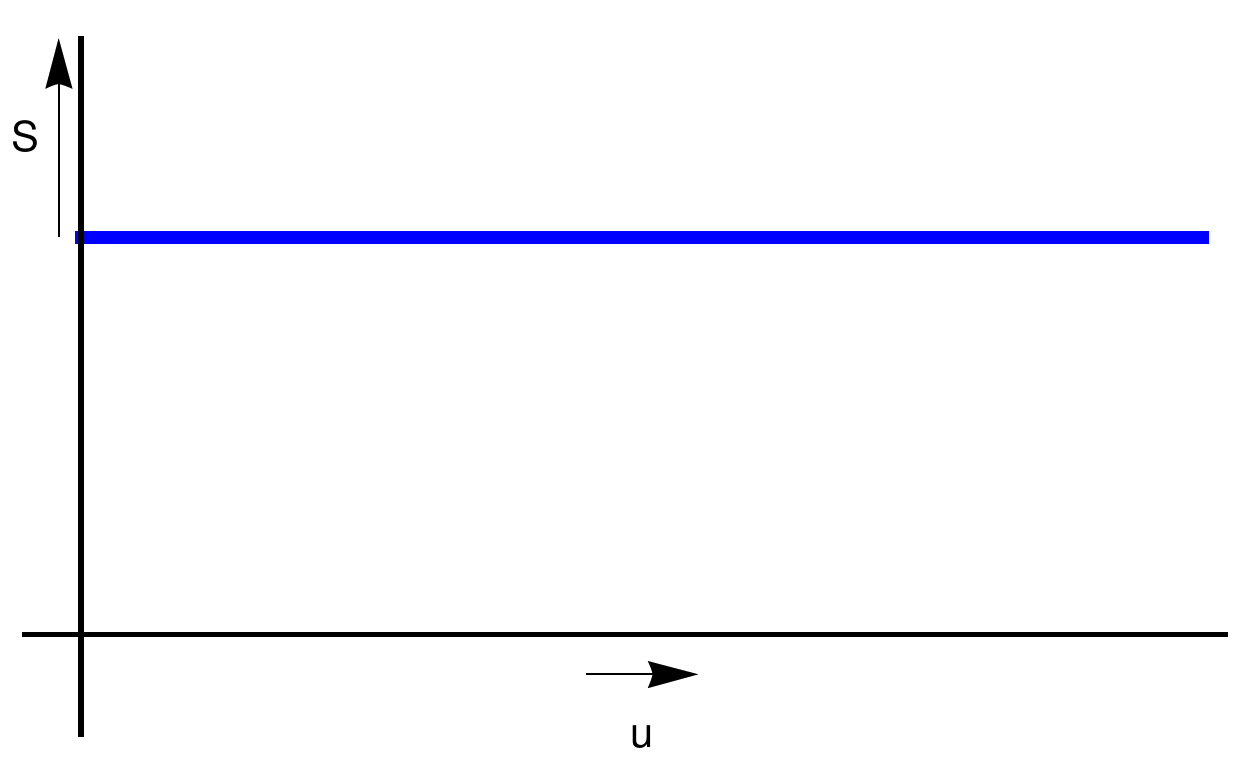}
\caption{\em The fine-grained von Neumann entropy of the segment $(-\infty, u)$ of $\scrip$ in a theory of gravity. The constancy of this quantity is one way to formalize the  statement that the information about the black-hole interior is always available ``outside''. \label{figvn}}
\end{center}
\end{figure}

The claim that the information is always ``outside'' can be formalized as the result that the von Neumann entropy of the state defined on a segment $(-\infty, u)$ of $\scrip$ is independent of the upper limit $u$. (See Figure \ref{figvn}.) From a physical perspective, observables on this segment can be thought of as the natural observables for a set of observers who can make measurements {\em outside} a large ball although we caution the reader that for any precise question, the description on $\scrip$ is the correct one.

The result is easy to prove as is done in section 4 of \cite{Laddha:2020kvp}. To compute the von Neumann entropy of a state with respect to an algebra, we first define a density matrix, $\rho$,  as that element of the algebra whose trace with any other observable in the algebra, $b$, yields the expectation value of that observable in the given state: 
\be
\tr(\rho b) = \langle b \rangle.
\ee
Then the entropy is $S = -\tr(\rho \log \rho)$. A little reflection should be sufficient to convince the reader that, in a LQFT, this coincides with the standard definition of the entropy of a region when the algebra is taken to comprise all the local operators in the region.

 As we extend the segment $(-\infty, u)$ towards $u \rightarrow +\infty$,   it naively appears that we have access to additional local operators and so the density matrix should change.  This is indeed what would happen in a LQFT. However, as we showed above, in  a theory of gravity, all the operators on $\scrip$ can be represented as operators near $\scrippast$. Therefore the density matrix $\rho$ can always be chosen to be an operator near $\scrippast$ regardless of the upper limit of the segment. It is clear that the corresponding entropy is independent of $u$.

Since the von Neumann entropy with respect to an algebra is a measure of the information about the state that is available through observables in the algebra, the physical interpretation of this result is precisely what we need.  The result tells us that we gain no further information by including operators from larger and larger values of $u$, and all the information about the state is already present near $\scrippast$. This is in contrast to what happens in a LQFT where information would have emerged ``gradually'' on null infinity.  The expected curve of the von Neumann entropy is shown in Figure \ref{figvn}. The fact that we have a finite intercept is because it is natural to account for a mixed state on  ${\cal H}$, which arises after tracing out massive particles.

The reader might be concerned that a black hole is a very complicated nonperturbative state, that is not incorporated in the analysis of section \ref{secholography}.  But the beauty of  the description on $\scrip$ is that these complications all go away. The Hilbert space on $\scrip$ is just a Fock space, and so the description of a black hole in this space is just in terms of the free gas of Hawking radiation that it emits.  Therefore the claim 
is just that any state in this Fock space can be identified by considering correlators of the Hamiltonian, supertranslation charges, and other dynamical fields in a small interval near $\scrippast$.   This seems to be a very robust statement. 

\paragraph{\bf Difficulty of extracting information.} By thinking in terms of correlators, one can also make some simple-minded estimates of the difficulty involved in extracting information from the black-hole interior.  In  section \ref{morelowenergy} we already explained how simple states could be identified using two-point correlators near $\scrippast$.  Of course, for the final state of a black hole, it is insufficient to consider only two-point correlators. We expect that identifying a black-hole microstate will require at least $S$-point correlators where $S$ is the black hole entropy. Moreover, to distinguish between different black-hole microstates may require an accuracy that scales exponentially with $S$. These are formidable obstructions if one is an experimentalist! But, from a theoretical perspective, there seems to be no difficulty in identifying arbitrary pure and mixed states on ${\cal H}$ by using physical correlators on $\scrip$ in the range $u \in (-\infty, -{1 \over \epsilon})$.

Of course, it is possible to also use operators at larger values of $u$ to obtain the same information. The discussion of section \ref{morelowenergy} is again relevant: the profile of the shock-wave state considered there can be extracted either by using gravitational effects near $u \rightarrow -\infty$ or by making direct measurements of the massless field operators near $u = 0$. The process of using correlators at larger values of $u$ to reconstruct the state is what is commonly referred to as ``collecting Hawking radiation.'' However, these measurements {\em also} require at least $S$-point correlators and the same accuracy $e^{-S}$. So it is not a priori clear that it is any easier to extract information from a black hole by waiting and collecting the Hawking radiation than by using gravitational effects in the black hole exterior before it has evaporated. 

This also reveals  an important distinction between black holes and objects like burning coal. In the latter case, it is possible to take a limit $M_{\text{pl}} \rightarrow \infty$, while keeping the entropy of the object finite. In this limit, it becomes infinitely hard to use quantum gravity effects to reconstruct the microstate of the object. This should be clear from the discussion of section \ref{morelowenergy}: the correlators displayed there necessarily involve a measurement of the disturbance caused in the metric by the object that one is probing. This disturbance is proportional to $\gnewt$, and, in the limit where $\gnewt \rightarrow 0$, this disturbance vanishes. In this limit it is still possible to collect all the radiation from the coal, and make measurements on that radiation with exponential accuracy. Since the entropy is decoupled from $M_{\text{pl}}$ this difficulty remains fixed even
if gravity becomes very weak. And, in this limit, it is fair to say that information is ``within'' the coal when it has not evaporated, and only emerges with the radiation.

In the case of black holes, it is impossible to take $M_{\text{pl}} \rightarrow \infty$ while keeping $S$ finite since the entropy itself scales as $(M_{\text{pl}} r_h)^{d-2}$. So if we attempt to make gravity weaker, which makes
it harder to extract information by making measurements near $u \rightarrow -\infty$,  this also increases the difficulty of reconstructing information by ``collecting'' the Hawking radiation. This is the key difference between black holes and ordinary evaporating objects: it is impossible to take a limit where the effects of quantum gravity become unimportant for evaporating black holes, whereas such a limit does exist for ordinary objects.

\paragraph{Penrose diagram.}
There has also been some debate in the literature about the precise form that the Penrose diagram for an evaporating black hole should take. For instance \cite{Ashtekar:2020ifw} explained that the Penrose diagram of Figure \ref{backreactpenrose} should be replaced with the modified  Penrose diagram of Figure 2 of \cite{Ashtekar:2020ifw}.  One of the arguments advanced is that the Penrose diagram of Figure \ref{backreactpenrose} suggests that the interior of the black hole
is clearly spacelike to $\scrip$. Therefore, this diagram, suggests that information is lost in the singularity.

We understand the discomfort with the Penrose diagram of Figure \ref{backreactpenrose}. In fact, this Penrose diagram was, as pointed out in  \cite{Ashtekar:2020ifw}, originally viewed as a {\em consequence} of Hawking's independent argument for information loss. But, since the diagram has now become so standard, it is now viewed as an argument {\em in favour} of information loss.

We do not enter into this discussion in this article. Our perspective is that the Penrose diagram is useful for classical {\em analyses} of how information propagates but should not be taken overly seriously at the quantum level.  
We have repeatedly stressed that in quantum gravity, information about a region may be available outside its light cone. 
In the context of Figure \ref{backreactpenrose}, a copy of the information in the interior is always available in the exterior. So even as the interior ``disappears'' (i.e ceases to exist after the singularity of Figure \ref{backreactpenrose}), this copy continues to be available near $\scrippast$, and so there is no loss of information. 

On the other hand, an appealing aspect of the diagram of Figure 2 of \cite{Ashtekar:2020ifw}  is that it makes it clear that the structure of asymptotic infinity is not changed. Since we  use the Penrose diagram only as a guide for building intuition, and not as a precise indicator
of how information is localized, there are some advantages to thinking in terms of such a diagram.  But, for the rest of the article, to avoid confusing the reader, we will continue to draw the Penrose diagram of Figure \ref{backreactpenrose}. We remind the reader to keep two important caveats in mind: (a) the structure of asymptotic null infinity is still that of Minkowski space (b) information on one part of a Cauchy slice may also be available on another part of the same slice.

\subsection{Small black holes in AdS}

It is also interesting to consider the  evaporation of a ``small black hole'' in AdS. This is a black hole whose horizon radius is much smaller than the cosmological scale. Such a black hole is expected to behave qualitatively similarly to a black hole in asymptotically flat space. In particular, it is expected to evaporate into a gas of particles, unlike large black holes in AdS, which do not evaporate.

The principle of holography of information, as we formulated it in \ref{holinfads}, tells us that information about the black-hole microstate is completely contained in any asymptotic time band of an infinitesimal width $[0, \epsilon]$. In section 
 \ref{secholography}, we restricted ourselves to excitations using asymptotic operators at arbitrary times. But such excitations can be used to create black holes \cite{Bhattacharyya:2009uu}, including small black holes. So our results apply to that setting.  It is also natural to consider a pure state since our AdS analysis was general and did not exclude massive particles. Now consider the algebra of observables in the time band $[t, t + \epsilon]$.  The arguments above suggest that the von Neumann entropy of a pure state with respect to this algebra is just $0$, which is the same curve as Figure \ref{figvn} except that it does not even have an intercept.

This is not surprising when considered from the point of view of AdS/CFT, especially when combined with the extrapolate dictionary \cite{Banks:1998dd}. The extrapolate dictionary tells us that CFT operators can be matched to the boundary limit of bulk operators. Since CFT operators on a given time slice clearly have all possible information about the state, it is clear that correlators of bulk operators near the boundary in a small time band also contain all information about the state.

\paragraph{\bf Difficulty of extracting information.} 
It is interesting to again compare --- in the context of a small black hole --- the relative difficulty of identifying the microstate using the quantum gravity effects that we have described,  and by simply waiting for the black hole to evaporate and ``collecting'' Hawking radiation. 

\begin{figure}[!ht]
\begin{center}
\includegraphics[height=0.5\textheight]{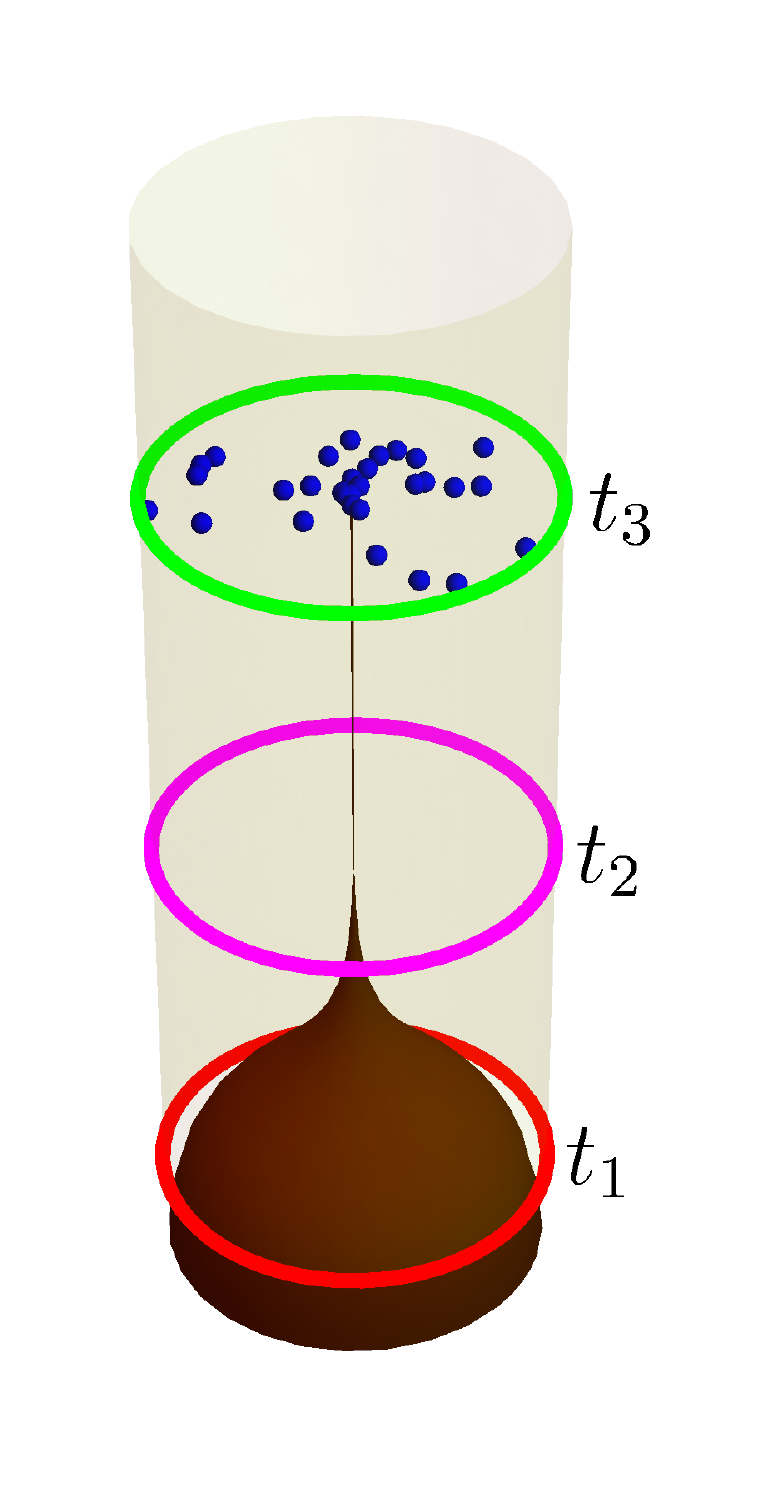}
\caption{\em A cartoon of a small black hole forming and evaporating in AdS. At time $t_1$, the black hole has not yet formed and the bulk state is well described by a cloud of dust. At time $t_2$, the black hole has formed whereas at $t_3$ it has evaporated completely into a gas of Hawking radiation. \label{figadsbh}}
\end{center}
\end{figure}

In Figure \ref{figadsbh}, we have shown a cartoon of a state that at time $t_1$ is well described by a gas of dust, which then collapses to a black hole at time $t_2$ and gradually becomes a gas of Hawking radiation again by time $t_3$.  Let us call this state, $|\Psi \rangle$.  We can now reconstruct the state by evaluating correlators of asymptotic operators in the vicinity of  any of the times $t_1, t_2, t_3$. For instance, if $O(\tau)$ is a boundary operator, then the set of all correlators of the form
\be
\langle \Psi | O(\tau_{1}) O(\tau_{2}) \ldots O(\tau_{n}) |\Psi \rangle,
\ee
with $\tau_i$ in the vicinity of $t_1$ give sufficient information to reconstruct the state. The same is true if $\tau_i$ are taken to be near $t_2$ or $t_3$.  If the reader prefers, this reconstruction can be phrased as a question purely within the boundary CFT. 

Of these three possibilities, it is clear that it is easiest to determine the state at time $t_1$. At this point of time, even simple correlators of the stress tensor and the fields that make up the dust may be sufficient to reconstruct the state without resorting to high accuracy. At time $t_2$, when the black hole has formed, we expect that it will still be possible to reconstruct the state but this will require the measurement of correlators with up to $\Or[S]$ insertions, and to an accuracy that scales exponentially with $S$.  This is an estimate of the difficulty required to extract information from the interior using quantum gravity effects in the exterior.

Now the important point is that at $t_3$, when the black hole has evaporated, it is still difficult to reconstruct the state. In fact, the entropy of the final Hawking radiation is {\em larger} than the black hole. If we call this entropy $S'$, with $S' > S$, the naive estimate of the difficulty of extracting information at time $t_3$ is that it will require correlators with up to $\Or[S']$ insertions with an accuracy that scales exponentially with $S'$. This means that it may be {\em harder} to reconstruct the state at time $t_3$, by collecting the radiation, than it is to reconstruct the state at time $t_2$ when the black hole is present.

This can also be seen by viewing the whole process from a CFT perspective. If the gravity theory is dual to a gauge theory, then the initial state is well described as en excitation of some energetic glueballs. At the intermediate time, $t_2$, the state has turned into a lump of quark-gluon plasma. At late times, $t_3$, the state has again evaporated into a number of glueballs.  But the second law of thermodynamics suggests that as time advances, it becomes more and more difficult to reconstruct the initial state. In other words, the state continues to thermalize and does not ``un-thermalize'' as the black hole evaporates.

The argument above is not watertight since small black holes are atypical states. And it is known that atypical states can sometimes violate generic considerations about the growth of complexity. So it may happen that, for such states, the  difficulty of reconstructing the initial state is lower at time $t_3$ than at time $t_2$. But our crude analysis at least strongly suggests that one cannot just assume that ``collecting'' the Hawking radiation provides an easier path to recovering information from evaporating black holes; it may, in fact, be easier to identify the black-hole microstate while the black hole exists.

\subsection{Revisiting the monogamy paradox \label{revisitmonogamy}}
We now turn to the monogamy paradox of section \ref{secmonogamy}. In this subsection we show that, even in the {\em absence} of black holes, ignoring the redundancy implied by the principle of holography of information leads to a similar paradox.  We will present a simplified version of the calculation in \cite{Raju:2018zpn}, and stress the main physical points. Let us consider a theory of gravity in AdS. This helps to avoid the  IR complications of flat space. Moreover, let us put the bulk theory in its vacuum state, $|0 \rangle$, about which we are confident that we understand the physics.  

\begin{figure}[!ht]
\begin{center}
\includegraphics[width=0.5\textwidth]{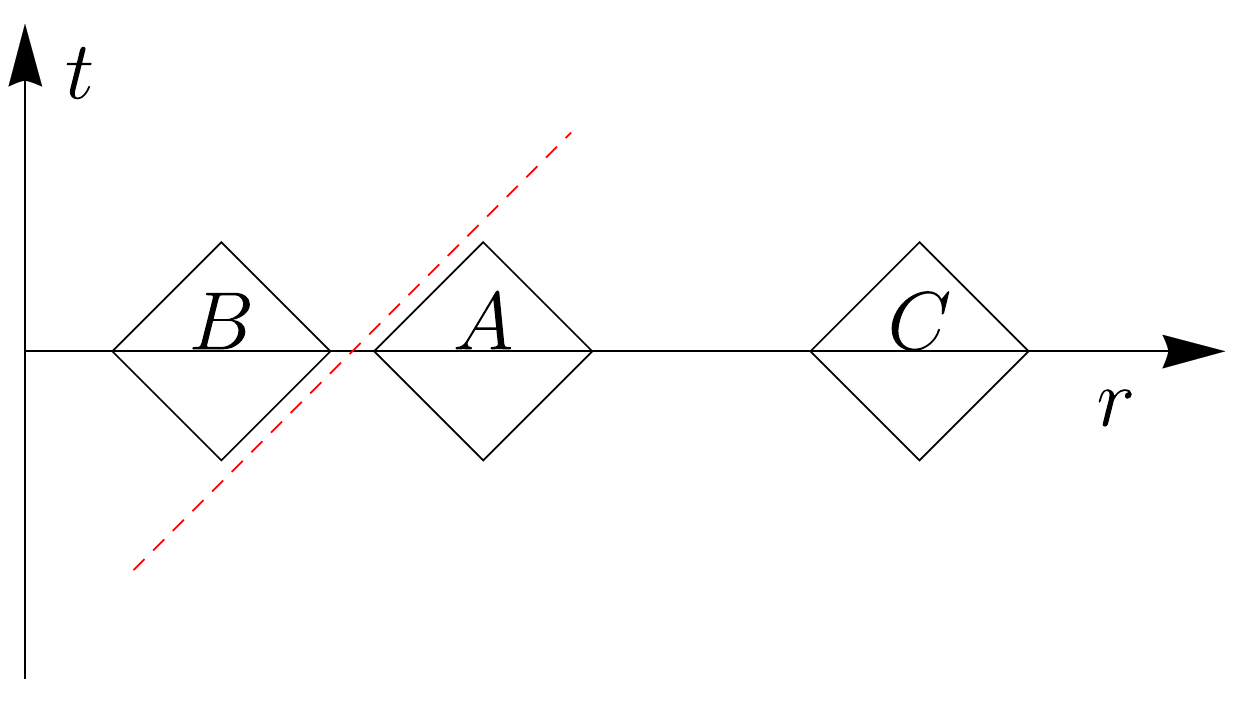}
\caption{\em The support of the operators A, B, C in section \ref{revisitmonogamy} in the r-t plane. A (d-1)-sphere has been suppressed at each point. This ``support'' only means that the gauge-fixed fields that enter these operators are located in these regions. In reality, the operators located in C can probe the information in A and B if they are cleverly constructed leading to a seeming paradox with the monogamy of entanglement. \label{monogsupport}}
\end{center}
\end{figure}

In the vacuum state, the classical metric in the bulk is simply
\be
\label{emptyglobalads}
ds^2 = -\left(1+r^2\right) dt^2  +{ dr^2 \over 1+r^2} + r^2 d\Omega_{d-1}^2 
\ee
Consider the $t = 0$ slice of the global AdS geometry, and consider a sphere at some radius $r_0$.  To make contact with the monogamy paradox of section \ref{secmonogamy}, let us consider a causal diamond just outside this sphere, which we denote by ``A'' and another causal diamond, just inside the sphere,  which we denote by  ``B''. We will use ``C'' to denote the region near the boundary. (See Figure \ref{monogsupport}.) Let us also consider the trajectory of an, imaginary, expanding spherical null shell that starts at $r_0$ at time $t = 0$. This trajectory  will play the role of the ``horizon''. 

 In section \ref{secoldinfo} we showed that if one considers a propagating scalar field then across {\em any} null surface, it is possible to define modes on the two sides that are entangled with each other. In this case by integrating the field within A and B one can get modes that we denote (using the notation of section \ref{secoldinfo}) by $\anh$ and $\tildanh$. There is also a well defined number operator corresponding to modes on both sides.
\be
N = \anh^{\dagger} \anh; \qquad \tilde{N} = \tildanh^{\dagger} \tildanh.
\ee
Here, when we refer to a field operator ``in A'', we are taking precisely the same perspective as in section \ref{secmonogamy}. This phrase simply means that, after a gauge-fixing or a relational procedure has been adopted, the coordinates that specify the operator are localized in this region.

We remind the reader that these modes implicitly depend on the choice of a frequency $\omega_0$. Denoting the eigenstates of $N$ and $\widetilde{N}$ by $|n \rangle_{A}$ and $|\tilde{n} \rangle_{B}$, 
we can further define operators that project onto these eigenstates or change the eigenvalue of $N$ or $\widetilde{N}$ by 1. (We hope that the notation will not lead the reader to confuse either $|0 \rangle_{A}$ or $|\widetilde{0} \rangle_{B}$ with the AdS vacuum.)
\be
\label{a1a2deftoymodel}
\begin{split}
A_1 = \sum_{n=0}^{\infty} \left( |2 n + 1 \rangle_{A} \, _{A}\langle 2 n + 1| - |2 n \rangle_{A} \, _{A}\langle 2 n| \right), \\
A_2 = \sum_{n=0}^{\infty} \left( |2 n + 1 \rangle_{A} \, _{A}\langle 2 n| +  |2 n \rangle_{A} \, _{A}\langle 2 n + 1| \right); \\
\widetilde{A}_1 = \sum_{\widetilde{n}=0}^{\infty} \left( |2 \widetilde{n} + 1 \rangle_{B} \, _{B}\langle 2 \widetilde{n} + 1| - |2 \widetilde{n} \rangle_{B} \, _{B}\langle 2 \widetilde{n}| \right), \\
\widetilde{A}_2 = \sum_{\widetilde{n}=0}^{\infty} \left( |2 \widetilde{n} + 1 \rangle_{B} \, _{B}\langle 2 \widetilde{n}| +  |2 \widetilde{n} \rangle_{B} \, _{B}\langle 2 \widetilde{n} + 1| \right); \\
\end{split}
\ee

We have use a bra-ket notation to define the operators above but this can be made precise as in equations \eqref{raiselowerdef} and \eqref{projectdef}. In fact the reader will recognize that these are very similar to the operators that appear in equation \eqref{a1a2def},  except that rather than eigenstates of global modes we are referring to eigenstates of the near-horizon modes.

Once again we set
\be
B_1 = \cos \theta \widetilde{A}_1 + \sin \theta \widetilde{A}_2  \quad \text{and}  \quad B_2 = \cos \theta \widetilde{A}_1 - \sin \theta \widetilde{A}_2,
\ee
and choose $\tan \theta = 2 {e^{-{\pi \omega_0}} \over 1 + e^{-{2 \pi \omega_0}}}$. All the operators, $A_1, A_2, B_1, B_2$ clearly have operator norm $1$ and so we can form a CHSH operator as in \eqref{cabdefn}, whose expectation value in the AdS vacuum is
\be
\label{maxentacrossvac}
\langle 0 | C_{AB} | 0  \rangle = {2 \over 1 + e^{-\pi \omega_0}} \left(1 + 6 e^{-{\pi \omega_0}} + e^{-2 \pi \omega_0} \right)^{1 \over 2}.
\ee

The easiest way to understand this result is to note that the relations \eqref{stateaction}, \eqref{commutmodes} and \eqref{samesidecorr} tell us that for correlators of simple operators made up of $\anh$ and $\tildanh$, one can just think of the state as being proportional to the  entangled state of two simple harmonic oscillators: $ \sum e^{-n \pi \omega_0} |n \rangle |\widetilde{n} \rangle$. But for the reader who desires a more careful treatment, we emphasize that the entire calculation of the CHSH correlators above can be done carefully in perturbative quantum field theory coupled to gravity in global AdS and we refer the reader to \cite{Raju:2018zpn} for details. The actual computations performed in \cite{Raju:2018zpn} were somewhat complicated precisely because every correlator was derived using the elementary Greens functions in empty AdS.  Provided gravity is weakly coupled, the result \eqref{maxentacrossvac} receives corrections that are proportional to the ratio of the Planck scale and the AdS scale but these are not relevant for the $\Or[1]$ results that we are interested in at the moment. Also, for simplicity, the paper \cite{Raju:2018zpn} kept only the terms with $n=0$ and $n=1$ above, but the techniques of that paper can be easily extended to evaluate the entire sum over $n$ as we have done here.

Now we can use the construction described in section \ref{secholography} to create a seeming paradox with the monogamy of entanglement. If we denote
\be
|B_1 \rangle = B_1 |0 \rangle; \qquad |B_2 \rangle = B_2 | 0 \rangle,
\ee
then recall that near the boundary of AdS, i.e. in region C, it is possible to find operators that have the property that
\be
Q_1 | 0 \rangle = B_1 |B_1 \rangle; \qquad Q_2 | 0 \rangle = |B_2 \rangle. 
\ee
(These operators are discussed further in \cite{Raju:2018zpn}.)
Recall that it is also possible to construct the projector on the vacuum from near the boundary of AdS as we described in detail above. By combining the operators $Q_i$, with $i=1,2$, with the projector on the vacuum, we can construct operators like
\be
|B_i \rangle \langle 0| = Q_i \projvac; \quad |0 \rangle \langle B_i| = \projvac Q_i^{\dagger}; \quad |B_i \rangle \langle B_i | = Q_i \projvac Q_i^{\dagger}
\ee

Now, lets take a specific linear combination of these operators
\be
\label{cicomb}
C_i = {\left(|B_i \rangle \langle 0| + |0 \rangle \langle B_i |  - \langle 0 | B_i | 0 \rangle |0 \rangle \langle 0|  - \langle 0 | B_i | 0 \rangle |B_i \rangle \langle B_i| \right) \over \langle 0 | B_i^2 | 0 \rangle - \langle 0| B_i| 0 \rangle^2}
\ee
where $i$ runs over 1 and 2. The point of taking this slightly complicated-looking linear combination that is that the {\em operator norm} of $C_i$ is now bounded.  In fact, $\|C_i \| = \langle 0 | B_i^2 | 0 \rangle \leq 1$.\footnote{As we emphasized many times above, the operators $Q_i$ exist independently of gravity. But they cannot be used directly inside a CHSH operator because they are unbounded. This is what ensures that it is impossible to obtain a monogamy violation in a LQFT. So, in this example, the significance of the projector on the vacuum is that it allows us to construct a pair of {\em bounded} operators in the region C that--- when applied to the vacuum --- produce the states $|B_1 \rangle$ and $|B_2 \rangle$ respectively.}  But we still have
\be
C_1 | 0 \rangle = |B_1\rangle \qquad C_2 | 0 \rangle = |B_2 \rangle.
\ee
Therefore it is clear that, with $C_{AC}$ defined as in \eqref{cacdefn},
\be
\langle 0 | C_{A C} | 0 \rangle = \langle 0 | C_{AB} | 0 \rangle.
\ee

Since $\langle 0 | C_{AB} | 0 \rangle > 2$, for any value of $\omega_0$ it is clear that
\be
\langle 0 | C_{AB} | 0 \rangle^2 + \langle 0| C_{AC} | 0 \rangle^2 > 8.
\ee
So the monogamy inequality is violated, even though the regions A,B,C are spacelike-separated.
This should hardly be a surprise by this point. The region near the boundary contains a redundant copy of the information in the interior and there are two ways to probe information about the bulk. One is by using simple quasi-local operators, which are the operators that comprise A,B. The other is by using gravitational effects from far away and the operators from C use such effects because they rely on the projector on the vacuum. If one neglects this redundancy it directly leads to a paradox.

The reader may possibly be concerned about the following subtlety.
\begin{enumerate}[qseries]
\item
{\em 
Even though the operators A and B were constructed by smearing field operators in certain geometric regions, they are not really ``local'' operators since they must be dressed to the asymptotic boundary to make them gauge invariant.  So why is there a paradox?}

A reader who has this question is absolutely right that there is no real paradox in the construction above. But this is also true of the monogamy paradox of section \ref{secmonogamy}. As we have emphasized repeatedly, since the Hilbert space does not factorize, operators in the interior of the black hole do not commute with operators in the exterior even though they may nominally be spatially separated. It is this phenomenon that underlies the principle of holography of information. So the construction above is simply meant to demonstrate that if one makes the error of forgetting this fact --- whether in the presence or the absence of black holes ---  one ends up with a paradox.
\end{enumerate}

\subsubsection{A monogamy paradox about black holes}
The paradox in empty space already shows that  neglecting the redundancy  implicit in the principle of holography
of information can lead to paradoxes. Using the construction of \ref{holinfads}, it is possible to construct a similar paradox about a black-hole microstate. However, an important difference with the empty-AdS scenario is that the black hole paradox necessarily requires the use of very complicated operators.  

The first part of the argument above goes through almost entirely unchanged. Given a black-hole microstate, $|\Psi \rangle$, let us adopt the same notation that we adopted in section \ref{corrnull}. We can then
construct operators in regions A (outside the horizon) and B (inside the horizon) so that (as in equation \eqref{maxentacross})
\be
\langle \Psi | C_{AB} |\Psi \rangle =  {2 \over 1 + e^{-\beta \omega}} \left(1 + 6 e^{-{\beta \omega}} + e^{-2{\beta \omega}} \right)^{1 \over 2}.
\ee

Let us again denote (through a slight abuse of the notation above),
\be
B_1 | \Psi \rangle = |B_1 \rangle; \qquad B_2 |\Psi \rangle = |B_2 \rangle.
\ee
Now the argument of section \ref{secholography} tells us that it is possible to find three asymptotic operators (in region C) that we denote by $Q, Q_{B_1}, Q_{B_2}$, and which have the property that
\be
Q | 0 \rangle = |\Psi \rangle; \qquad Q_{B_1} | 0 \rangle = |B_1 \rangle ; \qquad Q_{B_2} | 0 \rangle = |B_2 \rangle.
\ee
Clearly, all these three operators are very complicated operators. For all except a very simple class of microstates we cannot write them down explicitly and we only have an existence proof.  We can then also construct
operators
\be
\begin{split}
&|B_i \rangle \langle \Psi| = Q_{B_i} \projvac Q^{\dagger}; \qquad |\Psi \rangle \langle B_i| = Q \projvac Q_{B_i}^{\dagger}; \\
&|B_i \rangle \langle B_i | = Q_{B_i} \projvac Q_{B_i}^{\dagger}; \qquad |\Psi \rangle \langle \Psi| = Q \projvac Q^{\dagger}.
\end{split}
\ee
But using these operators we can against construct operators $C_1, C_2$ defined as
\be
C_i =  { |B_i \rangle \langle \Psi | + |\Psi \rangle \langle B_i |  - \langle \Psi | B_i | \Psi \rangle |\Psi \rangle \langle \Psi |  - \langle \Psi | B_i | \Psi \rangle |B_i \rangle \langle B_i|  \over  \langle \Psi | B_i^2 | \Psi \rangle - \langle \Psi | B_i| \Psi \rangle^2 }.
\ee
When these operators are inserted into the CHSH operator we again find
\be
\langle \Psi | C_{AC} |\Psi \rangle =  \langle \Psi | C_{AB} |\Psi \rangle
\ee
Since both the expectation values
\be
\langle \Psi | C_{AC} |\Psi \rangle = \langle \Psi | C_{AB} |\Psi \rangle > 2,
\ee
for any value of $\omega$, we clearly have a monogamy paradox about the black hole as well.

This analysis tells us that for the purpose of generating a monogamy paradox,  it is {\em not necessary} to go beyond the Page time. This is yet another indicator that the information is {\em always outside} a black hole. Just as in empty AdS, information can either be probed by complicated operators near infinity --- which are the operators that appear in $C$ above --- or by quasilocal field operators that appear in A or B.

\subsection{The Page curve and islands \label{secpage}}
There has been considerable recent interest in the computation of the Page curve of holographic CFTs coupled to nongravitational baths.  We refer the reader to \cite{Almheiri:2020cfm} and references there for a recent review.  Here, we would like to clarify some conceptual aspects of this computation and compare and contrast these results with the results that we have described above. We first start by clarifying the nature of the computations that have been performed. We then discuss some of the terminology that accompanies these computations. We then briefly describe some technical aspects of the computation, and then contrast these results with the story that we have described above.

\subsubsection{What has been computed}
Some of the initial computations of the Page curve \cite{Almheiri:2019hni,Almheiri:2019yqk,Penington:2019kki} were performed in $1+1$ dimensions, and the framework was later generalized to higher dimensions \cite{Almheiri:2019psy}. We will provide a broad description that, we believe, is applicable to all the precise computations that have been performed so far.

\begin{figure}[!ht]
\begin{center}
\includegraphics[height=0.4\textheight]{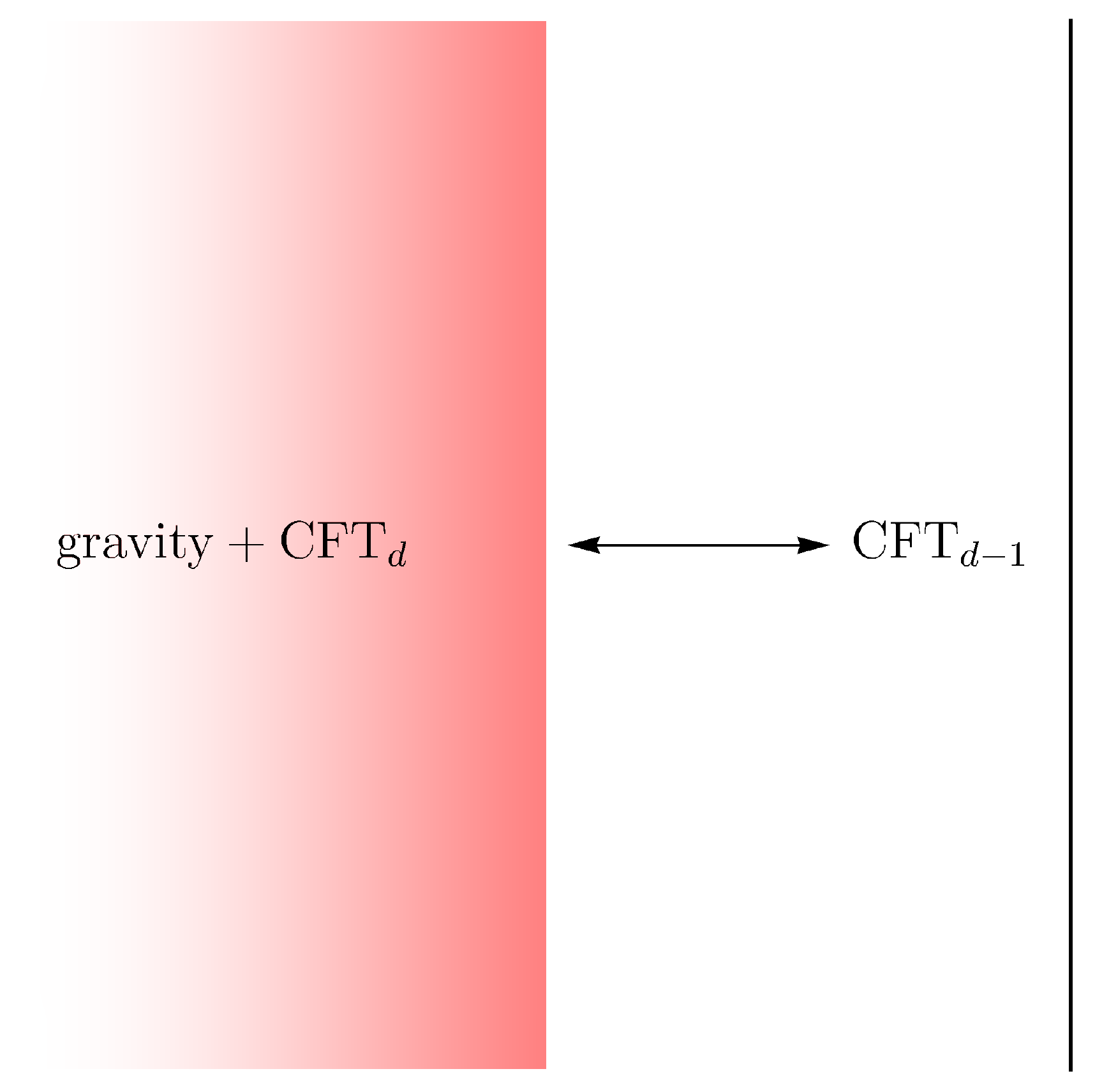}
\caption{\em We start by considering a theory of gravity in AdS$_d$, where the matter sector is itself a holographic CFT. The entire system has a dual description as a CFT$_{d-1}$. \label{figholographicrest1}} 
\end{center}
\end{figure}

Consider a theory of gravity in a $d$-dimensional asymptotically AdS spacetime, where the matter content itself is that of a holographic CFT \cite{Almheiri:2019psf}.   We will call this matter theory  ``CFT$_d$''. 
Since the entire system is asymptotically AdS, the standard arguments tell us that the combined gravity-matter system has a description in terms of a $(d-1)$-dimensional CFT that lives on the boundary of AdS$_d$. We will call this theory ``CFT$_{d-1}$.''  This reduction is displayed in Figure \ref{figholographicrest1}.

\begin{figure}[!ht]
\begin{center}
\includegraphics[height=0.4\textheight]{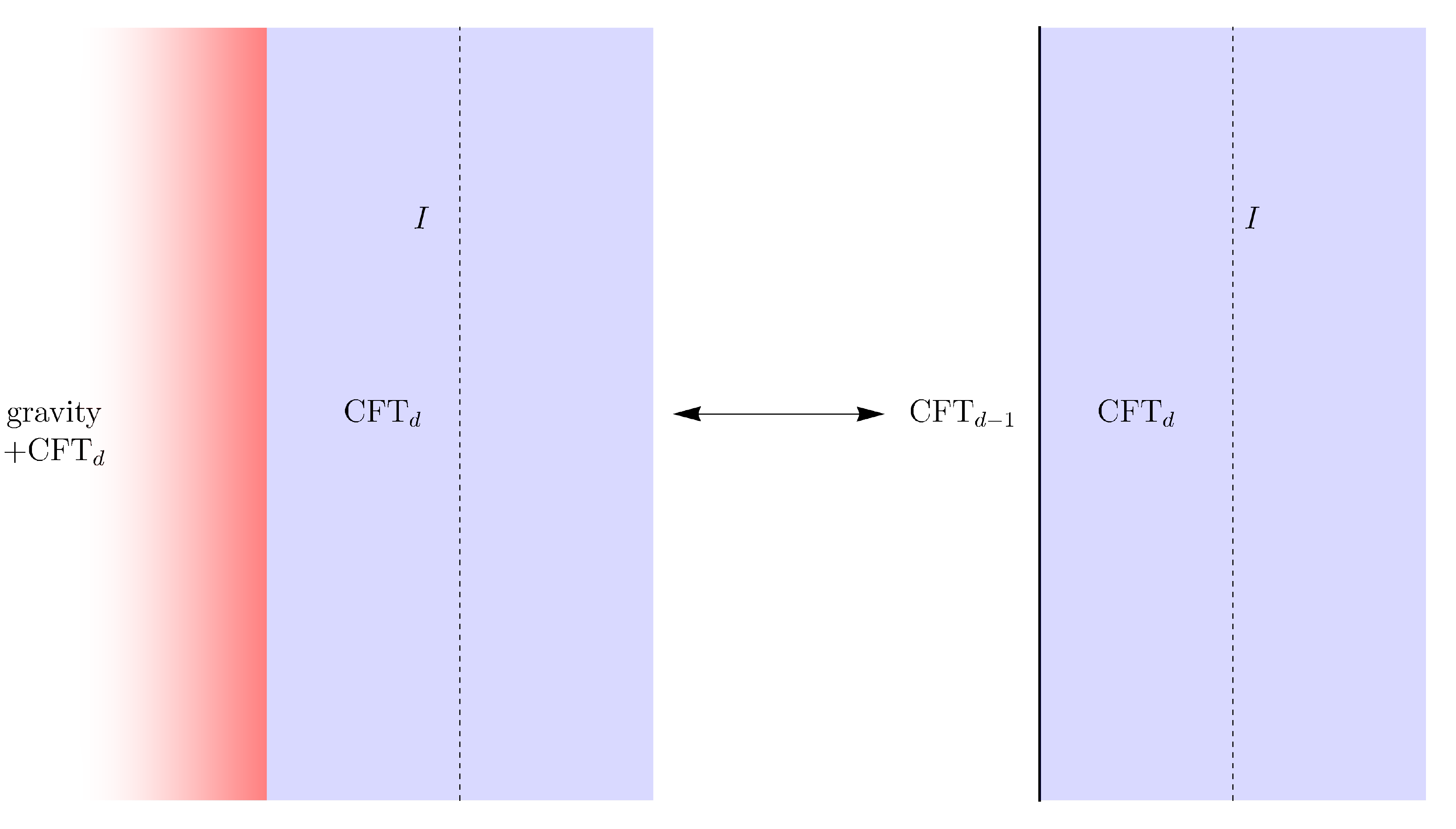}
\caption{\em We couple the CFT$_{d-1}$ to the CFT$_d$. So we now have a d-dimensional space that ends at a defect. This can also be described as coupling between the gravitational theory and the CFT$_d$ at the AdS boundary as shown on the left. We then divide the system into two parts across the imaginary interface I. \label{figholographiccoupled}} 
\end{center}
\end{figure}

Now, let us take this CFT$_{d-1}$ and couple it to another copy of CFT$_d$ in a fixed flat background that we will call the ``bath''. This can also be 
thought of a system where we have a CFT$_d$ coupled to gravity in AdS where the matter that lives in AdS is coupled near the AdS boundary to another copy of the matter theory on a  
flat background. The coupling is chosen to correspond to ``transparent boundary conditions'', which means that in the AdS picture we would like  matter excitations to propagate from AdS to flat space. This coupled system is shown in Figure \ref{figholographiccoupled}.  

This system has  has an entirely nongravitational description: the bath is a  $d$-dimensional quantum mechanical system that lives on a space that ends at a $(d-1)$-dimensional defect, where another nongravitational quantum mechanical system --- the CFT$_{d-1}$ --- lives.\footnote{Here we are using the word ``defect'' simply to mean that there are some additional degrees of freedom that are confined to a $(d-1)$ dimensional submanifold of the $d$-dimensional space.}
Now, let us divide this {\em nongravitational system} into two parts by placing an imaginary interface marked as ``I'' in  Fig \ref{figholographiccoupled}. So one part is the  union of the defect and some portion of the bath and the other part is the remainder of the bath.  Since these systems are nongravitational, it is safe to assume that the Hilbert space factorizes at the interface, I.

We can now consider almost any state with large entropy in this combined system. For instance, at $t = 0$,  we could place the initial defect CFT$_{d-1}$ in a highly excited state and the bath CFT$_d$ in its ground state. With the passage of time, energy will flow from the defect and across the interface, I, and go on to the asymptotic region in the bath. 
 By the very general arguments of section \ref{secavent}, we expect that the entanglement entropy between the defect and the rest of the system will obey a Page curve. This is the Page curve that has been computed, in all examples where a 
controlled and precise computation is possible.

\subsubsection{The terms ``black hole'' and ``radiation''}
The Page curve that we described above has been described as the entanglement entropy of the ``radiation'' with the ``black hole''. Of course, these are just names for subsystems that appear in the computation. But we feel that this terminology has caused confusion, since the words ``black hole'' and ``radiation'' are not used in their usual sense as we now explain.

If the original excited state of the defect CFT$_{d-1}$ had a dual description as a black hole then in the picture of Figure \ref{figholographiccoupled}, it looks like radiation from the black hole is entering the bath and crossing the imaginary interface, I. 
 This is the reason that \cite{Almheiri:2019hni,Almheiri:2019yqk} termed the subsystem to the left the ``black hole'' and the subsystem to the right ``the radiation.''   

But, one should remember that the precise division of the system into a ``black hole'' and ``radiation''  is always made across  the interface I in the nongravitational region, as depicted in Figure \ref{figholographiccoupled}. Therefore, at no point does the subsystem called the ``black hole'' describe only the degrees of freedom inside the black hole. Similarly, at no point does the subsystem called the ``radiation'' describe all the degrees of freedom in radiation.  Rather, even at $t = 0$, the subsystem called the ``black hole'' comprises the black hole,  {\em plus} the radiation extended all the way out to asymptotic infinity in AdS$_d$ and also some part of the nongravitational bath.\footnote{At later times, the bulk degrees of freedom corresponding to what is called the ``black hole'' and the ``bath'' are more subtle. The appearance of islands tell us that, at late times, the interior of the black hole in the bulk may really be described by operators from the ``bath''. And the system called the ``black hole'' may describe only part of the region outside the horizon, and either part or none of the region inside the horizon.}

This may seem like a technicality but it is important. In the perspective that we have advocated in previous sections, at $t = 0$, information about the black hole {\em already} resides in the operators near the asymptotic AdS$_d$ boundary. When the system is coupled to the nongravitational bath, part of the information moves from this asymptotic AdS$_d$ region into the bath. Therefore, the Page curve  never measures
the information ``emerging'' from the horizon; it is consistent with the idea that information about the black-hole interior is always available outside,  and it only tells us  about information transfer between two nongravitational regions outside the horizon.

It is also incorrect to think of the Page curve as telling us about the information emerging from some large but finite ball that contains the black hole but is in the region of spacetime where gravity is dynamical.  Indeed, if one were to attempt to move the interface I  even slightly into the region with dynamical gravity,  then we expect that the Page curve for the entropy of the ``radiation'' would immediately vanish. The principle of holography of information would then apply and tell us that the information in the region enclosed by I would then also be available outside I, and so the entropy of the region outside would always be 0.

For these reasons, we feel that that it is somewhat misleading to suggest that the Page curve that has been computed tells us about the information content of the radiation or its entanglement with the black hole.
It seems best,  from a conceptual point of view, to understand these computations from a boundary perspective. Here, we are simply computing the entanglement between two parts of an ordinary quantum mechanical system. This is in the spirit of many previous computations that have been carried out by applying the AdS/CFT duality to nongravitational systems including condensed matter systems. We ask a question that is conceptually best defined on the boundary and is not necessarily a particularly natural question to ask directly in the bulk. Then we use a holographic dual for the boundary theory only as a  calculational tool to answer the well defined boundary question. The same question can also be answered directly in the boundary theory \cite{Sully:2020pza} without making reference to the bulk, although the bulk computation may be calculationally simpler.

\subsubsection{Islands \label{secislands}}

We now briefly review the calculational techniques used in these analyses, that lead to the emergence of ``islands'' in the entanglement wedge.
Let us call the region to the right of the interface, $I$, as shown in Figure \ref{figholographiccoupled}, $R$.  Then, in $d = 1$, it is possible to carefully compute the entanglement entropy of $R$ using the replica trick in the Euclidean path integral \cite{Almheiri:2019qdq,Penington:2019kki}. The analysis is similar to the analysis of \cite{Lewkowycz:2013nqa}. The final result is an ``island rule'' for computing the entropy of the region $R$. The island
rule is that to compute the entropy of $R$, we should consider all possible ``island'' regions in the bulk gravitational region and compute
\be
\label{islandformula}
S(R) = \text{min}\left[ \text{ext} \left[{A(\partial(\text{island})) \over 4 \gnewt} + S_{\text{bulk}}(R \cup \text{island}) \right]\right].
\ee
Here we are instructed to look for extrema of the expression in the inner bracket, and take the global minimum of all such extremal configurations.
The entropy that appears on the right hand side above  is a free-field or a ``coarse-grained'' entropy of the quantum fields. In the time-dependent state that we have been considering, where we start with energy in the
gravitational region that flows into the bath,  the island emerges at late times because this bulk-entropy term  starts to compete with the area term, and moreover its variation under small deformations cancels the variation of the area term allowing the boundary of the island to be an extremal surface. So although the boundary of the island is expensive in terms of area, it may still be the minimal extremal surface  if its area contribution is offset by the reduction
that the island causes in $S_{\text{bulk}}$. 

Although the island rule can be derived independently, it can also be thought of as a generalization of the quantum extremal surface prescription in AdS/CFT \cite{Engelhardt:2014gca}. This prescription generalizes the Ryu-Takayanagi (RT)  \cite{Ryu:2006bv,Ryu:2006ef} and  Hubeny, Rangamani and Takayanagi  (HRT) \cite{Hubeny:2007xt} prescription for the entanglement entropy, $S(B_R)$, of a region, $B_R$, on the boundary of AdS  by including leading bulk quantum effects. The rule is that
\be
\label{qesformula}
S(B_R) = \text{min}\left[ \text{ext} \left[{A \over 4 \gnewt} + S_{\text{bulk}} \right]\right].
\ee
Here $A$ is the area of a surface in the bulk that is anchored at the boundary region $B_R$ and $S_{\text{bulk}}$ is the free-field entropy of quantum fields in the region between $B_R$ and the extremal surface. This surface must satisfy the ``homology constraint'' \cite{Headrick:2014cta,Headrick:2013zda}, which means that the region between the extremal surface and $R$ must have no {\em other} boundaries.   If there are multiple such surfaces, we pick the global minimum of the right hand side of equation \eqref{qesformula}. This formula can again be justified using a replica trick on the boundary \cite{Faulkner:2013ana}. The smallest bulk causal diamond that contains the region between the minimal surface and $B_R$ is called the {\em entanglement wedge} of $B_R$.

It is not, a priori, clear how one should apply equation \eqref{qesformula} to the setup of Figure \ref{figholographiccoupled}, where the region R is disconnected from the gravitational bulk. The island rule \eqref{islandformula} tells us that we can still use the prescription of \eqref{qesformula} but, while computing extremal surfaces,  we should also include disconnected surfaces in the gravitational bulk.

While the island rule and its derivation using replica wormholes in $d = 1$ is very interesting, we would like to describe, in a little more detail, how islands can be understood in higher dimensions. The key point, explained in \cite{Almheiri:2019hni} based on \cite{Almheiri:2019psf} is that we could take  the  setup shown in Figure \ref{figholographiccoupled}, to be {\em doubly holographic} by taking the $\text{CFT}_{d}$ to be itself a holographic theory. Then --- as first pointed out by Karch and Randall \cite{Karch:2000ct,Karch:2000gx} (see also \cite{Aharony:2003qf,Takayanagi:2011zk} for more details) --- it is believed that the entire system  has a dual gravitational description in terms of gravity propagating on a AdS$_{d+1}$ manifold that ends on a  Randall-Sundrum brane \cite{Randall:1999ee,Randall:1999vf} whose worldvolume is itself an AdS$_d$ manifold. This dual gravitational description geometrizes the quantum effects of the CFT.

Thinking in terms of this picture also led to the recognition \cite{Geng:2020qvw} that there is an interesting subtlety with the description in terms of the theory of gravity coupled to a CFT$_d$ shown on the left side of Figure \ref{figholographiccoupled}: since the stress tensor on the ($d-1$)-dimensional boundary is no longer conserved due to the coupling with the bath, it is free to pick up an anomalous dimension \cite{Aharony:2006hz}. Consequently, the graviton in the $d$-dimensional bulk is {\em massive}. Since the CFT$_d$ has a large number of degrees of freedom compared to the graviton, it is believed that the value of the graviton mass should not affect entanglement entropy computations. But, as emphasized in \cite{Geng:2020qvw}, it is not clear whether the lessons learned in this setup can be straightforwardly extrapolated to massless gravity.

Apart from this subtlety, this picture is rather general, and should apply, in principle,  to the time-dependent state that we discussed above, where we start with energy in the defect that flows into the bath. But, in practice,  it is difficult to find solutions with a brane embedded in a time-dependent geometry.  For this reason, the authors of  \cite{Almheiri:2019psy}  considered an eternal black hole in AdS$_{d+1}$ containing a brane. For a general brane configuration,  the metric for even this geometry cannot be written down analytically, and was found numerically for $d = 4$ in \cite{Almheiri:2019psy}. A specific configuration ---  where the brane is ``perpendicular'' to the boundary --- and where the metric can be analyzed explicitly was studied in \cite{Geng:2020qvw}.
  Islands can also be studied without black holes at all \cite{Chen:2020uac}.

Since the eternal black hole has two asymptotic regions, the boundary dual to this system comprises two copies of the setup described above. On each boundary, we have a $(d-1)$-dimensional theory that describes the brane coupled to a $d$-dimensional conformal field theory. At $t = 0$, the whole system is entangled, in a thermofield doubled state, with a copy of itself on the other asymptotic boundary. We will use the adjectives ``left'' and ``right'' to refer to
these copies. As usual, we examine the entanglement entropy between two regions on the boundary. The region  $R$, that we are interested in is now the union of some part of the $d$-dimensional theory on the right with some part of the $d$-dimensional theory on the left. The complement of this region, which we will call $\widetilde{R}$, is a union of four parts: the left defect theory, the right defect theory,  and parts of the left and the right $d$-dimensional theories.

\begin{figure}[!ht]
\begin{center}
\includegraphics[width=0.8\textwidth]{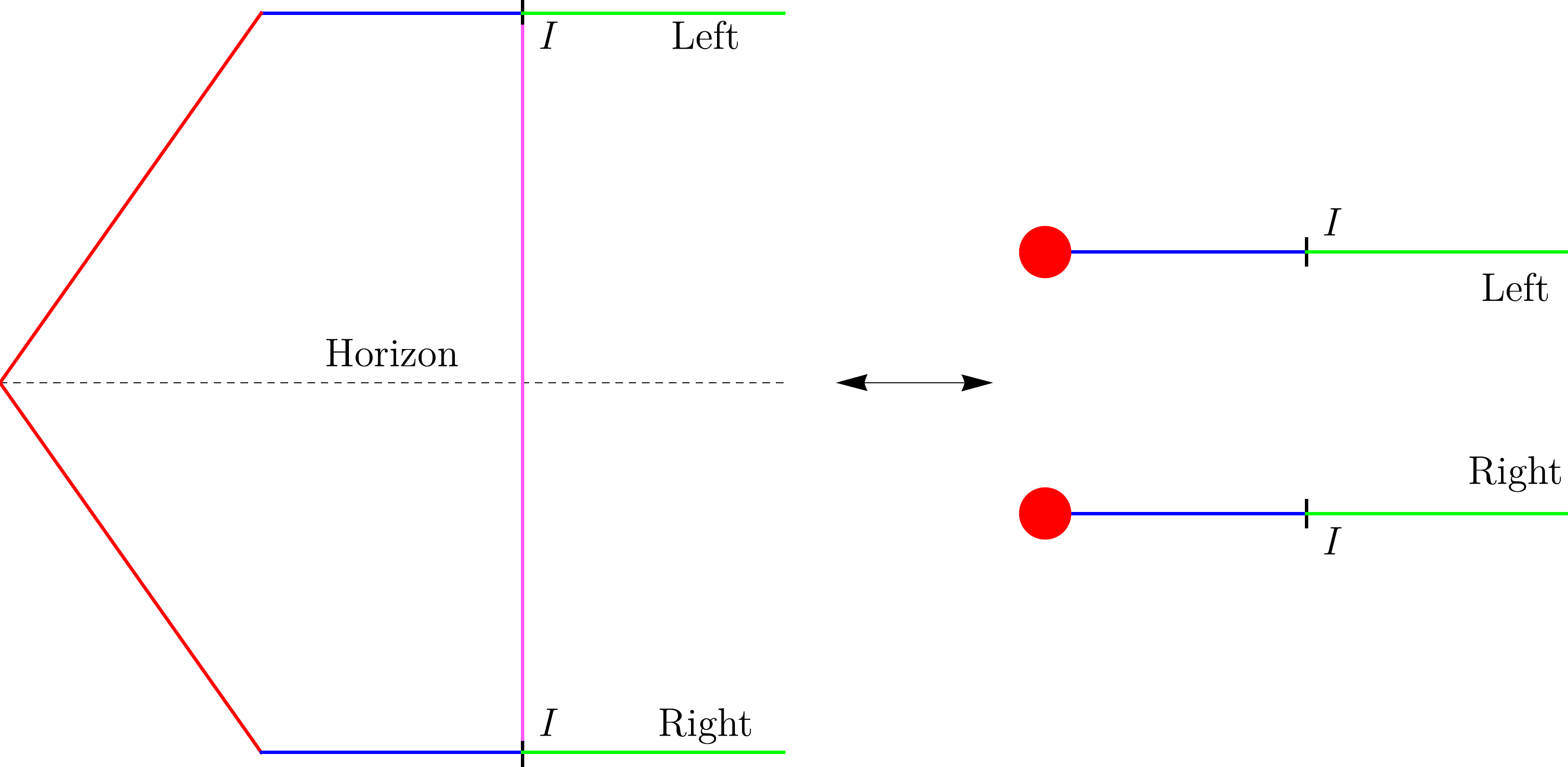}
\caption{\em We consider a thermofield doubled state of  a CFT$_{d-1}$ coupled to a CFT$_d$. The left and right systems (shown on top and bottom in the figure) are decoupled but entangled. This state has a dual description as a $d+1$-dimensional AdS spacetime,  with an AdS$_d$  brane (marked in red). The problem is to find the entropy of the union of the green regions on the top and the bottom boundary, which is denoted by $R$. A standard minimal surface calculation tells us that at $t = 0$, this is computed by a surface (purple curve) that runs from one boundary to the other through the bifurcation point. \label{rsinit}}
\end{center}
\end{figure}

The entanglement entropy between these regions, $S(R)$, can be computed using the standard RT/HRT formula in the bulk. So we simply use
\be
\label{rtformula}
S(R) = \text{min}\left[ \text{ext} \left[{A \over 4 \gnewt}  \right] \right] + \Or[1]
\ee
In this setup, the quantum corrections that appeared in equation \eqref{qesformula} are not important because the bulk geometry already captures them.

Near $t = 0$ we expect that the correct minimal surface is the one that cuts through the horizon and connects the part of $R$ on the left boundary with the part of $R$ on the right boundary as shown in Figure \ref{rsinit}. This is similar to the surface that was found in \cite{Hartman:2013qma}. However, the volume of this surface starts to grow with time. This feature can be seen even in the standard eternal black hole geometry. Maximum spatial volume slices that cut through the black-hole interior, and are anchored at the same value of $t$ on both sides, stretch with time due to the geometry of the interior. If this growth were to continue indefinitely, it would lead to a paradox. The region $\widetilde{R}$ is a finite region and it cannot have unbounded entanglement with any other system.  

\begin{figure}[!ht]
\begin{center}
\includegraphics[height=0.4\textheight]{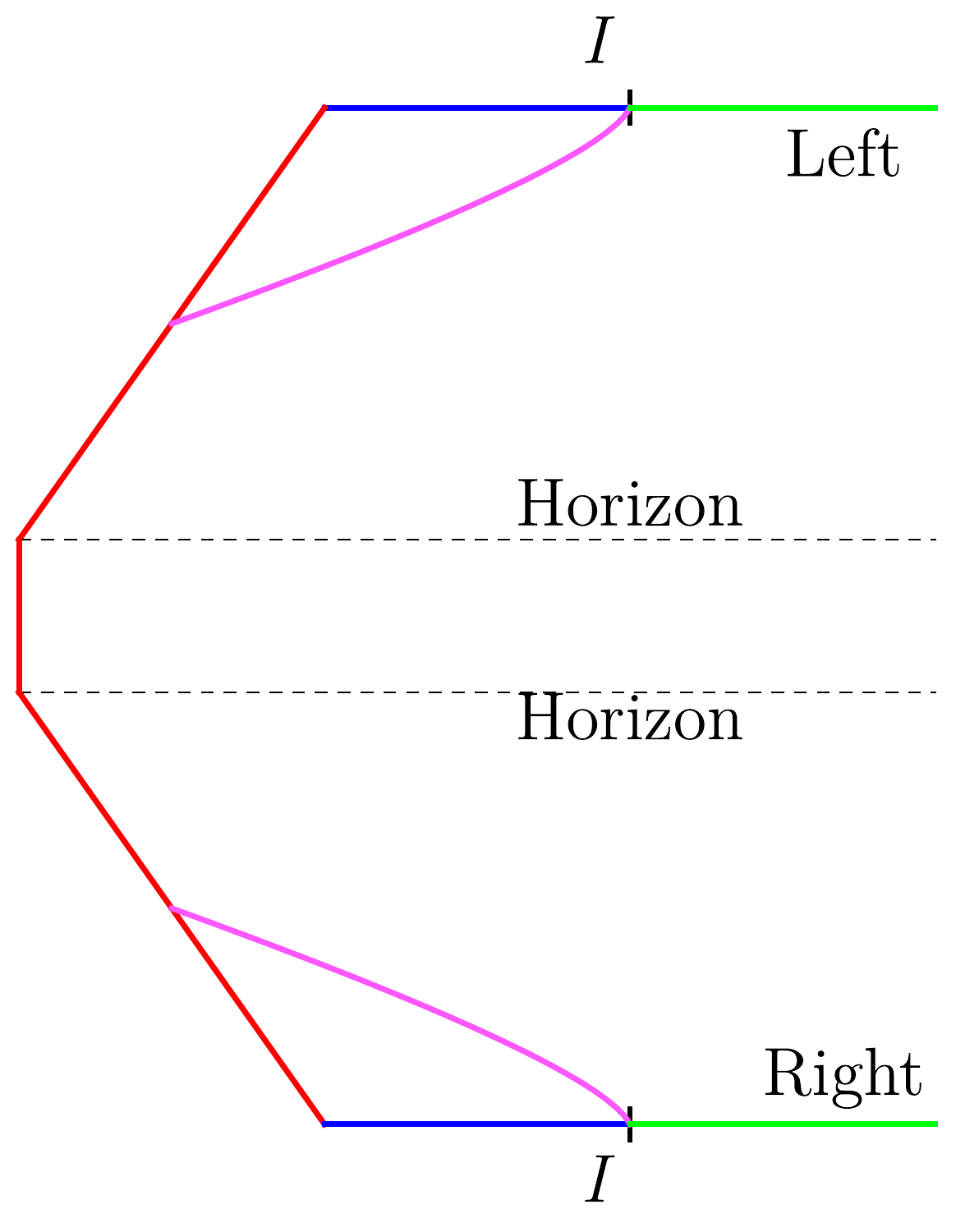}
\caption{\em If we evolve the system with time, then the original RT surface stretches for a while but is eventually beaten by a new RT surface shown in the figure. This is the phase transition at the heart of the ``Page curve.'' \label{rtisland}}
\end{center}
\end{figure}

The paradox is resolved because, at later times, a new surface of smaller area emerges.  The surface starts at the interface between $R$ and $\widetilde{R}$ and ends on the brane. This is the ``island'' surface. This surface is shown in Figure \ref{rtisland}. 

\begin{figure}[!ht]
\begin{center}
\includegraphics[width=0.7\textwidth]{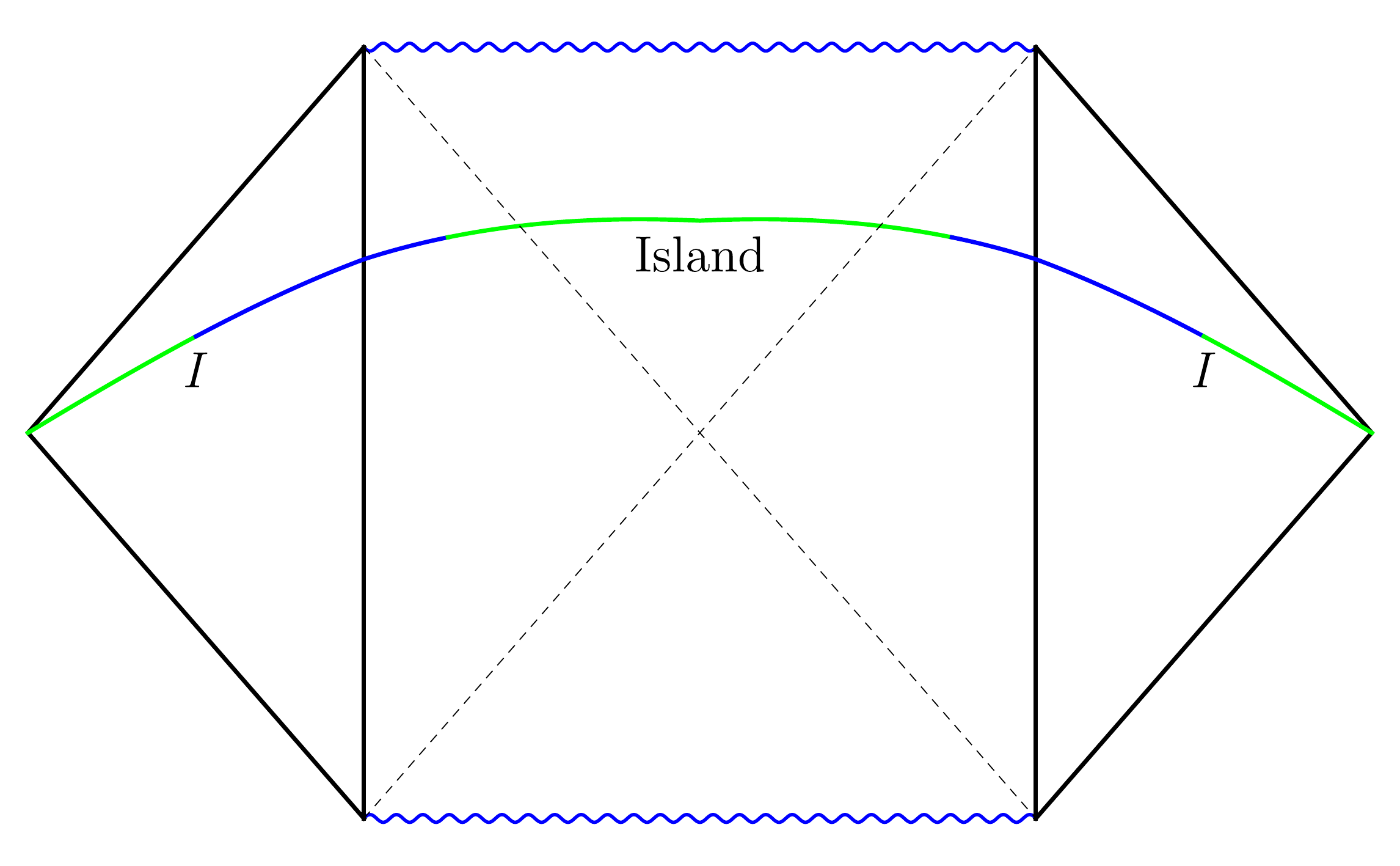}
\caption{\em A Penrose diagram showing the  same setup as Figure \ref{rtisland}. The island surface of Figure \ref{rtisland} tells us that the entanglement wedge of the green regions in the exterior includes a spatially disconnected region (also marked in green) in the middle of the Cauchy slice that runs through the geometry. \label{dualisland}}
\end{center}
\end{figure}

The term ``island'' can be understood by looking at the same picture from a  $d$-dimensional perspective, where only the brane and the boundary are plotted and we suppress the rest of the AdS$_{d+1}$ as shown in Figure \ref{dualisland}. In this picture, we see that the late-time entanglement wedge of R contains I, which is spatially disconnected from R. It is generally believed in AdS/CFT that operators in any region on the boundary describe the bulk degrees of freedom in its entanglement wedge \cite{Czech:2012bh,Dong:2016eik}. This tells us that the degrees of freedom in R also describe physics in a spatially disconnected island region.

\paragraph{\bf Gravitating baths \\}
It was recently pointed out in \cite{Geng:2020fxl} that the Karch-Randall setup described above allows for an elegant method of introducing gravity in the bath: one simply adds a second $AdS_{d}$ brane to the setup \cite{Kogan:2000vb}. This can be visualized as moving the boundary in Figure \ref{rsinit} into the bulk so that the bulk spacetime is now a ``wedge'' between two branes.   Then \cite{Geng:2020fxl} found that, for the purpose
of computing a fine-grained entropy, one can no longer meaningfully divide the degrees of freedom in the bath across an arbitrary interface,  as can be done when the bath is nongravitational.   Consequently, there is no analogue of the ``black hole''/``radiation'' Page curve in the presence of a gravitating bath. This is precisely what the arguments presented above would lead us to expect.

However, the entire system of two gravitating branes has a description in terms of a nongravitational CFT$_{d-1}$ \cite{Akal:2020wfl}. It turns out that it is possible to divide the internal degrees of freedom of this nongravitational CFT$_{d-1}$  into two specific subsystems, so that the entanglement entropy between these subsystems follows a Page curve when the brane configuration satisfies some geometric constraints. This entropy, and the associated Page curve, can be computed using the formula \eqref{rtformula} and, once again, islands play an important role. This is also consistent with our explanation above:  the Page curve is the answer to a well-defined 
question about a {\em nongravitational} system that can be obtained by analyzing  minimal-area surfaces in a gravitational dual.

\subsubsection{Contrast with the holography of information}
\begin{figure}[!ht]
\begin{center}
\includegraphics[height=0.5\textheight]{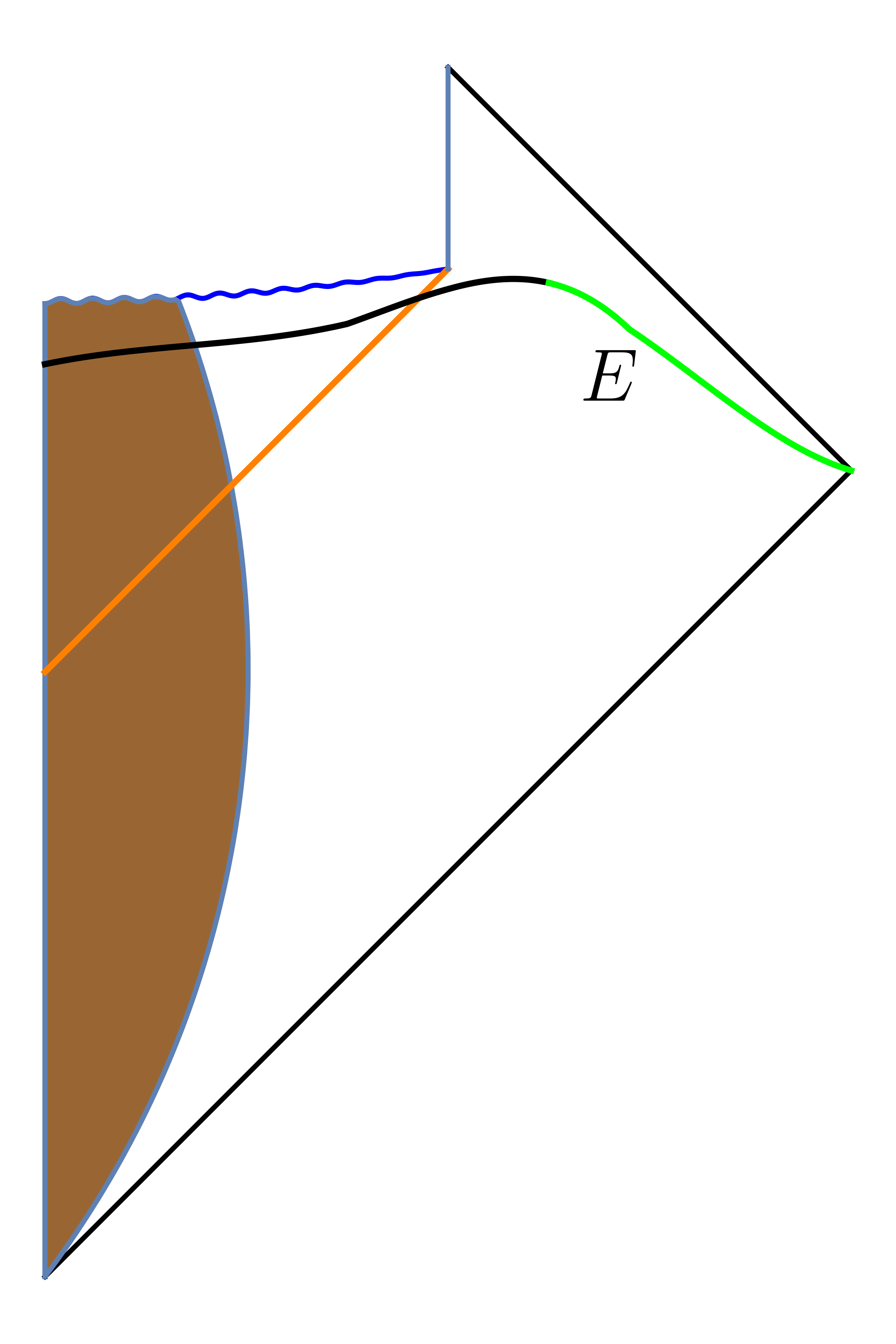}
\caption{\em Some papers in the literature suggest that one should apply the quantum extremal surface prescription to compute the entropy of the green region marked E. But this prescription cannot be justified since gravity is dynamical at the interface between $E$ and its complement. Our analysis suggests that the prescription leads to incorrect answers for the fine-grained entropy. \label{randomdivision}}
\end{center}
\end{figure}

To the extent that the recent developments compute the time dependence of the entropy of two parts of a nongravitational boundary using holographic techniques, we believe that they are both interesting and important. However, some papers in the literature have  attempted to apply the quantum extremal surface prescription to black holes in asymptotically flat space \cite{Krishnan:2020oun,Hashimoto:2020cas,Gautason:2020tmk,Alishahiha:2020qza}. For instance, the Penrose diagrams of   \cite{Almheiri:2020cfm} appear to suggest that one should consider an evaporating-black-hole spacetime as shown in Figure \ref{randomdivision}, and simply divide the spacetime into two parts: say, the exterior of a large ball of radius which we call E, and the interior. Then it is proposed in \cite{Almheiri:2020cfm} that we simply apply an analogue of equation \eqref{islandformula} to compute the entanglement entropy of E using an island rule
\be
\label{qesformulaflat}
S(E) = \text{min}\left[ \text{ext} \left[{A(\partial \text{island}) \over 4 \gnewt} + S_{\text{bulk}}(E \cup \text{island}) \right]\right],
\ee
where the island can be inside the horizon.
But note an important difference between equation \eqref{qesformulaflat} and equation \eqref{islandformula}. In equation \eqref{qesformulaflat}, the region E whose entropy we are attempting to compute is a region with dynamical gravity, which is not true in equation \eqref{islandformula}. 

In fact the quantum extremal surface formula \eqref{qesformula} that underpins  the ``island rule'' itself applies only when we are computing the entanglement entropy of
a part of the boundary in a nongravitational theory. There is no similar formula that can be applied to compute the entanglement entropy of some arbitrary region in a spacetime with dynamical gravity.

Once one assumes a formula of the form \eqref{qesformulaflat}, and allows the two terms on the right hand side to compete, the rest of the story of the Page curve follows. At later times, as $S_{\text{bulk}}$ increases, the minimal surface prescription demands that this growth be cut off.  This can be done by finding an island in the black-hole interior.  

The difficulty is not just that the formula \eqref{qesformulaflat} cannot be justified carefully. The difficulty is more serious because the analysis of sections \ref{secholography} and \ref{secresolveinfo} tells us that formula \eqref{qesformulaflat} is {\em incorrect} if $S(E)$ on the left hand side is the fine-grained gravitational entropy of the region E. As we have explained, in a region with dynamical gravity, information about the interior of a ball is always contained in its exterior. Therefore we expect that if one makes a division on a Cauchy slice as shown in Figure \ref{randomdivision}, the fine-grained entropy of E is zero and is {\em not} given by \eqref{qesformulaflat}. 

As we have emphasized above, the fact that formula \eqref{qesformulaflat} is incorrect can be seen already in perturbation theory. At early times, before the black hole has evaporated, there is no island in the black-hole interior. So if formula \eqref{qesformulaflat} and the associated entanglement-wedge conjecture had been correct, one would have expected to find an algebra of observables in the black-hole interior that commute with operators in $E$. But as we emphasized  in section \ref{resolvemonogamy} since the Hilbert space does not factorize, it is impossible  find even a {\em single} operator in the interior that commutes with all operators in the region $E$. Moreover, as we emphasized in section \ref{morelowenergy}, one can read off some features of an excitation in the interior even in perturbation theory by considering two point functions of the Hamiltonian with other dynamical fields.

Sometimes the formula \eqref{qesformulaflat} is justified using the following words:  
\begin{quote}
``If we go very far from the black-hole horizon, the effects of  gravity become weak. So it may be justified to treat the division between E and 
its complement as if gravity is simply non-dynamical in the region $E$.''
\end{quote}
We do not find this reasoning persuasive. It is true that, for simple observables, like low-point correlators, the effects of weak gravity can be ignored.  But 
the entanglement entropy is a fine-grained quantity that is sensitive to exponentially small effects. For such fine-grained quantities, it does not seem to be true that the effects of weak gravity can be neglected.

Apart from all the evidence for this conclusion that we have provided in the previous sections,  some additional evidence comes from applying the island rule itself. In these examples, as we explained above, at late times the  interior of the black hole is described by degrees of freedom that are very far away.  This is obviously a gravitational effect since such an effect would be absurd in a local quantum field theory. This itself tells us that the effects of weak gravity can be dramatic and extend over long distances for some questions.  Therefore, the island rule itself provides us with a warning that it should not be applied to black holes in asymptotically flat space, or other evaporating black holes in spacetimes  where gravity is everywhere dynamical.

\subsubsection{Is there a Page curve in flat space? \label{secispageflat}}
Can formula \eqref{qesformulaflat} be rescued so as to recover a Page curve in flat space?

Perhaps by restricting the set of observables in E in some way, one could come up with an algebra whose associated entropy obeys \eqref{qesformulaflat}. A natural guess might be to discard the ADM energy  --- which is an integrated component of the metric and clearly lives in the region E --- from the set of observables in E. So one could simply declare, by fiat, that the operator corresponding to the integral of a particular component of the metric cannot be measured, whereas other correlators of the metric can be measured.  But this is not so straightforward when one takes the region E to be the exterior of a large ball. Since gravity is still dynamical, in general, even if one removes a component of the metric from the algebra initially, this metric component reappears in the OPE of other operators.

There is one exception to this rule, and this arises at null infinity. It is simplest to restrict to four spacetime dimensions, where the asymptotic algebra at null infinity is best understood. If we interpret E to be the segment $(-\infty, u)$ of future null infinity, then one could choose to simply throw out the  Bondi mass aspect at each cut from the list of observables. This automatically also removes  the ADM Hamiltonian and the  supertranslation charges from the
set of observables.  This still leaves a large algebra at null infinity, which comprises the other components of the metric, and other dynamical fields including massless scalars and gauge fields. The algebra restricted to these remaining operators at one cut commutes exactly with the restricted algebra  at another cut. (For instance, the reader can consult equation 2.11 of \cite{Laddha:2020kvp}.) Moreover, this operator algebra is a trivial algebra, corresponding to operators in a Fock space. So the Bondi mass aspect does not reappear in the algebra once it has been discarded.   A reasonable hope is that if one computes the entanglement entropy of the segment of null infinity shown in Figure \ref{figvn}, with respect to this restricted algebra, it will obey a Page curve.\footnote{The possible obstruction is that 
this restricted algebra of the news and other dynamical fields does not have any information about the entanglement between the hard and soft degrees of freedom. The conventional Page curve in four dimensions will only emerge  if this hard-soft entanglement does not contain much information and is unimportant for computations of the von Neumann entropy. It appears to be commonly believed that it is indeed the case that the hard-soft entanglement does not carry much
information in a generic state but we are not aware of any quantitative check of this belief.} 

If this is the quantity that $S(E)$ on the left hand side of \eqref{qesformulaflat} is computing, then it is not a fine-grained entropy since it does not keep track of the component of the metric that is sensitive to the energy. But it is not a coarse-grained entropy since it keeps track of exponentially small correlations in some sectors. So it would appear to be some kind of ``intermediate'' entropy, between these two.

Moreover if this is what $S(E)$ is supposed to mean, it would appear to be a somewhat artificial quantity from a physical viewpoint. For instance, it would be natural to allow observers in E to measure physical quantities like the fluctuations in the Riemann tensor. But these fluctuations involve all components of the metric including the Bondi mass aspect.  So to observe a Page curve, the observers would have to ``close their eyes'' to this information, and keep track of other components  of metric. This may be mathematically well defined. But, the physically correct statement would then be that information about the black hole is always available outside, although there is a well-defined procedure for throwing away part of that information at each stage so as to recover a Page curve.

\begin{figure}[!ht]
\begin{center}
\includegraphics[height=0.5\textheight]{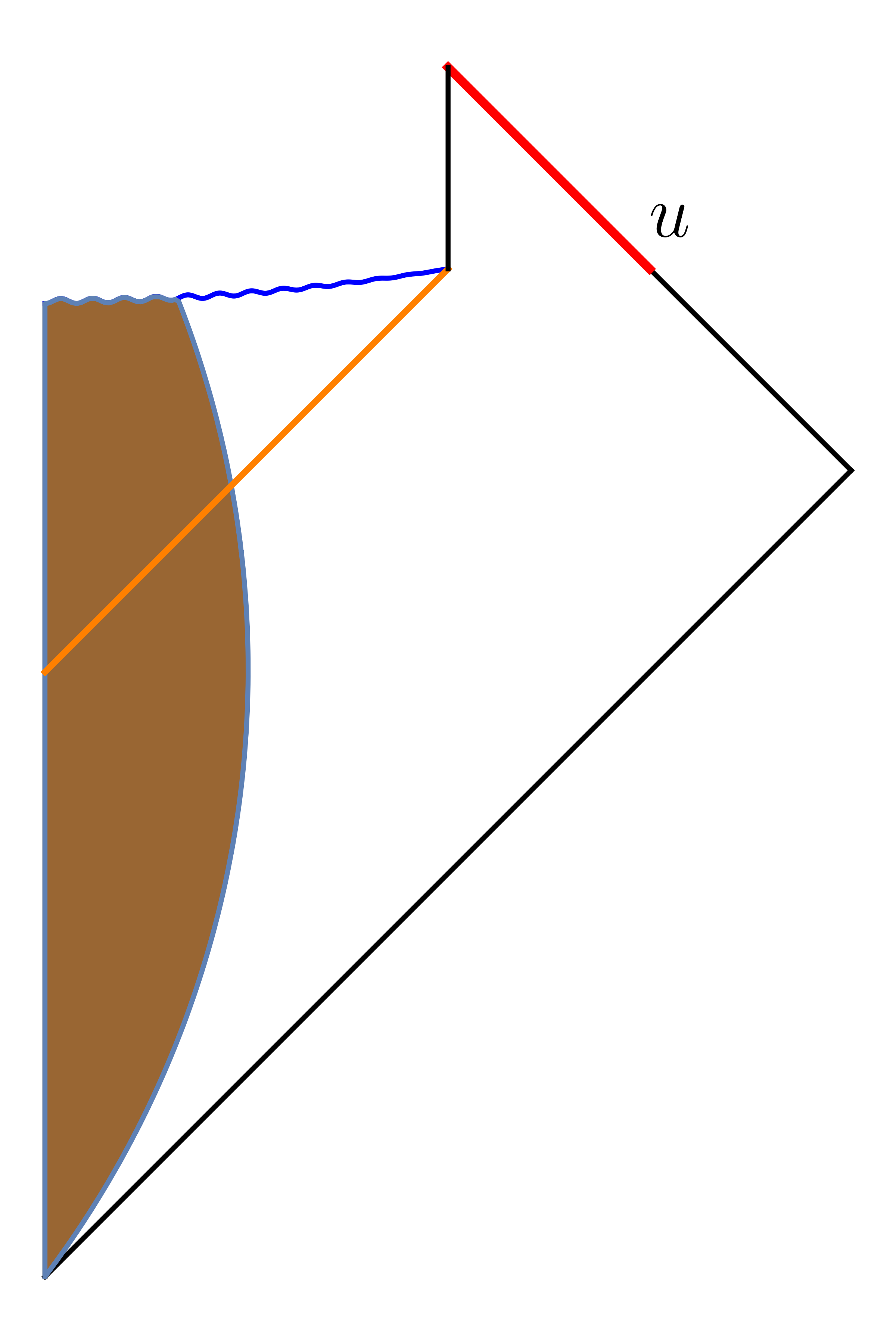}
\caption{\em The entropy of the segment $(u, \infty)$ (marked in red) may follow a Page curve as a function of $u$ if the entanglement between the soft and hard degrees of freedom is insignificant. But the physical interpretation of this Page curve is quite different from the conventional interpretation: for instance, the signature of unitarity is that entropy is a minimum near $u = -\infty$. \label{uppercut}}
\end{center}
\end{figure}

There is another way to think of the restriction of the algebra above. Instead of considering the von Neumann entropy of a segment of $\scrip$ that stretches from $(-\infty, u)$, one could consider the entropy of the segment $(u, \infty)$ of $\scrip$. See Figure \ref{uppercut}. It turns out that the commutant of the algebra of operators in the segment $(u, \infty)$ is precisely the restricted algebra above, where one considers the algebra of all operators from $(-\infty, u)$ but removes the mass aspect. Since the entropy of a pure state with respect to its algebra and with respect to its commutant is the same, we expect that this entropy may also obey a Page curve as a function of $u$. By result 2 of \cite{Laddha:2020kvp} (discussed near Figure \ref{cutinfo}), instead of considering the semi-infinite interval $(u, \infty)$  the same information can be extracted from the algebra of operators in the vicinity of the cut near $u$.

The physical interpretation of the Page curve for the algebra from $(u, \infty)$ is rather different from the conventional interpretation of the Page curve. In the conventional analysis of the Page curve of radiation, where one considers the state on $\scrip$ from $(-\infty, u)$,  the starting point of the curve at $u \rightarrow -\infty$ is considered to be trivial. The entropy is at a minimum there, because no radiation has emerged at all. Unitarity is usually thought to manifest itself in the fact that the entropy returns to its starting point.  But when we think of the algebra of operators from $(u, \infty)$, unitarity is manifested in the fact that the entropy is minimum at $u = -\infty$. And it is the entropy at late times, near $u =  \infty$, that is trivial.

Physically, one way to think of observables in the segment $(u, \infty)$ of  $\scrip$ is as follows.  Consider a Cauchy slice at very late times, $t$, after all black holes have evaporated. Observers who are constrained to explore retarded-time values larger than some $u$ correspond to observers who are constrained to explore the interior of a ball with radius $t - u$. In the limit $u \rightarrow -\infty$, such observers can explore the entire Cauchy slice. Increasing the lower limit on $u$ corresponds to contracting the ball that the observers can access: the observers still have information about the interior of the ball, but lose information about the exterior. As the ball contracts, the state of the interior starts to appear more-and-more mixed and so its entropy rises.  Eventually, the ball becomes small enough that the degrees of freedom in the interior cannot support a large entropy, and so the entropy starts decreasing. Finally, when the ball shrinks to zero size, it is clear that the entropy inside is 0.

We can summarize our discussion in terms of the following conclusion.
\begin{lesson} 
In asymptotically flat space, information about the microstate is accessible outside even before the black hole has evaporated, and so information does not ``emerge from the black hole'' according to the Page curve.
 But the Page curve may be the answer to other interesting and mathematically well-defined questions one can ask about evaporating black holes.
\end{lesson}

\section{Paradoxes for large AdS black holes \label{seclargeads}}
We now turn to large AdS black holes in AdS/CFT. An AdS black hole with a horizon radius that is much larger than the AdS radius has a positive specific heat \cite{Hawking:1982dh}. This means that it does not evaporate. For our purposes, the important distinction between small and large AdS black holes is that large AdS black holes dominate the microcanonical ensemble. This fact allows us to use several general 
results from statistical mechanics, and will be an important ingredient in all the paradoxes below. 

\subsection{Paradoxes for single-sided black holes \label{paradoxsingleside}}
We start by describing a number of paradoxes that appear for single-sided black holes. These paradoxes were emphasized in \cite{Almheiri:2013hfa} and in \cite{Marolf:2013dba}. Some of the arguments were clarified and elaborated in \cite{Papadodimas:2015jra} and, here, we will follow the treatment of section 5 of \cite{Papadodimas:2015jra}. 

\begin{figure}[!ht]
\begin{center}
\includegraphics[width=0.5\textwidth]{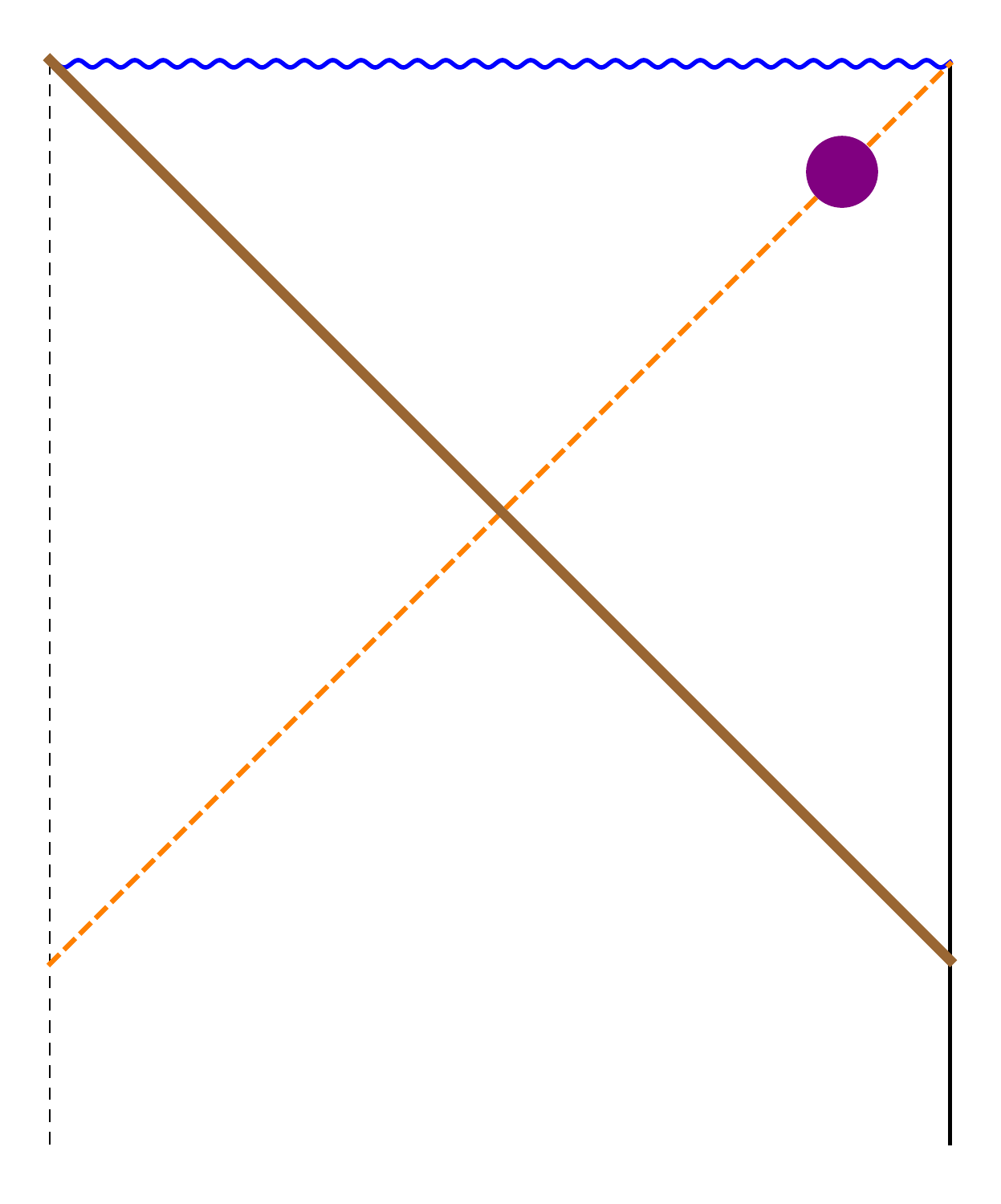}
\caption{\em A single-sided large black hole in AdS. The black hole is formed from the collapse of matter (brown line) and the state of the system is pure. We examine physics in the purple patch, at times, which straddles the horizon (marked in orange).\label{figsingleside}}
\end{center}
\end{figure}

The stage for this subsection is shown in Figure \ref{figsingleside}. We consider a single-sided black hole corresponding to some pure state in the theory. Perhaps this black hole was formed by the collapse of matter at an earlier time, but we are interested in the experience of an infalling observer who falls into the black hole at late times. The late-time classical metric is given by equation \eqref{latetimemetric} with $f(r)$ given by equation \eqref{fadsexplicit}.

The main takeaway from the paradoxes in this subsection is that the following two assumptions are mutually inconsistent.
\begin{enumerate}
\item\label{assumbig1}
Typical pure states drawn from the microcanonical ensemble have an empty horizon, where the standard rules of bulk effective field theory are applicable.
\item\label{assumbig2}
Degrees of freedom in the black-hole interior are described by the same
operators in the CFT for each black-hole microstate.
\end{enumerate}

\subsubsection{The negative occupancy paradox}

A simple paradox can be constructed using the two assumptions above. As we explained in section \ref{entangacrosshor}, a necessary condition for a smooth horizon is the existence of mirror operators $\ta_{\omega, \ell}$ behind the horizon
that satisfy the conditions outlined there. Combining equations \eqref{acrosshorcond} and \eqref{acrosshorcorr} we find that the mirror modes are also thermally occupied in a black-hole microstate.
\be
\label{mirroroccupancy}
\langle \Psi|  \ta_{\omega, \ell} \ta^{\dagger}_{\omega, \ell} |\Psi \rangle = {1 \over 1 - e^{-\beta \omega}}.
\ee
Now the mirror operators are slightly unusual as can be seen by examining the expansion \eqref{adsexpandbehindhor}.  The annihilation operators multiply a factor of $e^{i \omega t}$ in that expansion. This means that the 
commutation relations of these operators with the asymptotic Hamiltonian, $H$,  are\footnote{For a more careful
derivation, using a relational gauge fixing, we refer to section 6.3.1 of  \cite{Papadodimas:2015xma}. In brief, to specify the precise location of the operator in a diffeomorphism-invariant manner,  we shoot a geodesic from a given point on the right boundary, follow it for a certain proper distance and then consider the operator that probes the field at the end point of  the geodesic. The asymptotic Hamiltonian generates time translations near the right boundary. But the isometry of the geometry ensures that the entire geodesic is translated in the $t$-coordinate. Therefore, for operators that are gauge fixed in this manner, the asymptotic Hamiltonian generates translations of the $t$-coordinate even inside the horizon. See Figure \ref{dressedobservable} for an analogous procedure in the eternal geometry.
}
\be
\label{commuthamilt}
[H, \ta_{\omega, \ell}] = \omega \ta_{\omega, \ell}; \qquad [H, \ta_{\omega, \ell}^{\dagger}] = -\omega \ta_{\omega, \ell}^{\dagger}.
\ee
Note that the signs on the right hand sides above are the opposite of the usual signs. There are small corrections to the relations above since the operators are slightly smeared about a given frequency, but these corrections
will not be important.

By the result that we proved in section \ref{puremixed}, expectation values of operators in a typical state are exponentially close to their values in the microcanonical ensemble. But by the standard equivalence of ensembles,
expectation values in the microcanonical ensemble are close (up to power-law corrections in the entropy) to their values in the canonical ensemble. So \eqref{mirroroccupancy} implies that, up to negligible corrections,
\be
\label{occupexpected}
{1 \over Z(\beta)}\tr(e^{-\beta H} \ta_{\omega, \ell} \ta_{\omega, \ell}^{\dagger}) = {1 \over 1 - e^{-\beta \omega}},
\ee
where $Z(\beta) = \tr(e^{-\beta H})$ is the partition function.
But now a few simple manipulations lead to a paradox. First, using the cyclicity of the trace we find that
\be
\label{occupcyclicity}
\tr(e^{-\beta H} \ta_{\omega, \ell} \ta_{\omega, \ell}^{\dagger}) = \tr(\ta_{\omega, \ell}^{\dagger} e^{-\beta H} \ta_{\omega, \ell}).
\ee
Then using the commutator of the mirror modes with the Hamiltonian in equation \eqref{commuthamilt} we find that
\be
\tr(\ta_{\omega, \ell}^{\dagger} e^{-\beta H} \ta_{\omega, \ell}) = \tr( e^{-{\beta \omega}} e^{-\beta H} \ta_{\omega, \ell}^{\dagger} \ta_{\omega, \ell}).
\ee
Using the fact that the mirror modes are canonically normalized $[\ta_{\omega, \ell}, \ta_{\omega, \ell}^{\dagger}] = 1$ we find that
\be
{1 \over Z(\beta)} \tr( e^{-{\beta \omega}} e^{-\beta H} \ta_{\omega, \ell}^{\dagger} \ta_{\omega, \ell}) = e^{-{\beta \omega}} \left({1 \over Z(\beta)}\tr(e^{-\beta H} \ta_{\omega, \ell} \ta_{\omega, \ell}^{\dagger}) - 1 \right)
\ee
But putting all of this together we find that 
\be
{1 \over Z(\beta)}\tr(e^{-\beta H} \ta_{\omega, \ell} \ta_{\omega, \ell}^{\dagger}) = -{e^{-{\beta \omega}}  \over 1 - e^{-\beta \omega}},
\ee
which is not only not the same as \eqref{occupexpected}, it is manifestly absurd since the thermal expectation value of a positive operator cannot be negative. 

\subsubsection{The infalling number operator}
Another paradox involving the assumptions \ref{assumbig1} and \ref{assumbig2} was constructed in \cite{Marolf:2013dba}. Consider the operator 
\be
\label{nadef}
\begin{split}
N_a = (1 -e^{- \beta \omega})^{-1} \Big[ &\left(a_{\omega, \ell}^{\dagger} - e^{-{\beta \omega\over 2} }\ta_{\omega, \ell} \right) \left(a_{\omega, \ell}- e^{-{\beta \omega\over 2} } \ta_{\omega, \ell}^{\dagger} \right) \\ &+ \left( \ta_{\omega, \ell}^{\dagger} - e^{-{\beta \omega\over 2} } a_{\omega, \ell} \right) \left( \ta_{\omega, \ell}- e^{-{\beta \omega\over 2} }a_{\omega, \ell}^{\dagger} \right)  \Big]. 
\end{split}
\ee
Now applying the conditions \eqref{acrosshorcond} it is clear that for a smooth black hole state, we expect
\be
\label{naintypical}
\langle \Psi | N_a | \Psi \rangle = 0.
\ee
There is a simple physical reason for this. The operator $N_a$ can be thought of as the ``number operator'' for the infalling observer. In fact, the linear combinations of ordinary and mirror operators that appear above are the so-called Unruh combinations which are often used in understanding the relationship between Rindler and Minkowski quantizations of flat space. (See the operators $d_k$ discussed on p.116 of \cite{birrell1984quantum}.)

On the other hand, if one consider the eigenstates of the Schwarzschild number operator (defined as in \eqref{schwarzndef} by $\schwarzn_{\omega, \ell}=a_{\omega, \ell}^{\dagger} a_{\omega, \ell}$ ) that satisfy
\be
\schwarzn_{\omega, \ell} |n \rangle = n |n \rangle,
\ee
then it is clear that
\be
\label{nanineq}
\langle n| N_a | n \rangle \geq {2 e^{-\beta \omega} \over 1 - e^{-\beta \omega}}.
\ee
The inequality above is derived by noting that in a Schwarzschild number eigenstate there are no correlators between $a_{\omega, \ell}$ and $\ta_{\omega, \ell}$\footnote{Since $[N_{\omega, \ell}, \ta^{\dagger}_{\omega, \ell}] = 0$, we have $N_{\omega, \ell} \ta^{\dagger}_{\omega, \ell} |n \rangle = n \ta^{\dagger}_{\omega, \ell} | n \rangle$. This is clearly orthogonal to  $a_{\omega, \ell} |n \rangle$, which satisfies $N_{\omega, \ell} a_{\omega, \ell} |n \rangle = (n-1) a_{\omega, \ell} |n \rangle$.} and that
\be
a_{\omega, \ell} a_{\omega, \ell}^{\dagger} > 1, \qquad \ta_{\omega, \ell} \ta_{\omega, \ell}^{\dagger} > 1, \qquad a_{\omega, \ell}^{\dagger} a_{\omega, \ell} > 0 \qquad \ta_{\omega, \ell}^{\dagger} \ta_{\omega, \ell} > 0.
\ee

Now one can derive a contradiction as follows. If we think of the black-hole microstate drawn from the energy range $[E_0 - \Delta, E_0 + \Delta]$ that contains $e^{S}$ states then using the fact that typical states are exponentially close to the microcanonical trace we find from equation \eqref{naintypical} that
\be
\label{microtrace}
\langle \Psi | N_a | \Psi \rangle = {1 \over e^{S}} \sum_{E=E_0 - \Delta}^{E_0 + \Delta} \langle E | N_a | E \rangle  = 0,
\ee
up to small corrections that are irrelevant here. 

On the other hand, we also have
\be
\label{nomegcommut}
[\schwarzn_{\omega, \ell}, H] = 0,
\ee
which follows from the fact that $a_{\omega, \ell}$ and $a_{\omega, \ell}^{\dagger}$ have opposite energies. But then one can perform a change of basis in the trace in \eqref{microtrace}, and evaluate it in the basis of Schwarzschild number eigenstates.  These  number eigenstates are highly degenerate. Also, the change of basis may not be exact due to small corrections that come from interactions, but it was shown in \cite{Papadodimas:2015jra} that these edge effects do not matter. It is important that $N_a$ is a positive operator and therefore not subject to large cancellations between possible positive and negative contributions.

But, in the number eigenbasis basis, we find
\be
\label{numbertrace}
{1 \over e^{S}} \sum_{E} \langle E | N_a | E \rangle  =  {1 \over e^{S}} \sum_n \langle n | N_a | n \rangle \geq {2 e^{-\beta \omega} \over 1 - e^{-\beta \omega}}.
\ee
Equations \eqref{microtrace} and \eqref{numbertrace} are clearly in contradiction with each other.

\subsection{A paradox with the eternal black hole \label{paradoxeternal}}
The paradoxes  of subsection \ref{paradoxsingleside} might tempt the reader to decide that perhaps one should
simply give up on the assumption that typical states correspond to black holes with a smooth interior. Therefore, in this subsection, we would like to discuss a
paradox that pertains to the eternal black hole, and which we believe is both conceptually very important and under-appreciated.

\begin{figure}[!ht]
\begin{center}
\includegraphics[width=0.5\textwidth]{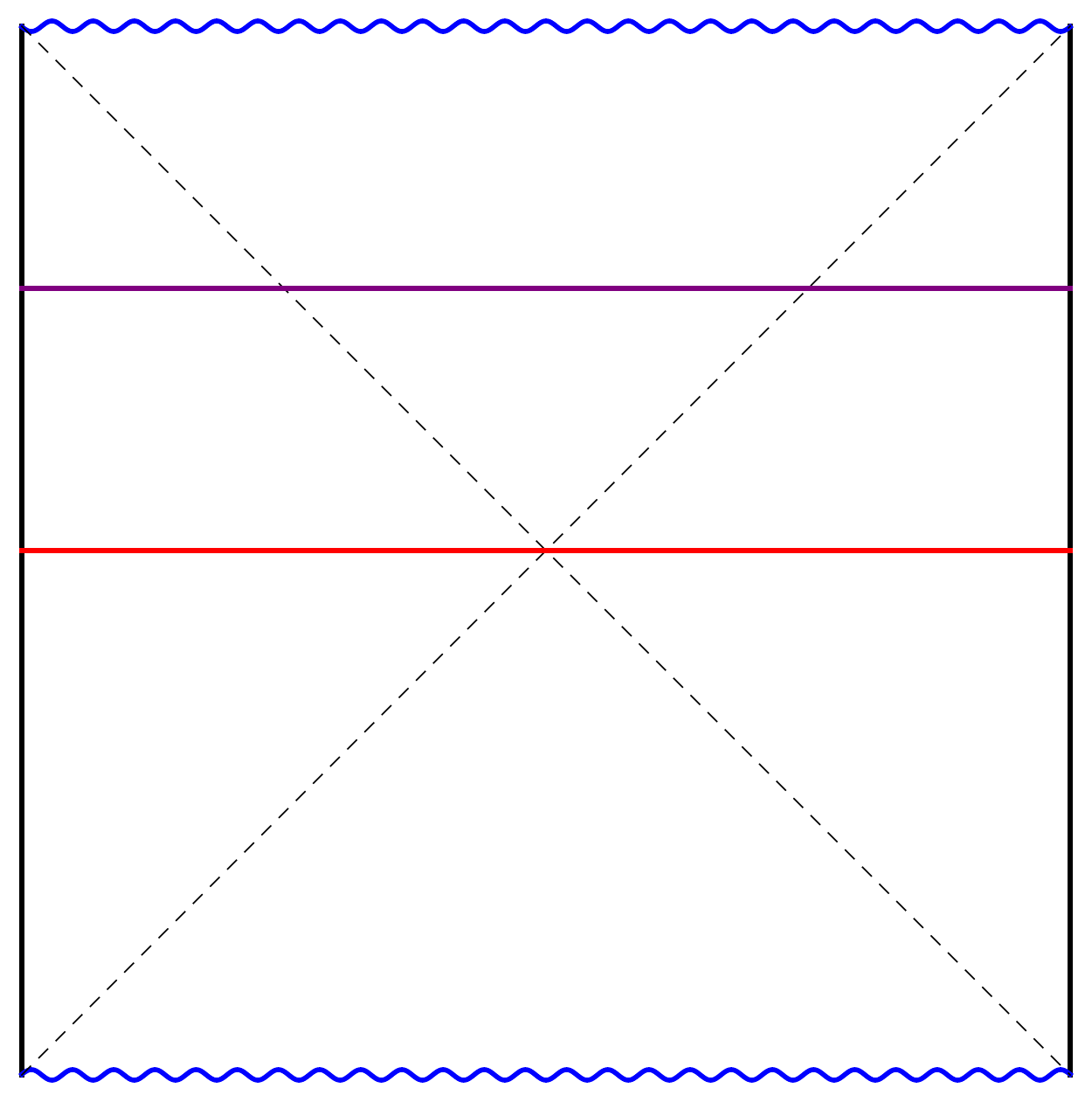}
\caption{\em An eternal AdS black hole. We have also marked two Cauchy slices: one (red) anchored at $t_{\text{left}} = t_{\text{right}} = 0$, and another one where time has been evolved forward on both sides. \label{figeternal}}
\end{center}
\end{figure}
It is believed that if one takes two identical holographic CFTs,  each with a holographic dual, and considers the state
\be
\label{thermofielddefn}
|\tfd \rangle = {1 \over Z(\beta)^{1 \over 2}} e^{-{\beta E \over 2}} |E, E \rangle,
\ee
then this state is dual to an eternal AdS black hole. (See Figure \ref{figeternal}.) We will distinguish the CFTs by the adjectives ``left'' and ``right'' and, in the equation above, the states $|E, E \rangle$ are simultaneous eigenstates of the left and right Hamiltonians, $H_L$ and $H_R$. This duality was first conjectured concretely in \cite{Maldacena:2001kr} based on earlier ideas \cite{Israel:1976ur}. This is also the prototypical example of the ER=EPR proposal \cite{Maldacena:2013xja}. 

The conjectured duality has a remarkable feature.  The two CFTs are entangled but noninteracting. In general when two systems are noninteracting, it is possible to perform a unitary operation on one side without affecting {\em any observable} on the other side. This is true even if the systems 
are entangled since a unitary on one system leaves the density matrix of the other system invariant. However, this duality suggests that the situation is subtly  different in gravity. Consider creating one excitation by applying a unitary to the right CFT, and a second excitation by applying a unitary to the left CFT. If both excitations fall into the black hole, they will then interact in the interior. So the claim is that excitation from the right CFT has a different ``experience'', depending on whether or not one had acted with a unitary on the left. 

This feature of the duality led \cite{Avery:2013bea,Mathur:2014dia} to object to this proposal.  Nevertheless, this  duality has been subjected to a number of checks. This includes the very nontrivial check of \cite{Gao:2016bin},  where a double trace deformation that acts on both boundaries was used to ``open up'' the wormhole and explore the interior. Moreover,  the geometry --- including the interior of the black hole --- appears to be under better control than the geometry corresponding to a typical pure state in a single CFT.

We now show how making the following three assumptions about this familiar system leads to a paradox. 
\begin{enumerate}
\item\label{eternalassumbig1}
The eternal black hole is dual to the thermofield doubled state.
\item\label{eternalassumbig2}
Degrees of freedom in the black-hole interior are described by the same
operators in the CFT for the thermofield doubled state and a class of states related to it by simple Hamiltonian evolution.
\item\label{eternalassumbig3}
Disentangled states, $|E,E \rangle$, are not connected by a wormhole.
\end{enumerate}
Arguably, these assumptions are even weaker than the assumptions \ref{assumbig1} and \ref{assumbig2}. The paradox that we will describe was first outlined in \cite{Papadodimas:2015xma}. A more lengthy description can be found in \cite{Papadodimas:2015jra}. (See section 6.) A similar paradox was also earlier described by \cite{Marolf:2012xe} although from a different perspective.

Consider the experience of an infalling observer who starts from the right boundary, at say $t = 0$,  and then falls into the black-hole interior. The experience of this observer
is related to the correlation functions
\[
\langle \tfd | \phi(x_1) \ldots \phi(x_n) |\tfd \rangle
\]
where $x_1, \ldots x_n$ are points along the observer's trajectory. Now, what if one were to translate the entire trajectory, including the starting point on the right boundary, in the $t$-coordinate? A translation in the $t$-coordinate
by an amount $\tau$ is generated by the operator $e^{-i H_R \tau}$. The experience of the observer would then be given by the correlators
\[
\langle \tfd | e^{i H_R \tau} \phi(x_1) \ldots \phi(x_n) e^{-i H_R \tau} |\tfd \rangle.
\]

Now the key physical point is that,  in the original undisturbed eternal black hole geometry, the experience of the observer must be the same for any value of $\tau$ including $\tau = 0$. This is clear from the bulk geometry but this is not just an approximate geometric statement. Rather it must be an exact statement because there is {\em no natural origin of time} in the eternal black hole.\footnote{Note that this would not be true if one were to excite the thermofield doubled state with an operator on the left or the right. This would break the time-translational symmetry.} This means that the correlators above must satisfy 
\be
\label{equalcorr}
\langle \tfd | e^{i H_R \tau} \phi(x_1) \ldots \phi(x_n) e^{-i H_R \tau} |\tfd \rangle =\langle \tfd | \phi(x_1) \ldots \phi(x_n) |\tfd \rangle, \quad \forall \tau \in (-\infty, \infty).
\ee

\begin{figure}[!ht]
\begin{center}
\includegraphics[width=0.5\textwidth]{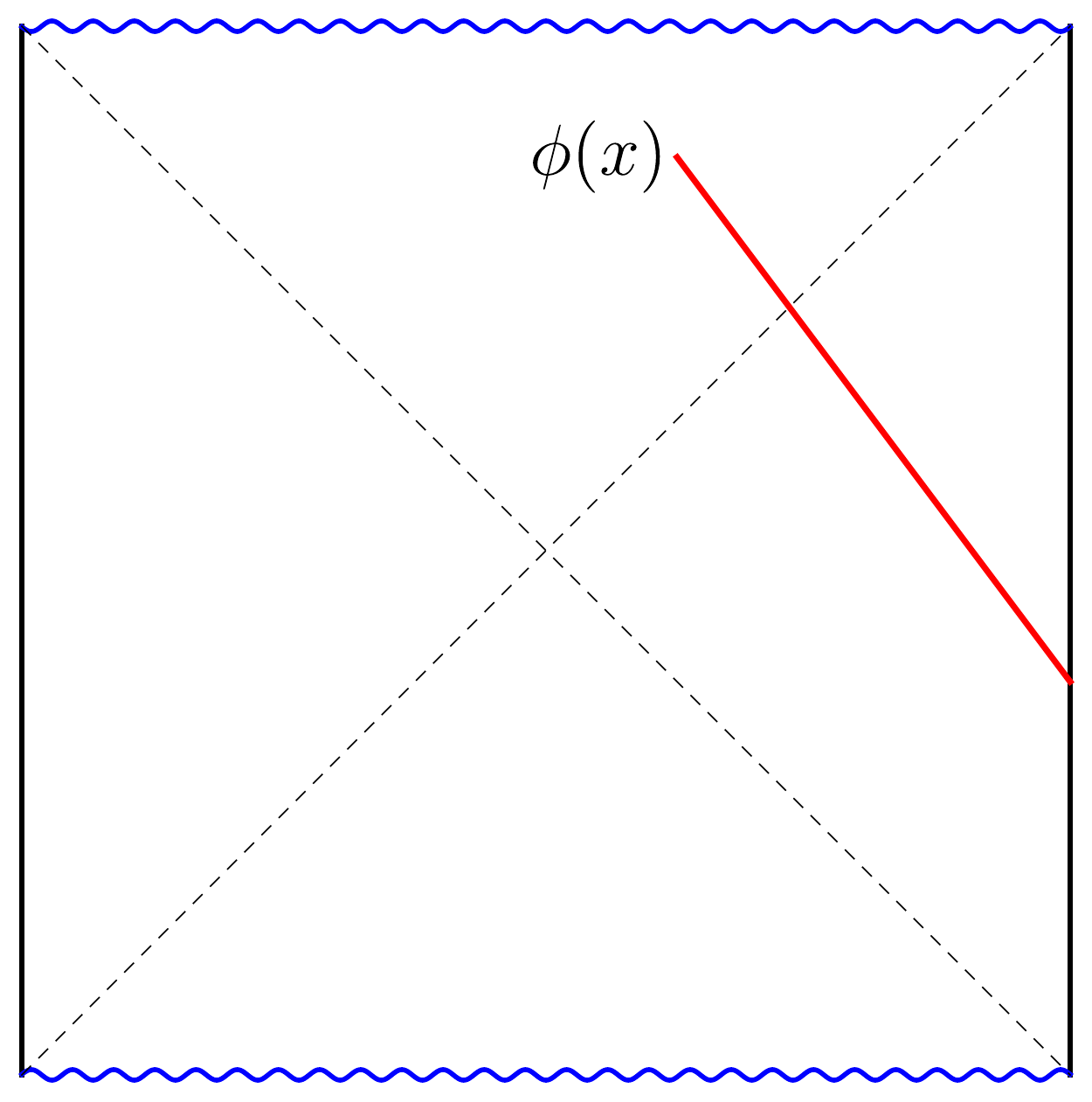}
\caption{\em The observables considered in the text, $\phi(x)$, are dressed to the right boundary. We prove that such observables must be state dependent in the eternal black hole geometry. \label{dressedobservable}}
\end{center}
\end{figure}
It is important that in the relation \eqref{equalcorr} the set of bulk observables we are looking at  --- including the observables that probe the interior --- are all dressed to the right boundary as explained in \cite{Papadodimas:2015xma,Papadodimas:2015jra}.    The relation \eqref{equalcorr} would {\em not} hold if, in place of $\phi(x_i)$, we considered cross-correlators of operators in the left CFT and operators in the right CFT. It would also not hold if we considered bulk correlators where some points were dressed to the left boundary, and others to the right boundary.

Now, let us evaluate the $\tau$-dependence of the correlator in
equation \eqref{equalcorr}. Using the expansion of the thermofield doubled
state, we find that 
\be
\label{corrtauexpansion}
\begin{split}
\langle \tfd | e^{i H_R \tau} \phi(x_1) \ldots \phi(x_n) e^{-i H_R \tau} |\tfd \rangle = {1 \over Z(\beta)} \sum_{E} e^{-\beta E} \langle E, E | \phi(x_1) \ldots \phi(x_n) | E, E \rangle& \\ + {1 \over Z(\beta)} \sum_{E \neq E'} e^{-{\beta(E+ E') \over 2}}  e^{i (E' - E) \tau} \langle E', E'| \phi(x_1) \ldots \phi(x_n) |E, E \rangle&
\end{split}
\ee
Therefore the condition \eqref{equalcorr}, which tells us that the correlator must be $\tau$-independent reduces to the condition that
\[
\sum_{E \neq E'} e^{-{\beta(E+ E') \over 2}}  e^{i (E' - E) \tau} \langle E', E'| \phi(x_1) \ldots \phi(x_n) |E, E \rangle
\]
is independent of $\tau$. But since we require this is to be true for arbitrary $\tau$, this can only happen if 
\be
\langle E', E'| \phi(x_1) \ldots \phi(x_n) |E, E \rangle = 0, \qquad \text{for} \quad  E' \neq E.
\ee
This tells us that the first term in equation \eqref{corrtauexpansion} is the only one that contributes to the correlator.

Now, we invoke assumption \ref{eternalassumbig3}, which tells us that in the {\em disentangled state}, $| E, E \rangle$ there is no geometric connection between the left and right asymptotic boundaries. So in this state, the experience of the infalling observer from the right cannot be affected by any unitary operator that acts only on the left:
\be
\langle E, E | \phi(x_1) \ldots \phi(x_n) | E, E \rangle = \langle E, E | U_L \phi(x_1) \ldots \phi(x_n) U_L^{\dagger}| E, E \rangle,
\ee
for arbitrary unitaries from the left CFT, $U_L$. But this means that the operators $\phi(x_i)$ must be operators {\em only} from the right theory since otherwise they could not commute with arbitrary unitaries from the left. 

This implies that, under all the assumptions above, the experience of the infalling observer from the right side in the eternal black hole is given by correlators of operators that purely come from the right CFT. Writing the operators that appear in these correlators as direct products of a right operator with the left identity, $\phi(x_i) = \phi_R(x_i) \otimes 1_L$, we find that the experience of the infalling observer is simply described by the correlators
\be
{1 \over Z(\beta)} \tr\left(e^{-\beta H_R} \phi_R(x_1) \ldots \phi_R(x_n) \right),
\ee
where the trace is only in the right CFT.

But in subsection \ref{paradoxsingleside} we showed that there were no state-independent operators in a single CFT whose correlators could correctly reproduce the correlators of effective field theory that describe a smooth horizon.  Therefore we find that there are no state-independent operators that can describe the interior of the eternal black hole either!

\subsection{Paradoxes about exponential decay \label{paradoxexponential}}

In section \ref{secstatedep}, we will explain how the paradoxes outlined in section \ref{paradoxsingleside} and \ref{paradoxeternal} can be resolved through state-dependent maps between the CFT and the black-hole interior. In this subsection, we describe paradoxes with a slightly different flavour, that can be resolved through exponentially small corrections to bulk computations, just as Hawking's original paradox was resolved in section \ref{resolvehawking}. For this reason, the paradoxes of this subsection  are not ``real'' paradoxes since exponentially suppressed effects are rarely under control in the bulk. Nevertheless, these paradoxes have attracted some attention because, on the boundary, 
these exponentially small effects can be computed. So these paradoxes --- while not conceptually particularly puzzling --- provide an arena where nonperturbative effects can be studied. 

One such paradox was outlined in \cite{Maldacena:2001kr}, together with the original proposal for the duality between the thermofield-double state and the eternal black hole. Consider exciting the eternal black hole geometry from the right boundary, and then detecting the strength of the disturbance that remains, near the boundary  after some time. This can be measured by a two-point function of an appropriate operator, $O_R(t)$, on the right boundary.   In the bulk, the excitation decays with time since the excitation tends to ``fall into'' the black hole. The decay of the perturbation near the boundary is given by the {\em quasinormal} modes \cite{Berti:2009kk,Kokkotas:1999bd,Horowitz:1999jd} of the black hole.  The bulk calculation tells us that the excitation falls off as $e^{-\kappa t}$ where $\kappa$ is a constant that is proportional to the surface gravity of the black hole. 

Now, let us parse the two-point function a little more. First note that 
\be
\langle \tfd | O_R(t) O_R(0) | \tfd \rangle = {1 \over Z(\beta)} \tr(e^{-\beta H_R} O_R(t) O_R(0)),
\ee
since the expectation value in the thermofield double just becomes the thermal expectation value
for operators from the right. By evaluating the trace explicitly in terms of a basis of energy eigenstates and also by inserting a complete
set of energy eigenstates between the operators we find that
\be
\label{thermalcorr}
\langle \tfd | O_R(t) O_R(0) | \tfd \rangle = {1 \over Z(\beta)} \sum_{E, E'} e^{-\beta E} e^{i (E - E') t} |\langle E| O_R(0) | E' \rangle|^2.
\ee

The expression \eqref{thermalcorr} can help us determine the late-time behaviour of this two-point function in the CFT.\footnote{For a more careful analysis of the early-time behaviour and approach to thermalization, we refer the reader to \cite{Banerjee:2019ilw,Mandal:2015kxi,Kulkarni:2018ahv}.} Even though the sum involves all energies from general thermodynamic considerations, we expect that it will receive its dominant contributions from a band of energies with $\Or[e^{S}]$ states.  Second, for a typical operator, the eigenstate thermalization hypothesis \cite{srednicki1994chaos,srednicki1999approach} tells us that we should expect
\be
|\langle E| O_R(0) | E' \rangle| = \Or[e^{-{S \over 2}}], \qquad \text{for}~~E \neq E'.
\ee
This estimate simply arises from an estimate of the overlap between two randomly chosen vectors in a space of dimension $e^{S}$.
It is also convenient to remove a constant contribution from the correlator by choosing a operator which has no diagonal elements
\be
\langle E| O_R(0) | E \rangle = 0.
\ee

Now, at late times there are $\Or[e^{2 S}]$ terms that contribute to \eqref{thermalcorr} due to the double sum over $E$ and $E'$. Each of these terms has typical size $\Or[e^{-{S \over 2}}]$ but they all contribute with random phases due to the leading time-varying exponential. Therefore, using a random phase approximation --- the sum of $N$ terms with random phases and size $a$ is typically of size $\sqrt{N} a$ --- we find that the sum itself yields a contribution that is $\Or[e^{S \over 2}]$.  The leading partition function and thermal suppression factor gives a contribution of size $e^{-S}$. Putting all these factors together we find that
\be
\label{longtimetwopt}
\langle \tfd | O_R(t) O_R(0) | \tfd \rangle  = \Or[e^{-{S \over 2}}],
\ee
for large $t$. Note that this estimate holds at ``generic'' late times. There are special values of $t$, where the phases in the sum over energies contribute coherently, for which the sum can become much larger. 

But this long-time estimate is in contradiction with the bulk computation for times $t = \Or[S]$. The bulk calculation predicts that the two-point function decays indefinitely in magnitude. This contradicts the CFT expectation that the  two-point function will develop a ``fat tail'' of size $\Or[e^{-S \over 2}]$ and not fall further.  Note that to obtain a sharp contradiction with interminable exponential decay, one must consider a boundary theory with a finite density of states, but this is naturally obtained when the boundary CFT is placed on a compact manifold. 

In  \cite{Maldacena:2001kr}, it was suggested that the required exponentially small correction to the bulk prediction  could be understood as a contribution of the ``other'' bulk saddle point:  a gas of gravitons in AdS that is a subdominant phase at high temperatures compared to the black hole, but will continue to contribute at exponentially small levels. In \cite{Barbon:2004ce,Barbon:2003aq}, it was pointed out that a more refined analysis was required to restore unitarity to the bulk theory. This issue has also been addressed using sophisticated CFT techniques, and we refer the reader to \cite{Fitzpatrick:2016mjq,Chen:2016cms,Anous:2016kss} for further details.

A similar paradox arises even without the two-point function. This was first pointed out in \cite{Papadodimas:2015xma}. Above, we considered a set of time shifted thermofield states
\be
|\tfdtau \rangle = e^{-i H_R \tau} |\tfd \rangle.
\ee
These states correspond to the original thermofield doubled state from equation \eqref{thermofielddefn} modified by adding $\tau$-dependent phases to each pair of energy eigenstates that appears in the sum. In fact, since 
there is no natural origin of time, these states should be treated on the same footing as the original thermofield state and it is only a convention to privilege the state where the phases are set to $1$ over other its time-shifted cousins. 

Now we see that the overlap of such a state with the original thermofield state is just the ratio of an analytically continued partition function to the original partition function.
\be
\label{overlapsum}
\langle \tfdtau | \tfd \rangle = {1 \over Z(\beta)} \sum_{E} e^{-i E \tau - {\beta E}} = {Z(\beta + i \tau) \over Z(\beta)}.
\ee
A simple random-phase estimate (see p. 77 of \cite{Papadodimas:2015jra}) tells us that this overlap initially dies off very rapidly. 
\be
\label{innerprod}
|\langle \tfdtau|  \tfd \rangle |^2 =  e^{-{\tau^2 C \over \beta^2}},
\ee
where $C$ is the specific heat of the boundary theory. But it is clear, once again, that this decay cannot persist for ever. Once $\tau$ becomes very large we expect that this overlap will saturate with a fat tail of magnitude $\Or[e^{-{S \over 2}}]$. This estimate arises just like the estimates above. We simply  note that at late times, a set of $\Or[e^{S}]$ random phases contribute to the sum in \eqref{overlapsum} and so the analytically continued partition function is of size $\Or[e^{{S \over 2}}]$. So its ratio with the original partition function at late times is $\Or[e^{-{S \over 2}}]$.

This analytically continued partition function, commonly called the ``spectral form factor'', has been the subject of several recent studies. It can be computed in the Sachdev-Ye-Kitaev model \cite{Cotler:2016fpe,Sonner:2017hxc},  where it turns out to have a rich structure at intermediate times that interpolates between the short-time behaviour of \eqref{innerprod} and the long-time fat tail.

\subsection{Bags of gold}

The eternal black hole allows us to make an interesting link with a simplified version of an old paradox, called the ``bags of gold'' paradox.

In the eternal black hole hole geometry of Figure \ref{figeternal}, in AdS$_{d+1}$, we can consider Cauchy slices that locally maximize volume and run from a time $t$ on one side to the same time $t$ on the other side. A simple calculation  shows that as $t$ increases the volume of these slices increases linearly at late times 
\be
{d V \over d t} \propto t,
\ee
for large $t$, where the constant of proportionality can be found in \cite{Hartman:2013qma} and is not relevant here.

This large volume leads to a paradox as pointed out in \cite{Mathur:2014dia}. At large $t$, we can consider a {\em dilute gas} of excitations on the slice.  As we produce excitations on the slice, we can simultaneously reduce the background mass in the black hole to keep the total energy constant. Using standard kinetic theory, the entropy of the gas grows as $S_{\text{gas}} \propto V$. But this means that for very large $t$ we can make $S_{\text{gas}} \gg S$. Since the black hole and its interior is supported in the union of two CFTs which, together, have only $e^{2 S}$ states at the relevant energy, it looks like the number of bulk states vastly exceeds the number of allowed states on the boundary.

This paradox is related to older puzzles about exotic classical solutions where the interior of a black hole contains an entire expanding cosmology \cite{Marolf:2008tx,Hsu:2008yi}. These solutions are commonly called ``bags of gold'', a terminology introduced by  Wheeler in a related but distinct context in \cite{wheeler1964relativity}. Usually, it is assumed that even though these classical solutions exist,  perhaps the states corresponding to these solutions belong to a separate superselection sector that does not contribute to the ordinary entropy of the black hole. Or else, perhaps that these solutions cease to exist in the quantum theory.

However, in the eternal black hole,  it is not hard to explicitly construct states that are 
indistinguishable outside the horizon for an observer on the right but correspond to excitations in the interior.  Let $U_L(\tau)$ be a unitary operator that creates an excitation on the left boundary at time $\tau$. If we now consider the state
\be
\label{niceexcite}
U_L(\tau_1) \ldots U_L(\tau_n) |\tfd \rangle,
\ee
this is a state that has $n$ left moving excitations on the late-time slice. One can separate the unitaries sufficiently on the boundary leading a set of widely-spaced excitations on the late-time slice.  These excitations can also be created by acting with a unitary from the right CFT, and then conjugating it with the operator $e{-{\beta H_R \over 2}}$ which also creates an excitation behind the horizon as explained in \cite{Papadodimas:2017qit,deBoer:2018ibj}. 

The resolution to this paradox also appears to lie in exponentially small corrections, as was explored in \cite{Chakravarty:2020wdm, essay2020}. Naively, one might believe that the states where the unitaries are very far separated in time have arbitrarily small inner product. But using precisely the same arguments that we used above, we expect that
\be
\langle \tfd | U_L(\tau_1) U_L(\tau_2)^{\dagger} | \tfd \rangle = \Or[e^{-{S \over 2}}],
\ee
for $|\tau_1 - \tau_2|$ very large. So however widely we separate the times in the state from \eqref{niceexcite} to get a dilute gas, we necessarily end up with states that are not quite orthogonal. 
What this implies is that a linear combination of a large number of seemingly independent excitations sometimes represents the {\em same} state as another excitation. This is similar to the phenomenon that we outlined in section \ref{secholography} where the degrees of freedom in one region had another description in terms of degrees of freedom in another region.

 One aspect of this proposal that requires further study is that even in an ordinary high-dimensional system --- such as the gas in a room --- it is possible to find a larger number of seemingly independent excitations of the form \eqref{niceexcite}.  But usually these states do not have any nice physical interpretation. In an ordinary system, an excitation that is created in the far past thermalizes. But, in gravity, on the other hand, the black-hole interior seems to  retain a record of very ``old'' excitations. It would be nice to understand this better.

\section{Interior reconstruction in AdS/CFT \label{secstatedep}}
In this section, we describe how the interior of a black hole can be related to boundary operators in AdS/CFT. This will also lead us to the phenomenon of state dependence. We will show how state dependence can be used
to resolve the paradoxes of sections \ref{paradoxsingleside} and \ref{paradoxeternal}.  The construction of interior operators that we describe here was developed in \cite{Papadodimas:2013wnh,Papadodimas:2013jku,Papadodimas:2015jra}. An earlier version of this construction was also provided in \cite{Papadodimas:2012aq}, but the analysis of \cite{Papadodimas:2012aq} is superseded by \cite{Papadodimas:2013jku} as explained in section 6.3.1 of \cite{Papadodimas:2013jku}. A related description of the black-hole interior was discussed in \cite{Verlinde:2013uja,Verlinde:2013qya,Verlinde:2012cy}.

\subsection{Review of the  mirror-operator construction \label{mirroropreview}}
Before we describe how CFT operators can be mapped to operators in the black-hole interior, we briefly remind the reader of the mapping in the black hole exterior.

Operators outside the black-hole horizon can be mapped to smeared CFT operators \cite{Bena:1999jv} using the standard HKLL method \cite{Hamilton:2007wj,Hamilton:2006fh,Hamilton:2005ju,Hamilton:2006az} or other equivalent methods \cite{Koch:2010cy,Jevicki:2015sla}. The idea of the construction is quite simple, and a recent review is  \cite{Kajuri:2020vxf}. Consider some propagating bulk field in the black hole spacetime. If one gives data for the field near the timelike boundary of AdS, then it is possible to solve the partial differential equations that control the field's dynamics, radially, and obtain the field configuration in the exterior of the black hole.  There are some subtleties in this process, which have to do with the fact that the mapping between CFT operators and bulk operators must be understood in the sense of a distribution.\footnote{This point was first explained in \cite{Papadodimas:2013jku} (see page 5) and subsequently elaborated in \cite{Morrison:2014jha}. See also \cite{Banerjee:2019kjh}.} 

However, if one works in Fourier space, this mapping  is quite simple as we have already described in section \ref{adshawkingderiv}. The bulk field can be written using equation \eqref{phiexpandads}, where the operators $A_{\omega, \ell}$ that appear are related to Fourier modes of boundary generalized free fields by equation \eqref{bulkbdryrelation}. (For a discussion of ``generalized free fields'', we refer the reader to \cite{ElShowk:2011ag} and to section 2 of \cite{Papadodimas:2012aq}.)  As in previous sections,  we will work with slightly smeared versions of the sharp modes that enter the field expansion. We denote the smeared modes by $a_{\omega, \ell}$ and we remind the reader that these modes are unit normalized: $[a_{\omega, \ell}, a_{\omega, \ell}^{\dagger}] = 1$.

To understand the black-hole interior we now need to find some operators $\ta_{\omega, \ell}$ that satisfy the conditions of section \ref{entangacrosshor}. In fact, the paradoxes described in section \ref{paradoxsingleside} tell us that if we demand that the conditions of section \ref{entangacrosshor} hold {\em exactly}, then it is impossible to find operators $\ta_{\omega, \ell}$. The insight in \cite{Papadodimas:2013jku} was that the conditions of section \ref{entangacrosshor} do {\em not} need to be satisfied exactly but only within  correlators that involve ``simple operators''. If we make this relaxation, then operators $\ta_{\omega, \ell}$ can indeed be found as we now show.

The first step in the  mirror-operator construction is to define what one means by  ``simple operators''.   The mirror-operator construction requires any definition of simple operators to satisfy a few conditions.  First, the set of simple operators must form a complex vector space so that linear combinations of simple operators are also simple. Second, the dimension of the set of simple operators must be much smaller than the number of states at the energy of interest.  Moreover, we will temporarily exclude the Hamiltonian, its powers and also polynomials of any other conserved charges in the theory from the set of simple operators. This is because these operators require special treatment as we explain below. 

One possible choice of simple operators is to consider polynomials of bounded order in the smeared modes of the generalized free-fields. This means that, given the modes $a_{\omega, \ell}$, monomials like
\[
a_{\omega_1, \ell_1}, \quad a_{\omega_1, \ell_1}^{\dagger}, \quad a_{\omega_1, \ell_1} a_{\omega_2, \ell_2}, \quad a_{\omega_1, \ell_1} a_{\omega_2, \ell_2}^{\dagger}, \quad a_{\omega_1, \ell_1} a_{\omega_2, \ell_2} a_{\omega_3, \ell_3}^{\dagger} a_{\omega_4, \ell_4}^{\dagger} \ldots,
\]
would be all be simple operators. Any linear combination of such monomials would also be a simple operator.  This is the perspective adopted in \cite{Papadodimas:2013jku}, where some precise cutoffs were also put on the polynomials.  We will work with this notion of simple operators here, but it is possible for the reader to adopt some other definition of simple operators that satisfies the conditions above.

Notice that, in general, the set of simple operators does {\em not} form an exact algebra since multiplying several of its elements together may take us outside of this set \cite{Ghosh:2017gtw}. This is particularly clear when we take simple operators to comprise polynomials of bounded order in some elementary modes. Evidently, by multiplying two such polynomials together, one may exceed the cutoff placed on the order. Nevertheless, in many cases, simple operators can be multiplied together to yield other simple operators. For instance, if we take the operator $a_{\omega, \ell}$ and also the operator $a_{\omega, \ell}^{\dagger}$ then the number operator $a_{\omega, \ell}^{\dagger} a_{\omega, \ell}$ is also simple and, in the equations below, we will assume that this product makes sense. We refer the reader to \cite{Papadodimas:2013jku,Papadodimas:2013wnh} for a more careful treatment.

The next step is to construct the {\em little Hilbert space.}  Let $\al_i$  be a linear basis for  a set of simple operators and let the CFT be in a microstate, $|\Psi \rangle$. Then the span of the  vectors
\be
\label{simplebasis}
|v_i \rangle = \al_i |\Psi \rangle,
\ee
is defined to be the little Hilbert space. This means that the little Hilbert space comprises all states in the Hilbert space obtained by acting on the CFT microstate with a simple operator.  In equation \eqref{simplebasis}, $i = 1 \ldots {\cal D}$, where ${\cal D}$ is the dimension of the set of simple operators. For a typical black-hole microstate, which is drawn from a space of dimension $e^{S}$, since ${\cal D} \ll e^{S}$, we expect that all the vectors above will be linearly independent and so the dimension of the little Hilbert space is also ${\cal D}$. 

The construction of the little Hilbert space was first described in \cite{Papadodimas:2013jku}. Subsequently, the construction of the little Hilbert space, as performed about the AdS vacuum,  was termed the ``code subspace'' in \cite{Almheiri:2014lwa}, following the terminology of \cite{Verlinde:2012cy}, and this latter terminology is also commonly adopted in the literature. This construction plays a key role in the relationship between quantum error correction and AdS/CFT \cite{Pastawski:2015qua}. 

We are now ready to define the mirror operators. It is important to distinguish between  equilibrium states and non-equilibrium states in this construction and we treat both in turn below.

\subsubsection{\bf Mirror operators in equilibrium states}
Say that the CFT is in a typical microstate, $|\Psi \rangle$, drawn from the set of states in the CFT at a given high energy so that the dual is expected to be a large black hole. As we discussed in section \ref{secoldinfo}, a  typical microstate has the property that correlators of all ordinary operators, such as the $\al_i$, are exponentially close to the microcanonical ensemble. In particular, this also means that such correlators are time-translationally invariant. We will call such states ``equilibrium states.''  The precise definition of an equilibrium state is discussed in some detail in section 5.1 of \cite{Papadodimas:2013jku}. Physically, an equilibrium state corresponds to a black hole, whose exterior has ``settled down.''

We then define the elementary mirror operators $\ta_{\omega, \ell}$ on the basis of the little Hilbert space to satisfy the following equations.
\be
\label{taelemdef}
\ta_{\omega, \ell} |v_m \rangle = |u_m \rangle,
\ee
where the vectors $|v_m \rangle$ are the same as in equation \eqref{simplebasis} and 
\be
\label{mirroraction}
|u_m \rangle = \al_m e^{-{\beta \omega \over 2}} a_{\omega, \ell}^{\dagger} |\Psi \rangle.
\ee
The action of the mirror operators on any other state in the little Hilbert space is now completely fixed by linearity.
More specifically, if we define the matrix $g^{m n}$  so that it satisfies $\sum_{n=1}^{{\cal D}} g^{m n} \langle v_n | v_p \rangle = \delta^m_p$, then the mirror operators are explicitly given by
\be
\label{taexplicit}
\ta_{\omega, \ell} = \sum_{m,n=1}^{{\cal D}}g^{m n} |u_m \rangle \langle v_n|.
\ee
The action of the elementary mirror operators $\ta_{\omega, \ell}^{\dagger}$ is defined similarly. 
\be
\label{tadagdef}
\ta_{\omega, \ell}^{\dagger} \al_m |\Psi \rangle =  e^{\beta \omega \over 2} \al_m a_{\omega, \ell} |\Psi \rangle.
\ee
The equations above also automatically specify the action of any polynomial in $\ta_{\omega, \ell}^{\dagger}$ and $\ta_{\omega, \ell}$. For instance applying the above rules in succession
\be
\label{numbermirror}
\ta_{\omega, \ell} \ta_{\omega, \ell}^{\dagger} |\Psi \rangle =  e^{\beta \omega \over 2} \ta_{\omega, \ell} a_{\omega, \ell} |\Psi \rangle = a_{\omega, \ell} a^{\dagger}_{\omega, \ell} |\Psi \rangle.
\ee

The equations above are designed so that the conditions of section \ref{secoldinfo} will be met within simple operators. In words, this works as follows. Consider some correlation function in the state $|\Psi \rangle$
with insertions of some ordinary operators and a mirror operator.  Then the equations above instruct us to first move the mirror operator to the extreme right, so that it is next to the state, and then convert it to an ordinary operator using the rule above. If there are insertions of multiple mirror operators, then the equations above instruct us to perform this procedure recursively, starting with the right-most mirror operator in the correlator. As shown above, this procedure suffices to define the mirror of a general polynomials in the modes of the elementary fields. But the mirror, $\widetilde{A}_n$, of a general simple operator, $A_n$, can also be defined directly through the equivalent equation:
\be
\label{generalmirror}
\widetilde{\al}_n \al_m |\Psi \rangle = \al_m e^{-{\beta H \over 2}} \al_n^{\dagger} e^{\beta H \over 2} |\Psi \rangle.
\ee

\paragraph{\bf Commutators and conserved charges \\}
Recall that he mirror operators not only need to be correctly entangled with the operators outside the horizon, they also need to  commute with those operators outside to a good approximation.  The equations above ensure that {\em within} correlations involving simple operators on the state, $|\Psi \rangle$,  any polynomial of the mirror operators commutes with simple operators.
\be
[\ta_{\omega, \ell}, \al_m] |\Psi \rangle = 0;  \qquad  [\ta_{\omega, \ell}^{\dagger}, \al_n] |\Psi \rangle = 0.
\ee
This is an important requirement and this requirement was often neglected in considerations of the interior prior to the above construction.

However this is not the end of the story since, as we have emphasized repeatedly above, a bulk operator cannot commute with the Hamiltonian or with other conserved charges. This issue is also carefully accounted for in the mirror-operator construction. Here, we just describe the treatment of the Hamiltonian, and direct the reader to section  3.2.4 of \cite{Papadodimas:2013jku} for a treatment of more general non Abelian charges.

Consider the states, $|\Psi \rangle, H |\Psi \rangle, H^2 |\Psi \rangle \ldots H^{\cal D'} |\Psi \rangle$ where ${\cal D}'$ is again some finite cutoff. About {\em each state}, $H^n |\Psi \rangle$ we can generate a little Hilbert space as above. For a generic state, $|\Psi \rangle$, which is not an energy eigenstate or a sum of a small number of energy eigenstates, the vectors in these little Hilbert spaces are linearly independent. Of course, the vectors are {\em not} orthogonal since, for instance, $\langle \Psi | H |\Psi \rangle \neq 0$. But here we are just assuming that $H |\Psi \rangle - \langle \Psi| H  | \Psi \rangle |\Psi \rangle$ gives us a new vector that cannot be generated by the action of another simple operator on $|\Psi \rangle$.\footnote{This condition can also be relaxed, and the mirror operators can be constructed for energy and charge eigenstates as is explained in section 3.2.4 of  \cite{Papadodimas:2013jku}. Here, we do not deal with those cases to keep the presentation simple.}
  
We now improve the definition of the mirror operators to
\be
\label{acthn}
\ta_{\omega, \ell} \al_m H^n |\Psi \rangle = \al_m e^{-\beta \omega \over 2} a_{\omega, \ell}^{\dagger} H^n |\Psi \rangle; \qquad \ta_{\omega, \ell}^{\dagger} \al_m H^n |\Psi \rangle = \al_m e^{\beta \omega \over 2} a_{\omega, \ell} H^n |\Psi \rangle.
\ee
The point of this definition is that it implements the commutators
\be
[H, \ta_{\omega, \ell}] = \omega \ta_{\omega, \ell}; \qquad [H, \ta_{\omega, \ell}^{\dagger}] = -\omega \ta_{\omega, \ell}.
\ee
The signs above are crucial since they are the opposite of the signs that we usually expect for creation and annihilation operators. This is just a manifestation of the fact discussed above. If one fixes the field in the black-hole interior by dressing it to the asymptotic AdS boundary then an analysis of large diffeomorphisms tells us that that mirror operators must transform as indicated above.

\subsubsection{\bf Mirror operators in near-equilibrium states}
The construction above was given for equilibrium states but it cannot be quite correct in all states. The simplest way to see this is to evaluate the expectation of the operator in equation \eqref{numbermirror}. We find that for an equilibrium state
\be
\label{notalwayseq}
\langle \Psi | \ta_{\omega, \ell} \ta_{\omega, \ell}^{\dagger} |\Psi \rangle = \langle \Psi | a_{\omega, \ell} a_{\omega, \ell}^{\dagger} |\Psi \rangle
\ee

But this equation cannot hold for a general state. For instance, one could simply excite a black hole by sending in a shock wave which disturbs the geometry outside differently from the geometry inside. In such a state, one would expect that the occupancy of the mirror modes is different from the occupancy of the outside-modes. If a construction were to predict that equation \eqref{notalwayseq} holds in all states --- which, we emphasize, the mirror-operator construction does not do --- then it would suffer from what is sometimes called the ``frozen vacuum'' problem \cite{Bousso:2013ifa}.

The main idea of \cite{Papadodimas:2013jku} was to account for this by considering ``near equilibrium'' states. A near equilibrium state, $|\Psi' \rangle$,  is an equilibrium state which has been excited by a simple unitary operator.
\be
|\Psi' \rangle = U |\Psi \rangle; \qquad U = e^{i \al_m},
\ee
where $\al_m$ is a {\em Hermitian} simple operator.   This kind of state can be used to model the process above, where one ``throws'' an excitation into a black hole.

Now the point is that because this excitation is created by a simple operator outside the black hole, it can also be {\em detected} by measuring correlators outside the black hole. So, if we are given the state, $|\Psi' \rangle$, there is an algorithmic procedure to back-calculate both the ``base state'', $|\Psi \rangle$ and also the unitary $U$. Details of this procedure are provided in section 5.2 of \cite{Papadodimas:2013jku}. 

For such a near-equilibrium state we  define the action of the mirror operators using the following equations.
\be
\label{tanoneqdef}
\ta_{\omega, \ell} \al_m |\Psi' \rangle = \al_m  e^{-\beta \omega \over 2} U  a_{\omega, \ell}^{\dagger}  U^{\dagger} |\Psi' \rangle; \qquad \ta_{\omega, \ell}^{\dagger} \al_m |\Psi' \rangle = \al_m  e^{\beta \omega \over 2} U  a_{\omega, \ell}  U^{\dagger} |\Psi' \rangle.
\ee
This definition is important to ensure that $\ta_{\omega, \ell}$ acts the same way on $\al_m |\Psi' \rangle$ as it would on $\al_m U |\Psi \rangle$. The factors of $U^{\dagger}$ and $U$ above simply  ``strip off'' the excitation created by $U$, move $\ta_{\omega, \ell}$ to the right so that it acts directly on $|\Psi \rangle$, and then reinsert the factor of $U$. More specifically, the definition in \eqref{tanoneqdef} is exactly the same as
\be
\ta_{\omega, \ell} \al_m U |\Psi \rangle = \al_m  e^{-\beta \omega \over 2} U  a_{\omega, \ell}^{\dagger}  |\Psi \rangle; \qquad \ta_{\omega, \ell}^{\dagger} \al_m U |\Psi \rangle = \al_m  e^{\beta \omega \over 2} U  a_{\omega, \ell} |\Psi\rangle.
\ee
This is, of course, what we would expect from our discussion of the construction of mirror operators in equilibrium states.

The construction presented above, which incorporates near-equilibrium states,  completely resolves the issue of the ``frozen vacuum''. But it would be nice, if one did not have to 
 distinguish near-equilibrium states from equilibrium states by hand and if there was a single formula that could be applied both in equilibrium and near-equilibrium states. There is a natural guess for such a formula that exploits
the analogy between the mirror-operator construction and the Tomita-Takesaki theory of modular automorphisms of von Neumann algebras \cite{takesaki2006tomita}. For a detailed discussion, we refer the reader to section 6 of \cite{Papadodimas:2013jku}. But, in brief, one starts by defining a ``modular operator'', $\Delta$, on the little Hilbert space. The quickest way to arrive at this operator is to define it through its matrix elements as
\be
\label{deltadef}
\langle \Psi | A_n \Delta A_m |\Psi \rangle = \langle \Psi | A_m A_n |\Psi \rangle
\ee
Then it is not difficult to see that for states, $|\Psi \rangle$, that comprise a {\em narrow range} of energy eigenstates around the energy, $E$, we can set $\Delta = e^{-\beta (H - E)}$. Then a natural guess is to attempt to make the following replacement in equation \eqref{generalmirror} 
\be
\label{possiblereplacement}
 e^{-{\beta H \over 2}} \al_n^{\dagger} e^{\beta H \over 2} \rightarrow \Delta^{1 \over 2} \al_n^{\dagger} \Delta^{-{1 \over 2}}.
\ee
One might hope that, in a near-equilibrium state,  inserting the right hand side of \eqref{possiblereplacement} would automatically lead to  \eqref{tanoneqdef}. Unfortunately, this does not happen. So while the definition of the mirror operators via the modular operator is tempting, it is insufficient to rid us of the necessity of distinguishing equilibrium and near-equilibrium states by hand. This issue was also recently discussed in \cite{Jafferis:2020ora}.

\subsection{State dependence \label{secorigstate}}
We now explain the notion of ``state dependence.'' So far we have specified how a mirror operator is constructed in a particular little Hilbert space.  To make this explicit we introduce an additional notational convention just for this subsection. Consider the little Hilbert space constructed about a state $|\Psi_1 \rangle$. Let us denote a mirror operator constructed in this little Hilbert space, using the procedure above,  by  $\tal^{\Psi_1}$.   This particular mirror operator correctly describes the experience of an infalling observer in the state, $|\Psi_1 \rangle$ and also in states obtained by making small excitations of $|\Psi_1 \rangle$ but it  simply annihilates states that are orthogonal to this little Hilbert space. This can be seen from the explicit formula for the elementary mirror mode in equation \eqref{taexplicit}.

Now consider another microstate, $|\Psi_2 \rangle$, corresponding to a black hole with the same energy. We can similarly define a mirror operator about the little Hilbert space constructed about this microstate. Using the same notation as above,  let us call this operator, $\tal^{\Psi_2}$.

Now we come to an important point. In general, if we pick two random microstates we expect that the overlap of the little Hilbert spaces constructed about $|\Psi_2 \rangle$ and $|\Psi_1 \rangle$ will be exponentially suppressed. This is because the overlap of two vectors, chosen at random from a Hilbert space of dimension $e^{S}$ is expected to be $\Or[e^{-{S\over 2}}]$.  So, to an excellent approximation, the operator $\tal^{\Psi_1}$ effectively annihilates  vectors from $\hilb[\Psi_2]$ and the operator  operator $\tal^{\Psi_2}$ effectively annihilates vectors from $\hilb[\Psi_1]$.  As a result,  we can now also think of the operator
\[
\tal^{\Psi_1} + \tal^{\Psi_2}.
\]
This operator now provides the correct description for an infalling observer both in states near $|\Psi_1 \rangle$ and in states about $|\Psi_2 \rangle$ since it acts correctly on the {\em direct sum} of $\hilb[\Psi_1]$ and $\hilb[\Psi_2]$.  Note that if we consider a state that is a linear combination of a state from $\hilb[\Psi_1]$ and another state from $\hilb[\Psi_2]$, then for correlators with insertions of the mirror operator above and of ordinary operators, this state simply behaves like a {\em classical mixture} because the interference terms are negligibly small. 

If, $|\Psi_3 \rangle, \ldots |\Psi_n \rangle$ are all equilibrium microstates then we can building larger and larger sums of mirror operators, and consider the operator
\[
\sum_{i=1}^n \tal^{\Psi_i}.
\]
For small values of $n$, this operator continues to work correctly on the entire subspace spanned by $\hilb[\Psi_1] \oplus \hilb[\Psi_2] \oplus \ldots \hilb[\Psi_n]$. But for {\em very large} values of $n$, we could potentially run into trouble. This is because the small interference terms that we mentioned above could ``conspire'' to create a problem so that the sum of operators above fails to act correctly even in one of the original states. To see this note that when the sum above is inserted in $|\Psi_1 \rangle$, we find
\be
\label{sumopinsertion}
\langle \Psi_{1} | \sum_i \tal^{\Psi_i} |\Psi_{1} \rangle
= \langle \Psi_1 | \tal^{\Psi_1} | \Psi_1 \rangle + \sum_{i=2}^n \langle \Psi_{1} |\tal^{\Psi_i} | \Psi_1 \rangle 
\ee 

\begin{figure}[!ht]
\begin{center}
\includegraphics[width=0.5\textwidth]{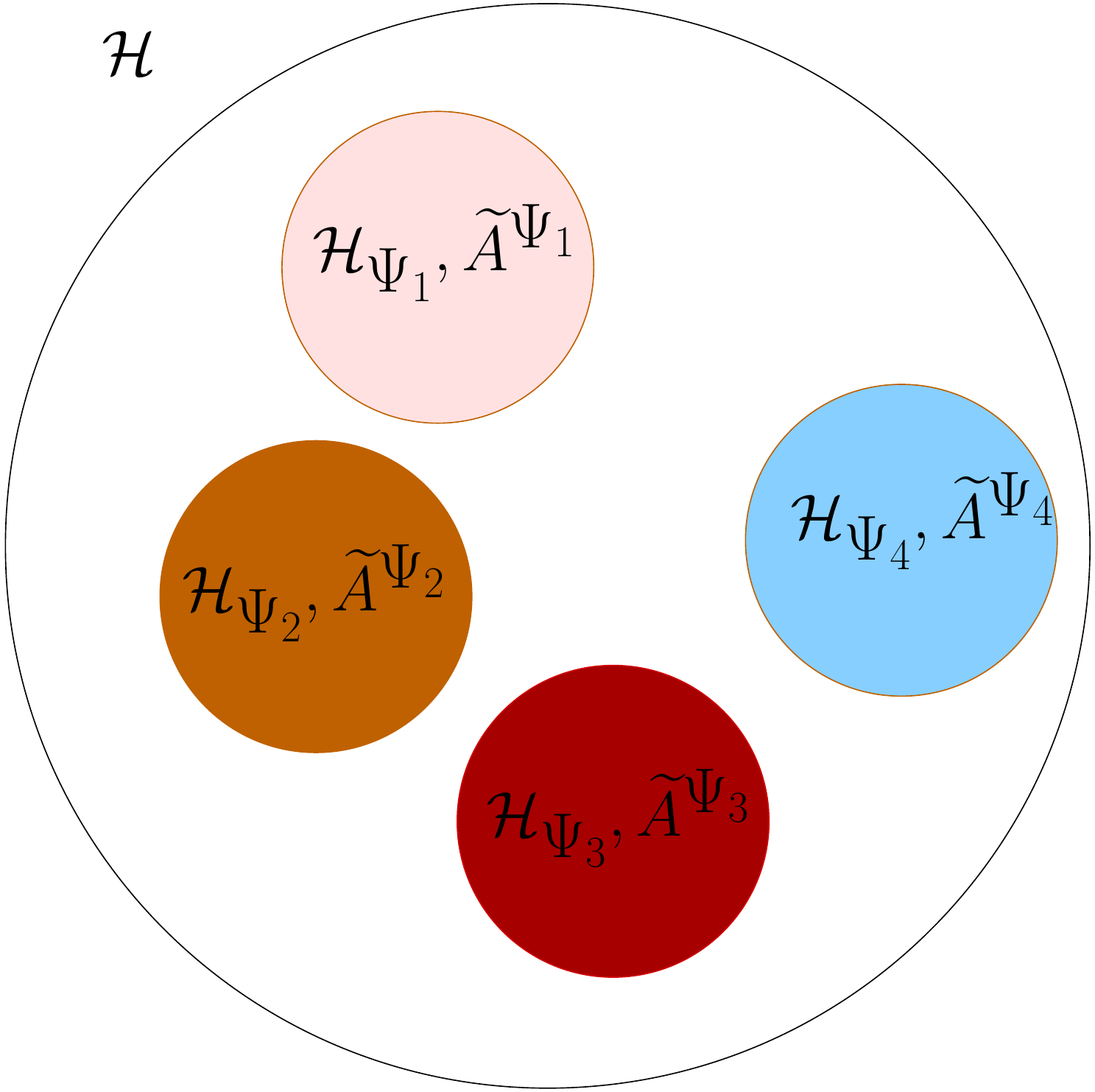}
\caption{\em A cartoon showing the origin of state dependence. We pick states, $|\Psi_1 \rangle \ldots |\Psi_n \rangle$, and define little Hilbert spaces, ${\cal H}_{\Psi_i}$  about these states. For a small number of states, the operator $\sum_i \tal^{\Psi_i}$ correctly describes the interior for the direct sum of all these little Hilbert spaces. But if we try and repeat this construction about a large number of states, the little Hilbert spaces start to overlap and no single operator has the correct properties in all these spaces. \label{multihpsi}}
\end{center}
\end{figure}

The first term above already gives us the desired behaviour for the mirror operator. But the second term may have a significant effect. This is because even though we expect the matrix elements of $\tal^{\Psi_i}$ to be very small in the state $|\Psi_1 \rangle$, this matrix element is not exactly 0. In fact, in a space of dimension $e^{S}$ we expect the second term to contribute $n$ terms of size $\Or[e^{-{S \over 2}}]$. Since these terms are incoherent, which means that they could have random phases, we expect that the second term in \eqref{sumopinsertion} will start interfering with the first term when  $n = \Or[e^{S}]$.  Therefore, if we try and  expand the sum above so as to make the mirror operator work on the entire Hilbert space, we might end up with an operator that does not work correctly even on the states that we started with. A cartoon depicting this issue is shown in Figure \ref{multihpsi}.

The idea of state dependence is that for correlators in a given microstate, say $|\Psi_1 \rangle$, one must select the mirror operators as constructed in $\hilb[\Psi_1]$. So one abandons the idea of finding a single operator that works correctly in all states and instead declares that the operators that the infalling observer uses depend on the state of the system.

The reason to believe that state dependence might be the correct framework to describe the experience of the infalling observer is that  usually, in a quantum mechanical system, there is a clear distinction between the system being observed and the observer. So the observer can choose a fixed set of operators to work with, independent of the state of the system. In this case, we are asking an unusual system where the infalling observer is part of the system that she is trying to observe. So the hypothesis is that perhaps the observables she chooses are influenced by the system.

We should also mention that if black holes with a smooth horizon had occupied only a small volume of the Hilbert space, then state dependence might not have been required. In such a situation, we might have been be able to get away by including a small number of terms in the sum above.   It is interesting --- and we will return below to this observation --- that  flat space black holes  and small black holes in anti-de Sitter space do occupy only an exponentially small volume of the Hilbert space. In both these cases, the entropy of the final Hawking radiation is always {\em larger} than that of the initial black hole. So a typical state at the energy of an evaporating black hole,  corresponds to  a gas of gravitons and not to a black hole.  So it is not clear whether state dependence is required for evaporating black holes.

\subsection{Resolving paradoxes with state dependence \label{secresolvestatedep}}
We now explain how the state-dependent construction described in previous subsections completely resolves the paradoxes for large single-sided black holes described in section \ref{paradoxsingleside} and 
the paradox for the eternal black hole described in \ref{paradoxeternal}.

The main point to note  is that these paradoxes all involve a trace over the Hilbert space of states at a given energy. But this is precisely what one cannot do with state-dependent operators. Let us start with the ``negative occupancy paradox.'' The step that fails with state-dependent operators is equation \eqref{occupcyclicity}. If we represent the trace as a sum over energy eigenstates then we find that
\be
\label{nocycstatedep}
\sum_{E} \langle E| e^{-\beta H} \ta_{\omega, \ell} \ta_{\omega, \ell}^{\dagger} | E \rangle  \neq \sum_{E} \langle E| \ta_{\omega, \ell}^{\dagger} e^{-\beta H} \ta_{\omega, \ell} | E \rangle.
\ee
Even for state-independent operators, the equality of the sums in equation \eqref{nocycstatedep} would not hold term by term but the differences would cancel out in the sum. This does not happen for state-dependent operators
since the operator $\ta_{\omega, \ell}$ depends on the state $|E \rangle$ within which it is inserted. It is, in fact, manifest that the mirror operators have the correct occupancy as can be seen from equation \eqref{numbermirror}. In any
equilibrium state (including an energy eigenstate) we have
\be
\langle E| \ta_{\omega, \ell} \ta_{\omega, \ell}^{\dagger} | E \rangle = \langle E| a_{\omega, \ell} a_{\omega, \ell}^{\dagger} | E \rangle = {1 \over 1 - e^{-\beta \omega}},
\ee
precisely as expected.

The resolution of the paradox about the number operator for the infalling observer  is the same. It is not permissible to change bases from energy eigenstates to number eigenstates. So with the number operator, $N_a$ as defined
in equation \eqref{nadef}, we  find that
\be
\sum_{E} \langle E | N_a | E \rangle  \neq \sum_n \langle n | N_a | n \rangle.
\ee
In fact, in an equilibrium state, using the definition of the mirror operators we find that
\be
\begin{split}
&\left(a_{\omega, \ell}- e^{-{\beta \omega\over 2} } \ta_{\omega, \ell}^{\dagger} \right) | \Psi \rangle = 0; \\
&\left( \ta_{\omega, \ell}- e^{-{\beta \omega\over 2} }a_{\omega, \ell}^{\dagger} \right) | \Psi \rangle = 0.
\end{split}
\ee
and therefore $\langle \Psi | N_a |\Psi \rangle = 0$ in any equilibrium state.

Let us now turn to the eternal black hole. In the case of the eternal black hole it is the equality \eqref{corrtauexpansion} that is invalid. Since the state-dependent operator does not act linearly, we find that its action
on a sum of $e^{S}$ states cannot be replaced with a sum of its action on individual states.
\be
 \phi(x_1) \ldots \phi(x_n) e^{-i H_R \tau} |\tfd \rangle \neq  {1 \over Z(\beta)} \sum_{E} e^{-{\beta E \over 2} - i E \tau}  \phi(x_1) \ldots \phi(x_n) | E, E \rangle,
\ee
when some of the $x_i$ are inside the horizon.

The mirror modes in the eternal black hole can, of course, be obtained using the construction above. But, in fact there is another way to construct the mirror modes. This second method arises from
the intuition that the mirror modes that are encountered by the infalling observer from the right should be related to modes from the left CFT. However, the modes from the left CFT, which we denote by $a^{L}_{\omega, \ell}$, cannot quite be the correct operators because they commute with the right Hamiltonian. But if we are looking to construct field operators that are dressed to the right boundary, the correct commutator is
\be
[H_R, \ta_{\omega, \ell}] = \omega \ta_{\omega, \ell},
\ee
which is not satisfied if we set $\ta_{\omega, \ell} = a^{L}_{\omega, \ell}$. However, the modes of the left CFT can be improved to give the correct state-dependent mirror operators as follows.
Let  $|\tfdtau \rangle$ be the time-shifted thermofield doubled states as above, and consider the operator
\be
\label{guessta}
\ta_{\omega, \ell} = {\sqrt{ C \over \pi \beta^2}} \int_{-\cutoffT}^{\cutoffT} d \tau  a^{L}_{\omega, \ell} e^{i \omega \tau} |\tfdtau \rangle \langle \tfdtau |, 
\ee
where we have introduced a cutoff, $\cutoffT$ whose significance will become evident momentarily. 

Now consider the action of this operator on the original thermofield state. Using the inner product \eqref{innerprod} we find that the integral in \eqref{guessta} can be written
\be
\label{actionontfd}
\ta_{\omega, \ell} |\tfd \rangle = {\sqrt{ C \over \pi \beta^2}} \int_{-\cutoffT}^{\cutoffT} a^L_{\omega, \ell} e^{i \omega \tau} | \tfdtau \rangle e^{-\tau^2 C \over 2 \beta^2} d \tau = a_{\omega, \ell}^L | \tfd \rangle = e^{-{\beta \omega \over 2}} a_{\omega, \ell}^{\dagger} | \tfd \rangle.
\ee
up to small corrections that can be neglected.\footnote{The modes $a_{\omega, \ell}$ that are relevant for the infalling observer from the right are, of course, related to modes from the right CFT.} Similarly when this operator acts on a time-shifted state then provided that $\tau < \cutoffT$, we find that
\be
\ta_{\omega, \ell} |\tfdtau \rangle = a^L_{\omega, \ell} e^{i \omega \tau} | \tfdtau \rangle =  e^{-{\beta \omega \over 2}} a_{\omega, \ell}^{\dagger} | \tfdtau \rangle,
\ee
which is also correct.

\begin{figure}[!ht]
\begin{center}
\includegraphics[width=\textwidth]{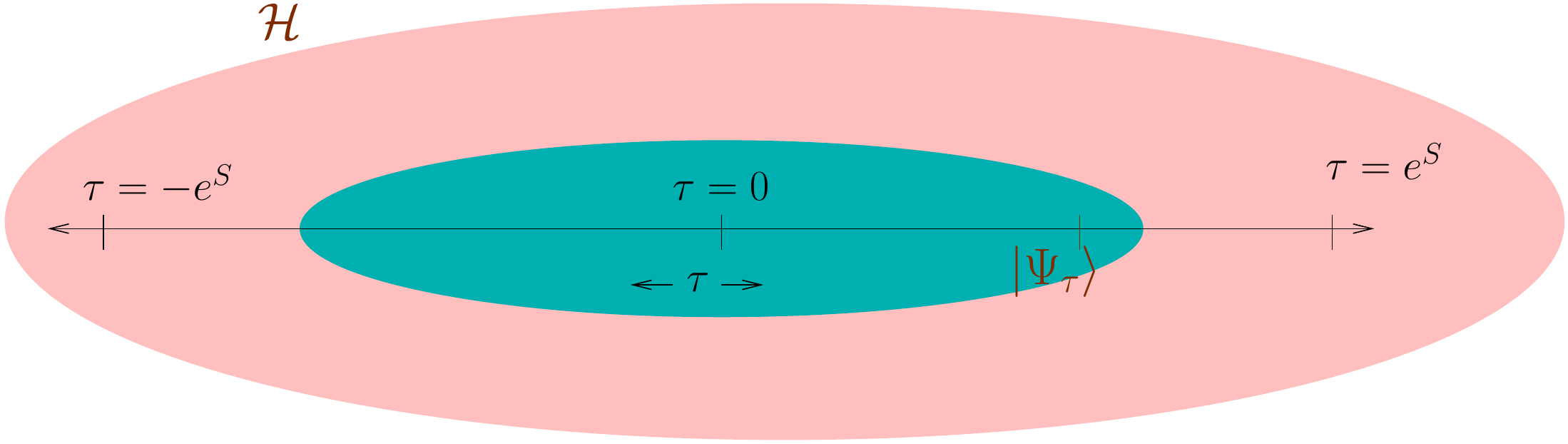}
\caption{\em We can write down a single operator in the CFT that describes the operators in the black-hole interior, not only in the original thermofield doubled state but also in ``nearby states'' including states $|\tfdtau \rangle$ for large values of $\tau$. But once we reach exponentially large values of $\tau$, we need to switch to a different operator. In this cartoon, the full Hilbert space is depicted in pink and the subset of the space on which this operator can be used in shaded in cyan. \label{tshiftpatches}}
\end{center}
\end{figure}

Why not simply take $\cutoffT \rightarrow \infty$ so that the action of $\ta_{\omega, \ell}$ becomes correct on all time-shifted states? The problem is that as we explained above the inner product \eqref{innerprod} is not correct for large $\tau$ and develops a fat tail of size $e^{-{S \over 2}}$. So if we take $\cutoffT \rightarrow \infty$ we would find that the integral in \eqref{actionontfd} would fail to converge. We can take the cutoff, $\cutoffT$ to be exponentially large, $\cutoffT = \Or[e^{S \over 2}]$, but we cannot take it to infinity. This is an explicit example of how it is possible to find
an ordinary  operator that acts correctly on a large range of states, but impossible to find an ordinary  operator that acts correctly on all states. Once we start considering states $|\tfdtau \rangle$ for very large $\tau$, it is necessary to switch to a {\em different} operator and the operator in equation \eqref{guessta} ceases to operate correctly. This picture is shown in Figure \ref{tshiftpatches}.

\paragraph{\bf Other interior constructions.}
Several other proposals have been made to understand the black-hole interior from the CFT. In this context, we would like to emphasize the following. In any microstate that has a smooth interior, any other construction of interior operators, if correct, {\em must agree} with the construction described in section \ref{mirroropreview} in its action on the little Hilbert space constructed about the microstate. This is because the behaviour of the mirror operators on a smooth microstate in the little Hilbert space is completely fixed by effective field theory. 

The construction of section \ref{mirroropreview} does not tell us, by itself, whether a given microstate has a smooth interior. And other constructions can disagree on microstates that do not have a smooth interior.  They can also differ in their action on states that are not part of the little Hilbert space. 

It is also possible to construct operators that are differently gauge fixed and so differ in their commutator with the Hamiltonian. In \cite{Harlow:2014yoa}, for instance,  it was proposed that one should use the construction of section \ref{mirroropreview}  but set $[H, \ta_{\omega, \ell}] = 0$. This would modify equation \eqref{acthn}, which implies $[H, \ta_{\omega, \ell}] = \omega \ta_{\omega, \ell}$. There are several difficulties with the proposal of \cite{Harlow:2014yoa} that are discussed in section 8.2 of \cite{Papadodimas:2015jra}. 

In some cases, one is looking for interior operators that, rather than being correctly entangled with exterior operators, have some other desired properties.  For instance, in the ``Petz map'' construction of \cite{Penington:2019kki}, one is looking for an operator in the interior whose action on a state reproduces the action of an operator in the exterior. Similarly, in \cite{Kourkoulou:2017zaj}, one is looking for operators that can introduce particles behind the horizon. But these operators are related to the mirror operators in a very simple way. One simply has to modify the right hand side of equation \eqref{mirroraction} to obtain the desired action of the interior operators, and then use the formula \eqref{taexplicit}. 

In the ``geon state'' \cite{Guica:2014dfa}, it is possible to study the construction of section \ref{mirroropreview} in detail. It is also possible to understand this construction in terms of deformations of the boundary, as explained in \cite{deBoer:2019kyr,deBoer:2018ibj}. For some special states, it may be possible to use bulk evolution or other techniques \cite{Heemskerk:2012mn,Almheiri:2017fbd} to write down specific formulas for interior operators. We also find the construction of  \cite{Nomura:2018kia,Nomura:2019dlz,Nomura:2019qps,Nomura:2020ska}  interesting.

Since some of the subtleties of interior reconstruction are not treated carefully in parts of the literature, we find the following set of questions useful to keep in mind while evaluating any proposal for the interior.
\begin{enumerate}
\item
Does the proposed interior operator have the correct commutator with exterior operators? For instance, if  we had simply equated $\ta_{\omega, \ell} = e^{-{\beta \omega \over 2}} a_{\omega, \ell}^{\dagger}$ this operator would still have had the correct action on an equilibrium microstate. But since this operator fails to commute with exterior operators even within simple correlators, it would not have yielded correct results when inserted in a middle of a correlation function that also contained insertions of other operators.
\item
Does the proposed operator have the right nonzero commutator with the Hamiltonian?
\item
Does the proposed operator act correctly on near-equilibrium states, or does it suffer from a ``frozen vacuum'' problem?
\item
Does the proposed operator act correctly in generic states, or is it constructed for a special class of states?
\end{enumerate}

\subsection{Consistency of  state-dependent maps \label{mpparadox}}
We showed in section \ref{secresolvestatedep} that state-dependent maps between the bulk and the boundary resolve several paradoxes associated with AdS black holes. Nevertheless, since state dependence is unusual from the point of view of quantum mechanics, it is important to check this hypothesis for consistency. In fact, state dependence itself gives rise to a new paradox pointed out in  \cite{Marolf:2015dia}. We describe this paradox and then a partial resolution.

Consider a large black hole in AdS, and let $\schwarzn_{\omega, \ell}$  be the Schwarzschild number operator for particles of frequency $\omega$ as defined in equation \eqref{schwarzndef}.  Then, if $|\Psi \rangle$ is a black-hole microstate with an empty interior,  consider the state $e^{i \theta \schwarzn_{\omega, \ell}} |\Psi \rangle$. The operator $e^{i \theta \schwarzn_{\omega, \ell}}$ is an operator of low energy in the boundary CFT due to equation \eqref{nomegcommut}.  But nevertheless from the bulk effective field theory point of view, it tends to ``rephase'' the operators outside the horizon while leaving the operators behind the horizon untouched. This disturbs the entanglement between the operators inside and the outside and is perceived by the infalling observer as an excitation. Therefore this operator generates another state with almost the same energy, but whose interior is excited. 

If the question of whether the interior is excited or not could be answered by a state-independent operator, and if one also assumed that almost all states at a given energy have a smooth interior, this would lead to a paradox.
This is because for every state with a smooth interior at a given energy, one can find a state with slightly higher energy that has an excitation. But these excited states now almost form a basis for the set of all states at higher energy. So it cannot be that almost all states at the slightly higher energy also have a smooth interior. 

With the use of state-dependent operators, there is no paradox since the behaviour of a state-dependent observable on a basis of states is insufficient to determine its behaviour on the rest of the Hilbert space. But, for state-dependent operators, the argument of \cite{Marolf:2015dia} does imply that if most states are smooth,  then there must be two states whose inner product is almost $1$ but with the property that one state corresponds to an excited interior and the other to an empty interior. This appears to violate the Born rule, which tells us that states that are almost parallel must be physically almost identical.

This can be framed as a mathematically more precise contradiction as was done in \cite{Raju:2016vsu}. It is possible to show that if $\al$ is a state-independent observable then the action of a low energy unitary, $U$ that changes the energy of a typical state, $|\Psi \rangle$ by only a small amount,
\be
\delta E \equiv \langle \Psi | U^{\dagger} H U |\Psi \rangle - \langle \Psi | H | \Psi \rangle, 
\ee
should not change the expectation value of $\al$ significantly.
\be
\label{bound}
\delta  \al  \equiv \big|\langle \Psi | U^{\dagger} \al U |\Psi \rangle - \langle \Psi | \al |\Psi \rangle \big| \leq 2 \sqrt{\beta \delta E} \sigma,
\ee
when $\beta \delta E \ll 1$ and where 
\be
\label{sigmadef}
\sigma^2 \equiv \langle \Psi| \al^{\dagger} \al | \Psi \rangle   - \left| \langle \Psi | \al | \Psi \rangle \right|^2,
\ee
The argument above suggests that the inequality \eqref{bound} could be violated for state-dependent operators.

A partial resolution to this paradox was provided in \cite{Raju:2016vsu}. In the discussion above, we have considered  deformations of the state by  ``simple operators'' that are low-order polynomials in {\em modes} of the field and are naturally defined in Fourier space. This definition has the advantage of leading to the algebraically simple results that we have obtained above.  However, \cite{Raju:2016vsu} pointed out that another notion of  ``simple'' is to consider deformations of the state by operators that are {\em local} in position space on the boundary, and ask how such deformations affect observations made by a {\em local} bulk observer. If one adopts this notion of simplicity --- that is tied to position space rather than frequency space ---  then \cite{Raju:2016vsu} pointed out that violations of \eqref{bound} cannot be observed using any simple physical process. 

\begin{figure}[!ht]
\begin{center}
\includegraphics[height=0.5\textheight]{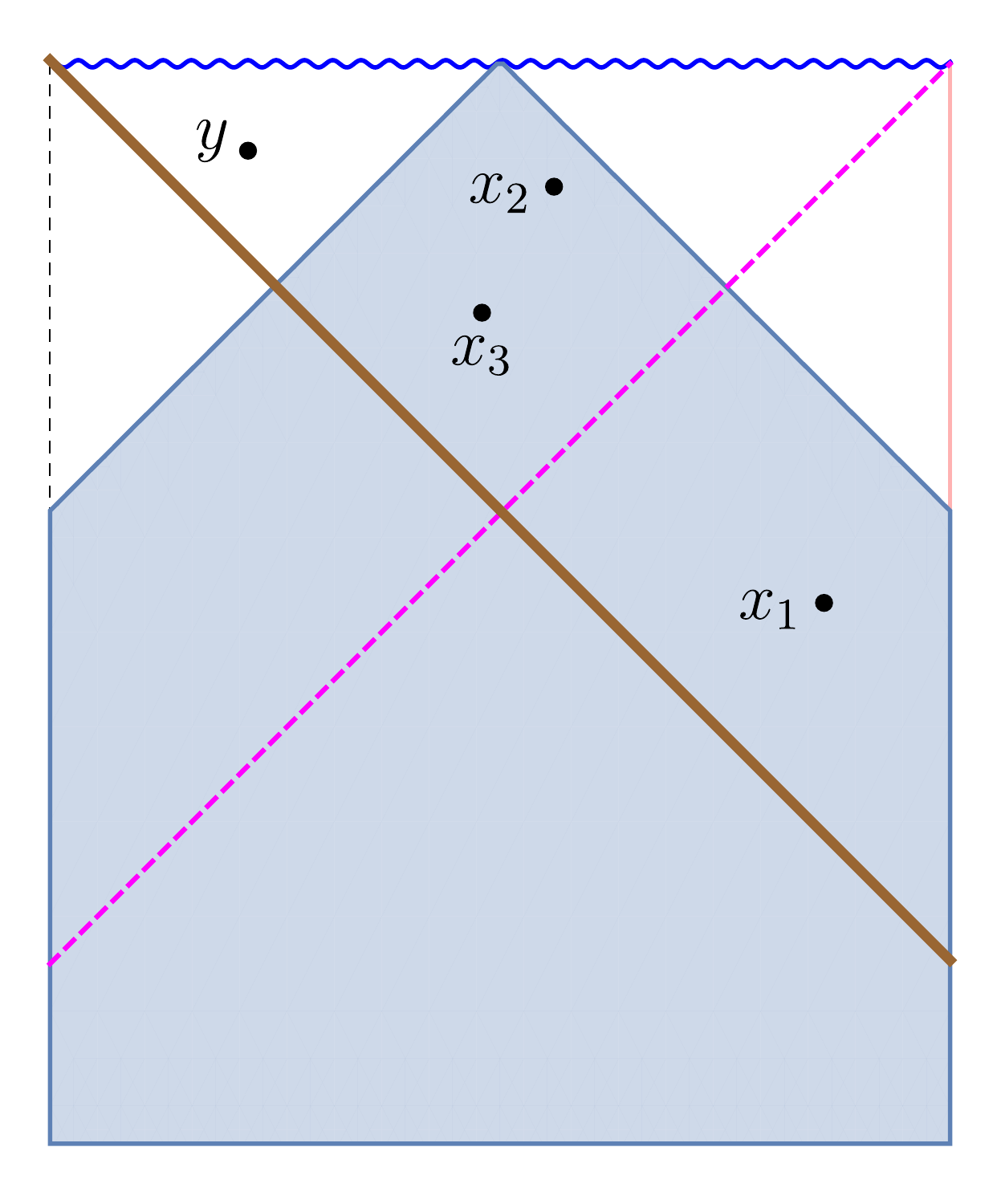}
\caption{\em The figure shows a causal patch (shaded region) in a single-sided large black hole in AdS. The points $x_1, x_2, x_3$ are in a single causal patch, whereas $y$ is outside the causal patch that contains these three points.  The black hole is formed by the collapse of a shell (shown in brown) and the dashed line on the left is the origin of polar coordinates, not another asymptotic region. \label{causalpatch}}
\end{center}
\end{figure}

More precisely, the main result of \cite{Raju:2016vsu} was as follows. Let $U$ be of the form
\be
\label{causalunitary}
U = \exp\big[-i \int_{-\infty}^{\tcaus} J(t) Q(t) d t \big],
\ee
where $Q(t)$ is a low-order polynomial in generalized free-field operators at time $t$ on the boundary. In the exponent, this polynomial is integrated from $-\infty$ to some upper limit $\tcaus$ with some weight function $J(t)$ to obtain the unitary excitation.   Let the observable $\al$ be of the form
\be
\label{causalobservable}
\al = \phi(x_1) \phi(x_2) \ldots \phi(x_n)
\ee
corresponding to bulk quantum-field observables in the {\em causal patch} that extends from the point $\tcaus$ on the boundary. (See Figure \ref{causalpatch}.) A causal patch is physically important \cite{Freivogel:2014fqa}  because correlators that do not belong to a single
causal patch cannot be determined by physical observers.\footnote{The discussion of \cite{Raju:2016vsu} is  within the context of quantum field theory on a fixed spacetime background. It does not account for  how observers
may use gravitational effects as described in section \ref{secholography} to extend the region from which they can obtain information. It is an important open problem to include such observations in the analysis.} Then, it is a remarkable property of position-space AdS correlators that
the inequality \eqref{bound} is satisfied for any unitary excitation, $U$, of the form \eqref{causalpatch} and observables $\al$ of the form \eqref{causalobservable}.

This result covers several physically important processes. This is because the usual physical way to obtain a unitary transformation of a state is by modifying the Hamiltonian by adding to it a simple polynomial of generalized free field operators on the boundary.
\be
\label{deformhamilt}
H(t) =  H + J(t) Q(t).
\ee
By the standard rules of quantum mechanics, the expectation value of the operator $\al$, in a state $|\Psi \rangle$ in the presence of this deformation should be computed through the expression
\be
\langle \Psi| \overline{{\cal T}} \{e^{i \int_{-\infty}^{\tcaus} J(t) Q(t) d t}  \} \phi(x_1) \phi(x_2) \ldots \phi(x_n) {\cal T}\{e^{-i \int_{-\infty}^{\tcaus} J(t) Q(t) d t} \} | \Psi \rangle
\ee
where ${\cal T}$ is the time-ordering symbol. The important aspect of the expression above is that the upper limit of the integral does not extend beyond $\tcaus$ because the modification of the Hamiltonian to the future does not
affect any correlator in the causal patch.  But then the result of \cite{Raju:2016vsu} applies and the bound \eqref{bound} is satisfied. Therefore the summary is that if observers are restricted (a) to the unitary transformations that can be carried out by deforming the Hamiltonian with local operators and (b) to the natural set of physically accessible observables, then they cannot observe the  violations of the Born rule described in \cite{Marolf:2015dia}. 

This is not a complete resolution of the paradox. This is because, even within effective field theory, we  do sometimes consider deformations of the Hamiltonian that are simple in terms of modes but may not be local in position and so are more general than the deformation displayed in equation \eqref{deformhamilt}.
For instance, to act with the operator $e^{i \theta \schwarzn_{\omega, \ell}}$ that we discussed above, we could consider the deformation
\be
\label{complexdeform}
H(t) = H + \theta \delta(t)  \schwarzn_{\omega, \ell}.
\ee
which adds the Schwarzschild number operator 
to the Hamiltonian for a short amount of time near $t = 0$. 
This is not of the form shown in equation \eqref{deformhamilt} because the number operator is not a local operator when written in terms of generalized free fields at time $t$.  In terms of such operators, we need to write $\schwarzn_{\omega, \ell} = a_{\omega, \ell}^{\dagger} a_{\omega, \ell}$, where $a_{\omega, \ell}$ itself is obtained by integrating the boundary operator for a long time period to isolate a particular frequency.

Note that, in the boundary theory, the deformation in equation \eqref{complexdeform} is allowed and can even be written in terms of a boundary operator at a single time. This is because the set of operators in the boundary theory on any time slice is complete.
So, generalized free field operators at a later time, and at an earlier time, are already present in the algebra of all operators at a time $t$. Of course, if one attempts to rewrite  these operators from the ``future'' and the ``past'' in terms of generalized free fields at time $t$, this leads to a very complicated expression.
The result of \cite{Raju:2016vsu} does not apply to deformations of the form \eqref{complexdeform}. 

We believe that the paradox of \cite{Marolf:2015dia} deserves more attention than it has received. It appears to us that the following possibilities are open.
\begin{enumerate}
\item
Perhaps the response of the bulk to deformations of the form \eqref{complexdeform} --- where one deforms the Hamiltonian with an operator that is not a simple polynomial of generalized free fields at the same time  ---  is not correctly described by QFT in curved spacetime. For instance, in the last section of \cite{Raju:2016vsu}, it was speculated that such a deformation might affect observers in different causal patches differently. 
\item
\label{possadditional}
Related to the possibility above, perhaps the boundary description must be supplemented with additional information before one can describe the black-hole interior unambiguously. For instance, \cite{vanBreukelen:2019zxq} explored the possibility that an external ``clock'' is required to complete the description of the interior. The addition of extra degrees of freedom can turn states that are almost parallel into orthogonal states, and thereby ameliorate the paradox above.
\item
Perhaps \eqref{bound} simply does not hold for observables inside the horizon. This corresponds to the possibility that state dependence is indeed severe enough for two states that are almost parallel to yield physically distinct observations.
\item
Perhaps this paradox  rules out state dependence. This corresponds to the possibility that typical states in AdS/CFT have a firewall just behind the horizon. In this case, the construction of mirror operators described in section \ref{mirroropreview} would still be valid for those microstates that have an empty interior but would not apply to typical microstates. 
\end{enumerate}
A new physical idea is required to choose between these possibilities. We discuss these possibilities some more in the next section.

\section{Is structure at the horizon required? \label{secisstructure}}

We would now like to take stock of what we have discussed so far. In the sections above, we have described various paradoxes about black holes and their resolutions.  But when we put all these discussions together, what do they tell us about the nature of the black-hole horizon? Do we expect that the horizon of a typical large black hole is just featureless and empty, or does it have a firewall or a fuzzball lurking just behind it. We frame this questions more precisely below and then describe our answers, first for large AdS black holes, and then for evaporating black holes.  We have tried to keep this section somewhat self contained, and so there is some overlap with points covered previously.

Our conclusion, which we state more succinctly in conclusions \ref{lessonlargebh} and \ref{lessonevap}, is as follows. For evaporating black holes, including small AdS black holes and flat-space black holes, all {\em currently known} versions of the information paradox can be resolved without appealing to a firewall or a fuzzball. For large black holes in AdS/CFT  the question  is more subtle and has to do with the consistency of state-dependent reconstructions of the black-hole interior.  State dependence is necessary also to reconstruct the interior of the eternal black hole, where a number of independent calculations suggest that the interior should be empty. State dependence is also used widely in many calculations in AdS/CFT. But some questions about state dependence have not yet been resolved. So, the somewhat unsatisfactory situation for typical large black holes in AdS/CFT appears to be that there is no clean way to definitively rule out firewalls and fuzzballs, and also no persuasive argument that implies their existence.

In the last part of this section, we also explain why the program of constructing  microstate geometries --- while interesting in its own right --- cannot yield evidence in favour of the fuzzball proposal and cannot even be used to conclude that such geometries  form a basis of all black-hole microstates. We also briefly touch on the proposal known as ``nonviolent nonlocality.''

\subsection{Framing the question}

We expect that any theory of quantum gravity will contain a number of states that correspond to black-hole microstates. Let us call a typical state from this class, $|\Psi \rangle$. We also expect that the theory will contain 
observables corresponding approximately to local physical quantities at a point.  The position of the point can be fixed relationally with respect to the asymptotic boundary or by means of a gauge-fixing procedure. Presumably, the precise method used to fix the location does not significantly  affect low-point correlators of this observable. Let $\qop(x)$ denote such an observable. This observable may be the value of a propagating field that we have considered previously, or it may correspond to some other geometric quantity like the square of the Riemann tensor. With this observable in hand, one can evaluate its correlators  at different points like
\[
\langle \Psi |  \qop(x_1) \ldots \qop(x_n) |\Psi \rangle,
\]
where the points $x_1, \ldots x_n$ could be inside or outside the horizon but are away from the singularity.

On the other hand, one could also consider the geometry of a black hole formed from collapse and consider late times in this geometry. Here by ``late time'' we  mean that the time difference to the collapsing matter is long  compared to the temperature of the black hole but much shorter than its evaporation time. Using effective field theory in this background gives a definite prediction for the same correlator that is displayed above.  Let us call this prediction $G_{\text{standard}}(x_1, \ldots x_n)$.  

Then we are interested in whether the correlator in the full theory agrees with this standard prediction or not.
\be
\label{structurequestion}
\text{Is}\quad \langle \Psi |  \qop(x_1) \ldots \qop(x_n) |\Psi \rangle = G_{\text{standard}}(x_1, \ldots x_n) \quad \text{in~a~typical~state~$|\Psi \rangle$?}
\ee
The fuzzball and firewall proposals suggest that the answer to this question should be ``no''. Since the horizon has ``structure'' in these proposals, when some points are taken to be behind the horizon, the correlator in a typical state deviates significantly from its value in a standard black hole geometry. The conventional idea is that the answer to the question above is ``yes''; this is the idea that horizons are empty.

Although we do not subscribe to the fuzzball or firewall proposals, we would like to start by pointing out that these proposals are not obviously incorrect, as is sometimes suggested.
For instance, one common criticism is that they ``violate the equivalence principle.'' Another common response is the  observation that our world could be crossing a horizon, just as the reader reaches this sentence,  since a huge shell of matter with a radius of a billion light years could be collapsing on us. But there is clearly no firewall or fuzzball near the Earth.

But the firewall and fuzzball proposals --- in their reasonable form ---  do {\em not} suggest that {\em all} horizons have structure. Rather they suggest that {\em typical} horizons or {\em very old horizons} have structure. It is important to remember that the textbook treatment of black holes as formed by the Vaidya solution \cite{vaidya1951gravitational}  or the Datt-Oppenheimer-Snyder solution \cite{datt1999class,oppenheimer1939continued} does not correspond to a typical black-hole microstate, and nor does it train our intuition to think about very ``old'' black holes.

One way to reach a typical state is to start with a black hole formed from collapse and then manipulate it over a long time scale. If one manipulates one qubit of information per Planck time,  then the time it will take to reach a typical state is approximately the entropy of the black hole multiplied with the Planck time. For a solar-mass black hole, in four dimensions, this works out to $3 \times 10^{33}$ years. The time scale to form an ``old black hole'' is even longer. For a solar-mass black hole, the Page time is about $10^{67}$ years. These time scales are outside the realm of our usual intuition and it is not beyond imagination that some conspiracy of quantum fluctuations in such long-time processes could generate boosted particles near the horizon or cause the geometry to tunnel into a fuzzball, or lead to other effects as explored in \cite{Hutchinson:2013kka}. Therefore, the firewall and fuzzball proposals deserve to examined seriously. We first discuss typical states in AdS/CFT and then discuss evaporating black holes.

\subsection{Typical states in AdS/CFT \label{typicalads}}

In AdS/CFT, we believe that any typical state drawn from the microcanonical ensemble at a high enough energy corresponds to a large black hole in the bulk. Even here, the question \eqref{structurequestion} is too difficult to answer. But, one can ask a somewhat easier question. Do there exist CFT operators, $\qop_{\text{CFT}}(x)$, so that 
\be
\label{cftcondition}
\langle \Psi |  \qop_{\text{CFT}}(x_1) \ldots \qop_{\text{CFT}}(x_n) |\Psi \rangle = G_{\text{standard}}(x_1, \ldots x_n),
\ee
where $|\Psi \rangle$ is a typical state drawn from the microcanonical ensemble.
If the answer to this existence question is ``yes'', one could declare that these CFT operators give us the right answer for bulk observations. One would still need to understand --- from a fundamental perspective --- why these operators constitute the right observables for a bulk infalling observer but that is a second-order question. 

On the other hand, if the answer to this existence question is ``no'', this suggests that the answer to the question \eqref{structurequestion} is also ``no'', and typical black-hole microstates in AdS have structure at their horizons. Another possibility, if \eqref{cftcondition} cannot be satisfied, is  that the CFT must be supplemented with additional degrees of freedom in order to describe the interior of typical black-hole microstates.

In section \ref{mirroropreview}, we have already addressed the existence question. The construction that we described in section \ref{mirroropreview} provides operators that, when inserted in a typical state, do yield the standard correlators as demanded by equation \eqref{cftcondition}. However, the price we pay for demanding that this condition hold is that the operators must be state dependent. As we explained in section \ref{mpparadox}, state dependence itself gives rise to new paradoxes. These paradoxes have been resolved partially but not completely. Therefore, it appears to us that the answer to question \eqref{cftcondition} depends on whether the origin of state dependence can be understood and whether the consistency of state-dependent interior reconstructions can be demonstrated cleanly.

As we already explained in section \ref{paradoxeternal}, this issue also afflicts the eternal black hole. 
So if one rules out state-dependent bulk-boundary maps altogether then one is forced to give up the picture that the thermofield doubled state corresponds to a geometric wormhole. But this would contradict several other calculations, including \cite{Gao:2016bin}, that suggest that the eternal black hole is most easily interpreted in terms of a wormhole in the bulk.

State dependence has sometimes been criticized for ``violating quantum mechanics.'' But, as we have discussed above, the kind of question that one is asking here --- about the experience of the infalling observer --- is unusual. Since the observer is necessarily part of the system, the observer cannot make an independent and fixed choice of observables, independent of the state of the system. The necessity of state-dependent maps has now been noted in a variety of  contexts in AdS/CFT  \cite{Berenstein:2016pcx,Berenstein:2017abm,Jafferis:2014lza,Guica:2015zpf,Jafferis:2017tiu,Jefferson:2018ksk,Bzowski:2018aiq,vanBreukelen:2019zxq}.  
Moreover, as pointed out in \cite{Papadodimas:2015jra}, the commonly-used RT/HRT formula that relates the entanglement entropy to an area is necessarily state dependent.  In the bulk, the area of a minimal surface appears to be an observable. But in the CFT, standard arguments tell us that there is no operator whose expectation value yields the von Neumann entropy. (See footnote \ref{footvn}.) Similarly, the reconstruction of operators in the entanglement wedge as proposed, for instance, in \cite{Faulkner:2017vdd} is explicitly state dependent because modular flow is state dependent.  A nice discussion of the necessity of state dependence for entanglement-wedge reconstruction in the presence of a black hole can be found in \cite{Hayden:2018khn}. This state dependence of geometric quantities is consistent with the philosophy that the bulk spacetime in AdS/CFT is built up from entanglement \cite{VanRaamsdonk:2010pw,VanRaamsdonk:2011zz}.

What we lack is a good ``theory of measurement'' for state-dependent operators. The standard von Neumann procedure of measurement involves adding an operator to the Hamiltonian to create entanglement between the system and a pointer. (See, for instance, section 3.1.1 of  \cite{preskill1998lecture}.) But clearly, one cannot add a state-dependent operator to the Hamiltonian. As discussed in  \cite{Kapustin:2013yda}, and also in some older papers  \cite{gisin1990weinberg,PhysRevLett.66.397}, deforming the Hamiltonian by a state-dependent observable leads to immediate difficulties the moment one considers entangled states. But it is clear, in any case, that deformations of the Hamiltonian cannot be the correct way to think of local measurement in gravity since local sources would violate the local conservation of energy. 

So there must be some other formalism, involving measurements happening autonomously due to the natural evolution of the system that describes measurement in this setting. It is ultimately this issue, of finding a description of the process of measurement when the observer is part of the system, that lies at the root of the ``Born rule'' paradox of section \ref{mpparadox}.  It may even be, as we mentioned at the end of section \ref{mpparadox}, that understanding this process of measurement forces us to introduce some additional degrees of freedom in the theory.  Since state dependence clearly appears to be an essential ingredient in AdS/CFT, it seems important to understand this issue better.

Our discussion can be summarized in terms of the following conclusion.
\begin{lesson}
\label{lessonlargebh}
The question of whether typical microstates in AdS/CFT correspond to black holes with an empty interior or not reduces to a question of the validity of the state-dependent interior construction of section \ref{mirroropreview}.
\end{lesson}

\subsection{Evaporating black holes \label{structureevap}}
Let us now consider evaporating black holes. By this, we mean small black holes in AdS/CFT or black holes in asymptotically flat space.\footnote{In particular, we are not referring to the case where one prepares a large black hole in AdS and then turns on a coupling to a nongravitational bath, for which the discussion of the previous subsection may be more relevant.} The statistical properties of such black holes are quite different from those of large black holes. This is because evaporating black holes {\em never dominate} the microcanonical ensemble. In fact, this is precisely why they evaporate. The gas of Hawking radiation, at the same energy as the black hole, has more entropy and is therefore thermodynamically favoured. This leads both to complications and to potential simplifications.
 
Let us first discuss the complications. The complication is that we do not know how to specify the ``ensemble of evaporating black holes.'' This is true even in four-dimensional asymptotically flat space, where the description of the Hilbert space on either past or future null infinity is very simple. As described in section \ref{holinfoflat}, the Hilbert space is just a direct sum of Fock spaces. Classically, there is some understanding of what kind of initial data on past null infinity will lead to the formation of black holes \cite{Gundlach:2007gc}. But, quantum mechanically, we do not know of any precise delineation of states on $\scrim$ that would evolve to form a single black hole, as opposed to states that would lead to multi-centered black holes or just a gas of particles. 

A similar comment holds in AdS, where there is no clear available description of the precise set of states that correspond to small AdS black holes. (See  \cite{Hanada:2016pwv} for some discussion.) This state of affairs should be contrasted with the good understanding that we have of large AdS black holes, where we expect that almost all states from the microcanonical ensemble at a sufficiently high energy correspond to large black holes.

On the other hand, the fact that the ensemble of evaporating black holes does not coincide with the microcanonical ensemble also leads to a potential interesting simplification. Note that each of the paradoxes of section \ref{paradoxsingleside} involved a trace over all states of a given energy. This is also true of the paradox of section \ref{mpparadox}. It was these paradoxes that taught us that state-independent operators cannot be used to describe the interior of typical large AdS black holes. Therefore this observation leads to the interesting {\em possibility} that perhaps the interior of evaporating black holes can be described using {\em state-independent} operators.

Let us follow this idea further. We expect that small black holes in AdS at a given energy can be described as a subspace of dimension $e^{S}$ that is embedded in a larger subspace of states at the same energy with dimension $e^{S'}$. Simple estimates --- obtained by comparing the entropy of Hawking radiation with the original Bekenstein-Hawking entropy --- suggest that ${S' \over S}$ is $\Or[1]$ \cite{Page:2013dx}, which implies that  $e^{S'} \gg e^{S}$. In asymptotically flat space, once we take care of the divergence caused by infinite volume, we expect a similar picture to hold: the Hilbert space of black hole states is embedded in a much larger space of states at the same energy.  

If $|\Psi_i \rangle$ is a given microstate for an evaporating black hole,  then it is clear that we can extend the construction of section \ref{mirroropreview} to find operators, $\qop^{\Psi_i}(x)$,  that satisfy equation \eqref{structurequestion} about  that specific microstate. We can also demand that $\qop^{\Psi_i}$ annihilate all vectors outside the little Hilbert space constructed on $|\Psi_i \rangle$. The question is whether it is possible to find a single operator that satisfies equation \eqref{structurequestion} about all evaporating black-hole microstates. A natural guess is to consider the summed operator 
\be
\label{sumopsmall}
\qop(x) = \sum_{i=1}^{e^{S}} \qop^{\Psi_i}(x),
\ee
where the sum runs over a basis of microstates.

We discussed a similar operator in section  \eqref{secorigstate}. We remind the reader that the key issue  has to do the size of the cross terms
\be
C_{i j} = \langle \Psi_i| \qop^{\Psi_j}(x_1) \ldots \qop^{\Psi_j}(x_n) |\Psi_i\rangle
\ee
for $i \neq j$. If the states $|\Psi_i \rangle$ are distributed generically inside the larger Hilbert space, then a natural estimate for the size of the cross term is $C_{i j} = \Or[e^{-{S' \over 2}}]$. And if so, then since $e^{S} \ll e^{S'}$, we find that the operator in \eqref{sumopsmall} may provide a single operator that correctly works in all black-hole microstates.

We immediately caution the reader that the argument here is too simple to be conclusive. In particular, since we do not understand the subspace of small black holes, we do not know if our estimate for the size of the cross term, $C_{i j}$ is correct. But it is accurate to state that, as of the writing of this article, there is no clear argument that the paradoxes that afflict large AdS black holes and force us to use state-dependent operators, are relevant for evaporating black holes.  

For evaporating black holes,  one can still construct a paradox, involving the monogamy of entanglement and ask questions about whether black hole formation and evaporation takes pure states to pure states. But, as we discussed in section \ref{secresolveinfo}, these paradoxes are nicely resolved by the principle of holography of information.  Therefore, in summary, it is safe to state our conclusion as follows.
\begin{lesson}
\label{lessonevap}
There is no currently known paradox about black holes in asymptotically flat space or about small AdS black holes that requires us to postulate that such black holes have structure at their horizons. 
\end{lesson}

\subsection{Can classical solutions tell us about horizon structure? \label{seccanclass}}

It is possible to find a number of horizon-free solutions of supergravity that have the same charges as a black hole. These solutions are called ``microstate geometries'', and large and interesting classes of these solutions have been found, starting with \cite{Lunin:2001fv,Lunin:2002bj}, and including the recent multi-charge solutions \cite{Bena:2016ypk,Bena:2015bea,Bena:2017xbt}. We refer the reader to  \cite{Mathur:2005zp,Bena:2007kg,Bena:2013dka} and references there for  further details.

While these solutions are interesting in their own right, we would like to explain the following conclusion. 
\begin{lesson}
\label{lessonfuzzballs}
Microstate geometries can neither represent typical microstates of black holes, and nor can they form a basis for the space of black holes. Therefore, they cannot be used to make reliable inferences about the nature of the black-hole horizon.
\end{lesson}
This issue was addressed in detail in  \cite{Raju:2018xue}, and we now review the argument briefly.

First, let us consider typical states.  The microstate geometry program is sometimes motivated using a qubit model, and so we start by reminding the reader of the meaning of a typical state. If one considers multiple spin-1/2 systems, and if we denote the spin-z eigenstates by $|0 \rangle$ and $|1 \rangle$ for each system, then the states
\be
\label{basis}
|0 0 \ldots 0 \rangle, \quad |1 0 \ldots 0 \rangle,  \quad |0 1 \ldots 0 \rangle \ldots 
\ee
form a basis for the Hilbert space of the combined system.  But, in fact, from the point of view of the z-spin, each of these states is an extremely {\em atypical} state. A typical state is a linear combination
\be
a_1 |0 0 \ldots 0 \rangle + a_2 |1 0 \ldots 0 \rangle + a_3 |0 1 \ldots 0 \rangle \ldots 
\ee
with some coefficients $a_i$ chosen according to the Haar measure of section \ref{puremixed}. 

The result \eqref{projectordeviation} tells us that the quantum probability distribution for any physical quantity cannot vary significantly from one typical state to another.  We earlier used this result to claim that typical pure states are exponentially close to mixed states. But this also implies that typical pure states are exponentially close to one another. Therefore, it is insufficient for the microstate-geometry program to answer question \eqref{structurequestion} in the negative. The program must also provide us with a {\em universal} replacement for the right hand side of \eqref{structurequestion}, in the form of some function $G_{\text{microstate}}(x_1, \ldots x_n)$, and make the positive assertion that
\be
\label{universalmicro}
\langle \Psi |  \qop(x_1) \ldots \qop(x_n) |\Psi \rangle = G_{\text{microstate}}(x_1, \ldots x_n),
\ee
for typical states, $|\Psi \rangle$.

But different microstate geometries would suggest different values for correlators of local observables that appear on the right hand side of equation \eqref{universalmicro}. So it cannot be that all of them correspond to typical states. And there must be some {\em privileged} microstate geometry, which describes the experience of an infalling observer in a typical state, and leads to the correct answer for the right hand side of equation \eqref{universalmicro}. What is this privileged geometry?

Euclidean quantum gravity would suggest that when the points $x_1, \ldots x_n$ are outside the horizon, this geometry should correspond {\em exactly} to the conventional black hole geometry. But we are not aware of any privileged geometry that has been proposed as a replacement of the conventional empty black-hole interior. This appears to us, to be a significant gap in the microstate-geometry program. 

Let us mention two possibilities. One possible proposal is that $G_{\text{microstate}}(x_1, \ldots x_n)$ continues to coincide with $G_{\text{standard}}(x_1, \ldots x_n)$ even when some of the points are taken inside the horizon. If so, then an infalling observer in almost all states would encounter a smooth horizon, and the microstate-geometry picture would start to coincide with the conventional picture of black holes.  Another logical possibility is that  space simply ends abruptly  at the horizon, and there is no meaning to observables in the interior. This is, in fact, the firewall proposal, which also suggests that the infalling observer has a uniform experience in almost all states: the observer meets a featureless singularity at the horizon.

Now, we turn to a different question. Even if the distinct microstate geometries cannot correspond to typical states, can they at least correspond to a {\em basis}, as is displayed in the list \eqref{basis}?  In fact this also turns out not be possible. The Euclidean theory not only allows us to compute the expectation value of observables like the metric, it also allows us to compute their {\em fluctuations}. These fluctuations are small since they correspond, in the Lorentzian setting, to Hawking radiation.  If the distinct reliable microstate geometries were to form a basis, it would lead to a prediction for fluctuations that is too large, and in contradiction with the Euclidean prediction.

We can see this more precisely as follows. Let us focus on some bounded observable with a {\em non-zero classical expectation value}. To emphasize this property, we denote this observable by $\qop^{\text{cl}}$. For instance, we could take $\qop^{\text{cl}}$ to be the square of the Riemann tensor smeared in a small region of spacetime.  Then a Euclidean calculation suggests that in a typical state, 
\be
{\langle (\qop^\text{cl})^2 \rangle - \langle \qop^{\text{cl}} \rangle^2 \over \langle \qop^{\text{cl}} \rangle^2} =\Or[{1 \over S}].
\ee
where $S$ is the entropy of the black hole. The reasoning is very simple, and simply relies on an estimate of the size of quantum fluctuations in the geometry. (See section 2.2 of \cite{Raju:2018xue}.)

Now, say that the different microstate geometries correspond, in the quantum theory, to a basis of states $|f_i \rangle$ with $i = 1 \ldots e^{S}$. Then we find that the quantity above can also be evaluated by taking the average of the fluctuations over all these microstates. Setting $\langle \qop^{\text{cl}} \rangle = {1 \over e^{S}} \sum_i \langle f_i | \qop^{\text{cl}} | f_i \rangle$ some simple algebra (see Result 2 of \cite{Raju:2018xue}) tells us that
\be
\langle (\qop^{\text{cl}})^2 \rangle - \langle \qop^{\text{cl}} \rangle^2  \geq {1 \over e^{S}} \sum_i (\langle f_i | \qop^{\text{cl}} | f_i \rangle - \langle \qop^{\text{cl}} \rangle)^2 = \langle d_i^2 \rangle
\ee
where we have introduced
\be
d_i \equiv \langle f_i | \qop^{\text{cl}} | f_i \rangle - \langle \qop^{\text{cl}}  \rangle,
\ee
which is the deviation of the expectation value of $\qop^{\text{cl}}$ in a particular fuzzball state from its mean expectation value. 

The estimate of fluctuations tells us  that $\langle d_i^2 \rangle$ is suppressed by $\Or[{1 \over S}]$ and the previous argument about typical states tells us that the mean value can be obtained from the conventional black hole geometry. Therefore, the inequality above leads us to the conclusion that, at most,  only a small fraction (suppressed by the entropy) of microstate geometries can differ from the conventional black hole geometry at $\Or[1]$.\footnote{A significant fraction of these geometries could differ from the conventional geometry at a level that is suppressed by the entropy. But recall that, at this level of accuracy,   quantum fluctuations in the geometry become important. So the correct description is then in terms of a wavefunctional over geometries rather than a single solution to the equations of motion.} 
In particular, the currently known  microstate geometries --- which all deviate by an $\Or[1]$ amount from the black hole solution --- must represent highly atypical basis elements.

The arguments above directly lead to conclusion \ref{lessonfuzzballs}. This cautions us to be conservative about the inferences that we draw about black holes, by studying a set of atypical states. For instance, it is sometimes suggested, from a study of microstate geometries, that  excitations about black-hole microstates should have energy gaps that are regulated by an inverse power of $S$ \cite{Tyukov:2017uig}.  But, in fact, we expect from general properties of statistical systems that the gap between successive energy eigenstates should be of size $\Or[e^{-S}]$.  
This is simply because, in an interacting system, all degeneracies are typically broken and since there are $e^{S}$ states in a band of energies of size $S$, the spacing between them must be $\Or[e^{-S}]$ \cite{casati1985energy,berry1977level}. Moreover, correlators of generic operators are expected to show signs of this small gap since we expect that the spectrum of such correlators in frequency space is effectively continuous.  This suggests that the larger energy gaps observed about microstate geometries are highly atypical.
We refer the reader to the discussion in section 2.1 of \cite{Raju:2018xue} for more details.

\subsection{Nonviolent nonlocality}
For completeness, we briefly mention a proposal known as ``nonviolent nonlocality'' \cite{Giddings:2012gc}. This proposal posits a modification to the conventional black hole geometry in the form of explicit nonlocal terms in the effective action. These terms are added in order to explicitly transfer information from the interior to the exterior. The terms are designed to be ``nonviolent'' so that the stress tensor encountered by the infalling observer is never divergent. It has been proposed that such terms could actually lead to {\em observable} signals that might be detected  with existing astrophysical techniques \cite{Giddings:2019jwy}.

We would like to make a few observations.  Similar issues were also discussed in \cite{vijaythesis}.
\begin{enumerate}
\item
The nonlocal terms introduced in this proposal not only modify observables inside the horizon, they even modify 
low-point correlation functions of quantum fields outside the horizon.  But, as above, general arguments from statistical mechanics tell us that, for typical black holes, such correlators are expected to be very close to their thermal values that can be reliably computed in the Euclidean theory.   The nonlocal terms introduced --- especially if they are posited to be large enough to be astrophysically observable ---  contradict this robust theoretical principle.
\item
It is {\em not necessary} to add  explicit nonlocal terms to the theory to ensure unitarity of evolution.  Gravitational effects automatically ensure that the information inside the horizon is available outside, as we have emphasized above. The argument that explicit nonlocal terms are required seems to flow  from an assumption that the Hilbert space should 
effectively factorize, which is described as a ``subsystem postulate'' in \cite{Giddings:2017mym}. But this assumption is incorrect in a theory of gravity.
\item
We do not understand the physical origin of the proposed nonlocal terms. If such terms are simply meant to model gravitational effects, it seems simpler to just treat propagating scalar fields interacting with gravity in the standard framework of effective field theory. On the other hand, if these terms are proposed as a genuinely new physical effect, they appear to be somewhat ad hoc.
\end{enumerate}

\section{Summary and open questions}
The information paradox is not one paradox but part of a web of interconnected puzzles about black holes. These puzzles are of interest, not just
in their own right, but also because they illuminate several general features of quantum gravity. In this article, we have surveyed some of these puzzles and reviewed the lessons that they teach us. These lessons are summarized in conclusions \ref{lessonone} -- \ref{lessonfuzzballs} that are distributed through the text. Here we provide a non-technical summary of the results that we have reviewed, and describe some  important open questions.
 
We started by discussing Hawking's original formulation of the information paradox. Based on a semiclassical computation, Hawking found that a black hole formed from collapse would emit radiation at a temperature determined by its surface gravity. He suggested that the radiation at $\scrip$ would be independent of the details of the collapse, except for a few macroscopic parameters. Hawking argued that black hole formation could lead a pure state to evolve into a mixed thermal state, which would contradict the unitarity of quantum mechanics. This argument is reviewed in section \ref{secoldinfo}.

This led us to our first physical lesson. From the point of view of 
any observable, almost all pure states in a large Hilbert space are exponentially close to a mixed state. This is the basic kinematical property that underlies
thermalization in quantum mechanical systems. Hawking's calculation, which is
equivalent to the computation of low-point correlators of field operators at $\scrip$, did not keep track of these exponentially small corrections. So it was simply not precise enough to lead to a paradox. 

We then considered more refined paradoxes in section \ref{secintpar}. These paradoxes
are based on some simple results from quantum-information theory. If one
divides a system into two parts then, in a generic state of the system, almost all observables in the smaller subsystem are maximally entangled with
observables in the larger subsystem. But this entanglement has to be monogamous. Both these properties can be expressed concisely in terms of Bell correlators. These results, when applied to evaporating black holes,  suggest that a paradox arises for old black holes. The late radiation must be entangled with the early radiation. But it must also be entangled
with mirror modes behind the horizon. This seems to violate the monogamy of entanglement. It even appears that this contradiction cannot
be resolved by small corrections.

The resolution to this paradox lies in the fact that gravity localizes 
quantum information very differently from quantum field theories. We devoted section \ref{secholography} to exploring this issue. The main physical lesson is that, in
a theory of gravity, the information available on the bulk of a 
Cauchy slice is also available near its boundary. More precisely, 
we showed that in asymptotically AdS spacetimes, all the information
in the bulk could be obtained from an infinitesimal time band near the boundary
of AdS. In asymptotically flat space, all information about massless particles
is available near either the past boundary of $\scrip$ or the future boundary of $\scrim$. Both the results in AdS and in flat space can be  proved if one assumes that the UV-complete theory of quantum gravity shares some simple properties of low-energy effective theory. It is important that neither result is
some formal result about operator algebras. Rather, these results are already
meaningful for low-energy states, where they can be verified in perturbation theory as we described in sections \ref{lowenergy} and \ref{morelowenergy}.

When applied to black holes, this property of gravity implies that a copy
of the information in the black-hole interior is already available in the exterior. There are, therefore, two ways to probe this information. One is by means of quasilocal operators, which correspond to propagating quantum fields.
But the other is by means of more complicated operators outside the black hole that exploit gravitational effects. Both procedures probe the same information. This feature of gravity implies that the common idea that the Hilbert space can be factorized into an interior part and an exterior part  fails as completely as possible in quantum gravity: the interior degrees of freedom are contained in the exterior.  If one insists on making an incorrect assumption about factorization,  then this immediately leads to a paradox with the monogamy of entanglement. 

We showed in section \ref{revisitmonogamy} that a similar paradox appears even in empty space if one considers an imaginary ball  and assumes that the Hilbert space factorizes into the degrees of freedom inside the ball and outside the ball. We then explicitly showed how one could construct a contradiction
with the monogamy of entanglement in black hole states by incorrectly assuming
that the interior degrees of freedom are separate from the exterior degrees of freedom. 

We also reviewed the recent computations of the Page curve of holographic
CFTs coupled to nongravitational baths. These computations are very interesting but they should be interpreted correctly.  What is called the
``radiation'' in these systems is not the region outside the horizon but
 part of an entirely separate nongravitational quantum mechanical system. Moreover, what is called the
black hole is not the region inside the horizon but a complete holographic system that, at late times, does not even contain the quasilocal operators that probe the black hole. So we argued that the correct interpretation of these
computations is to think of them purely on the boundary in terms of two nongravitational systems coupled to each other. We expect to see a Page curve in this setting. Moreover, when the nongravitational systems have 
a gravitational dual, this Page curve can be computed using holographic techniques. 

For black holes in asymptotically flat space, the Page curve does not describe how information ``emerges'' from the black hole.  Rather,  in asymptotically flat space and also in other settings where gravity is dynamical, a better
physical answer appears to be that the information is always outside. Nevertheless, even in such settings, the Page curve may be the answer to other kinds of questions that one can ask about black hole evaporation as we discussed in section \ref{secispageflat}.

We then turned to the issue of large black holes in AdS. Large black holes in AdS pose unique issues because they dominate the microcanonical ensemble. 
If one assumes that ``typical'' black-hole microstates --- where ``typical'' is defined using the Haar measure on the Hilbert space --- have a smooth horizon, and also assumes that the mapping between the black-hole interior and boundary operators is state independent, then this leads to paradoxes with general results from statistical mechanics. These paradoxes are not restricted to single-sided black holes. A similar paradox also arises for the eternal black hole, which is believed to be dual to the thermofield doubled state in the CFT.

In section \ref{secstatedep} we described a construction of the black-hole interior in AdS/CFT. This construction is based just on effective field theory. It is clear that in any
black-hole microstate, where the horizon is smooth, this construction 
correctly describes interior degrees of freedom. This construction is not intrinsically state dependent.  But if one demands that this construction of interior operators be valid in typical microstates, then this forces the interior operators to be state dependent. This was explained in section \ref{secorigstate}. State dependence is an unusual phenomenon, and it itself gives rise to paradoxes. We explained how these paradoxes could be partially resolved in section \ref{secresolvestatedep}.

One recent topic of discussion has been the question of whether
typical states are described by the conventional black hole geometry, or whether they instead correspond to fuzzballs or firewalls. In section \ref{structureevap}, we explained that for small black holes in AdS and black holes in asymptotically flat space, currently known  paradoxes can be resolved without any need to postulate structure at the horizon. 

For large black holes in AdS, the question is more subtle. One possibility is that  observables in the black-hole interior are state dependent so that the construction of section \ref{mirroropreview} is valid for typical microstates.  But another possibility that remains open is that typical microstates correspond to firewalls, in which case the construction of section \ref{mirroropreview} is valid only for a small subset of microstates that have a smooth interior. It is important to recognize that if state dependence is ruled out, one would also have to give up the picture of the thermofield double as being dual to a wormhole connecting two asymptotic regions. 

In section \ref{seccanclass} we also explained  why the construction of microstate geometries --- although interesting in its own right --- cannot be used to settle this question. General arguments from statistical mechanics suggest that these solutions can only be dual to a small class of microstates and not to typical microstates.

It seems to us that several very interesting questions remain open. The principle of holography of information, as we have formulated it, applies when one has access to the entire boundary of a Cauchy slice. But what if one is given access to only some part of the boundary? There have been several interesting developments in the field of subregion duality in the context of AdS/CFT. The principle there seems to be that if one is given access to a region on the boundary, this corresponds to information about the entanglement wedge of the region in the bulk. It would be very nice to have a direct Hamiltonian derivation of this principle. Such a derivation would also help us extend this idea to asymptotically flat spacetimes.

A second conceptual question is how the principle of holography of information should be extended to spatially compact spacetimes. It is tempting to conjecture that, in such a spacetime, any region has information about its complement and vice versa. This would be because, in such spacetimes, any region is always encompassed by its complement  and so it would appear that one cannot modify the state of the region without modifying the state of its complement. However, we do not know how to prove this idea. From a technical point of view, the tools that we used in section \ref{secholography}, including the projector on the vacuum, do not appear to generalize in a simple manner to this setting.

In this article, we have focused on how quantum information is localized in gravity. However, as AdS/CFT teaches us, it is not just information that is holographic. The dynamics of the bulk can also be reformulated in terms of the dynamics of holographic degrees of freedom. In asymptotically AdS spacetimes, this reformulation takes the elegant form of a conformal field theory. What is the analogue in asymptotically flat spacetimes? 

For large black holes in AdS, it appears to us that the question of 
the origins and implications of state dependence of the interior is
still not completely understood. State dependence
appears to be a feature of several aspects of AdS/CFT. But, on the other hand,
it also has the potential to give rise to paradoxes as in section \ref{mpparadox}. Moreover, from a deeper point of view, how does one operationally measure state-dependent operators? Usually, we think of the process of measuring an observable in 
terms of adding that observable to the Hamiltonian. But this is clearly impossible with state-dependent operators. So it may be that the description of the interior of black holes provides a gateway to some more fundamental questions about quantum mechanics and gravity.

\section*{Acknowledgments}
I am grateful to  Ahmed Almheiri, Abhay Ashtekar, Bidisha Chakrabarty, Tuneer Chakraborty, Joydeep Chakravarty, Chandramouli Chowdhury, Laurent Freidel,  Hao Geng,  Victor Godet, Thomas Hartman,  Simon Caron Huot, Daniel Jafferis, Andreas Karch, Vijay Kumar, Alok Laddha, R. Loganayagam, Raghu Mahajan, Juan Maldacena, Samir Mathur, Shiraz Minwalla, Ruchira Mishra,  Carlos Perez-Pardavila, Olga Papadoulaki, Priyadarshi Paul, Siddharth Prabhu, Lisa Randall, Marcos Riojas, Abhisek Sahu, Sudipta Sarkar, Sanjit Shashi, Pushkal Shrivastava, Sandip Trivedi, Xiaoliang Qi,  Madhavan Varadarajan, Edward Witten, and Zhenbin Yang for helpful discussions. I am especially grateful to Kyriakos Papadodimas for several helpful discussions. I am grateful to the organizers and participants of the Black Hole Microstructure conference, the Strings 2020 conference and the Island Hopping 2020 conference. This article draws on the talks and panel discussions that were part of those conferences. This work was partially supported by a Swarnajayanti fellowship,  DST/SJF/PSA-02/2016-17, of the Department of Science and Technology. 

\appendix
\section{A geometric-optics derivation of Hawking radiation \label{secraytracehawk}}
In this Appendix, we will present a simplified version of  Hawking's original derivation \cite{Hawking:1974sw} of the spectrum of the radiation emitted by a black hole, which used geometric optics. The geometric optics method is clever but it is difficult to  extend it beyond leading
order, and the approximations made in the calculation are not immediately apparent. It is for this reason that we presented the more straightforward derivation of Hawking radiation in section \ref{hawkingderiv}. This is perhaps
also the reason that this original method is not commonly covered in textbooks. Nevertheless, the geometric-optics analysis provides physical insight into the process of particle creation and so we review it here.

\begin{figure}[!ht]
\begin{center}
\includegraphics[width=0.4\textwidth]{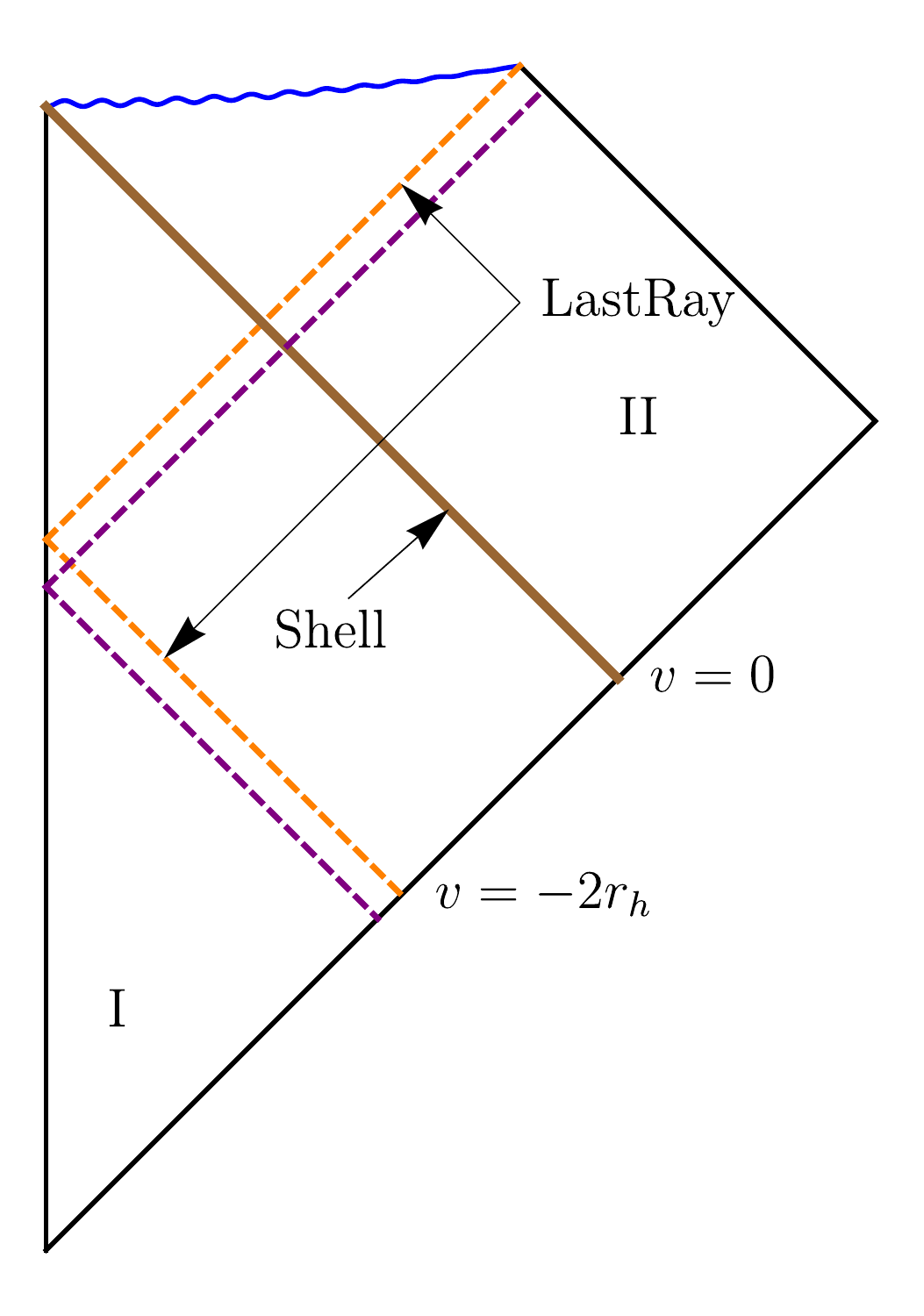}
\caption{\em Ray tracing in a collapsing geometry. A black hole is formed
by the collapse of a  shell (brown line), which marks the interface between
regions {\rm I} and {\rm II}. The last ray is marked in orange. We study the path of the purple ray, which starts from $\scrim$ slightly before the last ray.  \label{figraytracing}}
\end{center}
\end{figure}

Consider a Vaidya geometry, where a black hole is formed by the  collapse of a null shell, as shown in Figure \ref{figraytracing}. We will assume that the spacetime is spherically symmetric to simplify the notation. In the Vaidya
solution, the metric of Minkowski spacetime in region I of Figure \ref{figraytracing} is glued to a Schwarzschild spacetime in region II, which is given in \eqref{latetimemetric}. The metric, in arbitrary dimensions can be written in the form \cite{Iyer:1989nd} 
\be
ds^2 = -(1 - {\theta(v) \mu \over r^{d-2}}) d v^2 + 2 d v d r + r^2 d \Omega_{d-1}^2.
\ee
Here the collapsing shell is placed at $v = 0$.  Region I corresponds to $v < 0$ whereas region II corresponds to $v > 0$. For $v > 0$, the function that appears in the metric is just $f(r)$ as given in equation \eqref{fexplicit}, and the same relation holds between the parameter $\mu$ and the mass of the shell. After collapse, the horizon forms at $r_h = \mu^{1 \over d-2}$.

Now, let us recall the mathematical problem that we need to solve.  Consider a quantum field propagating on this background. For simplicity, we will take this field to be a scalar but this derivation can easily be generalized to fields with spin.   Then, there are two convenient expansions for this quantum field. First, the field can be expanded  in modes that have specified behaviour at $\scrim$. 
\be 
\phi = \sum_{\ell} \int d \omega  C_{\omega, \ell} \fpast_{\omega, \ell}(v, r) Y_{\ell}(\Omega) + \text{h.c}
\ee
where the asymptotic behaviour of $\fpast$ along past null infinity is given by
\be
\label{pastmodes}
\fpast_{\omega, \ell}(v,r) \underset{r,v \rightarrow {\cal I}^{-}}{\longrightarrow} {e^{-i \omega v} \over r}
\ee

The field can also be expanded in modes that have specified behaviour at $\scrip$ and at the horizon. We have already encountered this expansion  in section \ref{hawkingderiv}.
\be
\label{futuremodes}
\phi(v, r, \Omega) = \sum_{\ell} \int {d \omega}  \left(A_{\omega, \ell} \fout_{\omega, \ell}(v,r) + B_{\omega,\ell} \fin_{\omega,\ell}(v, r) \right) Y_{\ell}(\Omega) + \text{h.c}.
\ee
We remind the reader that, as specified in equation \eqref{foutfinexpansion}, the behaviour of $\fin$ at the future horizon at late times is
\be
\fin_{\omega, m}(t,r) \underset{r \rightarrow r_h}{\longrightarrow} \white_{\omega, \ell} e^{-i \omega v}.
\ee
 We specified the behaviour of $\fout$ at the horizon in \eqref{foutfinexpansion} but it is more convenient to now consider its behaviour at $\scrip$, which is
\be
\fout_{\omega, \ell}(v,r) \underset{r,v \rightarrow {\cal I}^{+}}{\longrightarrow} \white_{\omega, \ell} {r_h \over r} {e^{-i \omega (v - 2 r)}}
\ee
See equations 100--102 of \cite{dewitt1975quantum}. Note that $u = v - 2 r$ is also the natural coordinate on $\scrip$.

In principle, by solving the wave equation on the background of the collapsing geometry, one could obtain the Bogoliubov coefficients between the modes $C_{\omega, \ell}$ and the modes $A_{\omega, \ell}$ and $B_{\omega, \ell}$. However, Hawking pointed out that it is possible to derive an approximation to these coefficients by using geometric optics.

Let us understand the behaviour of light rays that start from $\scrim$ and end up on $\scrip$. In the diagram \ref{figraytracing} a sphere has been suppressed at each point. So, although we use the word ``rays'', the correct way to think of these rays is in terms of  spherical shells of light that are expanding or contracting. It is easy to work out the explicit trajectory of these rays. 
\be
\begin{array}{ll}
v = \text{constant} \qquad & \qquad \text{for~contracting~rays}; \\
v - 2 \rtor = \text{constant} \qquad &\qquad \text{for~expanding~rays~in~region~II}; \\
v - 2 r = \text{constant} \qquad &\qquad \text{for~expanding~rays~in~region~I}.
\end{array}
\ee
Here the tortoise coordinate is defined, as usual by ${d \rtor \over d r} = f^{-1}(r)$ and we choose the integration constant so that $\rtor \rightarrow r$ at large $r$.
Rays that start in the very far past travel along a constant value of $v$ till they reach $r = 0$. At this point, they stop contracting and start expanding instead. In the 2d diagram of Figure \ref{figraytracing}, it looks like the ray has ``bounced off the origin of polar coordinates.'' 

From Figure \ref{figraytracing} it is clear that there is a ``last ray'', which departs $\scrim$ as late as possible in order to be able to make it out to $\scrip$. Every ray that departs $\scrim$ at a later time ends up in the singularity.  If we denote the starting point of this ray on $\scrim$ by $v_{\text{last}}$ then, in the geometry under consideration, it is easy to see that $v_{\text{last}} = -2 r_h$.   After bouncing off $r = 0$, this ray starts traveling along the trajectory $v - 2 r = -2 r_h$. Therefore this last ray intersects the shell, which is at $v =0$, precisely at $r = r_h$ and so it gets ``stuck'' at this value of $r$.  

Now consider a ray that departs $\scrim$ at $-2 r_h - \delta$, just before the last ray. After bouncing off $r = 0$, this ray follows the trajectory $v - 2 r = -2 r_h - \delta$. It intersects the shell at 
\be
r = r_h + {\delta \over 2}.
\ee
Now recall that very close to $r = r_h$, we have $f(r) = 2 \kappa (r - r_h)$ where $\kappa = f'(r_h)/2$ is the surface gravity of the black hole. Therefore, for $\delta \ll r_h$, we see that the value of the tortoise coordinate when this ray crosses into region II is 
\be
\rtor = {1 \over 2 \kappa} \log({\delta \over r_h}),
\ee
where we have neglected some $\Or[1]$ terms that are unimportant when $\delta$ becomes very small. After crossing the shell this ray continues along a trajectory of constant $v - 2 \rtor$. Since $\rtor \rightarrow r$ for large $r$, this ray also reaches $\scrip$ at 
\be
u = {-1 \over 2 \kappa} \log({\delta \over r_h}).
\ee
Note that a very large coordinate range on ${\cal I}^+$ (all the way up to $u = \infty$) is generated by a very small coordinate range on ${\cal I}^{-}$ around the last ray. We will return to this point below.

In geometric optics, the {\em phase difference} between neighbouring rays is preserved. As a result if we consider a wave where the phase on ${\cal I}^+$ behaves as $e^{\pm i \omega u}$, then this can be traced back to a wave where the phase on ${\cal I}^{-}$ behaves as 
\be
\label{pastbehaviour}
\left({|v| \over  4 m}\right)^{\mp  i \omega \over 2 \kappa}
\ee
 for small {\em negative} values of  $v$. If we place the incoming state in the vacuum of the $C$ modes (which multiply $e^{-i \omega v}$) then it is not in the vacuum of the modes shown in \eqref{pastbehaviour}. In fact, the transformation between the modes $e^{-i \omega v}$ and the modes shown in \eqref{pastbehaviour} is precisely the transformation between Minkowski and Rindler modes. These Bogoliubov coefficients can be found explicitly as we have done in the main text of the article. Therefore we find that the vacuum of the $C$ modes translates into a thermal spectrum for the $A$-modes displayed in \eqref{futuremodes}. 

\paragraph{\bf The trans-Planckian problem \\}
The derivation above suggests that a semi-infinite interval on $\scrip$ is generated by a small range of coordinates on $\scrim$ near $v = v_{\text{last}}$. Therefore, the modes that emerge on $\scrip$ can be traced back to extremely blue-shifted modes on $\scrim$.  On the one hand, this feature is necessary for the geometric-optics approximation to hold since it is precisely because we are considering rapidly oscillating modes near the last ray that we can reliably ignore
scattering from the infalling matter. On the other hand this gives rise to an issue that is sometimes called the ``trans-Planckian problem'', since the geometric optics derivation suggests that the modes on $\scrim$ that lead to Hawking radiation are coming from ``beyond'' the Planck scale. The concern is that the vacuum may not be well described by quantum-field excitations at such short distances \cite{tHooft:1996rdg}.

The trans-Planckian problem is not really a problem, in that it does not invalidate the final result. It only reminds us that one has to be careful about this subtlety. One of the virtues of the derivation in section \ref{hawkingderiv} is that it does not suffer from the trans Planckian problem. It is indeed the case, that the approximately thermal nature of Hawking radiation has to do with short-range entanglement: even in section \ref{hawkingderiv}, we used the short-distance behaviour of the two-point function. But the reason for carefully introducing the smearing function in section \ref{hawkingderiv} was to smear this two-point function over a scale that is much larger than the Planck scale but much smaller than the characteristic curvature scale of the geometry. Provided the near-horizon region looks like empty space on this scale --- and the late time geometry has an approximate Killing symmetry ---  the final spectrum on $\scrip$  looks approximately thermal.

\paragraph{\bf Entanglement across the horizon \\}
The geometric-optics derivation also provides insight into why modes outside the horizon must be entangled with ``partners'' behind the horizon. The ray-tracing argument tells us that both the state of the fields outside the horizon and the state behind the horizon arise from a small vicinity of $v = v_{\text{last}}$ on $\scrim$:  the region outside the horizon depends on $v < v_{\text{last}}$ whereas the region inside the horizon depends on $v > v_{\text{last}}$. So the entanglement across the horizon is simply the usual short-distance entanglement of fields near $v = v_{\text{last}}$ on $\scrim$. 
 
\paragraph{\bf The information paradox \\}
It is also possible to  reframe the information paradox in this setting.  One paradox, of course, is that the spectrum on $\scrip$ appears to be thermal. As we already explained in section \ref{secoldinfo}, the difference between pure states
 and thermal states --- as measured by physical observations --- can be exponentially small and so this is not really a paradox. 

It is also possible to ask more refined questions. For instance, if one creates an excitation in the region $v > 0$, which fall into the black hole, how does that information emerge at $\scrip$. We see that the principle of holography of information, as established in section \ref{holinfoflat}, provides a clear answer. All the information available on $\scrim$ is also available on $\scrimfuture$ --- the future boundary of $\scrim$. Therefore, in a rather precise sense, the information never falls in and always remains outside.

\section{A primer on perturbation theory in global AdS \label{appadspert}}
In this section, we provide a very quick review of the behaviour of perturbative fields in global AdS. Many existing and excellent reviews of perturbation theory \cite{Aharony:1999ti,D'Hoker:2002aw} tend to focus on Poincare AdS, which provides a technically simpler arena for the computation of correlation functions that are of interest in AdS/CFT. However, global AdS is more important from a conceptual point of view since it provides a natural IR cutoff and leads to a theory with a finite density of states.

We remind the reader that the metric of empty global AdS$_{d+1}$, which we have already used above, is
\be
\label{emptyglobaladsapp}
ds^2 = -\left(1+r^2\right) dt^2 
+{ dr^2 \over 1+r^2} + r^2 d\Omega_{d-1}^2 
\ee
We have set the AdS radius to $1$.

Let us start by switching off gravity and also self-interactions and considering a minimally coupled scalar field propagating on this space. The field obeys the equation
\be
\left( \Box - m^2 \right) \phi(t, r, \Omega) = 0.
\ee
In this article, we have consistently imposed ``normalizable'' boundary conditions for all operators. This means that we set 
\be
\label{boundcond}
\phi \rightarrow 0, \quad \text{as} \quad r \rightarrow \infty.
\ee

With the choice of boundary conditions made above, the solutions to this equation are as follows. 
\be
\label{phiexpansion}
\begin{split}
&\phi(t, r, \Omega) = \sum_{n \geq 0} c_{n, \ell} a_{n, \ell} e^{-i (2 n + \ell + \Delta) t} Y_{\ell}(\Omega) \chi_{n, \ell}(r) + \text{h.c}, \\
&\chi_{n, \ell}(r) = r^\ell \left(r^2+1\right)^{-{\Delta +2n + \ell \over 2}} \, _2F_1\left(-n, - \Delta -n+{d\over 2} ;\frac{d}{2}+\ell;-r^2\right), \\
&c_{n, \ell} = \frac{\Gamma \left(\frac{1}{2} (d+2 \ell+2 n)\right) \Gamma \left(\frac{1}{2} (d-2 n-2 \Delta )\right) \sqrt{\frac{\Gamma \left(-\frac{d}{2}+n+\Delta +1\right) \Gamma (\ell+n+\Delta )}{\Gamma (n+1) \Gamma \left(\frac{d}{2}+\ell+n\right)}}}{\Gamma \left(\frac{d}{2}-\Delta \right) \Gamma \left(\frac{d}{2}+\ell\right) \Gamma
   \left(-\frac{d}{2}+\Delta +1\right)}.
\end{split}
\ee
Here, $a_{n, \ell}$ are mode operators, $Y_{\ell}$ are the spherical harmonics and 
\be
\Delta = {d \over 2} + \sqrt{\left({d \over 2}\right)^2 + m^2}.
\ee
It is clear that even though we imposed only \eqref{boundcond}, the solution above actually falls off as ${1 \over r^{\Delta}}$ near the boundary. The radial equation could have had another solution, which is excluded by our boundary conditions. An important aspect of the solution in \eqref{phiexpansion} is that the solutions obeying the boundary condition \eqref{boundcond} have quantized frequencies
\be
\omega_n = 2 n + \ell + \Delta.
\ee
This quantization condition comes from demanding that the solution be regular also near $r = 0$. Physically, the quantization of frequencies arises because global AdS with normalizable boundary conditions behaves like a ``box'',
which confines fields.

The normalization of the solutions is fixed by the equal-time commutation relations. In the metric \eqref{emptyglobaladsapp}, these commutation relations read
\be
[\phi(t, r, \Omega), {d \over d t} \phi(t, r', \Omega')] =  i {(1 + r^2) \over r^{d-1}} \hat{\delta}(\Omega, \Omega') \delta(r - r'), 
\ee
where $\hat{\delta}$ is the delta function on the sphere. We choose conventions for the spherical harmonics so that
\be
\sum_{\ell} Y_{\ell}(\Omega) Y^*_{\ell}(\Omega') = \hat{\delta}(\Omega, \Omega')
\ee
The normalization above ensures that the radial solutions are normalized as
\be
\int 2(\Delta + \ell + 2 n) \chi_{n, \ell}(r) \chi_{n', \ell}(r) {r^{(d-1)} \over 1 + r^2} = \delta_{n n'}
\ee
Therefore the equal-time commutation relations for the field are consistent
with canonical commutators for the modes that appear in the expansion of the field:
\be
[a_{n, \ell}, a_{n, \ell}^{\dagger}] = 1.
\ee

The modes that appear in this expansion can also be related to the modes that appear in the expansion of the boundary operator that is dual to this bulk field. Defining this operator by
\be
O(t, \Omega) = \lim_{r \rightarrow \infty} r^{\Delta} \phi(t, r, \Omega),
\ee
we find that
\be
O_{n, \ell} = \sqrt{G_{n, \ell}} a_{n, \ell},
\ee
where
\be
G_{n, \ell} = \frac{\Gamma (\ell+n+\Delta )^2}{\Gamma (n+1)^2 \Gamma \left(-\frac{d}{2}+\Delta +1\right)^2 \Gamma \left(\frac{d}{2}+\ell\right)^2}.
\ee

The vacuum is the unique state that satisfies
\be
a_{n, \ell} | 0 \rangle = 0, \quad \forall n \geq 0 \quad \text{and} \quad \forall \ell.
\ee
We can now construct a Fock space  by acting repeatedly on the vacuum
with creation operators to generate states like
\[
a_{n_1, \ell_1}^{\dagger} a_{n_2, \ell_2}^{\dagger} \ldots a_{n_q, \ell_q}^{\dagger} | 0 \rangle.
\]
Since we have switched off gravity and interactions, the energy of this state in AdS units is   $2 (n_1 + n_2 + \ldots n_q) + (\ell_1 + \ell_2 + \ldots \ell_q) + q \Delta$. This leads, schematically, to the energy levels displayed in Figure \ref{figadsspect}.

In CFT language, the vacuum state corresponds to the identity operator. The spectrum of single-particle states is just the spectrum of operator dimensions of a primary operator of dimension $\Delta$ and its descendants. The expansion above determines the two-point function of this primary operator
\be
\langle 0| O_{n, \ell} O_{n, \ell}^{\dagger} | 0 \rangle = G_{n, \ell}.
\ee
If we Fourier transform back to the time domain we find that
\be
\langle 0| O_{\ell}(t) O_{\ell}(t') | 0 \rangle = \frac{\Gamma (\ell+\Delta )^2 e^{i(\Delta +\ell)(t' - t)} }{\Gamma \left(\frac{1}{2} (-d+2 \Delta +2)\right)^2 \Gamma \left(\frac{1}{2} (d+2 \ell)\right)^2} \, _2F_1\left(\ell+\Delta ,\ell+\Delta ;1;e^{2 i (t' - t)}\right).
\ee

\paragraph{\bf Interacting fields \\}
Now, lets turn on weak gravity. When we say that gravity is weak, we mean that the parameter
\be
N = {\ell_{\text{ads}} \over \ell_{\text{planck}}} 
\ee
is large. Since we have chosen the convention that $\ell_{\text{ads}} = 1$, and since the energies above are all of the AdS scale, the natural perturbative parameter is ${1 \over N}$.  We can also turn on self-interactions in the field. It is conventional, in the AdS/CFT literature, to take the self-interactions to also be proportional to a power of ${1 \over N}$ so that there is no hierarchy between gravity and other forces. This is, however, not essential to the argument in the article.

The presence of gravity leads to a few changes in the picture above.  First, and most straightforward, there are now states that correspond to propagating graviton modes themselves. These modes can be understood as follows. We expand the metric as $g_{\mu \nu} + h_{\mu \nu}$ (where $g_{\mu \nu}$ is the empty AdS metric shown above) and expand the action to quadratic order in $h_{\mu \nu}$ \cite{Christensen:1979iy}. The quadratic action reads
\begin{equation}
\label{gravityquadratic}
S = {-1 \over 64 \pi \gnewt} \int \sqrt{-g} d^{d+1} \vec{x}  \left(\tilde{h}^{\mu \nu} \Box h_{\mu \nu} + 2 \tilde{h}^{\mu \nu} R_{\mu \rho \nu \sigma} h^{\rho \sigma}  + 2 \nabla^{\rho} \tilde{h}_{\rho \mu} \nabla^{\sigma} \tilde{h}_{\sigma}^{\mu} \right) + \ldots
\end{equation}
where $\tilde{h}^{\mu \nu} = h^{\mu \nu} - {1 \over 2} g^{\mu \nu} h^{\alpha \beta} g_{\alpha \beta}$,  all covariant 
derivatives are with respect to the background metric and $\ldots$ denote higher order terms that we have not displayed. We can quantize this quadratic action and this leads to the conclusion that the lowest graviton mode has energy $d$ and a spectrum that is qualitatively similar to the picture above. In CFT language, the graviton is dual to stress tensor. 

The second effect is that due to the interactions,  the energy levels themselves shift. The shift in the energies to leading order in ${1 \over N}$  can be computed through tree-level Feynman diagrams in AdS. In the CFT,  this shift corresponds to the anomalous dimension of the operator corresponding to the state.  In principle, loop-level shifts can also be computed but they have not been treated systematically in
the literature.

The third, and conceptually most important effect, is that the constraints of gravity ensure that the fluctuations of the metric near infinity are constrained by the excitations in the bulk. In particular, the energy of a state
in the bulk is determined by a surface integral near the boundary, which can be expressed covariantly \cite{Balasubramanian:1999re}. The expression for this integral becomes particularly simple   \cite{deHaro:2000vlm} if one chooses Fefferman-Graham gauge near the boundary --- which means that one sets $h_{r \mu} = 0$. Then the energy is simply measured by
\be
\label{energy}
H = {d \over 16 \pi \gnewt} \lim_{r \rightarrow \infty} r^{d-2} \int  d^{d-1} \Omega \, h_{t t}.
\ee
It is clear that the operator on the right hand side of \eqref{energy} is an observable near the boundary. If one considers a superposition of states of different energy--- say, chosen from the Fock space above --- then the operator in \eqref{energy} does not have a definite value but yields probabilistic results as mandated by quantum mechanics.

The perturbative formalism that we have developed here explicitly is a reliable guide to the low-energy spectrum. This spectrum corresponds to excitations that are of the cosmological scale.  Note that, in principle, one could proceed further systematically and attempt to include low-energy excitations of strings and branes that, in some cases, can be understood quite explicitly just by quantizing classical solutions \cite{Grant:2005qc,Mandal:2005wv,Maldacena:2000hw,mandal2008sgg,Ashok:2008fa}. At energies of order $N$, one encounters the smallest black holes.  This spectrum of excitations at higher energies is  evidently quite complicated. 

However, it is remarkable that in our argument for the holography of information in AdS in section \ref{holinfads}, we did not need any of the details of this high-energy spectrum. We only needed the assumption that all states in the Hilbert space could be obtained by applying asymptotic operators to the vacuum at arbitrary points of time. This is a weak assumption since this space of states clearly forms a superselection sector as explained in the text. Second, we needed the assumption that an asymptotic operator  could be used to accurately identify the vacuum. This is also a weak assumption, since it only requires that some of the {\em low-energy} properties of the UV theory coincide with the {\em low-energy} properties that one deduces from perturbation theory.

\bibliographystyle{JHEP}
\bibliography{references}
\end{document}